\documentclass{UCF_ETD}

\usepackage{times}
\usepackage{graphicx}
\usepackage{float}
\usepackage{multirow}
\usepackage{amsmath}
\usepackage{amsthm}
\usepackage{upgreek}
\usepackage[table]{xcolor}
\usepackage{textcomp}
\usepackage{aas_macros}
\usepackage{color}
\usepackage{wrapfig}
\usepackage{enumitem}
\usepackage{url}
\usepackage{caption}
\usepackage{subcaption}
\usepackage{amsfonts}
\usepackage{fixltx2e}
\usepackage[defaultlines=5,all]{nowidow}
\usepackage{placeins}

\usepackage[shortcuts]{extdash}

\usepackage{ifthen}
\usepackage{longtable}
\usepackage{etoolbox}
\usepackage{lscape}
\usepackage{rotating}

\usepackage{multibib}

\usepackage{setspace}
\usepackage{amssymb}
\usepackage[comma, sectionbib]{natbib}
\usepackage{chapterbib}
\usepackage{bibentry}

\newcommand\SST{{\em Spitzer Space Telescope}}
\newcommand\Spitzer{{\em Spitzer}}
\newcommand\chisq{\ifmmode{\chi\sp{2}}\else\math{\chi\sp{2}}\fi}
\newcommand\redchisq{\ifmmode{ \chi\sp{2}\sb{\rm red}}
                    \else\math{\chi\sp{2}\sb{\rm red}}\fi}
\newcommand\PT{$P$--$T$}

\newcommand\ms{m\,s$\sp{-1}$}

\newcommand\Teq{$T$\sb{eq}}
\newcommand\mjup{$M\sb{\rm Jup}$}
\newcommand\rjup{$R\sb{\rm Jup}$}
\newcommand\msun{$M\sb{\odot}$}
\newcommand\rsun{$R\sb{\odot}$}

\newcommand\ph{\phantom{1}}

\newcommand\waspo{{wa008bs11}}
\newcommand\waspt[1]{{wa008bs2{#1}}}
\newcommand\waspf[1]{{wa008bs4{#1}}}
\newcommand\mcc{MC\sp{3}}
\newcommand\molhyd{H$\sb{2}$}
\newcommand\methane{CH$\sb{4}$}
\newcommand\water{H$\sb{2}$O}
\newcommand\carbdiox{CO$\sb{2}$}

\newcommand\der{{\rm d}}
\newcommand\tnt{$\sp{-2}$}
\newcommand\arcsec{$\sp{\prime\prime}$}
\newcommand\arcmin{$\sp{\prime}$}
\DeclareOldFontCommand{\rm}{\normalfont\rmfamily}{\mathrm}
\DeclareOldFontCommand{\tt}{\normalfont\ttfamily}{\mathtt}
\DeclareOldFontCommand{\bf}{\normalfont\bfseries}{\mathbf}
\DeclareOldFontCommand{\it}{\normalfont\itshape}{\mathit}

\newcommand\degr{\sp{\circ}}
\newcommand\degree{\ifmmode\degr\else\math{\degr}\fi}

\newcommand\vs{\emph{vs.}}

\bibpunct[, ]{(}{)}{,}{a}{}{,}

\DeclareSymbolFont{UPM}{U}{eur}{m}{n}
\DeclareMathSymbol{\umu}{0}{UPM}{"16}
\let\oldumu=\umu
\renewcommand\umu{\ifmmode\oldumu\else\math{\oldumu}\fi}
\newcommand\micro{\umu}
\newcommand\micron{\micro m}
\newcommand\microns{\micron}

\let\oldsim=\sim
\renewcommand\sim{\ifmmode\oldsim\else\math{\oldsim}\fi}

\let\oldpm=\pm
\renewcommand\pm{\ifmmode\oldpm\else\math{\oldpm}\fi}

\newcommand\by{\ifmmode\times\else\math{\times}\fi}

\newcommand\ttt[1]{10\sp{#1}}
\newcommand\tttt[1]{\by\ttt{#1}}

\newbox{\wdbox}
\renewcommand\c{\setbox\wdbox=\hbox{,}\hspace{\wd\wdbox}}
\renewcommand\i{\setbox\wdbox=\hbox{i}\hspace{\wd\wdbox}}
\newcommand\n{\hspace{0.5em}}

\newcount\timect
\newcount\hourct
\newcount\minct
\newcommand\now{\timect=\time \divide\timect by 60
         \hourct=\timect \multiply\hourct by 60
         \minct=\time \advance\minct by -\hourct
         \number\timect:\ifnum \minct < 10 0\fi\number\minct}

\newcommand\mctc{\multicolumn{2}{c}}
\newcommand\mctl{\multicolumn{2}{l}}

\catcode`@=11

\newcommand\comment[1]{}
\newcommand\commenton{\catcode`\%=14}
\newcommand\commentoff{\catcode`\%=12}

\renewcommand\math[1]{$#1$}
\newcommand\mathshifton{\catcode`\$=3}
\newcommand\mathshiftoff{\catcode`\$=12}

\comment{the backslash is necessary}
\comment{alignment tab}
\let\atab=&

\newcommand\atabon{\catcode`\&=4}
\newcommand\ataboff{\catcode`\&=12}

\let\oldmsp=\sp
\let\oldmsb=\sb
\def\sp#1{\ifmmode
           \oldmsp{#1}%
         \else\strut\raise.85ex\hbox{\scriptsize #1}\fi}
\def\sb#1{\ifmmode
           \oldmsb{#1}%
         \else\strut\raise-.54ex\hbox{\scriptsize #1}\fi}
\newbox\@sp
\newbox\@sb
\def\sbp#1#2{\ifmmode%
           \oldmsb{#1}\oldmsp{#2}%
         \else
           \setbox\@sb=\hbox{\sb{#1}}%
           \setbox\@sp=\hbox{\sp{#2}}%
           \rlap{\copy\@sb}\copy\@sp
           \ifdim \wd\@sb >\wd\@sp
             \hskip -\wd\@sp \hskip \wd\@sb
           \fi
        \fi}
\def\msp#1{\ifmmode
           \oldmsp{#1}
         \else \math{\oldmsp{#1}}\fi}
\def\msb#1{\ifmmode
           \oldmsb{#1}
         \else \math{\oldmsb{#1}}\fi}

\def\supon{\catcode`\^=7}
\def\supoff{\catcode`\^=12}

\def\subon{\catcode`\_=8}
\def\suboff{\catcode`\_=12}

\def\supsubon{\supon \subon}
\def\supsuboff{\supoff \suboff}

\newcommand\actcharon{\catcode`\~=13}
\newcommand\actcharoff{\catcode`\~=12}

\newcommand\paramon{\catcode`\#=6}
\newcommand\paramoff{\catcode`\#=12}

\comment{And now to turn us totally on and off...}
\newcommand\reservedcharson{\commenton
                            \mathshifton
                            \atabon
                            \supsubon
                            \actcharon
                            \paramon}

\newcommand\reservedcharsoff{\commentoff
                             \mathshiftoff
                             \ataboff
                             \supsuboff
                             \actcharoff
                             \paramoff}

\catcode`@=12
\reservedcharsoff

\reservedcharson
\actcharon

\title{CHARACTERIZING EXOPLANET ATMOSPHERES: FROM LIGHT-CURVE OBSERVATIONS TO RADIATIVE-TRANSFER MODELING} 
\author{PATRICIO E. CUBILLOS} 

\prevdegreei {B.Sc. Universidad de Chile, 2006}
\prevdegreeii{M.Sc. Universidad de Chile, 2011}

\thesisname{dissertation}

\degreename{Doctor of Philosophy in Physics, Planetary Sciences Track}

\departmentsname{Physics}

\collegename{Sciences}

\termname{Fall}

\termyear{2015}

\advisorname{Joseph Harrington}

\begin{document}
\frontmatter
\maketitle
\copyrightpage{~Patricio E. Cubillos}

\begin{abstract}

Multi-wavelength transit and secondary-eclipse light-curve observations are some of the most powerful techniques to probe the thermo-chemical properties of exoplanets.  Although the large planet-to-star brightness contrast and few available spectral bands produce data with low signal-to-noise ratios, a Bayesian approach can robustly reveal what constraints we can set, without over-interpreting the data.  Here I performed an end-to-end analysis of transiting exoplanet data.  I analyzed space-telescope data for three planets to characterize their atmospheres and refine their orbits, investigated correlated noise estimators, and contributed to the development of the respective data-analysis pipelines.  Chapters 2 and 3 describe the Photometry for Orbits, Eclipses and Transits (POET) pipeline to model Spitzer Space Telescope light curves.  I analyzed secondary-eclipse observations of the Jupiter-sized planets WASP-8b and TrES-1, determining their day-side thermal emission in the infrared spectrum.  The emission data of WASP-8b indicated no thermal inversion, and an anomalously high 3.6 micron brightness.  Standard solar-abundance models, with or without a thermal inversion, can fit the thermal emission from TrES-1 well.  Chapter 4 describes the most commonly used correlated-noise estimators for exoplanet light-curve modeling, and assesses their applicability and limitations to estimate parameters uncertainties.  I show that the residual-permutation method is unsound for estimating parameter uncertainties.  The time-averaging and the wavelet-based likelihood methods improve the uncertainty estimations, being within 20 -- 50\% of the expected value.  Chapter 5 describes the open-source Bayesian Atmospheric Radiative Transfer (BART) code to characterize exoplanet atmospheres.  BART combines a thermochemical-equilibrium code, a one-dimensional line-by-line radiative-transfer code, and the Multi-core Markov-chain Monte Carlo statistical module to constrains the atmospheric temperature and chemical-abundance profiles of exoplanets.  I applied the BART code to the Hubble and Spitzer Space Telescope transit observations of the Neptune-sized planet HAT-P-11b.  BART finds an atmosphere enhanced in heavy elements, constraining the water abundance to ~100 times that of the solar abundance.

\end{abstract}

\acknowledgments{
  It's fascinating how these six years at UCF turned out to be much more
  than just a PhD degree.  It certainly went beyond expectations in all
  aspects, because not only did I had to be a student or researcher, I
  also had to play the role of a colleague, a teacher, a shrink, a
  patient, a friend, an much more.  This period has been a true life
  experience with all the ups and downs, it's been at times as hard,
  hopeless, and lonely as it's been fun, exciting, and fulfilling.
  Of course, I couldn't have accomplished everything I did without the
  contributions of many others.  First, I have to thank my advisor
  Joseph Harrington.  I really appreciate how much I have learned from
  you, not only your direct instruction but also your thoroughness and
  work ethics.  I can only regard you with the most respect.
  I also have to thank Jasmina Blecic.  I don't think there is any
  other person I have argued with so much and so many times, yet I'm still
  glad to see again.  I'm deeply grateful for your advice in the
  office and especially for your advice out of the office.  I really
  enjoyed having you as a friend and I'm sure I couldn't have
  graduated without your help.
  I thank Alejandra, you have been a terrific friend, thanks for your
  company and support when needed.
  Thanks to my family for their unconditional support, to all
  the friends and workmates I met while at UCF,  to Nicolas
  Massu for reminding us that ``nada es imposible'', and many thanks
  to Jennifer Parham for taking so much bureaucracy off of my
  shoulders.}

\notes{
  The work presented in this dissertation consists of published and
  in-preparation journal articles, each the result of a collaborative
  effort amongst the coauthors listed in each chapter.  Unless
  otherwise stated below, I have reworded contributed text from
  coauthors without changing the ideas or implications behind the
  text.  The paragraphs below describe the specific contributions made
  in each paper using the initials of each author.

In Chapter \ref{chap:wasp8b}
P.C. wrote the paper with contributions from J.H., N.M., K.S., R.A.H.,
J.B., D.R.A., M.H., C.J.C; P.C. reduced the data; N.M. produced the
atmospheric models.; R.A.H. produced the orbital parameter results;
P.C. and J.H. analyzed the results;
J.H. developed the POET analysis pipeline in IDL.
K.S., P.C., C.J.C, and R.A.H. adapted POET for Python.
P.C. implemented the orbital thermal variation model.

In Chapter \ref{chap:tres1}
P.C. wrote the paper with contributions from J.H., N.M., A.S.D.F.,
N.B.L., R.A.H., M.O.B.;
P.C. reduced the data; 
A.S.D.F. produced the orbital parameter results with contribution from
R.A.H.;
N.M. produced the atmospheric models; 
P.C. and J.H. analyzed the results;
P.C. implemented the DEMC algorithm,
N.B.L. implemented the least-asymmetry methods.

In Chapter \ref{chap:rednoise} P.C. wrote the paper with contributions
from J.H. and T.J.L., N.B.L., J.B., and M.S.;
T.J.L. provided the statistical background for the
residual-permutation method conclusions.
P.C. designed and carried out the simulations;
P.C. analyzed the results;
P.C. wrote the MC$\sp{3}$ module with contribution from N.B.L. and M.S.

In Chapter \ref{chap:BART},
P.C. wrote the paper with contribution from J.H., J.B., R.C.C. and
S.D.B.; P.C. analyzed the HAT-P-11b transit data and ran the HAT-P-11b
atmospheric analysis.

Regarding the BART code implementation:
J.H. is the architect of BART, specifying the components and the
information flow between them, identifying major areas of required
effort, and assigning personnel to programming and documentation
tasks.
P.C. reviewed and documented the original Transit code, implemented
the opacity-grid calculation, opacity-grid interpolation,
Voigt-function pre-calculation, filter-transmission integration;
P.C. wrote the MC$\sp{3}$ module documentation;
P.C., J.B., R.C.C wrote the Transit module documentation with
contribution from A.J.F.;
P.C. developed the code to read the HITRAN, P\&S, Schwenke, and Plez
line-transition files;
P.C. implemented the MC$\sp{3}$ module with contribution from
N.B.L. and M.S.;
P.C., M.M.S., N.B.L., and A.S.D.F. implemented the MPI communication
framework;
P.C. and M.M.S. implemented the three-channel Eddington-approximation model;
P.C. and M.M.S. reimplemented the C linereader into Python;
P.C. implemented the CTIPS module with contribution from A.J.F.;
P.C and R.C.C. implemented the C Simpson integration and spline
interpolation routines with contribution from S.D.B..

S.D.B. implemented the VO partition-function code with contribution
from P.C.;
S.D.B. tested Transit and provided feedback for P.C. to further
develop Transit;
D.B. implemented HITRAN CIA reader with contribution from P.C..

P.M.R. provided original version of Transit;
M.M.S. imported Transit from SVN into Github;
A.J.F. redesigned the Transit compilation scripts with contribution
from A.S.D.F.;
N.B.L. implemented SWIG wrapped to run Transit from Python and several
performance speed ups to Transit;
J.B. and M.O.B. implemented and wrote documentation for TEA with
contribution from M.M.S.;
C.M. computed independent radiative-transfer spectra for comparison
with Transit;
A.J.F. added support for shared memory in BART/Transit.
J.B. implemented the contribution-function, best-fit calculations,
eclipse-geometry radiative-transfer code.
R.C.C. implemented the HITRAN cross-section reader for Transit;
R.C.C., J.G., A.J.F., A.S.D.F. contributed in low-level tasks in the
development of BART/Transit.
}

\tableofcontents
\listoffigures
\listoftables
\mainmatter

\chapter{INTRODUCTION}
\label{chap:intro}

\setcitestyle{authoryear, round}

The study of planets orbiting stars other than the Sun, exoplanets,
has become one of the most exciting and fastest-growing fields of
astronomy.  Since the first exoplanet detection
\citep{MayorQueloz1995natJupiterExoplanet}, the discovery rate has
increased exponentially, reaching nearly two thousand confirmed
exoplanets to date.
There are multiple methods to detect exoplanets, the
radial-velocity and transit techniques being the most successful.  The
radial-velocity method uses high-precision spectrographs to measure
the radial velocity of stars.  Planet-hosting stars show periodic
oscillations around the center of mass of the system.  The
radial-velocity signal is proportional to the mass of the planet.
The transit method monitors the flux of stars as a function of time.  If
a planet's orbit is aligned such that the planet crosses in front of
the star as observed from Earth (a ``transit''), the telescope
detects the drop of the flux blocked by the planet.  The transit signal
is proportional to the size of the planet.
Given the observational biases of the detection methods, the first
objects detected were large and massive planets orbiting
extremely close to their host stars; they were named ``hot Jupiters''.

The further arrival of dedicated, state-of-the-art facilities allowed for
important breakthroughs.  High-precision spectrographs, like the High
Accuracy Radial Velocity Planet Searcher
\citep[HARPS,][]{MayorEtal2003HARPS} and photometers, like the Kepler
Space Telescope, yielded the first estimations of the exoplanetary
occurrence rate in the Milky Way.  Our galaxy hosts billions of planets,
with Earth-sized planets the most frequent.  To highlight some
findings, \citet{FressinEtal2013KeplerRate} estimated that 52\% of
stars host at least one exoplanet,
\citet{DressingCharbonneau2015apjOcurrenceHabitableMdwarfs} determined
a planet ocurrence rate of $2.5 \pm 0.2$ planets per M dwarf star, and
\citet{BonfilsEtal2011PlanetOccurrenceMdwarf} estimated the fraction
of habitable planets around M dwarf stars at $0.41\sp{+54}\sb{-13}$.

Building on these statistics, future dedicated missions, like the
Transiting Exoplanet Survey Satellite
\citep[TESS,][]{RickerEtal2014spieTESS}, the Next Generation Transit
Survey \citep[NGTS,][]{WheatleyEtal2013NGTS}, or the Characterizing
ExOPlanet Satellite (CHEOPS) mission \citep{BroegEtal2013CHEOPS}, are
expected to find thousands of exoplanets around the brightest stars in
the Solar neighborhood.  These planets will be the best-suited targets
for characterization.

\section{Exoplanet Characterization}

Although exoplanets can hardly be better characterized than any
Solar-System planet, their value resides in their much larger number
and diversity \citep{CowanEtal2015paspCharacterizingAtmospheres}.
Exoplanets provide a far more comprehensive view of planetary physics
than what we can learn from the Solar System alone.  However, to date,
no categorization has fully succeeded to describe the observed
physical properties of the aggregate of known exoplanets.
For example, \citet{FortneyEtal2008apjTwoClasses} proposed a
classification of hot Jupiters based on the incident irradiation.  The
most highly irradiated planets would show thermal inversions due to
TiO and VO absorption.  However, subsequent observations and studies
challenged the predictions and mechanism of this classification
\citep[e.g.,][]{KnutsonEtal2010ApJ-CorrStarPlanet,
  SpiegelEtal2009apjTiO, Madhusudhan2012COchemistry}.
\citet{CowanAgol2011Albedos} studied the Bond albedo from a sample of
24 exoplanets. They found low albedos, but detected no clear trends
between the estimated equilibrium temperatures (an estimation of the
planet's energy budget) and the observed effective temperatures.  The
observed population of exoplanets shows a large diversity of
properties; therefore, classifying them is a complex and multi-dimensional
problem.

Transiting exoplanets are the most favorable targets for
characterization.  These planets pass in front and behind their host
stars, as seen from Earth, leaving a characteristic signature in the
systems' light curves.  By combining photometric and radial-velocity
measurements, we can constrain their bulk size and mass, and hence,
constrain their average density.  A further spectral analysis of their
time-series photometry allows for atmospheric characterization.

When a planet passes in front of its host star, a transit event, the
fraction of light that is blocked, the transit depth, is proportional
to the planet-to-star area ratio, constraining the bulk size of the
planet.  In this configuration, the planetary atmospheric composition
modulates the stellar light that travels across the planet's limb (the
day--night terminator).  Opaque absorbing gasses make the planet look
larger.  Each atmospheric species imprints a characteristic absorbing
pattern as a function of wavelength.  Hence, the variation of the
transit depth across wavelengths (the modulation spectrum), constrains
the composition of the planetary atmosphere.  The magnitude of the
modulation variation is proportional to the atmospheric scale height,
making a transit observation sensitive to the bulk density of the
atmosphere as well.  Given the geometry of the light ray paths, even low
concentrations of an absorber can have a large impact in the
modulation spectrum.  For the same token, high-altitude haze layers
quickly flatten out a spectrum
\citep[e.g.,][]{KreidbergEtal2014natCloudsGJ1214b,
  KnutsonEtal2014natGJ436b}.  Nonetheless, cloudless counter-examples
have been observed \citep[e.g.,][]{FraineEtal2014natHATP11bH2O}.

When a planet passes behind its host star (a secondary-eclipse event),
the fraction of blocked light is proportional to the thermal
(infrared) or reflected (optical) planetary emission.  Similar to the
transit modulation spectrum, the eclipse depth as a function of
wavelength constrains the atmospheric composition and temperature of
the integrated day-side hemisphere.  The shape of the light curve
during ingress and egress constrains the two-dimensional planetary
emission pattern \citep{deWitEtal2012aapFacemap,
  MajeauEtal2012Facemap}.  As the projection of the stellar limb
progressively occults and uncovers fractions of the planetary disk, we can
relate the flux variation to specific regions on the planet.
Additionally, since the stellar limb at ingress crosses the planetary
disk at a different angle than at egress (for an imperfectly edge-on
orbit), the time-resolved occultation of the planet allows for a
two-dimensional mapping of the emission.  For an edge-on orbit, a
one-dimensional longitudinal map can still be recovered.  Furthermore,
a multi-wavelength face mapping would constrain the compositional and
temperature variation over the planet surface, putting to test
circulation \citep{ShowmanEtal2012DopplerSignatures} and chemical
kinetics \citep{AgundezEtal2012aaChem} theories.  Both transit and
eclipse spectra around the infrared--visible boundary are also
sensitive to Rayleigh scattering
\citep[e.g.,][]{BiddleEtal2014mnrasGJ3470b}.

Finally, measuring the flux of a system along a whole orbit, a
phase-curve, allows one to construct a longitudinal phase map of the
planetary emission.  Only a phase curve can break the degeneracy
between the albedo and the day--night energy redistribution.  By
measuring both the day- and night-side emission, the total emission
constrains the energy budget of a planet, whereas the day-to-night
temperature contrast constrains the heat redistribution.  The
longitudinal variation pattern will allow us to infer the general
dynamical circulation regime of the atmosphere.

Astronomers have used many ground-based and space facilities to
characterize exoplanets in the infrared.  Among these, the Spitzer and
Hubble space telescopes have provided the most fruitful results.  The
Spitzer Space Telescope had six broad wavebands covering the 3.6--24
{\micron} range during its cryogenic mission (2003 to 2009).  Since
then, only the two shortest-wavelength bands (3.6 and 4.5 {\microns})
remain in use.  On the other hand, the Hubble Space Telescope's (HST)
spectrograph, the Wide Field Camera 3 (installed in 2009), covers the
1.1--1.7 {\micron} range at a spectral resolution of
$\lambda/\Delta\lambda\sim 75$.  Note that most of the available
instrumentation was not conceived for exoplanet observations.  Thus,
to succeed, scientists have resorted to observing techniques and
post-acquisition calibrations, allowing them to extract signals even
fainter than the instruments' photometric design criteria.

\section{Atmospheric Modeling and Retrieval}

Modeling planetary atmospheres is a multi-disciplinary endeavor,
involving radiative processes, atmospheric chemistry, circulation
dynamics, cloud physics, etc.  The radiative-transfer equation
ultimately links the observed spectra (transit or eclipse depths) to
the atmospheric properties.  The temperature, pressure, and absorbing
species abundances are the main factors that determine the resulting
spectrum.  However, gases that are not spectroscopically active are
also important, as they contribute to the bulk density of the
atmosphere or modify the abundance of the absorbing gases through
chemical reactions.

In general, two data-modeling approaches prevail: forward modeling and
retrieval.  In the forward-modeling approach, the researcher
heuristically adjusts the parameters of a model until it resembles the
data well.  On the other hand, the retrieval approach uses algorithms
that explore and constrain the parameter phase space of a model, given
the existing data; for example, a Bayesian retrieval approach uses
Markov-chain Monte Carlo (MCMC) algorithms.
Currently, exoplanet data is sparse and of low signal-to-noise ratios,
often leading to large parameter uncertainties and degenerate
solutions.  Therefore, the retrieval approach is at an advantage over
the forward-modeling approach, since it provides an exhaustive
exploration, and statistically-robust constraints of the parameters'
phase space.
\citet{MadhusudhanSeager2010apjRetrieval} first applied the retrieval
approach to characterize exoplanet atmospheres.  Quickly after, others
groups developed their own retrieval tools
\citep[e.g.,][]{LeeEtal2012mnrasRetrieval, LineEtal2013apjRetrievalI,
  BennekeSeager2012apjRetrieval, WaldmannEtal2014TauRexI}.

In general, the radiative-transfer problem alone is well understood,
though there is a lack of opacity data at temperatures above
$\sim1000$ K.
The statistical treatment varies among the different groups, without
a clear superior algorithm so far.
More importantly, it is still little understood how radiative
processes, equilibrium and disequilibrium chemistry, circulation, and
cloud physics combine to determine the atmospheric abundances.  Hence,
the parameterization and priors of the atmospheric composition remain
one of the most open modeling aspects.  The priors (i.e.,
observational or theoretical constraints) guide ---and generally
improve--- the phase space exploration, limiting the MCMC to
physically-plausible regions.  How one parameterizes the models
affects the statistical treatment efficiency, for example, reducing
correlations between parameters.  However, the most efficient
parameterizations are not necessarily the easiest to interpret in
physical terms.

A clear obstacle in laying out significant constraints has been the
limited spectral coverage and coarse resolution of current technology
\citep{HansenEtal2014mnrasSpitzerFeatureless}.  However, in the future,
better instrumentation will help to overcome these limitations.
Expected to be launched in late 2018, the James Webb Space Telescope
(JWST) will be the first space telescope truly built for exoplanet
characterization.  JWST's unprecedent combination of collecting area, 
resolving power, and  spectral coverage will unveil new and more precise details of these
worlds.  From the ground, three Extremely-Large Telescopes are planned
to start their operations in the 2020s, the Thirty Meter Telescope
(TMT), the European Extremely Large Telescope (E-ELT), and the Giant
Magellan Telescope (GMT).  Together, they will cover both the northern
(TMT) and the southern (E-ELT and GMT) hemispheres.  Given their much
larger collecting area, these telescopes
will be optimized to observe fainter targets and perform high
resolving-power spectroscopy.

In this work, I analyzed exoplanet secondary eclipse observations and
developed an open-source Bayesian retrieval code to model exoplanet
atmospheres.  The code focuses on developing efficient statistical and
radiative-transfer routines.  The main goal of the code is to constrain
the temperature and composition of exoplanet atmospheres given the
available data.  To do so, the code explores the phase space of
parameterized models of the temperature and species abundances with
advanced MCMC algorithms.
In Chapters \ref{chap:wasp8b} and \ref{chap:tres1}, I describe the data
analysis for Spitzer secondary-eclipse observations for the cases of 
WASP-8b and TrES-1, respectively.
In Chapter \ref{chap:rednoise}, I further investigate the statistical
methods used in the field to analyze transiting exoplanet data.
Finally, in Chapter \ref{chap:BART}, I describe the Bayesian
Atmospheric Radiative Transfer model, provide validation tests, and
show the atmospheric analysis for transit observations of the
exoplanet HAT-P-11b.

\bibliographystyle{apj}
\bibliography{chap1-intro}

\chapter{WASP-8b: CHARACTERIZATION OF A COOL AND
 ECCENTRIC EXOPLANET WITH SPITZER}
\label{chap:wasp8b}

{\singlespacing 
\noindent{\bf Patricio Cubillos\sp{1,2},
  Joseph Harrington\sp{1,2},
  Nikku Madhusudhan\sp{3},
  Kevin B. Stevenson\sp{1},
  Ryan A. Hardy\sp{1},
  Jasmina Blecic\sp{1},
  David~R. Anderson\sp{4},
  Matthew Hardin\sp{1}, and
  Christopher J. Campo\sp{1}
}

\vspace{1cm}

\noindent{\em
\sp{1} Planetary Sciences Group, Department of Physics, University of Central
       Florida, Orlando, FL 32816-2385 \\
\sp{2} Max-Plank-Institut f\"ur Astronomie, K\"onigstuhl 17, D-69117,
       Heidelberg, Germany \\
\sp{3} Department of Physics and Department of Astronomy, Yale University,
       New Haven, CT 06511, USA \\
\sp{4} Astrophysics Group, Keele University, Staffordshire ST5 5BG, UK
}

\vspace{1cm}

\centerline{Received  6 July 2012.}
\centerline{Accepted  4 March 2013.}
\centerline{Published in {\em The Astrophisical Journal} 12 April 2013.}

\vspace{1cm}

\centering{Publication reference: \\
  Cubillos, P., Harrington, J., Madhusudhan, N., Stevenson, K., Hardy,
  R., Blecic, J., Anderson, D., Hardin, M., \& Campo, C. 2013, ApJ, 768, 42 \\
  http://arxiv.org/abs/1303.5468}

\vspace{1cm}

\centerline{\copyright AAS. Reproduced with permission}
}

\clearpage
\setcitestyle{authoryear, round}

\section[Abstract]{Abstract}

WASP-8b has 2.18 times Jupiter's mass and is on an eccentric
($e=0.31$) 8.16-day orbit. With a time-averaged equilibrium
temperature of 948 K, it is one of the least-irradiated hot Jupiters
observed with the \SST.  We have analyzed six photometric light curves
of WASP-8b during secondary eclipse observed in the 3.6, 4.5, and 8.0
\microns\ Infrared Array Camera bands.  The eclipse depths are
0.113\pm 0.018\%, 0.069\pm 0.007\%, and 0.093\pm 0.023\%,
respectively, giving respective brightness temperatures of 1552, 1131,
and 938 K.  We characterized the atmospheric thermal profile and
composition of the planet using a line-by-line radiative transfer code
and a MCMC sampler. The data indicated no thermal inversion,
independently of any assumption about chemical composition.  We noted
an anomalously high 3.6-\micron\ brightness temperature (1552~K); by
modeling the eccentricity-caused thermal variation, we found that this
temperature is plausible for radiative time scales less than
$\sim\ttt2$ hours.  However, as no model spectra fit all three data
points well, the temperature discrepancy remains as an open question.

\section{Introduction}
\label{introduction}

When transiting exoplanets pass behind their host stars (a secondary
eclipse), the observed flux drop provides a direct measurement of the
planet's thermal emission and reflected light. Today,
secondary-eclipse observations exist for nearly 30 exoplanets.  The
\SST\ \citep{WernerEtal2004apjsSpitzer} made most of these
observations, capturing broadband photometric light curves in six
near- and mid-infrared bands (3 -- 24 \microns).  Each band probes a
specific altitude range in a planet's atmosphere.  With Bayesian
fitting of model spectra, one can quantitatively constrain the
atmospheric chemical composition and thermal profile of the planet's
photosphere \citep{MadhusudhanSeager2010}.  WASP-8b, with a
time-averaged equilibrium temperature of 948 \pm\ 22 K (\Teq,
temperature at which blackbody emission balances absorbed energy,
assuming zero albedo and efficient heat redistribution), is one of the
coolest Jupiter-sized planets yet observed in eclipse, and thus serves
as an end member to the set of measured hot-Jupiter atmospheres.

To classify the hot-Jupiter population, \citet{Fortney2008} proposed a
separation between moderately and strongly irradiated planets.  The
higher atmospheric temperatures of the more strongly irradiated
planets allow the presence of highly opaque molecules (like TiO and
VO) at high altitudes.  These strong absorbers produce hot
stratospheres (thermal inversion layers).  In contrast, for the
moderately irradiated hot Jupiters, these absorbers condense and rain
out to altitudes below the photosphere, thus presenting no thermal
inversions.

In general, the observations agree with this hypothesis, but
exceptions indicate that the picture is not yet completely understood.
For example, secondary-eclipse observations of the highly irradiated
WASP-12b \citep{Madhusudhan2011Nat, CrossfieldEtal2012apjWASP12b},
WASP-14b \citep{BlecicEtal2011}, and TrES-3
\citep{FressinEtal2010ApJ-Tres3} do not show evidence of thermal
inversions.  Conversely, XO-1 has an inversion layer even though it
receives a much lower stellar irradiation \citep{Machalek2008-XO-1b}.
Photochemistry provides one explanation.  The non-equilibrium
atmospheric chemistry models of \citet{ZahnleEtal2009SulfurPhotochem}
suggested that heating from sulfur compounds in the upper atmospheres
of hot Jupiters could explain these inversions.  Alternatively,
\citet{Knutson2010ApJ-CorrStarPlanet} suggest that strong UV radiation
from active stars destroys the high-altitude absorbers.

The Wide-Angle Search for Planets (WASP) Consortium discovered WASP-8b
in 2008 \citep{Queloz2010Wasp8}.  The planet orbits the brighter
component (WASP-8A) of a binary stellar system.  The angular
separation (4.83{\arcsec}) with the secondary (WASP-8B) sets a minimum
separation of 440 AU between the stars.  WASP-8A is a G6 star, with
effective temperature $T$\sb{eff} = 5600 K.  Color and photometric
analyses indicate that WASP-8B is a colder M star
\citep{Queloz2010Wasp8}.  WASP-8b is a 2.18 Jupiter-mass ($M$\sb{Jup})
planet with 1.08 times Jupiter's radius ($R$\sb{Jup}) in a retrograde
8.16 day orbit.  Its large eccentricity ($e = 0.31$) should make the
planet's dayside temperature vary by hundreds of degrees along the
orbit, possibly forcing an unusual climate.

The age of the host star (4 Gyr) is shorter than the planet's orbital
circularization time \citep[\sim30 Gyr, see, e.g.,][]{Goldreich1966Q,
  Bodenheimer2001}; accordingly, WASP-8b has one of the most eccentric
orbits among the \sim10-day-period exoplanets
\citep{PontEtal2011MNRASeccentricities}.  The Kozai mechanism
\citep{WuMurray2003Kozai} may explain the combination of high
eccentricity and retrograde orbit orientation.  The radial-velocity
drift and the large eccentricity may also indicate a second planetary
companion \citep{Queloz2010Wasp8}.

We obtained six secondary-eclipse light curves of WASP-8b from four
visits of the \SST, observing in the 3.6, 4.5, and 8.0 \microns\ bands
of the Infrared Array Camera \citep[IRAC, ][]{FazioEtal2004apjsIRAC}.
The eclipse depths determine the planet's dayside infrared emission.
Our Markov-chain Monte Carlo-driven radiative-transfer code
constrained the molecular abundances and temperature profile of
WASP-8b's dayside atmosphere, testing for the expected absence of a
thermal inversion and estimating the energy redistribution over its
surface.  We constrained the orbit of WASP-8b by determining the
eclipse epochs and durations.  We also modeled the thermal variations
along the orbit of the planet to explore the effects of the
eccentricity.

Section \ref{sec:observations} presents the \Spitzer\
observations of the WASP-8 system.
Section \ref{sec:analysis} describes the photometric and modeling
analysis of our secondary eclipse observations.
Section \ref{sec:orbit} gives the orbital dynamical analysis.
Section \ref{sec:atmosphere} presents our constraints on WASP-8b's
atmospheric composition derived from the photometry.
Section \ref{sec:discussion} discusses the effects of
eccentricity on the orbital thermal variation of WASP-8b.
Finally, Section \ref{sec:conclusions} states our conclusions.

\section{Observations}
\label{sec:observations}

The {\em Spitzer Space Telescope} visited WASP-8 four times.  From two
consecutive eclipse observations, we obtained simultaneous light
curves at 4.5 and 8.0 \microns.  Later, from two more consecutive
eclipse observations during the {\em Warm} \Spitzer\ mission, we
obtained one light curve at 3.6 \microns\ and one at 4.5 \microns\
(see Table \ref{table:observations}).  The \Spitzer\ pipeline (version
18.18.0) processed the raw data, producing Basic Calibrated Data
(BCD).

\begin{table}[ht]
\centering
\caption[Observation Information]
        {Observation Information}
\label{table:observations}
\begin{tabular}{cccccc}
\hline
\hline
Label\sp{a} 
          & Wavel.     & Observation    & Duration  & Exp. time  & Cadence  \\ 
          & (\microns) & date           & (minutes) & (seconds)  & (seconds)\\
\hline                                                                  
\waspt2   & 4.5       &  2008  Dec  13 & 226       & \n 1.20    & \n 2.0    \\ 
\waspf2   & 8.0       &  2008  Dec  13 & 226       &   10.40    &   12.0    \\ 
\waspt1   & 4.5       &  2008  Dec  21 & 226       & \n 1.20    & \n 2.0    \\ 
\waspf1   & 8.0       &  2008  Dec  21 & 226       &   10.40    &   12.0    \\
\waspo    & 3.6       &  2010  Jul  23 & 458       & \n 0.36    & \n 0.4    \\ 
\waspt3   & 4.5       &  2010  Jul  31 & 458       & \n 0.36    & \n 0.4    \\ 
\hline
\multicolumn{6}{l}{\footnotesize {\bf Note}. \sp{a} {\it wa008b} designates the
  planet, {\it s} specifies secondary eclipse, and the two}  \\
\multicolumn{6}{l}{\footnotesize numbers indicate the wavelength channel and observation serial number (we} \\
\multicolumn{6}{l}{\footnotesize analyzed the 2008 December 21 data before the 2008 December 13 data and} \\
\multicolumn{6}{l}{\footnotesize inadvertently inverted the serial numbers).}
\end{tabular}
\end{table}

During the initial minutes of our observations, the telescope pointing
drifted \sim0.25 pixels before stabilizing.  Throughout the
observations, the pointing also jittered from frame to frame
($\sim0.01$ pixel) and oscillated in an hour-long periodic movement
($\sim0.1$ pixel amplitude).

\begin{figure}[tb]
\centering
\includegraphics[width=\linewidth, clip]{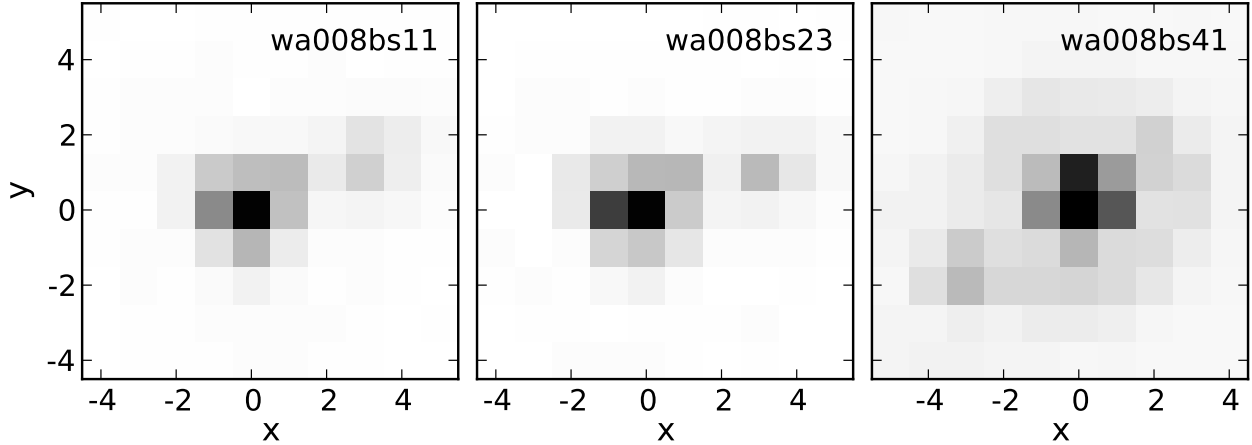}
\caption
[{\Spitzer} images of WASP-8b]
{{\Spitzer} images of WASP-8 at 3.6, 4.5, and 8.0 \microns,
  respectively. The brighter star, WASP-8A, is at the origin. The
  dimmer WASP-8B signal overlapped that of WASP-8A.}
\label{fig:SpitzerFOV}
\end{figure}

The separation between the centers of WASP-8A and WASP-8B in the IRAC
detectors is only 3.7 pixels. Consequently, the signal from the stars
overlapped, demanding special care during the data analysis (see
Figure \ref{fig:SpitzerFOV}).  Table \ref{table:WASPsystem} shows the
average and standard deviation of the flux ratio, separation, and
position angle (PA) of the secondary star with respect to WASP-8A
(derived from our centering routine, see Section
\ref{ss:center}). Our PA values agree with those of
\citet{Queloz2010Wasp8}, but our separation values are consistently
lower than theirs ($4.83 \pm 0.01${\arcsec}).

\begin{table}[ht]
\centering
\small
\caption{WASP-8 System}
\label{table:WASPsystem}

\begin{tabular}{cclclclcl}
\hline
\hline
Event   & \mctc{\underline{Flux\phantom{p}Ratio}}
        & \mctl{\underline{Separation (pix)}}
        & \mctl{\underline{Separation (\arcsec)}}
        & \mctl{\underline{Position Angle (deg)}}     \\
        & average & stddev  & average & stddev & average   & stddev  & average & stddev \\
\hline
\waspo  & 0.1420 & 0.0030  & 3.760 & 0.013  & 4.610 & 0.016   & 171.32 & 0.28 \\
\waspt1 & 0.1512 & 0.0017  & 3.734 & 0.007  & 4.541 & 0.009   & 171.00 & 0.15 \\
\waspt2 & 0.1600 & 0.0022  & 3.737 & 0.009  & 4.544 & 0.011   & 170.75 & 0.16 \\
\waspt3 & 0.1648 & 0.0039  & 3.726 & 0.017  & 4.497 & 0.020   & 170.78 & 0.29 \\
\waspf1 & 0.1718 & 0.0020  & 3.690 & 0.009  & 4.513 & 0.011   & 170.84 & 0.17 \\
\waspf2 & 0.1794 & 0.0023  & 3.686 & 0.010  & 4.506 & 0.012   & 170.96 & 0.16 \\
\hline
\end{tabular}
\end{table}

\section{Data Analysis}
\label{sec:analysis}

Our Photometry for Orbits, Eclipses, and Transits (POET) pipeline
produces light curves from BCD images.  Briefly, POET creates a bad
pixel mask for each image, finds the center position of the target,
executes interpolated aperture photometry, and fits a light curve
model that includes physical and systematic parameters.

\subsection{POET: Initial Reduction}

POET created bad pixel masks by discarding the flagged pixels from the
Spitzer BCD masks. Then, it discarded outlier pixels with a
sigma-rejection method.  At each pixel position and in sets of 64
consecutive images, POET calculated the median and standard deviation
of the unmasked pixels. Pixels diverging more than four times the
standard deviation from the median were masked. We iterated this
process twice.

We obtained the Julian Date of each frame from the UTCS\_OBS and
FRAMTIME entries of the files' headers.  We calculated the Barycentric
Julian Date (BJD) by correcting the projected light-travel time from
the telescope to the Solar System's barycenter using the Jet
Propulsion Laboratory (JPL) Horizons system.  We report the times in
both Coordinated Universal Time (UTC) and the Barycentric Dynamical
Time (TDB); the latter is unaffected by leap seconds
\citep{Eastman2010}.

\subsection{Centering}
\label{ss:center}

POET provides three routines to determine the center of the
point-spread function (PSF) in each image: center of light,
2D-Gaussian fitting, and least asymmetry \citep[Supplementary
Information]{StevensonEtal2010Natur}.  The proximity of WASP-8B
confuses these methods, so we added a double-PSF fit that shifts
supersampled PSFs to the target and secondary, bins them down, and
scales their amplitudes, as in \citet{Crossfield2010}.  For each
\Spitzer\ band we used Tiny
Tim\footnote{http://irsa.ipac.caltech.edu/data/SPITZER/docs/dataanaly-sistools/tools/contributed/general/stinytim/} (version 2.0) to create a stellar PSF
model with a 5600 K blackbody spectrum at 100{\by} finer resolution
than our images.  The double-PSF routine has seven free parameters:
the position of each star ($x\sb A$, $y\sb A$, $x\sb B$, $y\sb B$),
the integrated stellar fluxes ($F\sb A$, $F\sb B$), and the background
sky flux ($f$\sb{sky}).

To avoid interpolation when binning down, the PSF shifts are quantized
at the model's resolution, such that image and model pixel boundaries
coincide.  This quantization sets the position precision to 0.01
pixels.  It also excludes the position parameters from \chisq\
minimizers that assume a continuous function, such as
Levenberg-Marquardt.  So, we fit $F\sb A$, $F\sb B$, and $f$\sb{sky}
for a given position set \textbf{\emph{x}}~$ = \{x\sb A$, $y\sb A$,
$x\sb B$, $y\sb B\}$.

To avoid the computational challenge of performing a \chisq\
minimization for each \textbf{\emph{x}} in a 4D space at 0.01 pixel
resolution, we explored only specific coordinate positions.  Starting
at an initial guess position, and with an initial jump step of 100
positions (1 image pixel), we calculated \chisq\ at that position and
the 80 ($=3\sp4-1$) adjacent positions that are one jump step away
along all combinations of coordinate directions.  We either moved to
the lowest \chisq or, if already there, shrank the step by half.  We
repeated the procedure until the step was zero.

\subsection{Photometry}

Circular aperture photometry is unsuitable for this system, since any
flux from the secondary star (WASP-8B) contained in the aperture
dilutes the eclipse depth of WASP-8b.  Small pointing jitter would
also increase the light-curve dispersion for any aperture that
included much WASP-8B flux.  Apertures that are too large or small
both produce noisier light curves.  Thus, we modified the POET
interpolated aperture photometry \citep[Supplementary
Information]{HarringtonEtal2007natHD149026b} to remove the secondary
star two different ways.  In both methods, we subtracted the median
sky level prior to the stellar flux calculation.  The sky annulus
included values 7 -- 15 pixels from the target.

In our first method (B-Subtract), we subtracted the fitted, binned PSF
model of WASP-8B from each image. Then, we performed interpolated
aperture photometry centered on the target (A aperture).  In the
second method (B-Mask), we discarded the pixels within a circular
aperture centered at the position of the secondary before performing
aperture photometry.  The mask's aperture must encompass most of the
contribution from WASP-8B, but not from the target. Therefore, we
tested mask apertures with 1.6, 1.8, and 2.0 pixel radii.  For each
photometry method we tested a broad range of A-aperture radii in 0.25
pixel intervals.

The B-Mask method has less residual dispersion when the mask is
located at a fixed vector separation from WASP-8A (using the median of
all the measured separations in an event), than when its position is
determined for each individual frame.  This can be explained by the
dimmer signal of WASP-8B, which lowers the accuracy of its position
estimation.  So, within each dataset using B-Mask, we used the median
vector separation of the two objects.  For the B-subtract method, the
SDNR and eclipse-depths differences are marginal.

\subsection{Light-curve Modeling}

The eclipse depths of WASP-8b are on the order of 0.1\% of the
system's flux, well below {\em Spitzer's}\/ photometric stability
criteria \citep{FazioEtal2004apjsIRAC}. Thus, the eclipse light-curve
modeling requires a thorough characterization of the detector
systematics.  Systematic effects have been largely observed and
documented; they can have both temporal and spatial components, and
vary in strength and behavior for each dataset.

The main systematic at 3.6 and 4.5 {\microns} is intrapixel
sensitivity variation, $M(x,y)$, where the measured flux depends on
the precise position of the target on the array
\citep{StevensonEtal2012apjHD149026b, CharbonneauEtal2005apjTrES1}.
In addition, the detectors show a time-dependent sensitivity variation
called the \emph{ramp} effect, $R(t)$, suspected to be caused by
charge trapping \citep{Agol2010ApjHD189} at 8.0 \microns, but there
are also reports of a ramp in the 3.6 and 4.5 \micron\ bands
\citep[e.g., ][]{Campo2011, NymeyerEtal2011, Knutson2011gj436,
  Deming2011Corot, BlecicEtal2011, StevensonEtal2010Natur,
  StevensonEtal2012apjHD149026b}.  The eclipse and both systematic
variations entangle to produce the observed light curve.  To account
for their contributions, we modeled the light curves as
\begin{equation}
F(x,y,t) = F\sb s M(x,y) R(t) E(t),
\label{eq:fire}
\end{equation}
where $F\sb s$ is the out-of-eclipse system flux.  We used the eclipse
model, $E(t)$, from \citet{MandelAgol2002apjLightcurves}.  The eclipse
is parametrized by the eclipse depth, the mid-point phase, the
duration, and the ingress and egress times.  For the ingress/egress
times we adopted a value of 18.8 min, derived from the orbital
parameters of the planet. We used this value in all of our
eclipse-model fits.

The strength and behavior of the ramp variations are specific to each
dataset. Many formulae have been applied in the literature
\citep[e.g.,][]{DemingEtal2007, HarringtonEtal2007natHD149026b,
  Knutson2011gj436, StevensonEtal2012apjHD149026b}. The models are
formed with combinations of exponential, logarithmic, and polynomial
functions.  We tested dozens of equations; the best were:
\begin{eqnarray}
\label{eqnre}
{\rm risingexp}: \quad R(t) & = &  1 - e\sp{-r\sb{0}(t-t\sb{0})} \\
\label{eqnlog}
{\rm logramp}\sb q:  \quad R(t) & = & 1 + r\sb{q} [\ln(t-t\sb{0})]\sp q \\
\label{eqnlin}
{\rm linramp}:      \quad R(t) & = & 1 + r\sb{1}(t-t\sb c) \\
\label{eqnquad}
{\rm quadramp}\sb 1: \quad R(t) & = & 1 + r\sb{1}(t-t\sb c) + r\sb{2}(t-t\sb c)\sp{2} \\
\label{eqnquad2}
{\rm quadramp}\sb 2: \quad R(t) & = & 1 + r\sb{2}(t-t\sb c)\sp{2} \\
\label{eqnll}
{\rm loglinear}:     \quad R(t) & = & 1 + r\sb{1}(t-t\sb c) + r\sb{4} \ln(t-t\sb{0})
\end{eqnarray}
where $t\sb c$ is a constant value at the approximated mid-point
phase of the eclipse ($t\sb c = 0.515$ for this planet). Slight
changes in $t\sb c$ do not significantly affect the fitted eclipse
parameters.

We used our Bi-Linearly Interpolated Subpixel Sensitivity (BLISS)
mapping technique \citep{StevensonEtal2012apjHD149026b} to calculate
$M(x,y)$.  The BLISS method has been found to return a better result
than a polynomial fit \citep{StevensonEtal2012apjHD149026b,
  BlecicEtal2011}.

To determine the best-fitting parameters of our model (Eq.\
\ref{eq:fire}), we used a $\chi\sp2$ minimizer with the
Levenberg-Marquardt algorithm.  We used Bayesian posterior sampling
via a Markov-chain Monte Carlo (MCMC) algorithm to explore the phase
space and estimate the uncertainties of the free parameters of the
light-curve models.  Our code implements the Metropolis random walk,
which proposes parameter sets from a multivariate normal distribution
centered at the current position in the chain, computes
\math{\chi\sp{2}}, and accepts (or rejects) the new set with greater
probability for a lower (higher) \math{\chi\sp{2}}.  By generating
millions of parameters sets, the algorithm samples the posterior
distribution of the model parameters.  As a necessary condition for
chain convergence, we require the Gelman-Rubin statistic
\citep{Gelman1992} to be within 1\% of unity for each free parameter
between four MCMC chains.

The photometry routine uses the BCD uncertainty images to estimate the
uncertainties of the light-curve data points, $\sigma\sb i$.  However,
since the \Spitzer\ pipeline in general overestimates these
uncertainties (it is designed for absolute photometry), we multiply by
a constant factor ($\sigma\sb i \rightarrow f \cdot \sigma \sb{i}$),
such that the reduced $\chi\sp2 = 1$ in the light-curve fit.  This is
equivalent to estimating a single $\sigma$ from the scatter of model
residuals.  Both methods account for red noise, but ours retains the
(usually small) $\sigma\sb i$ variations due to aberrant frames.

To determine the best raw light curve (i.e., by selection of
photometry method and aperture radius), we calculated the standard
deviation of the normalized residuals (SDNR) of the light curve fit
\citep{StevensonEtal2012apjHD149026b, Campo2011}. Poor fits or data
with high dispersion increase SDNR; the optimum data set minimizes the
SDNR value.  Once we chose the best light curve, we compared the
different ramp models according to the Bayesian Information Criterion
\citep{Liddle2007},
\begin{equation}
{\rm BIC} = \chi\sp2 + k\ln N,
\end{equation}
where $k$ is the number of free parameters and $N$ the number of data
points.  The best model minimizes the BIC.  The probability ratio
favoring one model over a second one is $\exp(-\Delta{\rm BIC}/2)$.

\subsubsection{wa008bs11 Analysis} 
\label{sss:wa11}

This observation started 2.9 hours before the eclipse's first contact.
The telescope observed the target in sub-array mode, allowing a high
cadence (Table \ref{table:observations}).  We discarded the initial 15
minutes of observation while the telescope pointing settled.  Our data
present both intrapixel and weak ramp systematics.

\begin{table}[ht]
\centering
\caption{\waspo\ Ramp Model Fits}
\label{table:wa008bs11ramps}
\begin{tabular}{lccc}
\hline
\hline
$R(t)$       & SDNR       & $\Delta$BIC       & Ecl. Depth (\%) \\
\hline
quadramp\sb1 &  0.0061141 & \phantom{1}0.00   & 0.119 \\
risingexp    &  0.0061148 & \phantom{1}1.73   & 0.106 \\
logramp\sb1  &  0.0061153 & \phantom{1}2.96   & 0.096 \\
linramp      &  0.0061201 & \phantom{1}6.34   & 0.063 \\
loglinear    &  0.0061141 &           11.10   & 0.119 \\
\hline
\end{tabular}
\end{table}

The 2.25 pixel A aperture with B-subtract photometry minimized SDNR.
Table \ref{table:wa008bs11ramps} shows the five best-fitting models to
the best \waspo\ light curve. $\Delta$BIC is with respect to the
lowest BIC value.  The quadramp\sb{1} model is 2.4 times more probable
than, and consistent with, the second-best model. The linear
($11\sigma$) and and quadratic ($4\sigma$) terms of the quadramp\sb{1}
model (see Table \ref{table:fits}) confirm the need for a ramp model.
As a general remark, we noted that all the logramp\sb{q} models
produce similar BIC and eclipse parameter values; therefore, we will
refer only to the logramp\sb{1} model in the future. Models with more
free parameters do not improve BIC.  Following
\citet{StevensonEtal2012apjHD149026b}, we vary the bin size and the
minimum number of data points per bin ($mnp$) of the BLISS map to
minimize the dispersion of the residuals. We required at least 4
points per bin for any dataset.  The PSF-fitting position precision of
0.01 pixels sets our lower limit for the binsize.  For \waspo, $mnp =
5$ and a bin size of 0.015 pixels optimized the fit.

\begin{figure}[tb]
\includegraphics[width=0.315\textwidth]{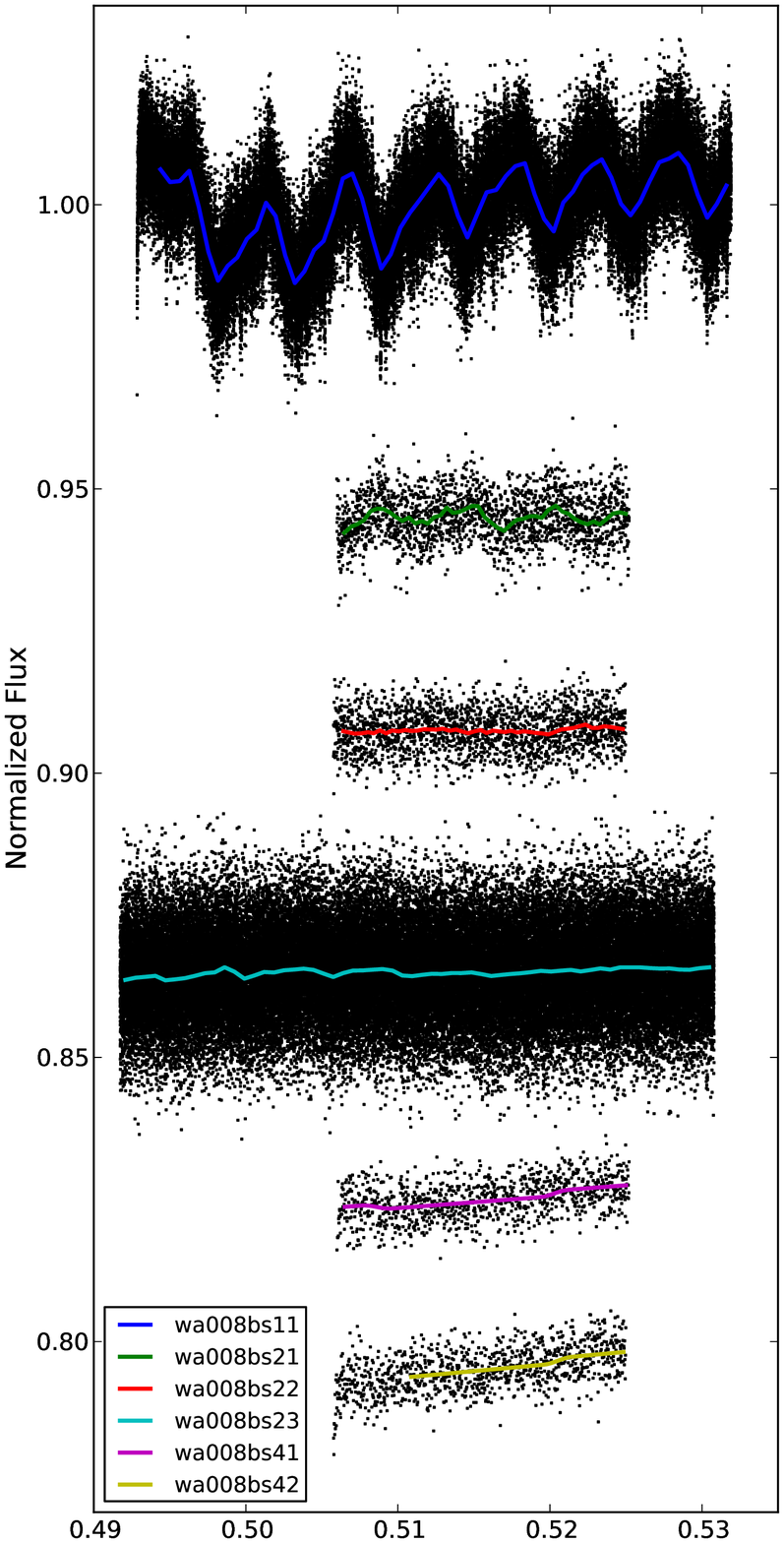}\hfill
\includegraphics[width=0.315\textwidth]{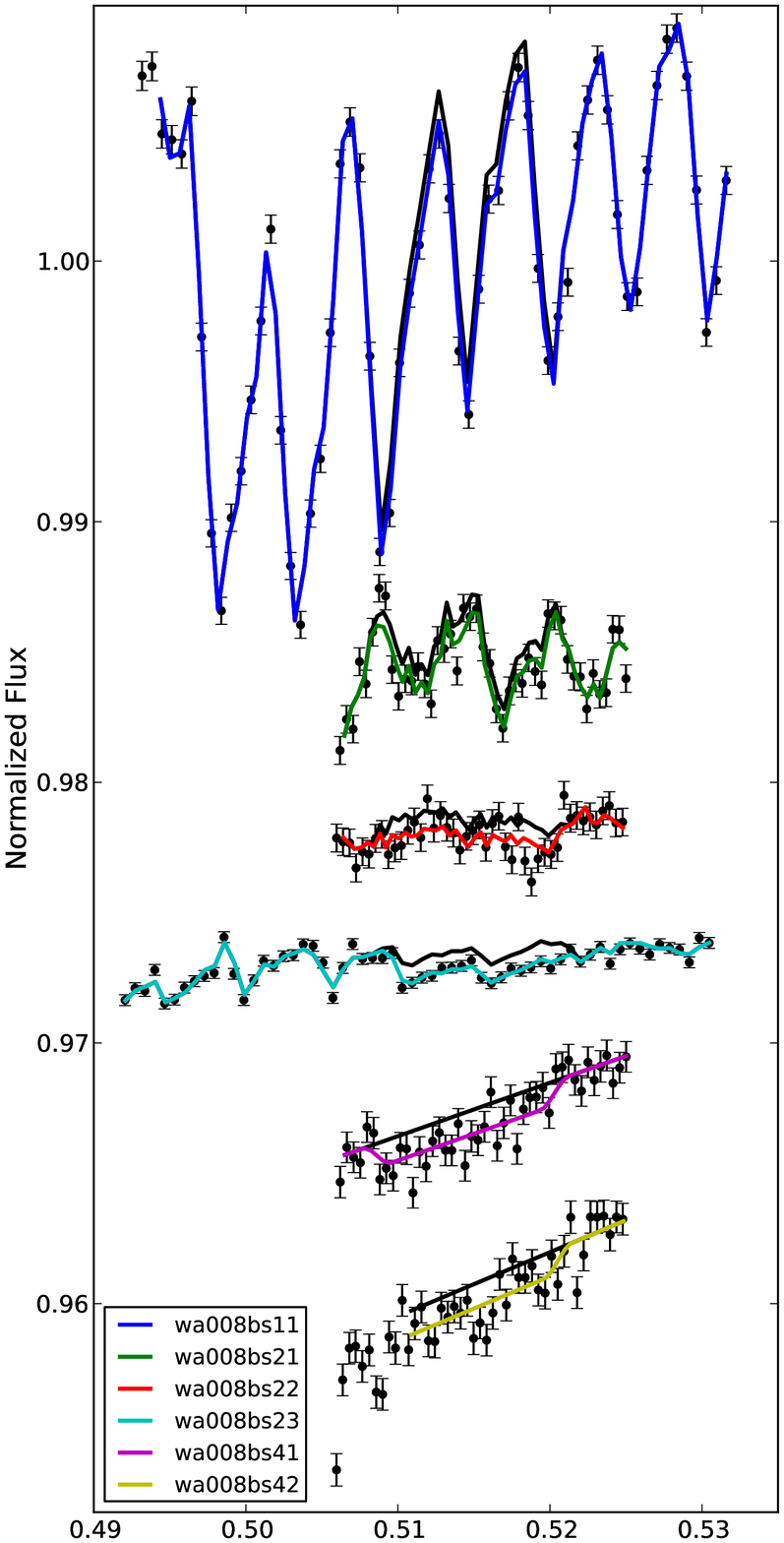}\hfill
\includegraphics[width=0.32\textwidth ]{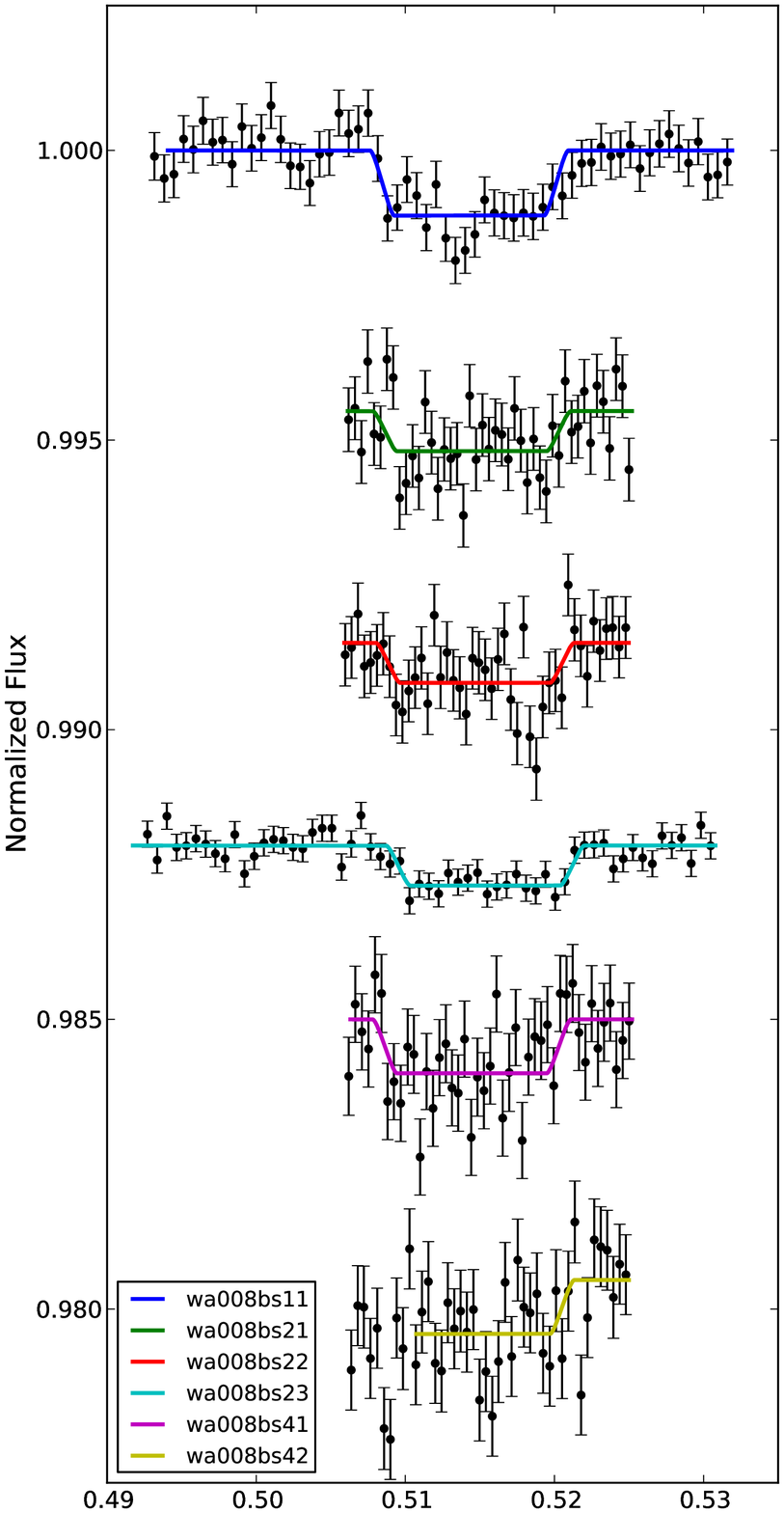}\hfill\strut
\put (-436, -8) {{\scriptsize Orbital Phase (8.2 day period)}}
\put (-277, -8) {{\scriptsize Orbital Phase (8.2 day period)}}
\put (-120, -8) {{\scriptsize Orbital Phase (8.2 day period)}}
\caption
  [Secondary-eclipse light curves of WASP-8b]
  {Raw (left), binned (center) and systematics-corrected (right)
  secondary-eclipse light curves of WASP-8b at 3.6, 4.5, and 8.0
  {\microns}.  The system flux is normalized to unity and the points
  are shifted vertically for clarity.  The colored curves are the
  best-fit models (see legend).  The black curves are the best-fit
  models excluding the eclipse component. The error bars in the center
  and right panels are the 1\math{\sigma} uncertainties.
  \label{fig:lightcurves}}
\end{figure}

\begin{figure}[tb]
\centering
\includegraphics[width=0.95\linewidth, clip]{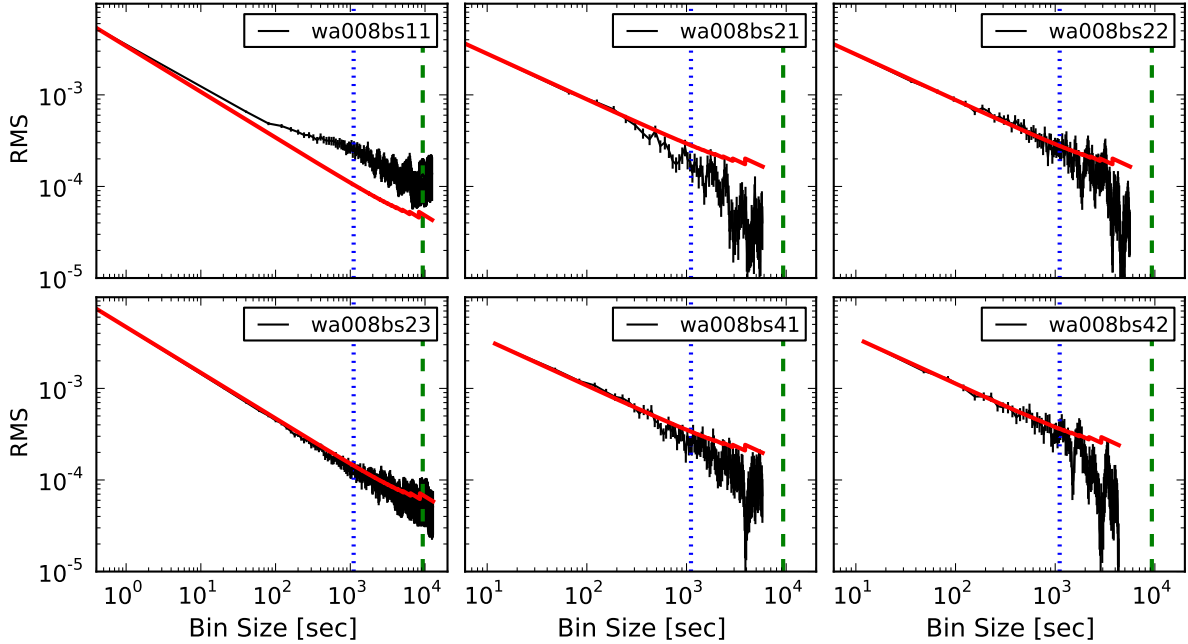}
\caption
  [Residuals RMS {\vs} bin size]
  {RMS of the fit residuals (black curves with 1$\sigma$
  uncertainties) {\vs} bin size of the WASP-8b light curves. The red
  curves are the expected RMS for Gaussian noise (extrapolated
  unbinned RMS scaling by the inverse square root of the bin
  size). The blue dotted and green dashed vertical lines mark the bin
  size corresponding to the eclipse ingress and duration time,
  respectively. \waspo's excess above the red line indicate correlated
  noise at bin sizes larger than the ingress time.}
\label{fig:rms}
\end{figure}

Figure \ref{fig:lightcurves} shows the raw, binned and
systematics-corrected \waspo\ light curves with their best-fitting
model.  We considered the correlated noise in the residuals as well
\citep{Pont2006Rednoise}.  Figure \ref{fig:rms} shows the
root-mean-square (RMS) of the residuals \vs bin size. The \waspo\ RMS
curve deviates above the expected RMS for pure Gaussian noise.
Following \citet{WinnEtal2008Rednoise}, to account for the correlated
noise, we weighted the light curve uncertainties by the factor $\beta$
(the fractional RMS excess above the pure Gaussian RMS at the bin size
corresponding to the eclipse duration). For \waspo\ we found $\beta
=2.4$.  We inspected all the pairwise correlation plots and histograms
and found only unimodal Gaussian distributions.

Alternatively, the residual-permutation (also known as {\em prayer
  bead}) algorithm is sometimes used to assess correlated noise in a
fit.  In this method, we cyclically shift the residuals from the best
model by one frame, add them back to that model, and re-fit, repeating
until we shift the residuals back to their original positions.  This
generates a distribution of values for each parameter, from which we
estimate the parameter uncertainties.  The eclipse-depth uncertainty
is 0.021\%, similar to the value found with the
\citet{WinnEtal2008Rednoise} method (see Table \ref{table:fits}).
However, we are cautious.  Although prayer bead has been broadly used
for the analysis of exoplanet lightcurve and radial-velocity fits
\citep[e.g.,][]{Southworth2008HomogTransit, BouchyEtal2005CorrNoise,
  PontEtal2005, GillonEtal2007Gj436, Knutson2008HD209,
  CowanEtal2012WASP12b}, we have found no detailed description of its
statistical properties in the literature.

\subsubsection{wa008bs23 Analysis} 
\label{sss:wa23}

\begin{figure}[htb]
\centering
\includegraphics[width=0.75\linewidth, clip]{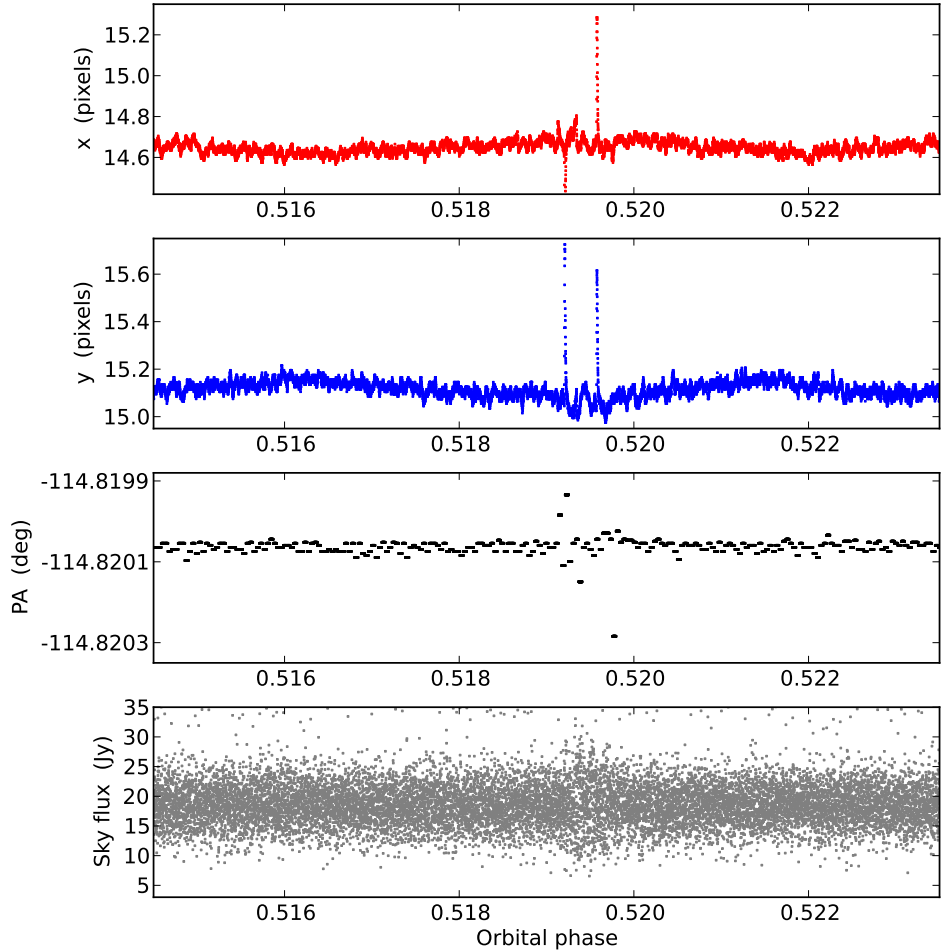}
\caption
  [\waspt3 target pointing, position angle, and sky flux]
  {\waspt3 target pointing, position angle and background sky
  flux near phase 0.519. We observed two sudden position shifts
  coincident with increases in the background flux. All values
  returned to normal almost instantly.}
\label{fig:impact}
\end{figure}

With the same observing setup as \waspo, this observation started 3.3
hours prior to the eclipse's first contact.  This dataset also
presented both intrapixel and ramp variations. Note that the
intrapixel systematic is weaker than at 3.6 \microns, attributed to
the smaller degree of undersampling at larger wavelengths by
\citet{CharbonneauEtal2005apjTrES1} and
\citet{Morales-CalderonEtal2006apjIntrapixel}. Even though the
pointing stabilized only after the initial 20 minutes, the light curve
did not deviate significantly; therefore, we included all data points
in the analysis. We noted two sudden pointing and PA deviations near
phase 0.519. After each incident, the telescope resumed its position
within \sim10 seconds (Figure \ref{fig:impact}).  Micrometeorite
impacts on the telescope can explain the abrupt deviations.
Simultaneously, we measured a slight increase in the background sky
flux dispersion, which returned to normality shortly after.  The
target flux did not show any extraordinary fluctuations during these
incidents.  However, the points outside the normal pointing range were
eliminated by the BLISS map's {$mnp$} criterion.

The analysis is analogous to \waspo. The SDNR indicated clearly that
the 2.25 pixel A aperture with B-Subtract photometry produced the
lowest dispersion.  The best-fitting ramp model is logramp\sb{1},
which is 21 times more probable than the rising exponential ramp
(Table \ref{table:wa008bs23ramps}).  The BLISS map is optimized at
$mnp=4$ and a bin size of 0.025 pixels.

\begin{table}[ht]
\centering
\caption{\waspt3 Ramp Model Fits}
\label{table:wa008bs23ramps} 
\begin{tabular}{lccc}
\hline
\hline
$R(t)$        &  SDNR        & $\Delta$BIC   & Ecl. Depth (\%) \\
\hline
logramp\sb1   &  0.0073830   & \phantom{1}0.00    & 0.0677   \\
risingexp     &  0.0073832   & \phantom{1}6.10    & 0.0730   \\
quadramp\sb1  &  0.0073833   & \phantom{1}9.40    & 0.0777   \\
loglinear     &  0.0073830   &           10.84    & 0.0685   \\
linramp       &  0.0073846   &           12.19    & 0.0564   \\
\hline
\end{tabular}
\end{table}

An initial MCMC run showed a significant linear correlation between
the system flux and the $r\sb{1}$ parameter of the logramp, which
prevented the MCMC chain from converging.  We solved this problem by
transforming the correlated parameters into an orthogonal set of
parameters, rerunning the MCMC chain, and inverting the transformation
on the resulting parameter values
\citep{StevensonEtal2012apjHD149026b}.  Figures \ref{fig:lightcurves}
and \ref{fig:rms} show the \waspt3 light curves with the best-fitting
model and RMS of the residuals \vs bin size.

\subsubsection{wa008bs21 \& wa008bs41 Analysis} 

We simultaneously observed \waspt1 and \waspf1 in full-array mode.
Prior to the eclipse observation, we exposed the detector (a
``preflash'' observation, \citealp{KnutsonEtal2009apjHD149026bphase})
for 25 minutes to a bright HII region, with coordinates
$\alpha~=~20$\sp{h} 21\sp{m} 39\sp{s}.28 and
$\delta~=~+37$\sp{\degree}
 31{\arcmin}
 03.6{\arcsec}, to minimize the
ramp systematic variation.  The secondary-eclipse observation started
only 26 minutes before the first contact. The telescope pointing
stabilized quickly, so fortunately we needed to remove only the
initial four minutes of observation.  Every 12 seconds, the detector
recorded two consecutive images (two-second exposures) at 4.5
{\microns} and one image at 8.0 {\microns}. (Table
\ref{table:observations}).

\begin{figure}[ht]
\centering
\includegraphics[width=0.8\linewidth, clip]{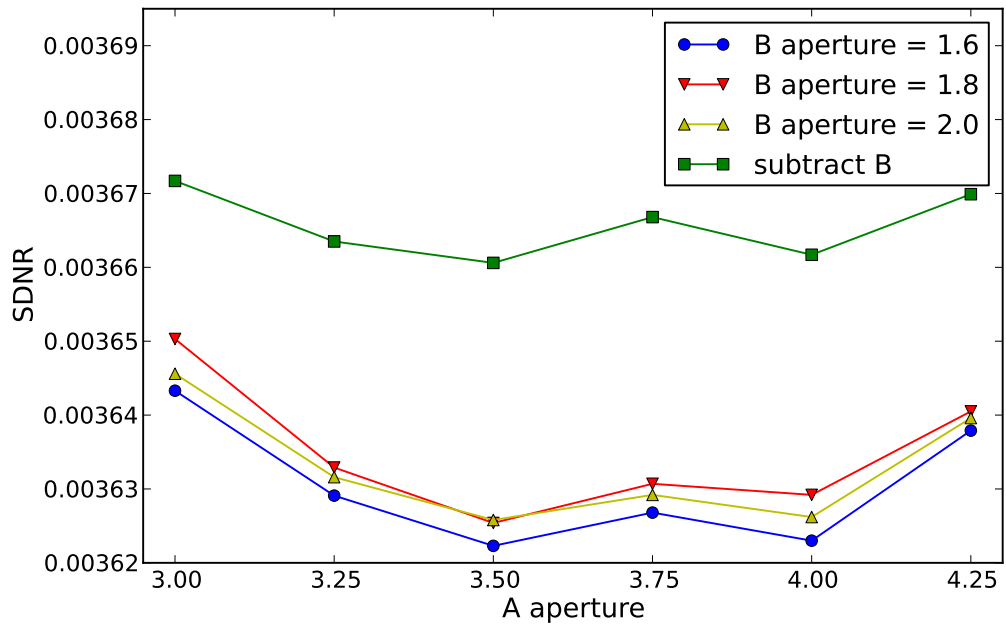}
\caption
  [\waspt1 standard deviation of the normalized residuals {\vs} aperture]
  {\waspt1 standard deviation of the normalized residuals
  \emph{vs.}\ aperture (in pixels). The SDNR curves use the best ramp
  model in Table \ref{table:wa008bs21ramps}. The legend indicates the
  photometry method.  SDNR increases at 3.75 pixels (coincident with
  the stars' separation).  The eclipse parameters are consistent over
  the 3.0--4.25 aperture range.  The optimum dataset uses 1.6 pixel
  B-Mask photometry with a 3.5 pixel A aperture.}
\label{fig:sdnr21}
\end{figure}

The SDNR analysis of \waspt1 showed that a 3.5 pixel A aperture with
1.6 pixel B-Mask photometry minimizes the dispersion (Figure
\ref{fig:sdnr21}).  The ramp models indicated a negligible ramp
variation.  Accordingly, a fit without a ramp model yielded the lowest
BIC.  Table \ref{table:wa008bs21ramps} shows the four best-fitting
models for the best \waspt1 data set.  The no-ramp model is 15 times
more probable than the quadramp\sb2 model.

\begin{table}[ht]
\centering
\caption{\waspt1 Ramp Model Fits}
\label{table:wa008bs21ramps} 
\begin{tabular}{lccc}
\hline
\hline
$R(t)$         & SDNR      & $\Delta$BIC  & Ecl. Depth (\%) \\
\hline
no ramp        & 0.0036223 & \n 0.00      & 0.0718   \\
quadramp\sb{2} & 0.0036195 & \n 5.76      & 0.1189   \\
linramp        & 0.0036223 & \n 7.56      & 0.0714   \\
quadramp\sb{1} & 0.0036197 &   13.14      & 0.1170   \\
\hline
\end{tabular}
\end{table}

\begin{figure}[htb]
\centering
\includegraphics[width=0.8\linewidth, clip]{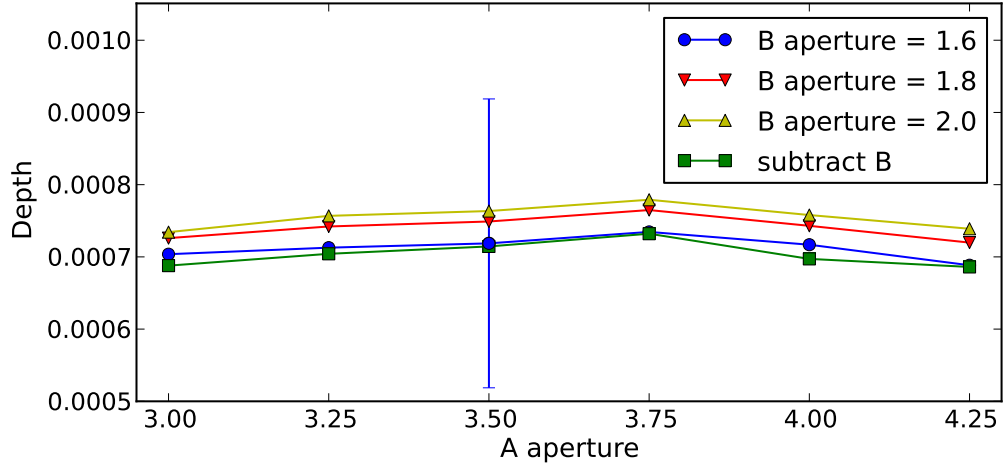}
\caption
  [Eclipse depth {\vs} A aperture for \waspt1]
  {Eclipse depth \emph{vs.} A aperture for \waspt1. Each color
    represent a different photometry method as in Figure
    \ref{fig:sdnr21}.  The blue error bar corresponds to the
    1-$\sigma$ uncertainty of the best model.  The eclipse duration
    and mid-point phase follow a similar trend.}
\label{fig:depth21}
\end{figure}

Because of the shorter out-of-eclipse observation, the system flux is
less-constrained for \waspt1\ than for the \waspo\ or \waspt3 events.
Combined with a correlation between the eclipse depth and system flux
(revealed by MCMC), the lower precision of the system flux translates
into a larger eclipse depth uncertainty.  Nevertheless, the \waspt1
fit parameters were consistent among the different apertures (Figure
\ref{fig:depth21}). The optimum parameters of the 
BLISS map are $mnp=5$ and a bin size of 0.025 pixels.

\begin{figure}[tb]
\centering
\includegraphics[width=0.8\linewidth, clip]{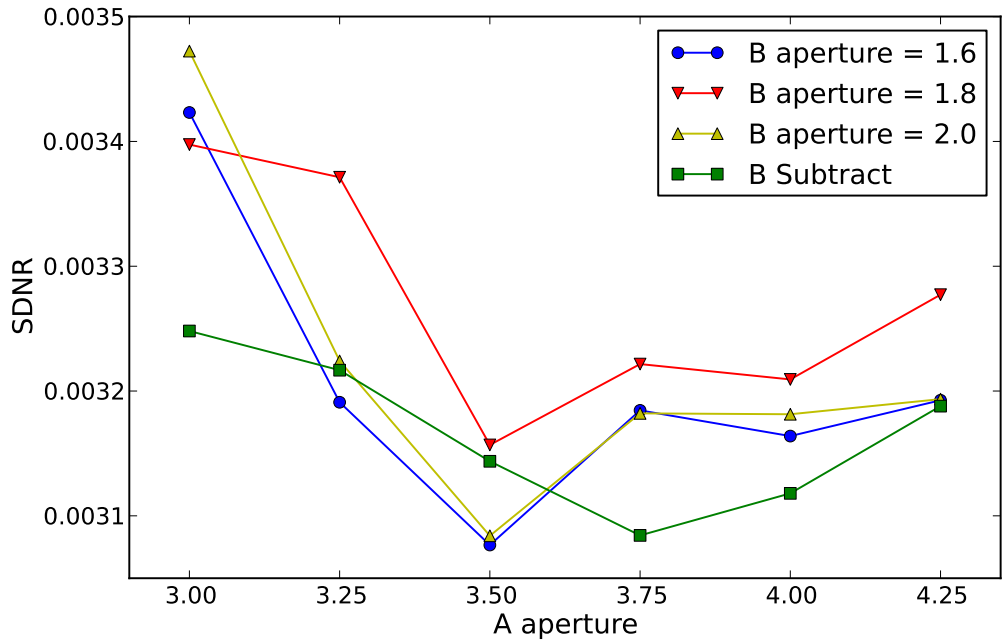}
\caption
  [Joint \waspt1 and \waspf1 standard deviation of the normalized
   residuals {\vs} aperture radius]
  {Joint \waspt1+\waspf1 standard deviation of the normalized
  residuals \emph{vs.}\ aperture radius (in pixels) of \waspf1, for
  different photometry methods.  All 24 light curves use the best ramp
  model from Table \ref{table:wa008bs41ramps}.  Light curves using 1.6
  and 2.0 pixel B-Mask photometry at 3.5 pixel A aperture produce
  consistent eclipse parameters and outperform the best B-subtract
  method (with a 3.75 pixel A aperture). The best B-subtract also
  yields a more scattered raw light curve. Hence, the optimum dataset
  uses 1.6 pixel B-mask photometry with a 3.5 pixel A aperture.}
\label{fig:sdnr41}
\end{figure}

The 8.0 \microns\ detector did not present an intrapixel pattern like
the 3.6 or 4.5 \microns\ detectors.  However, some of the raw light
curves for different apertures and photometry methods showed large
scatter and presented strong oscillations, producing implausible fit
parameters.  A pixelation effect \citep{Anderson2011MNRASWasp17b,
  StevensonEtal2012apjHD149026b} might be responsible.  As a
consequence, we were unable to fit the eclipse parameters
unambiguously for this data set alone.  Normally we study the events
individually to select the best aperture and photometry method, but in
this case we used a joint fit with the best \waspt1 dataset and model
to help constrain the 8.0 {\micron} eclipse curve, sharing the eclipse
duration and mid-point parameters.  The 3.5 pixel A aperture with 1.6
pixel B-mask photometry for \waspf1 minimized the joint SDNR (Figure
\ref{fig:sdnr41}).

Table \ref{table:wa008bs41ramps} compares the four best-fitting ramp
models for the best \waspf1 light curve. A linear ramp minimizes BIC,
and is 20 times more probable than the next-best model.  Figures
\ref{fig:lightcurves} and \ref{fig:rms} show the \waspt1 and \waspf1
light curves with their best-fitting models and RMS of the residuals
\vs bin size, respectively.

\begin{table}[ht]
\centering
\caption{\waspf1 Ramp Model Fits}
\label{table:wa008bs41ramps} 
\begin{tabular}{lccc}
\hline
\hline
$R(t)$         & SDNR 4\&2 & $\Delta$BIC & Ecl. Depth (\%)\\
\hline
linramp        & 0.0030766 &  0.00       & 0.0931   \\
quadramp\sb{1} & 0.0030744 &  5.94       & 0.1308   \\
risingexp      & 0.0030754 &  6.33       & 0.1150   \\
logramp\sb{1}  & 0.0030758 &  6.50       & 0.1078   \\
\hline
\end{tabular}
\end{table}

\subsubsection{wa008bs22 \& wa008bs42 Analysis} 

The observing setup of these events was identical to \waspt1 and
\waspf1, including the preflash observation.  The pointing of this
observation drifted noticeably more than in the other observations,
moving more than $0.4$ pixels during the initial 30 minutes and
stabilizing only during the eclipse.  As a consequence, the
illumination level of the individual pixels changed during the
beginning of the eclipse. The ramp variation, which depends on the
illumination \citep{Knutson2008HD209}, was disrupted.

The \waspt2 event, having a negligible ramp variation, was little
affected by the telescope pointing shift.  The SDNR calculation for
\waspt2 indicated the 1.8 pixel B-Mask photometry with 3.75 pixel A
aperture as the best dataset.  A light-curve model without a ramp
(Table \ref{table:wa008bs22ramps}) is 639 times more probable than the
quadramp\sb2 model. The optimal BLISS map has $mnp=4$ and a bin size
of 0.02 pixels.

\begin{table}[ht]
\centering
\caption{\waspt2 Ramp Model Fits}
\label{table:wa008bs22ramps}
\begin{tabular}{lccc}
\hline
\hline
$R(t)$         & SDNR      & $\Delta$BIC  & Ecl. Depth (\%) \\
\hline
no ramp        & 0.0025274 & n 0.00       & 0.0814  \\
quadramp\sb{2} & 0.0025253 &  12.94       & 0.1224  \\
logramp\sb{1}  & 0.0025267 &  14.22       & 0.0921  \\
risingexp      & 0.0025274 &  15.04       & 0.0814  \\
\hline
\end{tabular}
\end{table}

In contrast, we discarded the initial \waspf2 light curve past the
eclipse ingress due to the disrupted ramp variation.  The eclipse
model parameters are thus less constrained.  By this point, we already
had single-channel fits for the rest of the data, so we tuned the
\waspf2 analysis in a joint fit with all the other events, sharing the
eclipse duration and mid-time.  SDNR indicates the B-subtract method
with 4.00 pixel A aperture as the best dataset.  Table
\ref{table:wa008bs42ramps} presents the four best-fitting models. The
eclipse depth is consistent with the \waspf1 depth.

\begin{table}[ht]
\centering
\caption{\waspf2 Ramp Model Fits}
\label{table:wa008bs42ramps} 
\begin{tabular}{lccc}
\hline
\hline
$R(t)$         & SDNR       & $\Delta$BIC & Ecl. Depth (\%)\\
\hline
linramp        & 0.0032320  & 0.00        & 0.0932   \\
quadramp\sb{1} & 0.0032312  & 6.19        & 0.0892   \\
logramp\sb{1}  & 0.0032321  & 6.67        & 0.0938   \\
risingexp      & 0.0032330  & 7.08        & 0.0961   \\
\hline
\end{tabular}
\end{table}

\subsubsection{Final Joint-fit Analysis} 
\label{secjoint}

From the three individual fits to the 4.5 \microns\ observations we
found eclipse depths of $0.072\% \pm 0.021\%$, $0.086\% \pm 0.022\%$,
and $0.068\% \pm 0.007\%$ for \waspt1, \waspt2, and \waspt3,
respectively.  The weighted mean of the depths is $0.0692\% \pm
0.0065\%$. With a dispersion of 0.0062\% around the mean, this is not
larger than the individual uncertainties, thus we found no evidence
for temporal variability. This dispersion corresponds to 10\% of the
mean eclipse depth.  The consistency permitted a joint analysis of all
observations. We used the best light curves and models found in the
individual fits, where all events shared the eclipse duration, the
three 4.5-\micron\ events shared their eclipse depth, the two
8.0-\micron\ events shared their eclipse depth, the simultaneous
\waspt1 and \waspf1 events shared their eclipse mid-point phases, and
the \waspt2 and the \waspf2 events shared their eclipse mid-point
phases. Table \ref{table:fits} shows the light-curve modeling setup
and results.  We used these joint-fit results for the orbital and
atmospheric analysis. An electronic supplement contains the best light
curves, including centering, photometry, and the joint fit.

\begin{table}[htb]
\scriptsize
\centering
\caption{Best-fit Eclipse Light-curve Parameters}
\begin{tabular}{l@{\hspace{0.32cm}}
                c@{\hspace{0.32cm}}
                c@{\hspace{0.32cm}}
                c@{\hspace{0.32cm}}
                c@{\hspace{0.32cm}}
                c@{\hspace{0.32cm}}
                c}
\label{table:fits} \\
\hline
\hline
Parameter                                       & wa008bs11       & wa008bs21     & wa008bs22     & wa008bs23     & wa008bs41     & wa008bs42 \\
\hline
Array Position ($\bar x$, pix)                  & 14.74           & \ph20.76      & \ph20.39      & 14.65         & \ph19.11      & \ph18.72     \\
Array Position ($\bar y$, pix)                  & 15.07           & 233.30        & 233.30        & 15.12         & 230.27        & 230.20       \\ 
Position Consistency\sp{a} ($\delta\sb x$, pix) & 0.0072          & 0.0223        & 0.0220        & 0.0097        & 0.0273        & 0.0254       \\ 
Position Consistency\sp{a} ($\delta\sb y$, pix) & 0.0118          & 0.0228        & 0.0236        & 0.0101        & 0.0272        & 0.0274       \\ 
A Aperture Size (pix)                           & 2.25            & 3.5           & 3.75          & 2.25          & 3.5           & 4.0          \\ 
WASP-8B photometric Correction                  & subtract        & 1.6 mask      & 1.8 mask      & subtract      & 1.6 mask      & subtract     \\ 
System Flux \math{F\sb{s}} (\micro Jy)          & 144555.0(21.0)  & 91369.9(8.5)  & 90850.3(8.5)  & 87473.0(21.0) & 32892.5(6.6)  & 34949.8(8.9) \\ 
Eclipse Depth (\%)                              & 0.113(18)       & 0.0692(68)    & 0.0692(68)    & 0.0692(68)    & 0.093(23)     & 0.093(23)    \\ 
Brightness Temperature (K)                      & 1552(85)        & 1131(35)      & 1131(35)      & 1131(35)      & 938(99)       & 938(99)      \\ 
Eclipse Mid-point (orbits)                      & 0.51428(34)     & 0.51446(37)   & 0.51468(41)   & 0.51536(28)   & 0.51446(37)   & 0.51468(41)  \\ 
Eclipse Mid-point (MJD\sb{UTC})\sp{b}           & 5401.4981(28)   & 4822.2301(31) & 4814.0732(33) & 5409.6656(23) & 4822.2301(31) & 4814.0732(33)\\ 
Eclipse Mid-point (MJD\sb{TDB})\sp{b}           & 5401.4989(28)   & 4822.2309(31) & 4814.0739(33) & 5409.6663(23) & 4822.2309(31) & 4814.0739(33)\\ 
Eclipse Duration (\math{t\sb{\rm 4-1}}, hrs)    & 2.600(78)       & 2.600(78)     & 2.600(78)     & 2.600(78)     & 2.600(78)     & 2.600(78)    \\ 
Ingress/Egress Time (\math{t\sb{\rm 2-1}}, hrs) & 0.314           & 0.314         & 0.314         & 0.314         & 0.314         & 0.314        \\ 
Ramp Equation (\math{R(t)})                     & quadramp$_1$    & None          & None          & logramp$_1$   & linramp       & linramp      \\ 
Ramp, Linear Term (\math{r\sb{1}})              & 0.0707(70)      & $\cdots$      & $\cdots$      & 0.000504(45)  & 0.205(22)     & 0.246(37)    \\ 
Ramp, Quadratic Term (\math{r\sb{2}})           & $-$3.17(75)     & $\cdots$      & $\cdots$      & $\cdots$      & $\cdots$      & $\cdots$     \\ 
Ramp, Phase Offset (\math{t\sb{0}})             & $\cdots$        & $\cdots$      & $\cdots$      & 0.4917        & $\cdots$      & $\cdots$     \\ 
BLISS Map (\math{M(x,y)})                       &  Yes            & Yes           & Yes           & Yes           &  No           &  No          \\ 
Minimum Num. of Points Per Bin                &   5             &  5            &  4            &  4            &  $\cdots$     &  $\cdots$    \\ 
Total Frames                                    & 64320           & 2024          & 2024          & 64320         & 1012          & 1012         \\ 
Frames Used\sp{c}                               & 62203           & 1936          & 1879          & 64072         & 966           & 725          \\ 
Rejected Frames (\%)                            & 3.29            & 4.35          & 7.16          & 0.39          & 4.54          & 28.36        \\ 
Free Parameters\sp{d}                           & 6               & 4             & 4             & 5             & 5             & 5            \\ 
BIC Value                                       & 80444.5         & 80444.5       & 80444.5       & 80444.5       & 80444.5       & 80444.5      \\  
SDNR                                            & 0.0053772       & 0.0036250     & 0.0035698     & 0.0073926     & 0.0030768     & 0.0032320    \\ 
Uncertainty Scaling Factor                      & 0.3075          & 1.0280        &  1.0077       &  1.0902       & 1.1187        & 1.1382       \\ 
$\beta$ correction                              & 2.4             & $\cdots$      & $\cdots$      & $\cdots$      & $\cdots$      & $\cdots$     \\ 
Photon-limited S/N (\%)                         & 37.00           & 94.71         & 96.59         & 89.66         & 76.94         & 71.00        \\ 

\hline
\multicolumn{7}{l}{\scriptsize {\bf Notes.} The values quoted in parenthes are the 1$\sigma$ uncertainties.}  \\
\multicolumn{7}{l}{\scriptsize \sp{a} rms frame-to-frame position difference.} \\
\multicolumn{7}{l}{\scriptsize \sp{b} MJD = BJD - 2,450,000.} \\
\multicolumn{7}{l}{\scriptsize \sp{c} Frames excluded during instrument settling, for
       insufficient points at a BLISS knot, and for bad
       pixels in the photometry aperture.} \\
\multicolumn{7}{l}{\scriptsize \sp{d} In the individual fits.  Joint fit had 19 free parameters.}
\end{tabular}
\end{table}

\section{Orbital Dynamics}
\label{sec:orbit}

WASP-8b's high eccentricity ($e = 0.31$) implies that its separation
from WASP-8A at periapsis (0.055 AU) is about half that at apoapsis.
Given the argument of periapsis ($\omega=-85.86\degree$), the
secondary eclipse nearly coincides with the periapsis.  The planet,
therefore, receives over twice as much flux at eclipse as it would if
the orbit were circular, explaining in part our high brightness
temperature (see Table \ref{table:fits}).

Secondary-eclipse times can refine estimates of $e \cos \omega$ from
radial-velocity (RV) data.  The four eclipse events occurred at an
average eclipse phase of $0.514695 \pm 0.00018$.  After subtracting a
coarse light-time correction of $2a/c = 80$ s from this average phase,
we calculated $e \cos \omega = 0.02290 \pm 0.00028$ \citep[see Eq.\ 3
of][]{CharbonneauEtal2005apjTrES1}.  This is consistent with
\citet{Queloz2010Wasp8}, and photometrically confirms the nonzero
eccentricity of the planet's orbit (we fit $e \cos \omega$ below
without relying on the low $e$ approximation).

The eclipse timings were combined with 130 available RV data points
and with transit data from \citet{Queloz2010Wasp8} using the method
described by \citet{Campo2011} and \citet{NymeyerEtal2011}.
Forty-eight in-transit RV points were removed due to the
Rossiter-McLaughlin effect.

Our fit presented a moderate improvement to the orbital parameters of
WASP-8b (Table \ref{tab:orbit}), except for the period.  While
\citet{Queloz2010Wasp8} used several transits to measure the period,
we used their published mid-point epoch (a single date); hence, our
period is constrained mostly by our eclipses and the RV data, and thus
have a larger uncertainty.  By themselves, the secondary eclipses have
a period of $8.158774 \pm 0.00040$ days and a midpoint epoch of BJD
$2455409.6629 \pm 0.0017$ (TDB), not significantly ([$5.9 \pm 4.3]
\tttt{-5}$ days) longer than the period found by
\citet{Queloz2010Wasp8}.  The transit and eclipse periods place a
$9.8\tttt{-5}$ \degree\,day$\sp{-1}$ ($3\sigma$) upper limit on
possible apsidal precession, nearly three orders of magnitude larger
than the theoretical expectation for tidal effects
\citep{RagozzineWolf2009apjPlanetInteriors}.

\begin{table}[ht]
\centering
\caption{Eccentric Orbital Model}
\label{tab:orbit}
\begin{tabular}{lr@{\,{\pm}\,}lr@{\,{\pm}\,}l}
\hline
\hline
Parameter                                 & \mctc{This Work}           & \mctc{\citet{Queloz2010Wasp8}} \\
\hline
\math{e \sin \omega}                      & \math{-0.3078}  & 0.0020   & \math{-0.3092}   & 0.0029  \\
\math{e \cos \omega}                      & \math{ 0.02219} & 0.00046  & \math{ 0.023}    & 0.001   \\
\math{e}                                  & 0.309           & 0.002    & 0.310            & 0.0029  \\
\math{\omega} (\degree)                   & \math{-85.00}   & 0.08     & \math{-85.73}    & 0.18    \\
\math{P} (days)                           & 8.158719        & 0.000034 & 8.158715         & 0.000016\\
\math{T\sb{0}} (MJD\sb{TDB})              & 4679.33486      & 0.00057  & 4679.33509       & 0.00050 \\
\math{K} (ms\sp{-1})                      & 221.9           & 0.6      & 222.23           & 0.8     \\
\math{\gamma_C} (ms\sp{-1})               & \math{-1\,565.9} & 0.6     & \math{-1\,565.8} & 0.21    \\
\math{\gamma_H} (ms\sp{-1})               & \math{-1\,547.4} & 0.4     & \math{-1\,548.1} & 0.6     \\
\math{\dot{\gamma}} (ms\sp{-1}yr\sp{-1})  & \math{58.1}     & 1.2      & \math{58.1}      & 1.3     \\
Reduced \math{\chi^2}                     & \mctc{4.1}                 & \mctc{0.86}                \\
\hline
\end{tabular}
\end{table}

\section{Atmospheric Analysis}
\label{sec:atmosphere}

We use our IRAC observations of thermal emission from WASP-8b to
constrain the thermal structure and composition of the day-side
atmosphere of the planet.  The \Spitzer\ bandpasses at 3.6, 4.5, and
8.0 \microns\ contain strong spectral features due to several carbon
and oxygen-based molecules that are expected in hot-Jupiter
atmospheres. Methane (CH\sb4) has strong spectral features in the 3.6
and 8.0 \microns\ bands, carbon monoxide (CO) and carbon dioxide
(CO\sb2) have features at 4.5 \microns, while water vapor (H\sb2O) has
features in all three bands \citep{MadhusudhanSeager2010}. The
spectral features of the various molecules appear as absorption
troughs or emission peaks in the emergent spectrum depending on
whether the temperature decreases or increases with altitude,
respectively. Consequently, strong degeneracies exist between the
temperature structure and molecular composition derived from a
spectral dataset \citep[e.g.,][]{MadhusudhanSeager2010}. Nevertheless,
photometric observations made with \Spitzer\ have been successfully
used to constrain chemical compositions and temperature structures in
many exoplanetary atmospheres \citep[e.g.,][]{Barman2005,
  Burrows2007HD209, Knutson2008HD209, MadhusudanSeager2009apj,
  StevensonEtal2010Natur, Madhusudhan2011Nat}.

We model the dayside emergent spectrum of WASP-8b using the
atmospheric modeling and retrieval method of
\citet{MadhusudanSeager2009apj, MadhusudhanSeager2010}. The model
computes line-by-line radiative transfer in a plane-parallel
atmosphere assuming hydrostatic equilibrium, local thermodynamic
equilibrium, and global energy balance. We assume a Kurucz model for
the stellar spectrum \citep{CastelliKurucz2004} given the stellar
parameters.  The pressure-temperature (\PT) profile and molecular
mixing ratios are free parameters in the model, which can be
constrained from the data. The \PT\ profile comprises of six free
parameters and the mixing ratio of each molecular species constitutes
an additional free parameter.  Following
\citet{MadhusudanSeager2009apj}, we parametrize the mixing ratio of
each species as deviations from thermochemical equilibrium assuming
solar elemental abundances \citep{Burrows1999ChemEquilibrium}. We
include the dominant sources of opacity expected in hot Jupiter
atmospheres, namely molecular absorption due to H\sb2O, CO, CH\sb4 and
CO\sb2 \citep[Freedman, personal communication
2009]{Freedman2008Opacities}, and H$_2$-H$_2$ collision induced
absorption \citep{Borysow2002H2H2}. We explore the model parameter
space in a Bayesian way using an MCMC sampler
\citep{MadhusudhanSeager2010, MadhusudhanSeager2011}.  Given the
limited number of observations ($N_{\rm obs}$=3), our goal is not to
find a unique model fit to the data; instead, we intend to constrain
the region of atmospheric parameter space that is allowed or ruled out
by the data.

\begin{figure}[tb]
\centering
\strut\hfill
\includegraphics[width=0.6\textwidth,  clip]{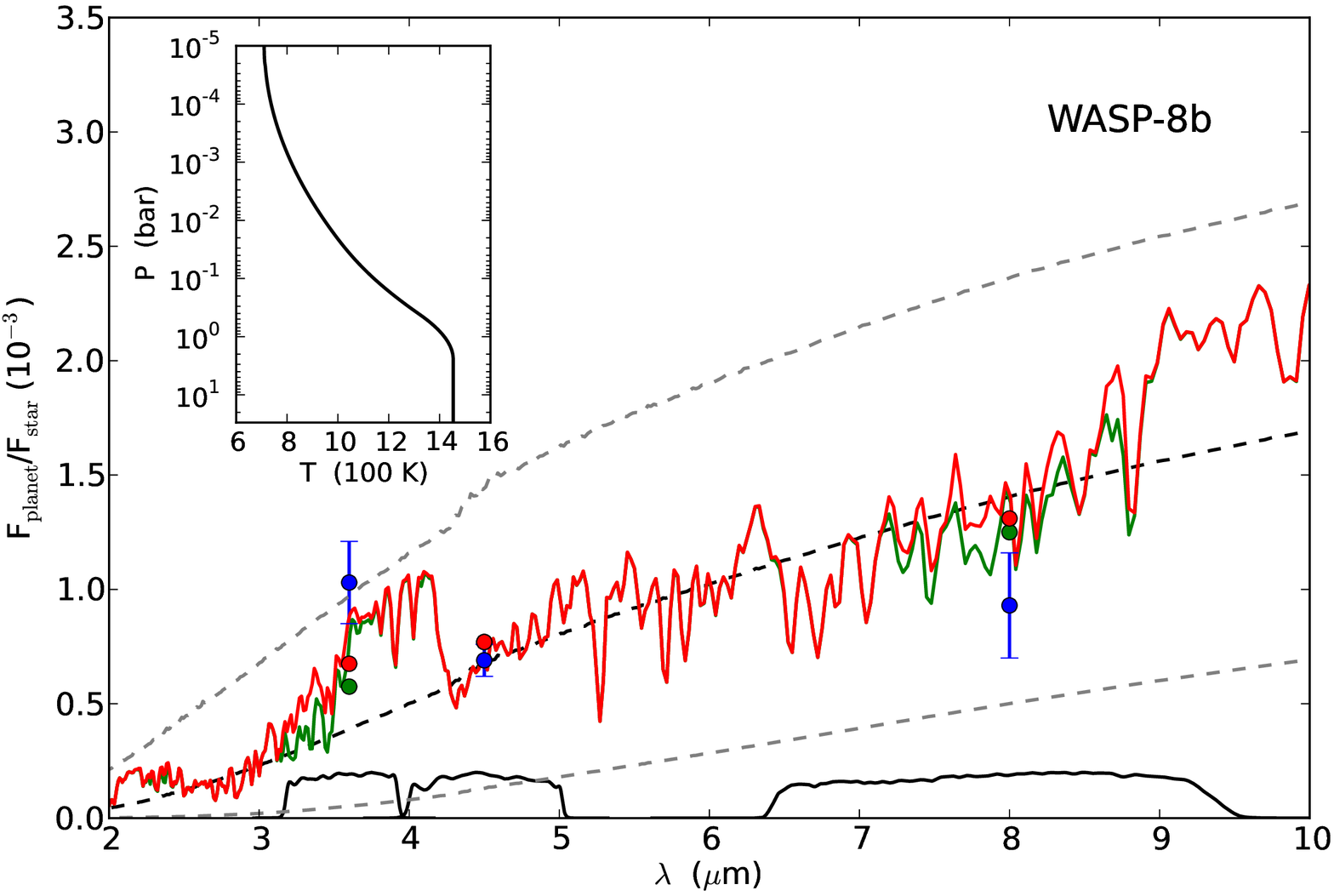}\hfill
\includegraphics[width=0.323\textwidth, clip]{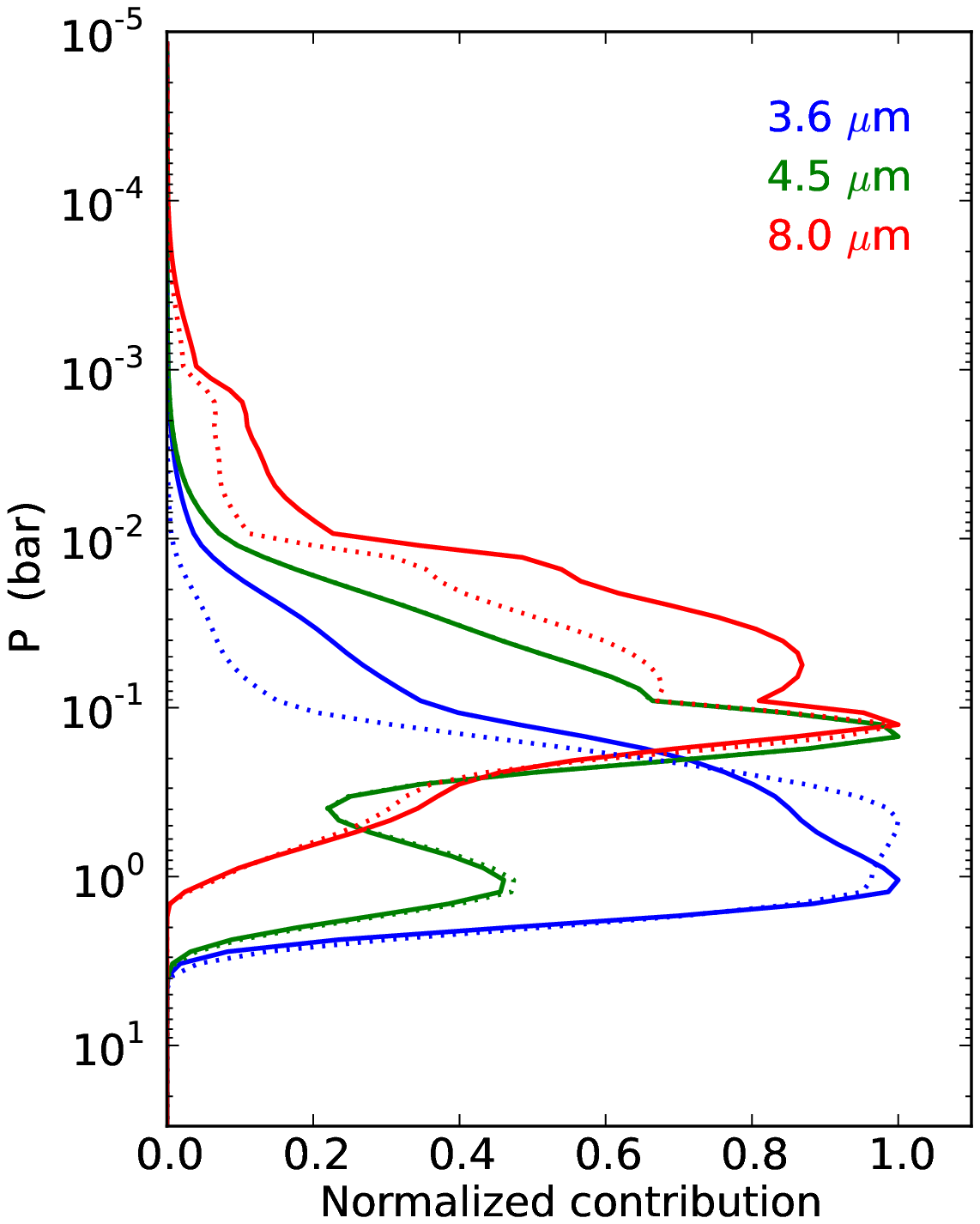}
\hfill\strut
\caption
  [Atmospheric models of WASP-8b]
  {Atmospheric models of WASP-8b.
   {\bf Left:} Atmospheric spectral emission of the dayside of
  WASP-8b.  The blue circles with error bars are the measured eclipse
  depths, or equivalently, the planet-star flux ratios.  The red and
  green curves show two model spectra with the same temperature
  profile (shown in the inset) but with different compositions. The
  green curve shows a model assuming chemical equilibrium with solar
  elemental abundances. The red curve shows a model with $10\sp{3}$ times
  lower methane abundance compared to the green model, but the
  abundances of the remaining molecules are identical to those in the
  green model.  The red and green filled circles are the corresponding
  model fluxes integrated over the \Spitzer\ bands (bottom solid
  lines).  The black dashed lines represent planetary blackbody
  spectra with $T = 710, 1100$, and 1450 K.  {\bf Right:} Normalized
  contribution functions of the solar-composition model (solid lines)
  and the low-CH\sb4-abundance model (dotted lines) in each \Spitzer\
  band (see legend).  The effective pressures of the contribution
  functions are 0.63, 0.35, and 0.12 bar at 3.6, 4.5, and 8.0
  \microns, respectively.}
\label{fig:atm}
\end{figure}

Our observations rule out a thermal inversion in the day-side
atmosphere of WASP-8b. This is evident from the planet-star flux
contrasts in the three IRAC bands at 3.6, 4.5, and 8.0 \microns. In
the presence of a thermal inversion, the planet-star flux contrasts in
the 4.5 and 8.0 \micron\ bands are both expected to be greater than
the flux contrast in the 3.6 \micron\ band \citep{Burrows2008,
  Fortney2008, MadhusudhanSeager2010}, due to spectral features of the
dominant molecules appearing as emission peaks as opposed to
absorption troughs. However, the low 4.5 and 8.0 \micron\ flux
contrasts relative to the 3.6 \micron\ contrast requires significant
absorption due to H$_2$O and CO across the spectrum, and hence the
lack of a thermal inversion in the atmosphere. Figure \ref{fig:atm}
shows model spectra of WASP-8b with no thermal inversion in the
temperature profile. The observed 4.5 and 8.0 \micron\ flux contrasts
are explained to a good level of fit by a model without a thermal
inversion and with solar abundance composition, as shown by the green
curve in Fig.~\ref{fig:atm}. Our inference of the lack of a thermal
inversion in WASP-8b is independent of any assumption about chemical
composition or C/O ratio \citep[e.g.][]{MadhusudhanSeager2011}.  The
lack of a thermal inversion in WASP-8b is not surprising, since it is
amongst the cooler population of irradiated hot Jupiters, which are
not expected to host inversion-causing species such as TiO or VO in
their upper atmosphere \citep{Fortney2008, Spiegel2009TiO}.

Our models are unable to reproduce the high planet-star flux contrast
observed in the 3.6 \micron\ IRAC band, independent of the
composition. The major sources of absorption in the 3.6 \micron\ band
are H\sb2O and CH\sb4. In principle, decreasing the CH\sb4 and/or
H\sb2O abundances can lead to a higher 3.6 \micron\ contrast. However,
as shown by the red curve in Fig.~\ref{fig:atm}, such an increase also
simultaneously increases the contrast in the 8.0 \micron\ band,
thereby worsening the fit overall. Another hindrance to fitting the
observed 3.6 \micron\ contrast is that it requires a hotter \PT\
profile, with $T \gtrsim$ 1550 K in the lower atmosphere, predicts
much higher fluxes in the 4.5 and 8.0 \micron\ bands than observed. On
the other hand, a cooler \PT\ profile than shown in Fig.~\ref{fig:atm}
would provide a better fit in the 4.5 and 8.0 \micron\ bands, but
would further worsen the fit in the 3.6 \micron\ band.  Consequently,
we choose an intermediate \PT\ profile that provides a compromise fit
to all three data points.

Although the 1D models shown in Fig.~\ref{fig:atm} output less energy
than the instantaneous incident irradiation during the eclipse
(concurrent with periastron passage), they output $\sim$ 20\% higher
energy compared to the time-averaged incident irradiation received at
the substellar point.  Considering that a pseudo-synchronous rotation
should facilitate the redistribution of energy to the night side, the
high emission measured suggests that WASP-8b is quickly reradiating
the incident irradiation on its day-side hemisphere, i.e. nearly zero
day-night redistribution.  Such a scenario would lead to a large
day-night temperature contrast in the planet which can be confirmed by
thermal phase curves of the planet observed using warm Spitzer
\citep[e.g.,][]{KnutsonEtal2009apjHD149026bphase}. The high emergent
flux also implies a very low albedo, as with most hot-Jupiter planets
\citep{CowanAgol2011Albedos}.

\section{The Unexpected Brightness Temperature 
                of WASP-8b}
\label{sec:discussion}

As seen in the previous section, the 3.6-\micron\ brightness
temperature is anomalously higher than expected.  The
hemisphere-averaged equilibrium temperature for instantaneous
re-radiation (time-averaged around the orbit) is only 948 K; even the
instantaneous equilibrium temperature at periapsis, 1128 K, is far
lower than this observation.  Thus, we modeled the orbital thermal
variation due to the eccentricity to determine if such a high
temperature is possible from irradiation alone.

Following \citet{CowanAgol2011TVariation}, we solved the energy
balance equation in a one-layer latitude--longitude grid over the
planetary surface.  The change in temperature of a cell with time,
${{\mathrm d} T}/{{\mathrm d} t}$, is determined by the difference
between the absorbed flux from the star and the re-emitted blackbody
flux,
\begin{equation}
\frac{{\rm d} T}{{\rm d} t} = \frac{1}{c\sb h} \left[ (1-A) \sigma T\sp4\sb{\rm eff} \left(\frac{R\sb{*}}{r(t)} \right)\sp2\cos\psi(t)\ -\ \sigma T\sp4 \right],
\label{eq:ebalance}
\end{equation}
where $c\sb h$ is the heat capacity per unit area; $T\sb{\rm eff}$ and
$R\sb*$ are the star's effective temperature and radius, respectively;
$r(t)$ is the planet-star separation; $\cos \psi(t) =
\sin\lambda\,{\rm max}(\cos\Phi(t),0)$ is the cosine of the angle
between the vectors normal to the planet surface and the incident
radiation, with $\lambda$ the latitude of the cell and $\Phi(t)$ the
longitude from the sub-stellar meridian; $\sigma$ is the
Stefan-Boltzmann constant.

Tidal interactions drive the planet's rotational angular velocity
($\omega\sb{\rm rot}$) toward synchronization with the orbital angular
velocity ($\omega\sb{\rm orb}$).  Hence, if the spin synchronization
timescale \citep[e.g.,][]{SeagerHui2002, Goldreich1966Q} is shorter
than the system age, we expect $\omega\sb{\rm rot} =\omega\sb{\rm
  orb}$.  In the case of WASP-8b, the timescale for tidal
synchronization is on the order of 0.05 Gyr, much shorter than the age
of the star.  However, a planet in an eccentric orbit, where
$\omega\sb{\rm orb}$ changes in time, is actually expected to reach a
pseudo-synchronization state \citep[e.g.,][]{Langton2008, Hut1981}, in
which the planet does not exchange net angular momentum with its
orbit. The planet acquires then a constant rotational angular velocity
close to the orbital angular velocity at periastron ($\omega\sb{\rm
  orb,p}$). In the literature we found different predictions for this
equilibrium angular velocity, from $0.8\, \omega\sb{\rm orb,p}$
\citep{Hut1981} to $1.55\, \omega\sb{\rm orb,p}$ \citep{Ivanov2007}.

The tidal evolution drives the orbit of a planet toward zero obliquity
in a timescale similar to the spin synchronization \citep{Peale1999}.
We thus adopted zero obliquity for our simulations.  We also assumed
$A=0$, supported by the atmospheric analysis (Section
\ref{sec:atmosphere}).  Beyond these assumptions, the parameters of
interest that control Equation (\ref{eq:ebalance}) are the radiative
time $\tau_{\rm rad}=c\sb h/\sigma T\sp3\sb0$ (where $T\sb0$ is the
sub-stellar equilibrium temperature at periastron) and the rotational
angular velocity of the planet $\omega_{\rm rot}$ (which determines
the sub-stellar longitude of a cell through the equation ${\rm
  d}{\Phi}(t) /{\rm d}t = \omega\sb{\rm rot} - \omega\sb{\rm
  orb}(t)$).  With these definitions Equation (\ref{eq:ebalance}) can
be re-written as:
\begin{equation}
\frac{{\rm d} T}{{\rm d} t} = \frac{T\sb0}{\tau\sb{\rm rad}} \left[ \left(\frac{a(1-e)}{r(t)} \right)\sp2 {\rm max}(\cos\Phi(t),0)\  -\ \left(\frac{T}{T\sb0}\right)\sp4 \right].
\label{eq:ebalance2}
\end{equation}

We derived the temperature of each cell as a function of time to study
its thermal evolution.  Assuming that each cell emits as a blackbody,
we calculated the photometric phase curve of the planet by integrating
over the hemisphere observable from Earth, weighted by the viewing
geometry.  Our simulations were for planets nearly in
pseudo-synchronous rotation ($\omega\sb{\rm rot} =$ 0.8, 1.0, and
$1.5\ \omega\sb{\rm orb,p}$). We tested values of $\tau$\sb{rad}
between 1 and \ttt3 hours.

\begin{figure}[tb]
\centering
\includegraphics[width=0.8\linewidth, clip]{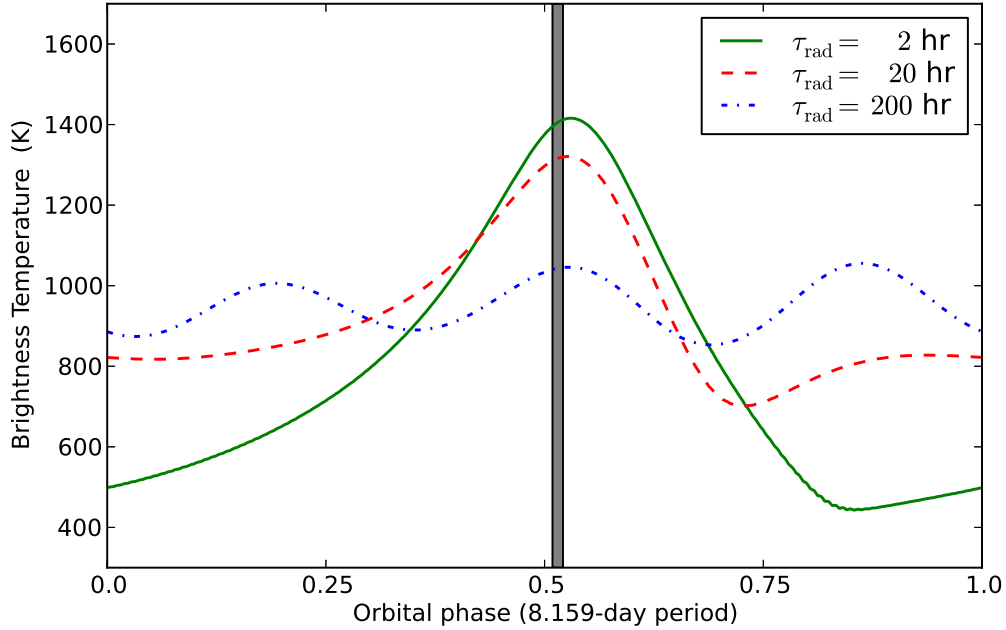}
\caption
  [Model brightness-temperature lightcurves of WASP-8b]
  {Model brightness-temperature lightcurves of WASP-8b as observed from
  Earth during one orbital period. Phase zero indicates the mid-transit
  time. The gray region indicates the secondary-eclipse interval, with
  periastron at phase 0.52. The models simulate super-rotating winds
  ($\omega\sb{\rm rot} = 1.5\ \omega\sb{\rm orb,p}$) for different
  radiative times (see legend). The curves with smaller $\tau$\sb{rad}
  show larger amplitudes.  For $\tau$\sb{rad} comparable to the orbit
  period, and since $\omega\sb{\rm rot} = 1.5\ \omega\sb{\rm orb,p}$,
  opposite sides will face the star during successive periastron 
  passages, leading to two bright spots and hence three periodic peaks
  per orbit.}
\label{fig:model}
\end{figure}

Figure \ref{fig:model} shows simulated brightness-temperature
lightcurves of WASP-8b after reaching a periodic stationary state
(after a few $\tau$\sb{rad}).  We noted that the higher irradiation at
periastron is not the only contribution to a higher temperature.  For
$\omega\sb{\rm rot} \geq \omega\sb{\rm orb,p}$, the sub-stellar
angular velocity (${\rm d}{\Phi} /{\rm d}t$) is minimum during
periastron, allowing the temperature to increase due to the longer
exposure to the irradiation.  For $\omega\sb{\rm rot} < \omega\sb{\rm
  orb,p}$, the sub-stellar angular velocity is negative for an instant
around periastron. Later, when the planet emerges from secondary
eclipse the over-heated region becomes observable from Earth. As a
result, the lightcurve shows a delayed maximum.

Our models show that for large radiative timescales, the temperatures
at secondary eclipse are lower than 1150 K, regardless of
$\omega\sb{\rm rot}$.  For radiative times shorter than $\sim \ttt2$
hr, the temperatures can be as high as 1400 K, similar to the
3.6-\microns\ measurement (Fig.\ \ref{fig:model}, top panel).
However, these models still cannot explain the observed
brightness-temperature discrepancy with wavelength.

The study of eccentric hot-Jupiter atmospheric circulation by
\citet{KatariaEtal2012CircEccentric} hints at a resolution to this
discrepancy.  Their Fig.\ 4 (top panel) shows that, as the planet
passes through periapsis, the time that the peak temperature is
reached varies as a function of pressure. This is typical of their
simulations (personal communication).  If this differential response
is significant in WASP-8b, it would introduce a discrepancy in the
observations since the \Spitzer\ bands sample different altitudes (see
Fig. \ref{fig:model} right panel).

Another possibility is to compare the radiative and advective
timescales at the altitudes sampled by each band.  Evaluating equation
1 of \citet{Fortney2008} using WASP-8b's {\PT} profile, indicates that
$\tau$\sb{rad} increases with depth between 0.1 and 1.0 bar, so there
should be less longitudinal temperature contrast at depth.  On the
other hand, models of \citet{KatariaEtal2012CircEccentric} show that
wind speeds decrease with depth, and thus $\tau$\sb{adv} also
increases with depth.  If the increase of $\tau$\sb{adv} with depth is
sharper than that of $\tau$\sb{rad}, then one would expect
less-homogenized temperatures at depth (but still above the
photosphere).  Hence, the rise in temperature (due to the increasing
incident irradiation) near periapsis could be more pronounced at 3.6
{\microns} than at longer wavelengths, given the weighting functions
of Figure \ref{fig:atm}.

\section{Conclusions}
\label{sec:conclusions}

\Spitzer\ observed secondary eclipses of WASP-8b in the 3.6, 4.5, and
8.0 \microns\ IRAC wavebands.  In our joint-fit model, we estimate
eclipse depths of $0.113\% \pm 0.018\%$, $0.069\% \pm 0.007\%$, and
$0.093\% \pm 0.023\%$ at 3.6, 4.5, and 8.0 \microns,
respectively. These depths correspond to brightness temperatures of
1552, 1131, and 938 K, respectively.  Although the 3.6-\microns\
eclipse depth is unexpectedly large, most of the ramp models had
consistent depths (within 1$\sigma$), while those with inconsistent
depths fit the data poorly.

Considering the \PT\ profile of WASP-8b, KCl, ZnS, Li\sb{$2$}, LiF, or
Na\sb{$2$}S clouds could form \citep[see Fig.\ 2a
of][]{LoddersFegley2006Chem}.  In analogy to brown dwarfs, partial
cloud coverage can cause photometric variability
\citep{Artigau2009BrownDwarfs}; however, our three 4.5 \micron\
observations, spanning 1.5 years, have consistent eclipse depths,
suggesting no temporal variation at secondary eclipse above a
hemispheric-mean level of $\sim 35$ K (1$\sigma$).  A moderate cloud
layer at altitudes higher than those probed by \Spitzer\ would produce
a featureless planetary spectrum at wavelengths shorter than 2
\microns\ \citep{Pont2008HD189Haze, Miller-Ricci2012Gj1214clouds} and
would block some of the stellar flux, decreasing the temperatures at
levels probed by Spitzer.  Yet, the observed temperatures, which
exceed the time-averaged equilibrium temperature, challenge this idea.

Given the high eccentricity, spin-orbit misalignment, and observed
radial-velocity drift in the of WASP-8 system, \citet{Queloz2010Wasp8}
suggested the existence of an additional, unseen body in the system.
Our orbital analysis is consistent with theirs.  It also improves the
orbital parameters and extends the baseline of sampled epochs.  This
constrains the long-term evolution of the orbit and aids the search
for a second planet, for example through the study of timing
variations \citep[e.g.,][]{AgolEtal2005TTV}.

The eclipse depths probe the day-side atmosphere of WASP-8b.  Our
results rule out the presence of a thermal inversion layer, as
expected, given the irradiation level from the host star.  A model
with solar-abundance composition explains the 4.5 and 8.0 \microns\
planet-star flux contrast; however, including the high 3.6 \microns\
flux contrast requires models that output nearly 20\% of the
orbit-averaged incident irradiation, independent of the atmospheric
composition.  If the orbit were circular (and thus the irradiation
steady-state), the high brightness temperatures would indicate a very
low energy redistribution to the night side of the planet.  For an
eccentric planet, it at least indicates a short $\tau$\sb{rad} (Figure
\ref{fig:model}).

By modeling the orbital thermal variations due to the eccentricity of
the orbit, we determined that it is possible for WASP-8b to achieve
temperatures as high as the 3.6 \microns\ brightness temperature.
However, the differing brightness temperatures in the other two bands
remain puzzling.  Neither the radiative-transfer model (Section
\ref{sec:atmosphere}) nor the phase-variation model (Section
\ref{sec:discussion}) embraces all the physics of the problem. The
radiative transfer code is a 1D, steady-state model representing
typical day-side conditions.  The phase-variation model describes
emission as a blackbody on a single-layer grid; it does not consider
absorption or emission features from the species in the atmosphere.
Clouds \citep[e.g.,][]{CushingEtal2008apjLTdwarfs}, atmospheric
dynamics \citep[e.g.,][]{Showman2009circulation}, and photochemistry
\citep[e.g.,][]{Moses2011Disequilibrium} are not directly considered
by these models.

What we can say for certain is that the assumptions of our simple
models have been violated, which is not surprising for this eccentric
planet.  While it may be possible to construct consistent, realistic
models, model uniqueness may be elusive until more and better data are
available.

Relatively few exoplanets with equilibrium temperatures below 1500~K
have been observed at secondary eclipse \citep{CowanAgol2011Albedos}.
The same is true for eccentric planets.  The characterization of
WASP-8b in this work thus addresses a particularly interesting, if
challenging, region of the exoplanet phase space.  Observation of
other planets with similar equilibrium temperatures or eccentricities
will help discover the physics that drive these unusual atmospheres.

\section{Acknowledgments}
We thank Ian Crossfield for his help with the Tiny Tim software.  We
thank contributors to SciPy, Matplotlib, and the Python Programming
Language; the free and open-source community; the NASA Astrophysics
Data System; and the JPL Solar System Dynamics group for software and
services.  PC is supported by the Fulbright Program for Foreign
Students.  NM acknowledges support from the Yale Center for Astronomy
and Astrophysics through the YCAA postdoctoral Fellowship.  This work
is based on observations made with the {\em Spitzer Space Telescope},
which is operated by the Jet Propulsion Laboratory, California
Institute of Technology, under a contract with NASA.  Support for this
work was provided by NASA through an award issued by JPL/Caltech.

\bibliographystyle{apj}
\bibliography{chap2-WASP8b}

\chapter{A SPITZER FIVE-BAND ANALYSIS OF THE JUPITER-SIZED PLANET TrES-1}
\label{chap:tres1}

{\singlespacing
\noindent{\bf
Patricio Cubillos\sp{1,2},
Joseph Harrington\sp{1,2},
Nikku Madhusudhan\sp{3},
Andrew S. D. Foster\sp{1},
Nate B. Lust\sp{1},
Ryan A. Hardy\sp{1}, and
M. Oliver Bowman\sp{1}
}

\vspace{1cm}

\noindent{\em
\sp{1} Planetary Sciences Group, Department of Physics,
       University of Central Florida, Orlando, FL 32816-2385 \\
\sp{2} Max-Plank-Institut f\"ur Astronomie, K\"onigstuhl 17,
  D-69117, Heidelberg, Germany \\
\sp{3} Department of Physics and Department of Astronomy, Yale
  University, New Haven, CT 06511, USA
}

\vspace{1cm}

\centerline{Received   6 May     2014.}
\centerline{Accepted  22 October 2014.}
\centerline{Published in {\em The Astrophisical Journal} 24 November 2014.}

\vspace{1cm}

\bibliographystyle{apj}
\centering{Publication reference: \\
  Cubillos, P., Harrington, J., Madhusudhan, N., Foster, A. S. D.,
  Lust, N. B., Hardy, R. A., \& Bowman, M. O. 2014, ApJ, 797, 42 \\
  http://arxiv.org/abs/1411.3093}

\vspace{1cm}
\centerline{\copyright AAS. Reproduced with permission}
}

\clearpage
\comment{
\usepackage{apjfonts}
\usepackage{ifthen}
\usepackage{natbib}
\usepackage{amssymb, amsmath}
\usepackage{appendix}
\usepackage{etoolbox}
\usepackage{longtable}
\usepackage{tablefootnote}
}





\setcitestyle{authoryear,round}

\section{Abstract}

  With an equilibrium temperature of 1200~K, TrES-1 is one of the
  coolest hot Jupiters observed by {\Spitzer}.  It was also the first
  planet discovered by any transit survey and one of the first
  exoplanets from which thermal emission was directly observed.  We
  analyzed all {\Spitzer} eclipse and transit data for TrES-1 and
  obtained its eclipse depths and brightness temperatures in the 3.6
  {\micron} (0.083\% {\pm} 0.024\%, 1270 {\pm} 110 K), 4.5 {\micron}
  (0.094\% {\pm} 0.024\%, 1126 {\pm} 90 K), 5.8 {\micron} (0.162\%
  {\pm} 0.042\%, 1205 {\pm} 130 K), 8.0 {\micron} (0.0213\% {\pm}
  0.042\%, 1190 {\pm} 130 K), and 16 {\micron} (0.33\% {\pm} 0.12\%,
  1270 {\pm} 310 K) bands.  The eclipse depths can be explained,
  within 1$\sigma$ errors, by a standard atmospheric model with solar
  abundance composition in chemical equilibrium, with or without a
  thermal inversion.  The combined analysis of the transit, eclipse,
  and radial-velocity ephemerides gives an eccentricity $e =
  0.033\sp{+0.015}\sb{-0.031}$, consistent with a circular orbit.
  Since TrES-1's eclipses have low signal-to-noise ratios, we
  implemented optimal photometry and differential-evolution
  Markov-chain Monte Carlo (MCMC) algorithms in our Photometry for
  Orbits, Eclipses, and Transits (POET) pipeline.  Benefits include
  higher photometric precision and \sim10{\by} faster MCMC
  convergence, with better exploration of the phase space and no
  manual parameter tuning.

\section{Introduction}
\label{sec:c3introduction}

Transiting exoplanets offer the valuable chance to measure the light
emitted from the planet directly.  In the infrared, the eclipse depth
of an occultation light curve (when the planet passes behind its host
star) constrains the thermal emission from the planet.  Furthermore,
multiple-band detections allow us to characterize the atmosphere of
the planet \citep[e.g.,][]{SeagerDeming2010AnnualRev}.  Since the
first detections of exoplanet occultations---TrES-1
\citep{CharbonneauEtal2005apjTrES1} and HD\,298458b
\citep{Deming2005Nat}---there have been several dozen occultations
observed.  However, to detect an occultation requires an exhaustive
data analysis, since the the planet-to-star flux ratios typically lie
below $\ttt{-3}$.  For example, for the {\SST}, these flux ratios are
lower than the instrument's photometric stability criteria
\citep{FazioEtal2004apjsIRAC}.  In this paper we analyze {\Spitzer}
follow-up observations of TrES-1, highlighting improvement in
light-curve data analysis over the past decade.

TrES-1 was the first exoplanet discovered by a wide-field transit
survey \citep{AlonsoEtal2004apjTrES1disc}.
Its host is a typical K0 thin-disk star
\citep{SantosEtal2006TrES1chemAbundances} with solar metallicity
\citep{LaughlinEtal2005TrES1followup, SantosEtal2006TrES1spectroscopy,
  SozzettiEtal2006TrES1starChemComp}, effective temperature $T\sb{\rm
  eff} = 5230 \pm 50$ K, mass $M\sb{*} = 0.878 \pm 0.040$ solar masses
({\msun}), and radius $R\sb{*} = 0.807 \pm 0.017$ solar radii
({\rsun}, \citealp{TorresEtal2008reanalyses}).
\citet{SteffenAgol2005TrES1transitTimes} dismissed additional
companions (with $M > M\sb{\oplus}$).
\citet{CharbonneauEtal2005apjTrES1} detected the secondary eclipse in
the 4.5 and 8.0 {\micron} {\Spitzer} bands.
\citet{KnutsonEtal2007TrES1GroundThEmission} attempted ground-based
eclipse observations in the L band (2.9 to 4.3 {\microns}), but
did not detect the eclipse.

The TrES-1 system has been repeatedly observed during transit from
ground-based telescopes \citep{NaritaEtal2007TrES1RLmeasurements,
  RaetzEtal2009TrES1Transits, Vanko2009TrES1Transits,
  RabusEtal2009TrEStransits, HrudkovaEtal2009TTVsearch,
  SadaEtal2012Transits} and from the {\em Hubble Space Telescope}
\citep{CharbonneauEtal2007TransitsReview}.  The analyses of the
cumulative data \citep{ButlerEtal2006Catalog,
  Southworth2008HomogeneousStudyI, Southworth2009HomogeneousStudyII,
  TorresEtal2008reanalyses} agree (within error bars) that the planet
has a mass of $M\sb{p} = 0.752 \pm 0.047$ Jupiter masses ({\mjup}), a
radius $R\sb{p} = 1.067 \pm 0.022$ Jupiter radii ({\rjup}), and a
circular, 3.03 day orbit, whereas \citet{WinnEtal2007apjTres1}
provided accurate details of the transit light-curve shape.  Recently,
an adaptive-optics imaging survey
\citep{AdamsEtal2013AOImagingCompanions} revealed that TrES-1 has a
faint background stellar companion ($\Delta$mag = 7.68 in the Ks band,
or 0.08\% of the host's flux) separated by 2.31{\arcsec} (1.9 and 1.3
{\Spitzer} pixels at 3.6--8~{\microns} and at 16 {\microns},
repectively).  The companion's type is unknown.

This paper analyzes all {\Spitzer} eclipse and transit data for TrES-1
to constrain the planet's orbit, atmospheric thermal profile, and
chemical abundances.  TrES-1's eclipse has an inherently low
signal-to-noise ratio (S/N).  Additionally, as one of the earliest
{\Spitzer} observations, the data did not follow the best observing
practices developed over the years.  We take this opportunity to
present the latest developments in our Photometry for Orbits,
Eclipses, and Transits (POET) pipeline
\citep{StevensonEtal2010natGJ436b, StevensonEtal2012apjHD149026b,
  StevensonEtal2012apjGJ436c, CampoEtal2011apjWASP12b,
  NymeyerEtal2011apjWASP18b, CubillosEtal2013ApjWASP8b} and
demonstrate its robustness on low S/N data.  We have implemented the
differential-evolution Markov-chain Monte Carlo algorithm
\citep[DEMC,][]{Braak2006DifferentialEvolution}, which explores the
parameter phase space more efficiently than the typically-used
Metropolis Random Walk with a multivariate Gaussian distribution as
the proposal distribution.  We also test and compare multiple
centering (Gaussian fit, center of light, PSF fit, and least
asymmetry) and photometry (aperture and optimal) routines.

Section \ref{sec:c3observations} describes the {\Spitzer} observations.
Section \ref{sec:c3analysis} outlines our data analysis pipeline.
Section \ref{sec:c3orbit} presents our orbital analysis.  Section
\ref{sec:c3atmosphere} shows the constraints that our eclipse
measurements place on TrES-1's atmospheric properties.  Finally,
section \ref{sec:c3conclusions} states our conclusions.

\section{Observations}
\label{sec:c3observations}

We analyzed eight light curves of TrES-1 from six {\Spitzer} visits
(obtained during the cryogenic mission): a simultaneous eclipse
observation in the 4.5 and 8.0 {\micron} Infrared Array Camera (IRAC)
bands (PI Charbonneau, program ID 227, full-array mode), a
simultaneous eclipse observation in the 3.6 and 5.8 {\micron} IRAC
bands (PI Charbonneau, program ID 20523, full-array), three
consecutive eclipses in the 16 {\micron} Infrared Spectrograph (IRS)
blue peak-up array, and one transit visit at 16 {\microns} (PI
Harrington, program ID 20605).  Table \ref{table:c3observations} shows
the {\Spitzer} band, date, total duration, frame exposure time, and
{\Spitzer} pipeline of each observation.

\begin{table}[ht]
\centering
\caption{Observation Information}
\label{table:c3observations}
\begin{tabular}{cccccc}
\hline
\hline
Event            & Band    & Observation & Duration & Exp. time & \Spitzer  \\
                 & \microns  & date      &  hours   & seconds   & pipeline  \\
\hline
Eclipse          & \n3.6     & 2005 Sep 17 & 7.27   &    \n1.2  & S18.18.0  \\ 
Eclipse          & \n4.5     & 2004 Oct 30 & 5.56   &     10.4  & S18.18.0  \\ 
Eclipse          & \n5.8     & 2005 Sep 17 & 7.27   &     10.4  & S18.18.0  \\ 
Eclipse          & \n8.0     & 2004 Oct 30 & 5.56   &     10.4  & S18.18.0  \\ 
Ecl. visit 1     &  16.0     & 2006 May 17 & 5.60   &     31.5  & S18.7.0   \\ 
Ecl. visit 2     &  16.0     & 2006 May 20 & 5.60   &     31.5  & S18.7.0   \\ 
Ecl. visit 3     &  16.0     & 2006 May 23 & 5.60   &     31.5  & S18.7.0   \\ 
Transit          &  16.0     & 2006 May 15 & 5.77   &     31.5  & S18.18.0  \\
\hline
\end{tabular}
\end{table}

\begin{figure}[htb]
\centering
\includegraphics[width=0.49\linewidth, clip]{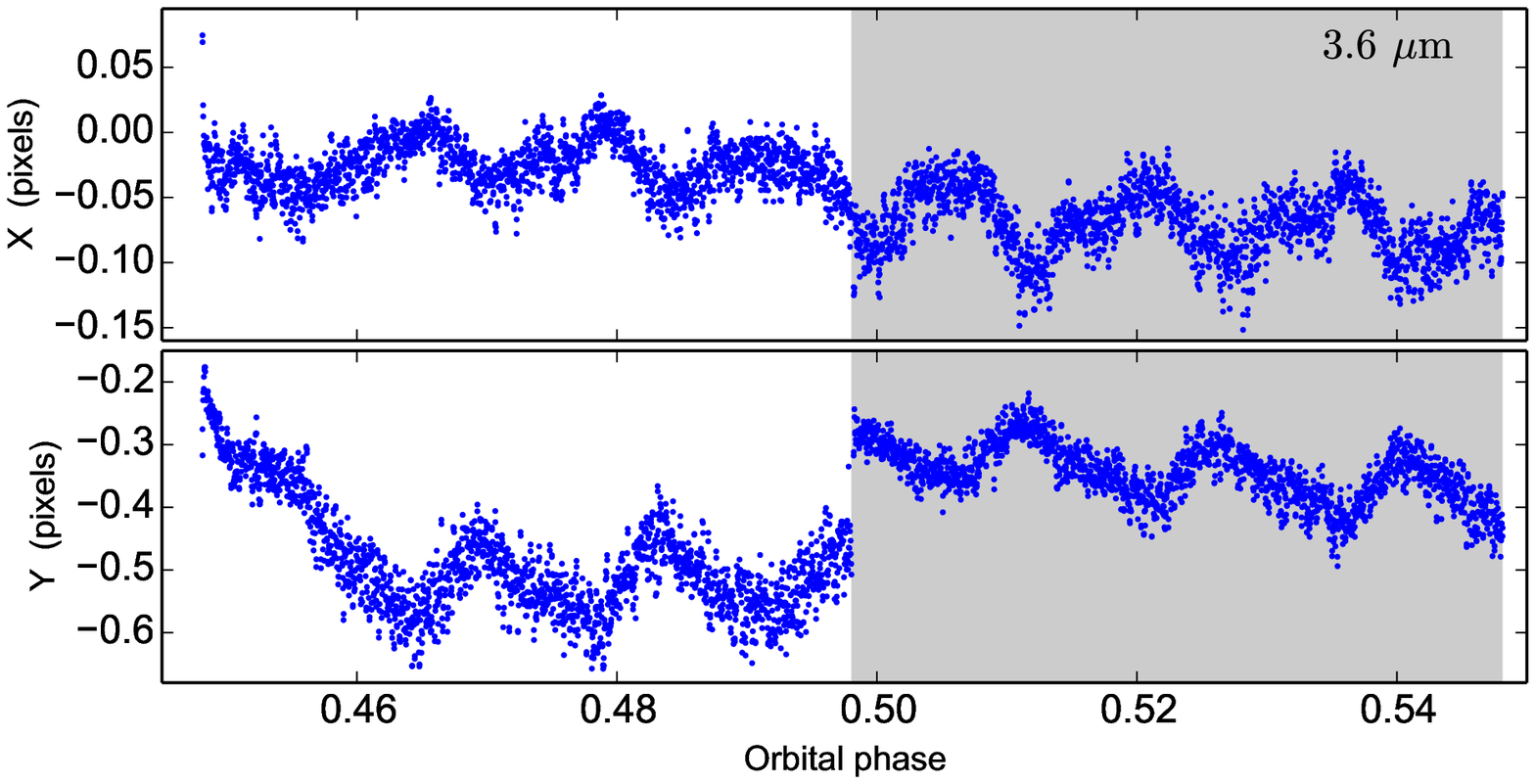}\hfill
\includegraphics[width=0.49\linewidth, clip]{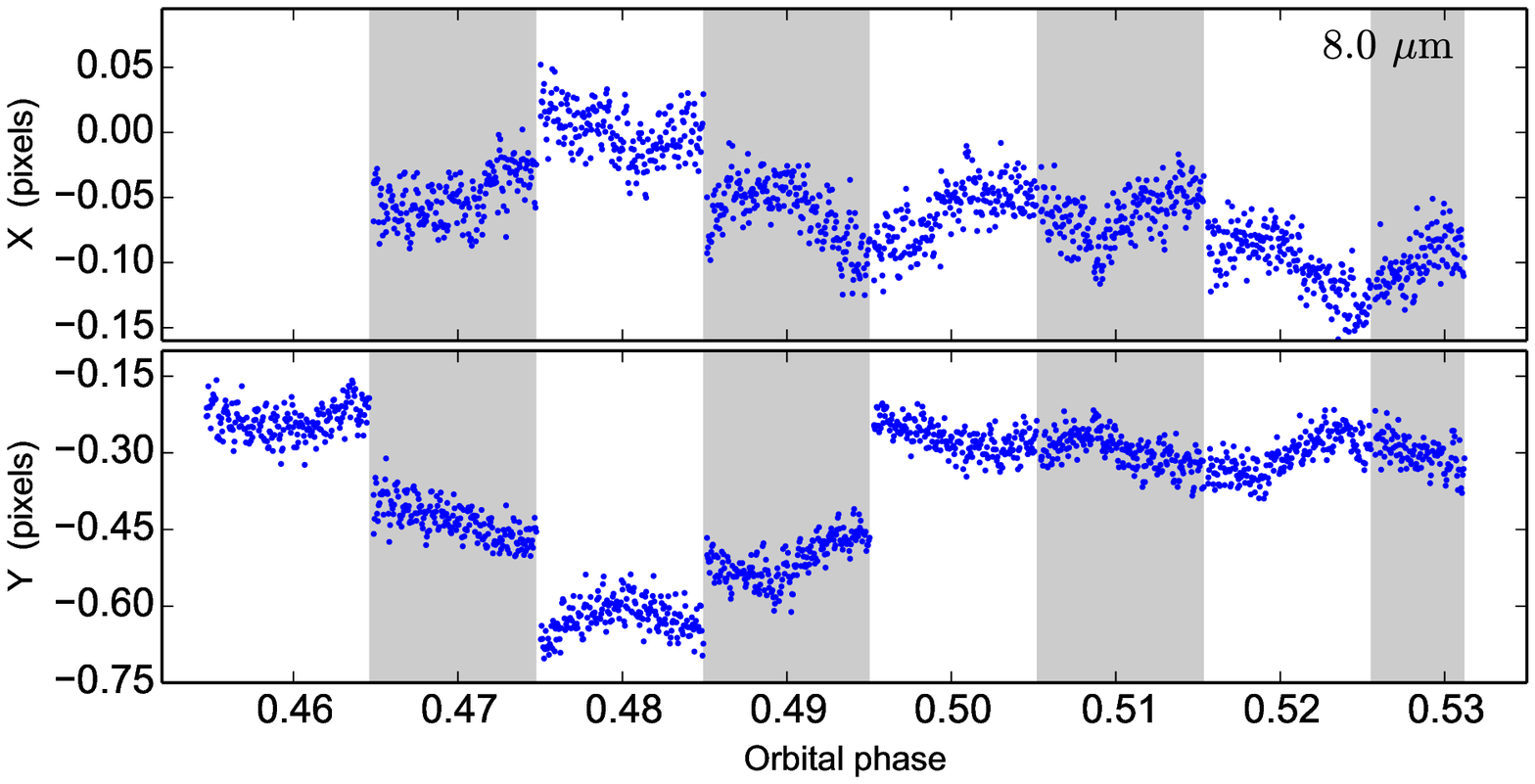}
\caption[Spitzer detector pointing]{{\bf Left:} TrES-1's $x$ (top) and
  $y$ (bottom) position on the detector at 3.6 {\microns} {\vs}\
  orbital phase.  The coordinate origin denotes the center of the
  nearest pixel. The shaded/unshaded areas mark different AORs.  The
  (\sim 0.1 pixels) pointing offsets are clear, as well as the usual
  hour-long pointing oscillation and point-to-point jitter (\sim 0.01
  pixels). {\bf Right:} Same as the left panel, but for the 8.0
  {\micron} light curve. The 5.8 and 4.5 {\micron} datasets were
  observed simultaneously with the 3.6 and 8.0 {\micron} bands,
  respectively; hence, their pointing correlates with the ones shown.}
\label{fig:positionAOR}
\end{figure}

In 2004, the telescope's Astronomical Observation Request (AOR)
allowed only a maximum of 200 frames
\citep{CharbonneauEtal2005apjTrES1}, dividing the 4.8 and 8.0
{\micron} events into eight AORs (Figure \ref{fig:positionAOR}).  The
later 3.6 and 5.8 {\micron} events consisted of two AORs.  The
repointings between AORs ($\sim 0.1$-pixel offsets) caused systematic
flux variations, because of IRAC's well-known position-dependent
sensitivity variations \citep{CharbonneauEtal2005apjTrES1}.  On the
other hand, the pointing of the IRS observations (a single AOR) cycled
among four nodding positions every five acquisitions, producing flux
variations between the positions.

\pagebreak

\section{Data Analysis}
\label{sec:c3analysis}

Our POET pipeline processes {Spitzer} Basic Calibrated Data to produce
light curves, modeling the systematics and eclipse (or transit)
signals.  Initially, POET flags bad pixels and calculates the frames'
Barycentric Julian Dates (BJD), reporting the frame mid-times in both
Coordinated Universal Time (UTC) and Barycentric Dynamical Time (TDB).
Next, it estimates the target's center position using any of four
methods: fitting a two-dimensional, elliptical, non-rotating Gaussian
with constant background
\citep[][Supplementary Information]{StevensonEtal2010natGJ436b};
fitting a 100x oversampled point spread function
\citep[PSF,][]{CubillosEtal2013ApjWASP8b}; calculating the center of
light \citep[][]{StevensonEtal2010natGJ436b}; or calculating the least
asymmetry (Lust et al.\ 2014, submitted).  The Gaussian-fit, PSF-fit,
and center-of-light methods considered a 15 pixel square window
centered on the target's peak pixel.  The least-asymmetry method used
a nine pixel square window.

\subsection{Optimal Photometry}
\label{sec:optimal}

POET generates raw light curves either from interpolated aperture
photometry \citep[][sampling a range of aperture radii in 0.25 pixel
  increments]{HarringtonEtal2007natHD149026b} or using an optimal
photometry algorithm (following \citealp{Horne1986Optimal}), which
improves S/N over aperture photometry for low-S/N data sets.  Optimal
photometry has been implemented by others to extract light curves
during stellar occultations by Saturn's rings
\citep{HarringtonEtal2010SatOcculatation} or exoplanets
\citep{Deming2005Nat, StevensonEtal2010natGJ436b}.  This algorithm
uses a PSF model, $P\sb{i}$, to estimate the expected fraction of the
sky-subtracted flux, $F\sb{i}$, falling on each pixel, $i$; divides it
out of $F\sb{i}$ so that each pixel becomes an estimate of the full
flux (with radially increasing uncertainty); and uses a mean with
weights $W\sb{i}$ to give an unbiased estimate of the target flux:
\begin{equation}
f = \frac{\sum\sb{i}W\sb{i}\ F\sb{i}/P\sb{i}}{\sum\sb{i} W\sb{i}}.
\label{eq:opphot1}
\end{equation}
Here, $W\sb{i}=P\sp{2}\sb{i}/V\sb{i}$, with $V\sb{i}$ the variance of
$F\sb{i}$.  Thus,
\begin{equation}
f\sb{\rm opt} = \frac{\sum\sb{i} P\sb{i}\ F\sb{i}/V\sb{i}}
                     {\sum\sb{i} P\sp{2}\sb{i}/V\sb{i}   }.
\label{eq:opphot2}
\end{equation}
We used the Tiny-Tim
program\footnote{\href{http://irsa.ipac.caltech.edu/data/SPITZER/docs/dataanalysistools/tools/contributed/general/stinytim/}{http://irsa.ipac.caltech.edu/data/SPITZER/docs/dataanalysistools/contributed/general/stinytim/}}
(ver. 2.0) to generate a super-sampled PSF model ($100\times$ finer
pixel scale than the {\Spitzer} data).  We shifted the position,
binned down the resolution, and scaled the PSF flux to fit the data.

\subsection{Light Curve Modeling}

Considering the position-dependent (intrapixel) and time-dependent
(ramp) {\Spitzer} systematics \citep{CharbonneauEtal2005apjTrES1}, we
modeled the raw light-curve flux, $F$, as a function of pixel position
$(x,y)$ and time $t$ (in orbital phase units):
\begin{equation}
F(x,y,t) = F\sb{s}\,E(t)\,M(x,y)\,R(t)\,A(a),
\label{eq:lcmodel}
\end{equation}
where $F\sb s$ is the out-of-eclipse system flux (fitting
parameter). $E(t)$ is an eclipse or transit (small-planet
approximation) \citet{MandelAgol2002ApJtransits} model.
$M(x,y)$ is a Bi-Linearly Interpolated Subpixel Sensitivity (BLISS)
map \citep{StevensonEtal2012apjHD149026b}.
$R(t)$ is a ramp model and $A(a)$ a per-AOR flux scaling factor.
The intrapixel effect is believed to originate from non-uniform
quantum efficiency across the pixels
\citep{ReachEtal2005paspIRACcalib}, being more significant at 3.6 and 4.5 {\microns}.  At the longer wavebands, the
intrapixel effect is usually negligible
\citep[e.g.,][]{KnutsonEtal2008apjHD209, KnutsonEtal2011apjGJ436b,
  StevensonEtal2012apjHD149026b}.  The BLISS map outperforms
polynomial fits for removing {\Spitzer}'s position-dependent
sensitivity variations \citep{StevensonEtal2012apjHD149026b,
  BlecicEtal2013apjWASP14b}.

For the ramp systematic, we tested several equations, $R(t)$, from the
literature \citep[e.g.,][]{StevensonEtal2012apjHD149026b,
  CubillosEtal2013ApjWASP8b}.  The data did not support models more
complex than:
\begin{eqnarray}
\label{eq:lin}
{\rm linramp:}  \quad R(t) & = & 1 + r\sb{1}(t-t\sb{c}) \\
\label{eq:quad}
{\rm quadramp:} \quad R(t) & = & 1 + r\sb{1}(t-t\sb{c}) + r\sb{2}(t-t\sb{c})\sp{2} \\
\label{eq:log}
{\rm logramp:}  \quad R(t) & = & 1 + r\sb{1}[\ln(t-t\sb{0})] \\
\label{eq:re}
{\rm risingexp:}\quad R(t) & = & 1 - e\sp{-r\sb{1}(t-t\sb{0})}
\end{eqnarray}
where $t\sb{c}$ is a constant, fixed at orbital phase 0 (for transits)
or 0.5 (for eclipses); $r\sb{1}$ and $r\sb{2}$ are a linear and
quadratic free parameters, respectively; and $t\sb{0}$ is a
time-offset free parameter.

\begin{figure}[htb]
\centering
\includegraphics[width=0.7\linewidth, clip]{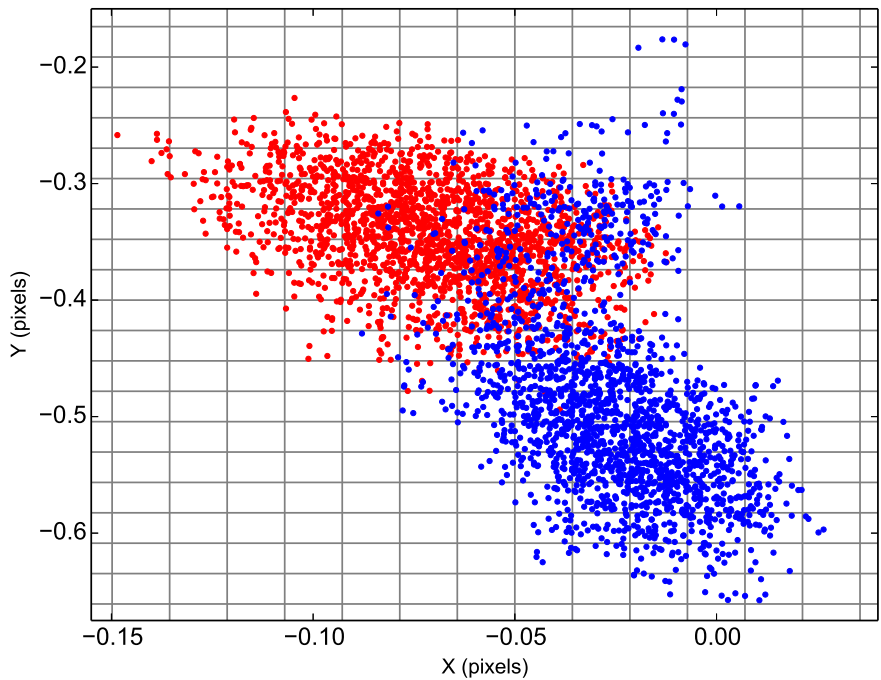}
\caption[3.6-{\micron} detector pointing.]{3.6-{\micron} detector
  pointing.  The blue and red points denote the data point from the
  first and second AOR, respectively.  The coordinate origin denotes
  the center of the nearest pixel.  The grid delimits the BLISS-map
  bin boundaries.}
\label{fig:BLISSmapch1}
\end{figure}

Additionally, the telescope pointing settled at slightly different
locations for each AOR, resulting in significant non-overlapping regions between
the sets of positions from each AOR (Figure \ref{fig:BLISSmapch1}).
Furthermore, the overlaping region is mostly composed of data points
taken during the telescope settling (when the temporal variation is
stronger).
The pointing offsets provided a weak link between the non-overlapping
regions of the detector, complicating the construction of the pixel
sensitivity map at 3.6 and 4.5 {\microns}.  We attempted the
correction of \citet{StevensonEtal2012apjHD149026b}, $A(a\sb{i})$,
which scales the flux from each AOR, $a\sb{i}$, by a constant factor.
To avoid degeneracy, we set $A(a\sb{1})=1$ and free subsequent
factors.  This can be regarded as a further refinement to the
intrapixel map for 3.6 and 4.5 {\microns}.  Just like the ramp models,
the AOR-scaling model works as an ad-hoc model that corrects for the
{\Spitzer} systematic variations.

Note that introducing
parameters that relate only to a portion of the data violates an
assumption of the Bayesian Information Criterion (BIC) used below; the
same violation occurs for the BLISS map (see Appendix A of
\citealp{StevensonEtal2012apjHD149026b}).  We have not found an
information criterion that handles such parameters, so we ranked these
fits with the others, being aware that BIC penalizes them too harshly.
It turned out that the AOR-scaling model made a significan improvement
only at 3.6~{\microns}; see Section \ref{sec:joint}.

\comment{For this data set, we discarded the other
models by contrasting the estimated orbital parameters to the rest of
the datasets.  At longer wavelengths, the
intrapixel effect is negligible, hence there are no solid grounds to
include the AOR model.}

To determine the best-fitting parameters, ${\bf x}$, of a model, ${\cal M}$
(Equation \ref{eq:lcmodel} in this case), given the data, ${\bf D}$,
we maximize the Bayesian posterior probability \citep[probability of the
model parameters given the data and modeling framework,][]{Gregory2005BayesianBook}:
\begin{equation}
  P({\bf x}|{\bf D},{\cal M}) = P({\bf x}|{\cal M})\, P({\bf D}|{\bf x},{\cal M}) / P({\bf D}|{\cal M}),
\label{eq:bayes}
\end{equation}
where $P({\bf D}|{\bf x},{\cal M})$ is the usual likelihood of the data given the
model and
$P({\bf x}|{\cal M})$ is any prior information on the parameters.  Assuming
Gaussian-distributed priors, maximizing Equation (\ref{eq:bayes}) can
be turned into a problem of minimization:
\begin{equation}
\min \bigg\{
  \sum\sb{j} \left(\frac{{\bf x}\sb{j}   -p\sb{j}}{\sigma\sb{j}}\right)\sp{2} +
  \sum\sb{i} \left(\frac{{\cal M}\sb{i}({\bf x})-{\bf D}\sb{i}}{\sigma\sb{i}}\right)\sp{2} 
     \bigg\},
\label{eq:minimization}
\end{equation}
with $p\sb{j}$ a prior estimation (with standard deviation
$\sigma\sb{j}$).  The second term in Equation (\ref{eq:minimization}) corresponds to $\chi\sp2$.  We used
the Levenberg-Marquardt minimizer to find ${\bf x}\sb{j}$ \citep{Levenberg1944, Marquardt1963}.  Next we sampled the
parameters' posterior distribution through a Markov-chain Monte Carlo
(MCMC) algorithm to estimate the parameter uncertainties, requiring
the Gelman-Rubin statistic \citep{Gelman1992} to be
within 1\% of unity for each free parameter before declaring convergence.

\subsection{Differential Evolution  Markov Chain}
\label{sec:demc}

The MCMC's performance depends crucially on having good proposal
distributions to efficiently explore the parameter space. Previous
POET versions used the Metropolis random walk, where new parameter
sets are proposed from a multivariate normal distribution.
The algorithm's efficiency was limited by the heuristic tuning of the
characteristic jump sizes for each parameter. Too-large values yielded
low acceptance rates, while too-small values wasted computational
power.  Furthermore, highly correlated parameter spaces required
additional orthogonalization techniques
\citep{StevensonEtal2012apjHD149026b} to achieve reasonable acceptance
ratios, and even then did not always converge.

We eliminated the need for manual tuning and orthogonalization
by implementing the differential-evolution Markov-chain algorithm \citep[DEMC,
][]{Braak2006DifferentialEvolution}, which automatically adjusts the
jumps' scales and orientations.  Consider ${\bf x}\sb{n}\sp{i}$ as the set of
free parameters of a chain $i$ at iteration $n$. DEMC runs several
chains in parallel, drawing the parameter values for the next
iteration from the difference between the current parameter states of two
other randomly-selected chains, $j$ and $k$:
\begin{equation}
\label{eq:demc}
{\bf x}\sb{n+1}\sp{i} = {\bf x}\sb{n}\sp{i} + \gamma \left( {\bf x}\sb{n}\sp{j} - {\bf x}\sb{n}\sp{k}\right) + \gamma\sb{2}{\bf e}\sb{n}\sp{i},
\end{equation}
where $\gamma$ is a scaling factor of the proposal jump.  Following
\citet{Braak2006DifferentialEvolution}, we selected $\gamma =
2.38/\sqrt{2d}$ (with $d$ being the number of free parameters) to
optimize the acceptance probability \citep[$\gtrsim $25\%,][]{RobertsEtal1997}.
The last term, $\gamma\sb{2}{\bf e}$,
is a random distribution (of smaller scale than the posterior
distribution) that ensures a complete exploration of posterior
parameter space.  We chose a multivariate normal distribution for {\bf e}, scaled by the factor $\gamma\sb{2}$.

As noted by \citet{EastmanEtal2013paspEXOFAST}, each parameter of $\bf
e$ requires a specific jump scale.  One way to estimate the
scales is to calculate the standard deviation of the parameters in
a sample chain run.
In a second method \citep[similar to that of
][]{EastmanEtal2013paspEXOFAST}, we searched for the limits around the
best-fitting value where $\chi\sp{2}$ increased by 1 along the
parameter axes.  We varied each parameter separately, keeping the
other parameters fixed.  Then, we calculated the jump scale from the
difference between the upper and lower limits, $({\bf x}\sp{\rm
  up}-{\bf x}\sp{\rm lo})/2$.  Both methods yielded similar results in
our tests.
By testing different values for $\gamma\sb{2}$, provided that
$|\gamma\sb{2}{\bf e}\sb{n}\sp{i}| < |\gamma ( {\bf x}\sb{n}\sp{j} -
{\bf x}\sb{n}\sp{k})|$, we found that each trial returned identical
posterior distributions and acceptance rates, so we arbitrarily set
$\gamma\sb{2}=0.1$.

\subsection{Data Set and Model Selection}
\label{sec:modelselec}

To determine the best raw light curve (i.e., the selection of
centering and photometry method), we minimized the standard deviation of
the normalized residuals (SDNR) of the light-curve fit
\citep{CampoEtal2011apjWASP12b}.  This naturally prefers good fits and
low-dispersion data.

We use Bayesian hypothesis testing to select the model best supported
by the data. Following \citet{Raftery1995BIC}, when comparing two
models ${\cal M}\sb{1}$ and ${\cal M}\sb{2}$ on a data set ${\bf D}$, the posterior odds
($B\sb{21}$, also known as Bayes factor) indicates the model preferred
by the data and the extent to which it is preferred.  Assuming that
either model is, a priori, equally probable, the posterior odds are
given by:
\begin{equation}
\label{eq:postodds}
 B\sb{21} =       \frac{p({\bf D}|{\cal M}\sb{2})}{p({\bf D}|{\cal M}\sb{1})}
          =       \frac{p({\cal M}\sb{2}|{\bf D})}{p({\cal M}\sb{1}|{\bf D})}
          \approx \exp\left(-\frac{{\rm BIC}\sb{2}-{\rm BIC}\sb{1}}{2}\right).
\end{equation}
This is the ${\cal M}\sb{2}$-to-${\cal M}\sb{1}$ probability ratio for
the models (given the data), with BIC = $\chi\sp{2} + k\ln{N}$ the
Bayesian Information Criterion \citep{Liddle2007mnrasBIC}, $k$ the
number of free parameters, and $N$ the number of points.  Hence,
${\cal M}\sb{2}$ has a fractional probability of
\begin{equation}
\label{eq:fracprob}
p({\cal M}\sb{2}|{\bf D}) = \frac{1}{1+ 1/B\sb{21}}.
\end{equation}
We selected the best models as those with the lowest BIC, and assessed
the fractional probability of the others (with respect to the best
one) using Equation (\ref{eq:fracprob}).

Recently, \citet{Gibson2014mnrasInferenceSystematics} proposed to
marginalize over systematics models rather than use model selection.
Although this process is still subjected to the researcher's choice of
systematics models to test, it is a more robust method.
Unfortunately, unless we understand the true nature of the systematics
to provide a physically motivated model, the modeling process will
continue to be an arbitrary procedure.  Most of our analyses prefer
one of the models over the others.  When a second model shows a
significant fractional probability ($>0.2$) we reinforce our selection
based on additional evidence (is the model physically plausible? or
how do the competing models perform in a joint fit?).  We are
evaluating to include the methods of
\citet{Gibson2014mnrasInferenceSystematics} to our pipeline in the
future.

\subsection{Light Curve Analyses}

We initially fit the eclipse light curves individually to determine
the best data sets (centering and photometry methods) and systematics
models.  Then, we determined the definitive parameters from a final
joint fit (Section \ref{sec:joint}) with shared eclipse parameters.
For the eclipse model we fit the midpoint, depth, duration, and
ingress time (while keeping the egress time equal to the ingress
time).  Given the low S/N of the data, the individual events do not
constrain all the eclipse parameters well.  However, the final joint
fit includes enough data to do the job.  For the individual fits, we
assumed a negligible orbital eccentricity, as indicated by transit and
radial-velocity (RV) data, and used the transit duration ($2.497 \pm
0.012$ hr) and transit ingress/egress time ($18.51 \pm 0.63$ min) from
\citet{WinnEtal2007apjTres1} as priors on the eclipse duration and
ingress/egress time.  In the final joint-fit experiments, we freed
these parameters.

\subsubsection[IRAC-3.6 {\micron} Eclipse]{IRAC-3.6 $\mu$m Eclipse}
\label{sec:tr001bs11}

\begin{figure}[htb]
\centering
\includegraphics[width=0.8\linewidth, clip]{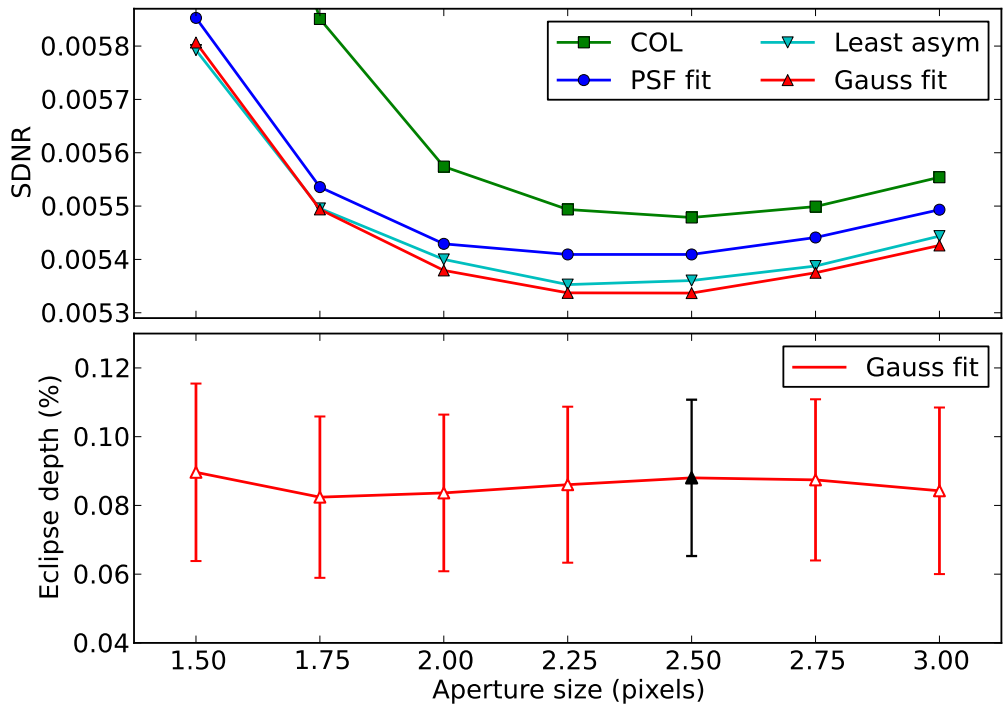}
\caption[Light curve's SNDR and eclipse depth at 3.6 {\microns}]{{\bf
    Top:} 3.6 {\micron} eclipse light-curve SDNR {\vs}\ aperture.  The
  legend indicates the centering method used.  All curves used the
  best ramp model from Table \ref{table:tr001bs11ramps}. {\bf Bottom:}
  Eclipse depth {\vs}\ aperture for Gaussian-fit centering, with the
  best aperture (2.5 pixels) in black.}
\label{fig:sdnr11}
\end{figure}

This observation is divided into two AORs at phase 0.498, causing a
systematic flux offset due to IRAC's intrapixel sensitivity
variations.
We tested aperture photometry between 1.5 and 3.0 pixels.
The eclipse depth is consistent among the apertures, and the
minimum SDNR occurs for the 2.5 pixel aperture with Gaussian-fit
centering (Figure \ref{fig:sdnr11}).

Table \ref{table:tr001bs11ramps} shows the best four model fits at the
best aperture; $\Delta$BIC is the BIC difference with respect to the
lowest BIC.  Given the relatively large uncertainties, more-complex
models are not supported, due to the penalty of the additional free
parameters.  The Bayesian Information Criterion favors the AOR-scaling
model (Table \ref{fig:sdnr11}, last column).

\begin{table}[ht]
\centering
\caption[3.6-{\microns} Eclipse - Ramp Model Fits]
        {3.6 {\microns} Eclipse - Ramp Model Fits\sp{a}}
\label{table:tr001bs11ramps}
\begin{tabular}{cccccc}
\hline
\hline
$R(t)\,A({\rm a})$ & Ecl. Depth\sp{b} & Midpoint & SDNR  & $\Delta$BIC & $p({\cal M}\sb{2}|D)$ \\
          & (\%)   & (phase)  &       &             &      \\
\hline
$A(a)$    & 0.083(24)  & 0.501(4)   & 0.0053763 &   0.0  & $\cdots$     \\
quadramp  & 0.158(29)  & 0.492(2)   & 0.0053712 &   2.8  & 0.19         \\
risingexp & 0.146(25)  & 0.492(2)   & 0.0053715 &   2.9  & 0.19         \\
linramp   & 0.093(23)  & 0.492(3)   & 0.0053814 &   7.4  & 0.02         \\
\hline
\multicolumn{6}{l}{\footnotesize {\bf Notes}.}  \\
\multicolumn{6}{l}{\footnotesize \sp{a} Fits for Gaussian-fit centering
                   and 2.5 pixel aperture photometry.} \\
\multicolumn{6}{l}{\footnotesize \sp{b} For this and the following tables,
                   the values quoted in parenthesis indicate the} \\
\multicolumn{6}{l}{\footnotesize 1$\sigma$ uncertainty
                   corresponding to the least significant digits.}
\end{tabular}
\end{table}

Although the fractional probabilities of the quadratic and exponential
ramp models are not negligible, we discard them based on the estimated
midpoints, which differ from a circular orbit by 0.008 (twice the
ingress/egress duration).
It is possible that a non-uniform brightness distribution can induce
offsets in the eclipse midpoint \citep{WilliamsEtal2006apj}, and these
offsets can be wavelength dependent.  However, this relative offset can be
at most the duration of the ingress/egress.
Therefore, disregarding non-uniform brightness offsets, considering
the lack of evidence for transit-timing variations and that all other
data predict a midpoint consistent with a circular orbit, the
3.6~{\micron} offset must be caused by systematic effects.
The AOR-scaling model is the only one that yields a midpoint
consistent with the rest of the data.  Our joint-fit analysis (Section
\ref{sec:joint}) will provide further support to our model selection.

\begin{figure*}[p]
\strut\hfill 
\includegraphics[width=0.28\textwidth, clip]{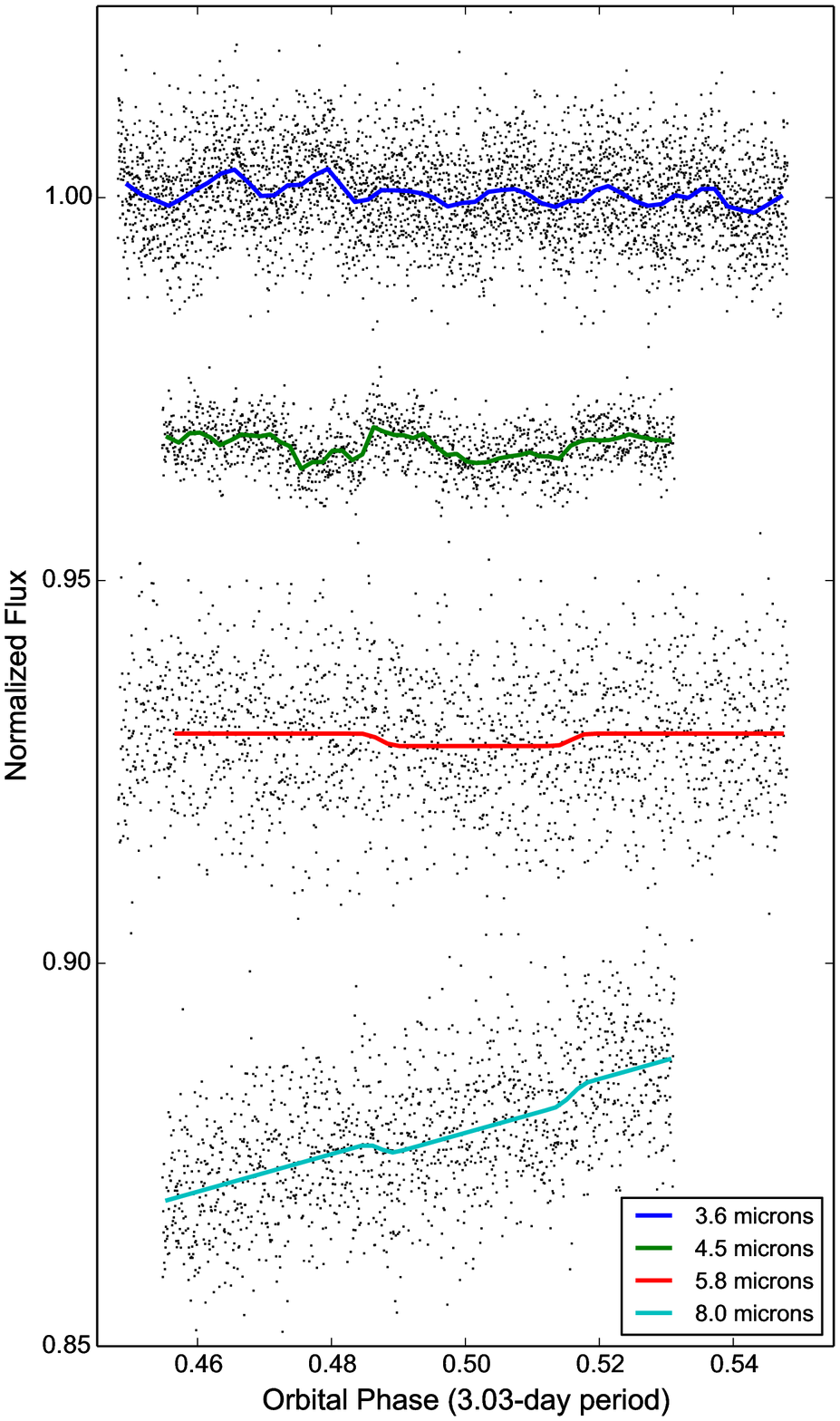}
\hfill       
\includegraphics[width=0.28\textwidth, clip]{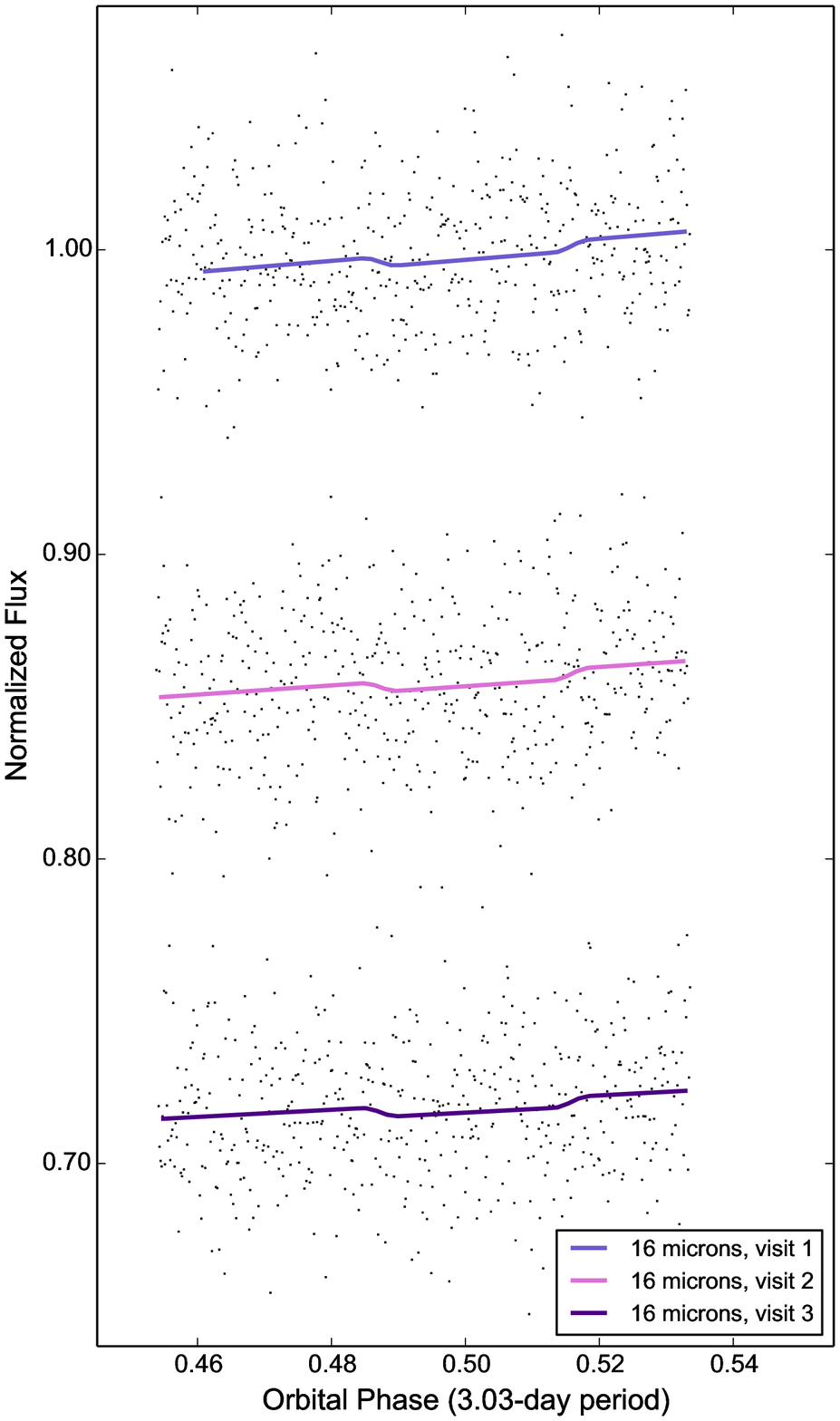}
\hfill       
\includegraphics[width=0.28\textwidth, clip]{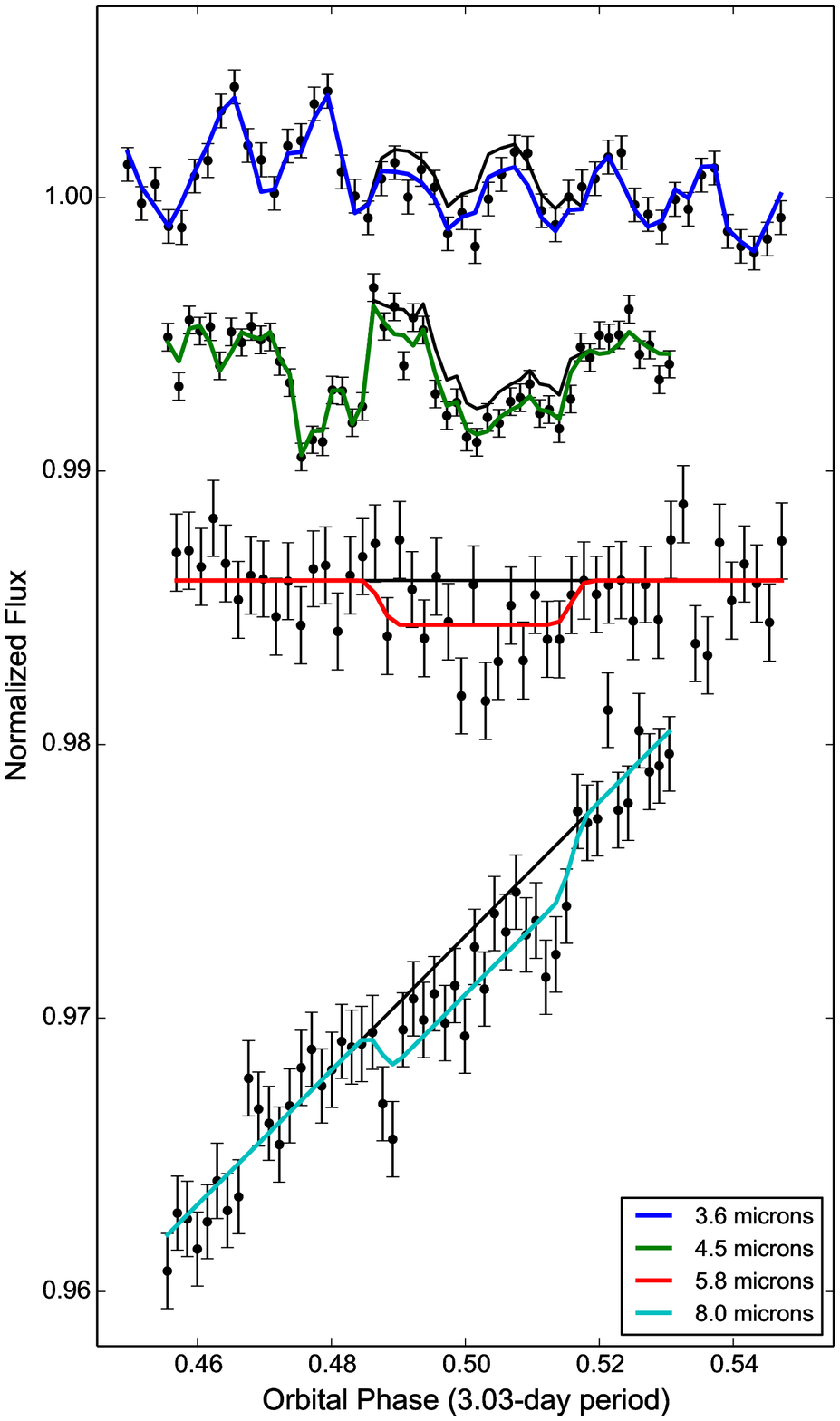}
\hfill\strut
\newline
\strut\hfill 
\includegraphics[width=0.28\textwidth, clip]{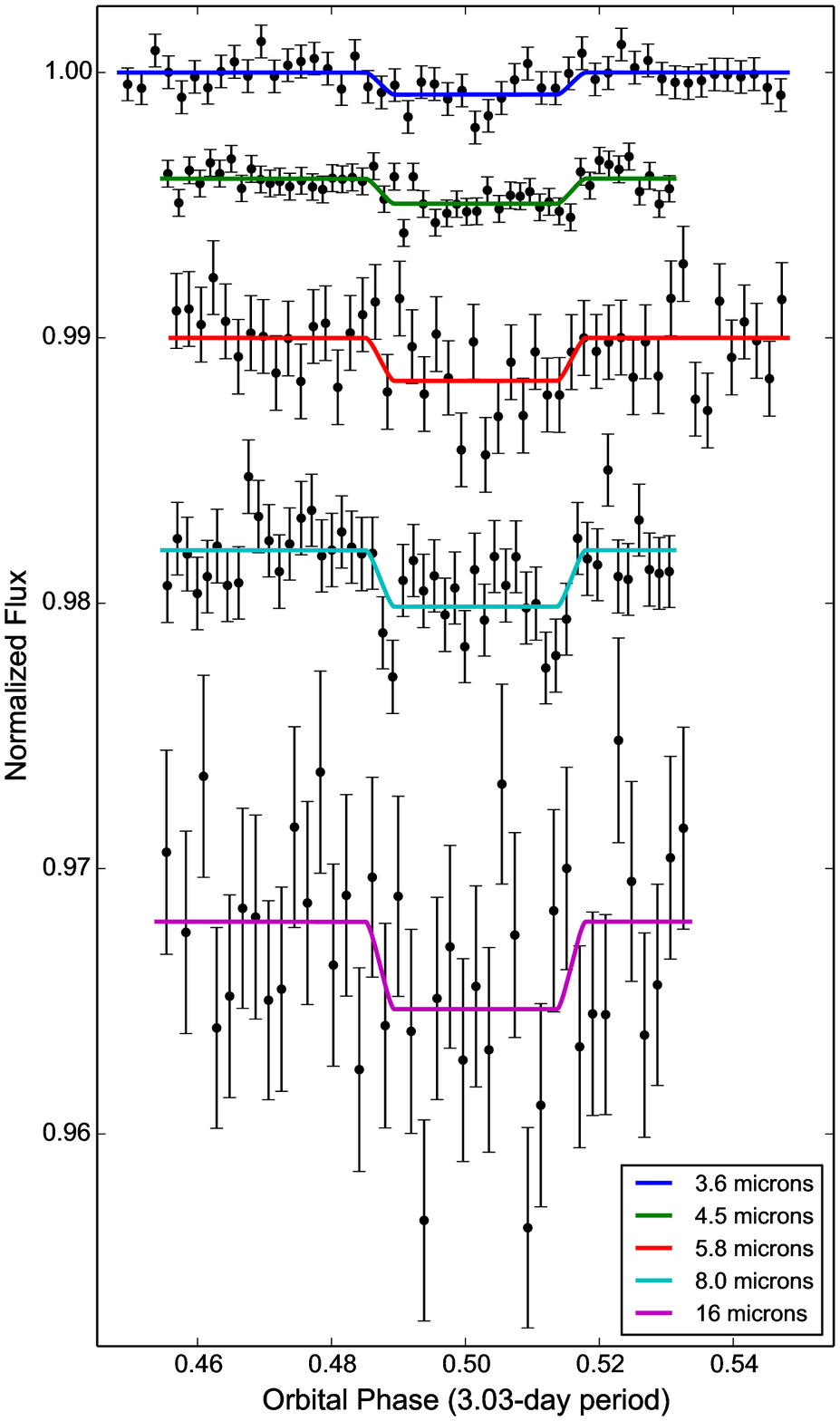}
\hfill       
\includegraphics[width=0.28\textwidth, clip]{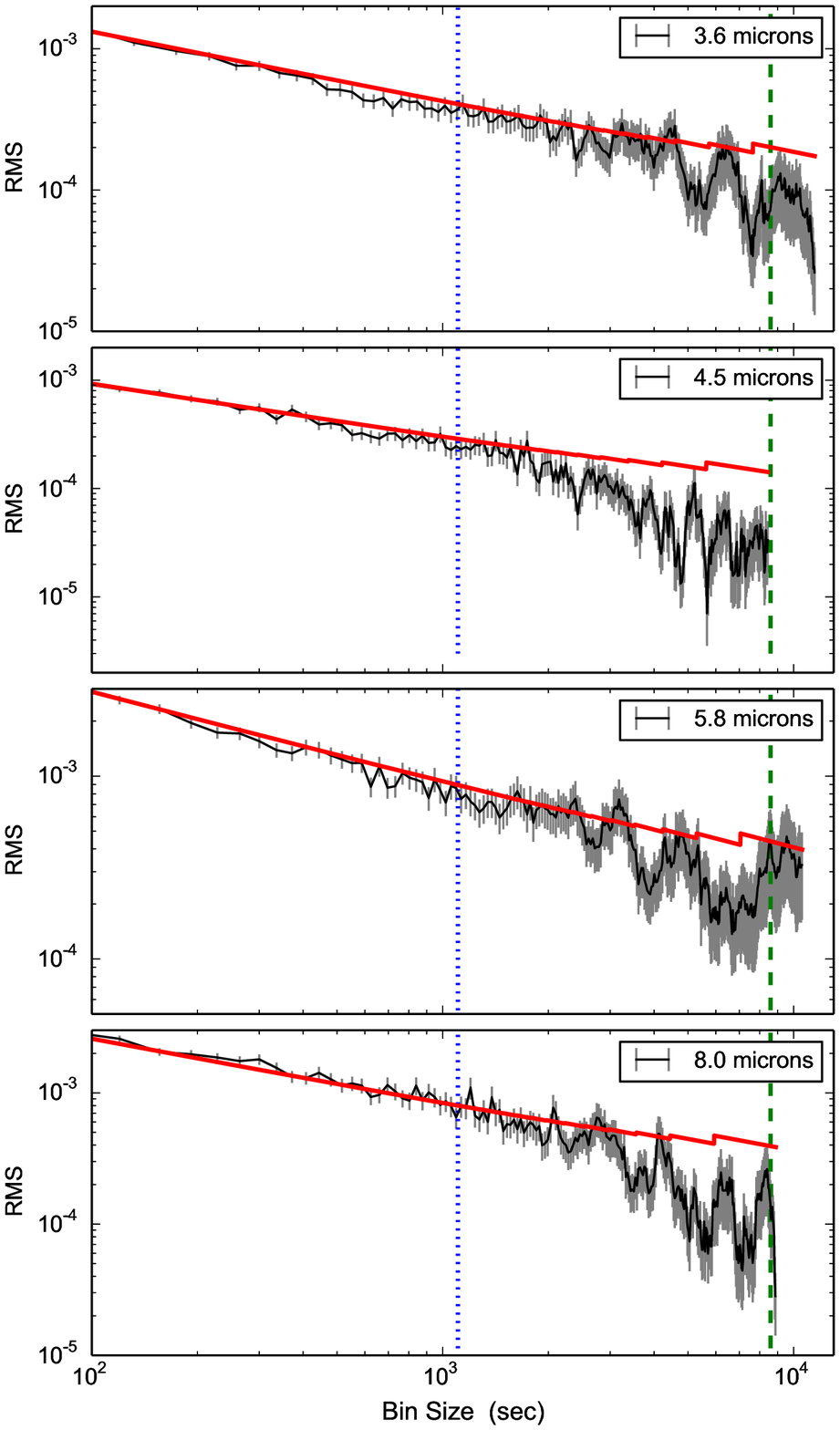}
\hfill       
\includegraphics[width=0.28\textwidth, clip]{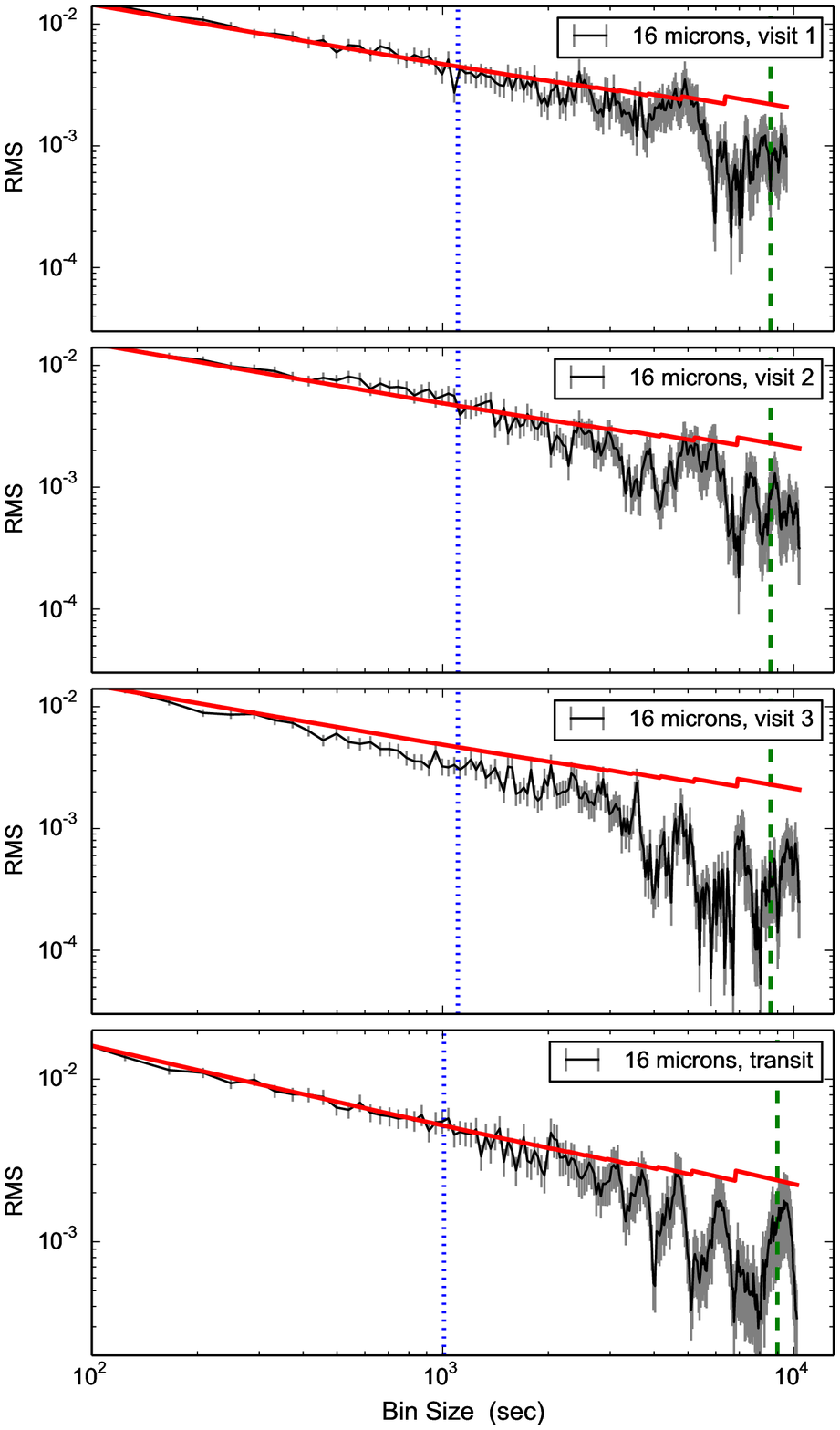}
\hfill\strut
\caption[TrES-1 secondary-eclipse light curves and rms-{\vs}-bin size
  plots]{TrES-1 secondary-eclipse light curves and rms-{\vs}-bin size
  plots.  Raw light curves are in the top-left and top-center panels.
  Binned IRAC data are in the top-right panel, and
  systematics-corrected traces are in the bottom-left panel.  The
  system flux is normalized and the curves are shifted vertically for
  clarity.  The colored solid curves are the best-fit models, while
  the black solid curves are the best-fit models excluding the eclipse
  component.  The error bars give the 1$\sigma$ uncertainties.  The
  bottom-center and bottom-right panels show the fit residuals' rms
  (black curves with 1$\sigma$ uncertainties) {\vs} bin size.  The red
  curves are the expected rms for Gaussian noise.  The blue dotted and
  green dashed vertical lines mark the ingress/egress time and eclipse
  duration, respectively.}
\label{fig:c3lightcurves}
\end{figure*}

We adjusted the BLISS map model following
\citet{StevensonEtal2012apjHD149026b}.  For a minimum of 4 points per
bin, the eclipse depth remained constant for BLISS bin sizes similar
to the rms of the frame-to-frame position difference (0.014 and 0.026
pixels in $x$ and $y$, respectively).  
Figure \ref{fig:c3lightcurves}
shows the raw, binned, and systematics-corrected light curves with
their best-fitting models.  

To estimate the contribution from time-correlated residuals we
calculated the time-averaging rms-vs.-bin-size curves
\citep{PontEtal2006mnrasRednoise, WinnEtal2008apjRednoise}.  This
method compares the binned-residuals rms to the uncorrelated-noise
(Gaussian noise) rms.  An excess rms over the Gaussian rms would
indicate a significant contribution from time-correlated residuals.
Figure \ref{fig:c3lightcurves} (bottom-center and bottom-right panels)
indicates that time-correlated noise is not significant at any
time scale, for any of our fits.

\subsubsection[IRAC-4.5 {\micron} Eclipse]{IRAC-4.5 $\mu$m Eclipse}
\label{sec:tr001bs21}

\begin{figure}[htb]
\centering
\includegraphics[width=0.3\linewidth]{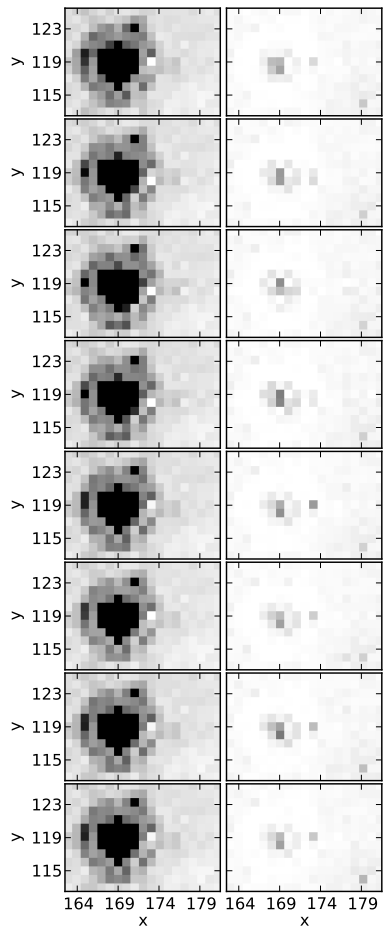}
\caption[Per-AOR mean {\Spitzer} frames at 4.5 {\microns}]{{\bf Left:}
  Per-AOR mean of the {\Spitzer} BCD frames at 4.5 {\microns} around
  TrES-1.  {\bf Right:} Per-AOR rms divided by the square root of the
  mean BCD frames at 4.5 {\microns}.  The flux is in electron counts,
  the color scales range from the 2.5th (white) to the 97.5th (black)
  percentile of the flux distribution.  TrES-1's center is located
  near $x=169, y=119$.  The miscalculated muxbleed-corrected pixels
  stand out at 4 and (sometimes) 8 pixels to the right of the target
  center.  The excess in the scaled rms confirms that the muxbleed
  correction is not linearly scaled with the flux.  The pixels around
  the target center also show high rms values, which might be due to
  the 0.2 pix motion of the PSF centers.}
\label{fig:bleed}
\end{figure}

Our analysis of the archival data revealed that the 4.5 {\micron} data
suffered from multiplexer bleed, or ``muxbleed'', indicated by flagged
pixels near the target in the mask frames and data-frame headers
indicating a muxbleed correction.
Muxbleed is an effect observed in the IRAC InSb arrays (3.6 and 4.5
{\microns}) wherein a bright star trails in the fast-read direction
for a large number of consecutive readouts.  Since there are 4 readout
channels, the trail appears every 4 pixels, induced by one or more bright pixels\footnote{
\href{http://irsa.ipac.caltech.edu/data/SPITZER/docs/irac/iracinstrumenthandbook/59/}{http://irsa.ipac.caltech.edu/data/SPITZER/docs/irac/iracinstrumenthandbook/59/}} (Figure \ref{fig:bleed}).

\pagebreak

TrES-1 (whose flux was slightly below the nominal saturation limit at 4.5
{\microns}) and a second star that is similarly bright fit the muxbleed
description.  We noted the same feature in the BCD frames used by
\citet[][{\Spitzer} pipeline version
S10.5.0]{CharbonneauEtal2005apjTrES1}.  Their headers indicated a
muxbleed correction as well, but did not clarify whether or not a
pixel was corrected.
\comment{From communications at that time, the PI and the IRAC team
concluded that the flagged data were not invalid.}

Since the signal is about $\ttt{-3}$ times the stellar flux level,
every pixel in the aperture is significant and any imperfectly made
local correction raises concern (this is why we do not interpolate bad
pixels in the aperture, but rather discard frames that have them).
Nevertheless, we analyzed the data, ignoring the muxbleed flags, to
compare it to the results of \citet{CharbonneauEtal2005apjTrES1}.  In
the atmospheric analysis that follows, we model the planet both with
and without this data set.

\begin{figure}[ht]
\centering
\includegraphics[width=0.8\linewidth, clip]{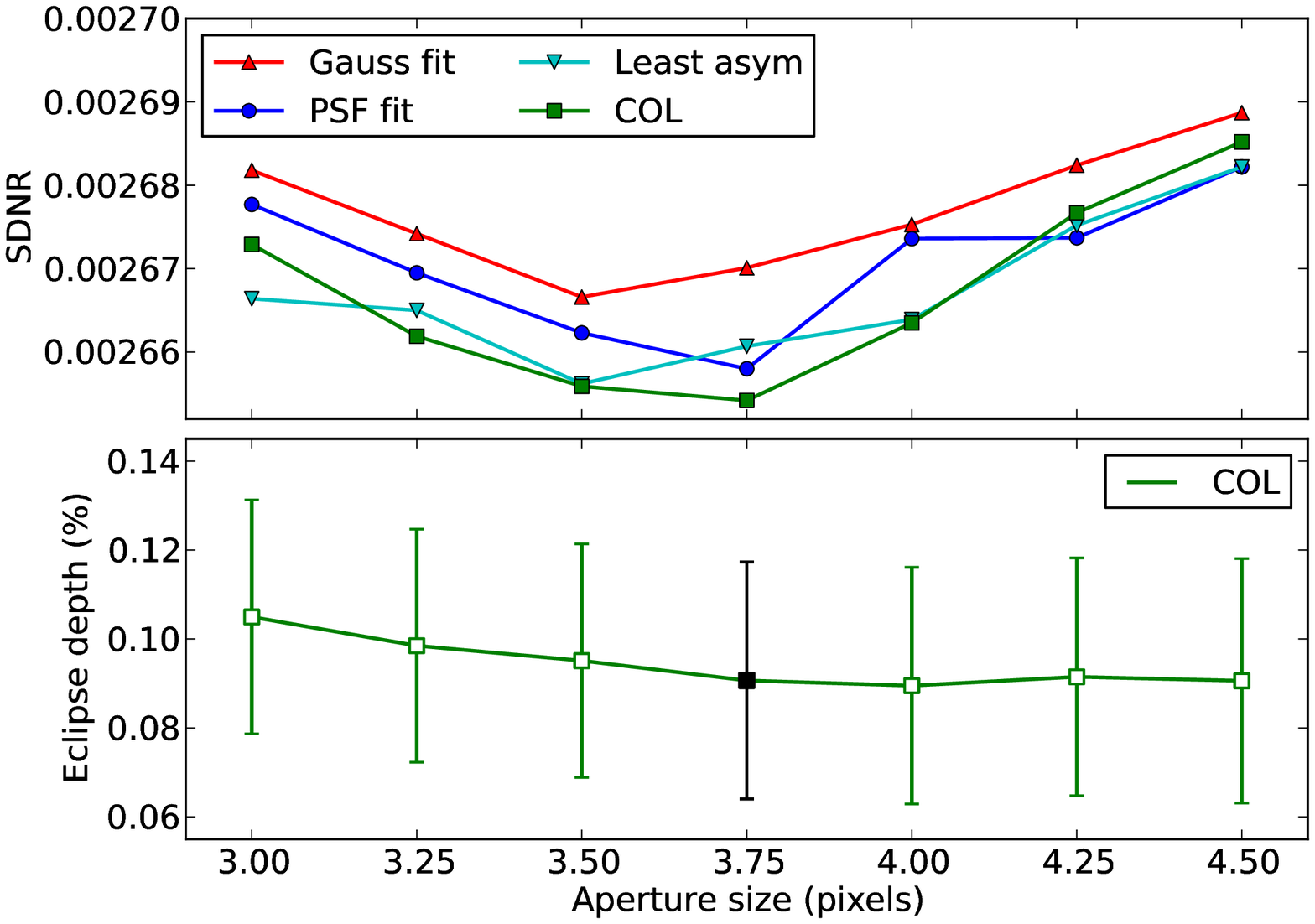}
\caption[Light curve's SNDR and eclipse depth at 4.5 {\microns}]{{\bf
    Top:} 4.5 {\micron} eclipse light-curve SDNR {\vs} aperture.  The
  legend indicates the centering method used.  All curves used the
  best ramp model from Table \ref{table:tr001bs21ramps}. {\bf Bottom:}
  Eclipse depth {\vs} aperture for center-of-light centering, with the
  best aperture (3.75 pixels) in black.}
\label{fig:sdnrdepth21}
\end{figure}

This light curve is also mainly affected by the intrapixel effect.
Since the 4.5~{\microns} light curve consisted of 8 AORs, some of
which are entirely in- or out-of-eclipse, making the AOR-scale model
to overfit the data. We tested apertures
between 2.5 and 4.5 pixels, finding the lowest SDNR for the
center-of-light centering method at the 3.75-pixel aperture (Figure
\ref{fig:sdnrdepth21}).  This alone is surprising, as it may be the
first time in our experience that center of light is the best method.
In the same manner as for the 3.6~{\micron} data, we selected BLISS
bin sizes of 0.018 ($x$) and 0.025 ($y$) pixels, for 4 minimum points
per bin.  A fit with no ramp model minimized BIC (Table
\ref{table:tr001bs21ramps}). Figure \ref{fig:c3lightcurves} shows the
data and best-fitting light curves and the rms-{\vs}-bin size plot.

\begin{table}[ht]
\centering
\caption[4.5-{\micron} Eclipse - Ramp Model Fits]
        {4.5 {\micron} Eclipse - Ramp Model Fits\sp{a}}
\label{table:tr001bs21ramps}
\begin{tabular}{cccccc}
\hline
\hline
$R(t)\,A({\rm a})$  & Ecl. Depth (\%) & SDNR  & $\Delta$BIC & $p({\cal M}\sb{2}|D)$ \\
\hline
no-model            & 0.090(28)       & 0.0026543 & \n0.0   & $\cdots$     \\
linramp             & 0.091(27)       & 0.0026531 & \n6.0   & 0.05         \\
risingexp           & 0.131(32)       & 0.0026469 & \n7.1   & 0.03         \\
quadamp             & 0.153(39)       & 0.0026481 & \n8.7   & 0.01         \\
logramp             & 0.090(22)       & 0.0026532 &  13.3   & $1\tttt{-3}$ \\
$A(a)$              & 0.140(43)       & 0.0026474 &  38.5   & $4\tttt{-9}$ \\
\hline
\multicolumn{6}{l}{\footnotesize {\bf Note.} \sp{a} Fits for center-of-light centering and 3.75-pixel
  aperture photometry.}
\end{tabular}
\end{table}

\subsubsection[IRAC-5.8 {\micron} Eclipse]{IRAC-5.8 $\mu$m Eclipse}
\label{sec:tr001bs31}

\begin{figure}[ht]
\centering
\includegraphics[width=0.8\linewidth, clip]{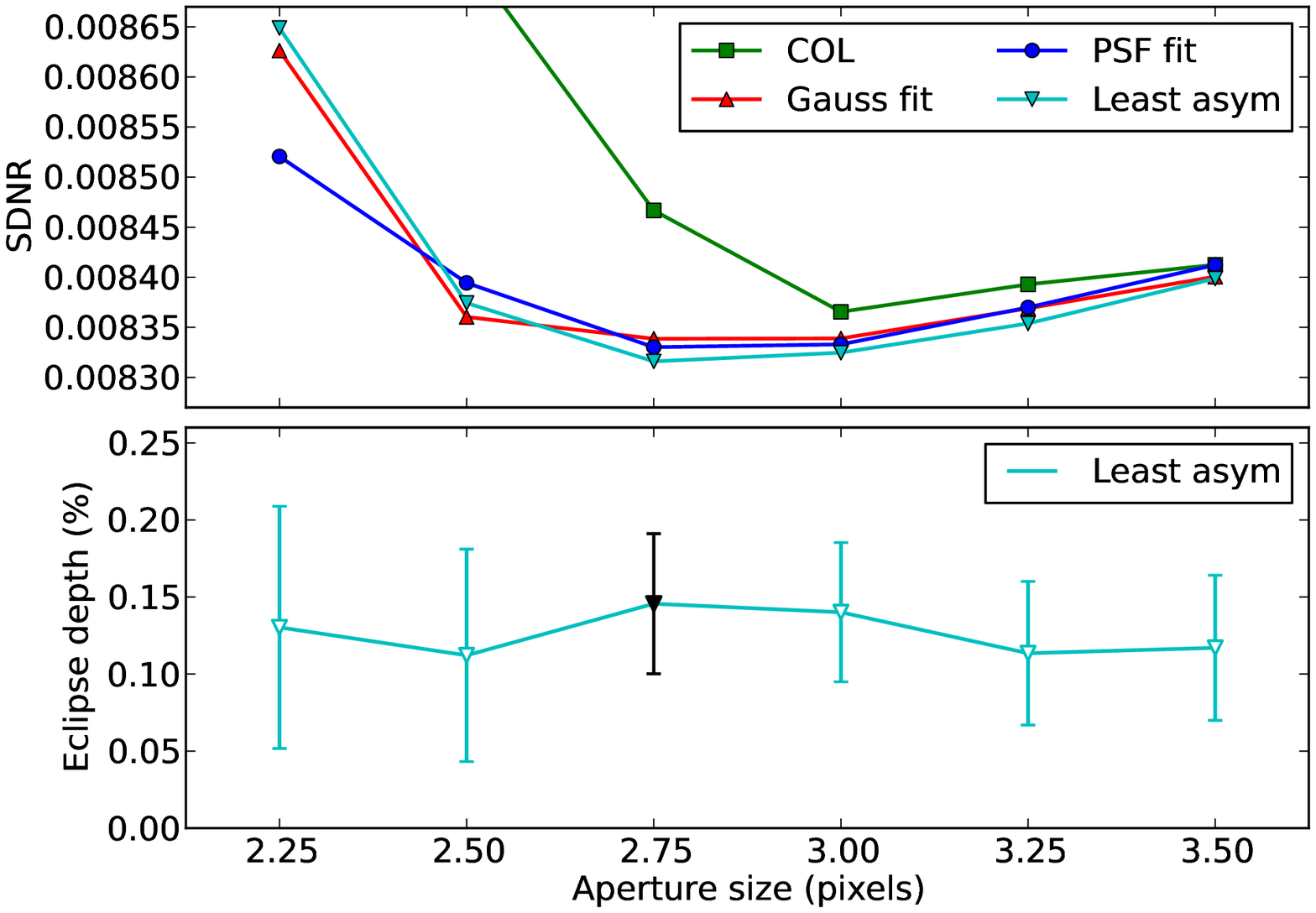}
\caption[Light curve's SNDR and eclipse depth at 5.8 {\microns}]{{\bf
    Top:} 5.8 {\micron} eclipse light-curve SDNR {\vs} aperture.  The
  legend indicates the centering method used.  All curves used the
  best ramp model from Table \ref{table:tr001bs31ramps}. {\bf Bottom:}
  Eclipse depth {\vs} aperture for least-asymmetry centering, with the
  best aperture (2.75 pixels) in black.}
\label{fig:sdnrdepth31}
\end{figure}

These data are not affected by the intrapixel effect.  We sampled
apertures between 2.25 and 3.5 pixels.  Least-asymmetry centering
minimized the SDNR at 2.75 pixels,
with all apertures returning consistent eclipse depths (Figure
\ref{fig:sdnrdepth31}).
The BIC comparison favors a fit without AOR-scale nor ramp models,
although, at some apertures the midpoint posterior distributions
showed a hint of bi-modality.  The eclipse depth, however, remained
consistent for all tested models (Table \ref{table:tr001bs31ramps}).
Figure \ref{fig:c3lightcurves} shows the data and best-fitting light
curves and rms-{\vs}-bin size plot.

\begin{table}[ht]
\centering
\caption[5.8-{\micron} Eclipse - Ramp Model Fits]
        {5.8 {\micron} Eclipse - Ramp Model Fits\sp{a}}
\label{table:tr001bs31ramps}
\begin{tabular}{cccccc}
\hline
\hline
$R(t)\,A({a})$  & Ecl. Depth (\%) & SDNR       & $\Delta$BIC & $p({\cal M}\sb{2}|D)$ \\
\hline
no-model        &   0.158(44)     &  0.0083287 &  \n0.0    & $\cdots$     \\
$A(a)$          &   0.142(45)     &  0.0083220 &  \n4.4    & 0.10         \\
linramp         &   0.154(44)     &  0.0083281 &  \n7.2    & 0.03         \\
quadramp        &   0.100(54)     &  0.0083259 &   13.0    & $2\tttt{-3}$ \\
risingexp       &   0.158(44)     &  0.0083287 &   14.9    & $6\tttt{-4}$ \\
\hline
\multicolumn{6}{l}{\footnotesize {\bf Note.} \sp{a} Fits for least-asymmetry
   centering and 2.75-pixel aperture photometry.}
\end{tabular}
\end{table}

\subsubsection[IRAC-8.0 {\microns} Eclipse]{IRAC-8.0 $\mu$m Eclipse}
\label{sec:tr001bs41}

\begin{figure}[th]
\centering
\includegraphics[width=0.8\linewidth, clip]{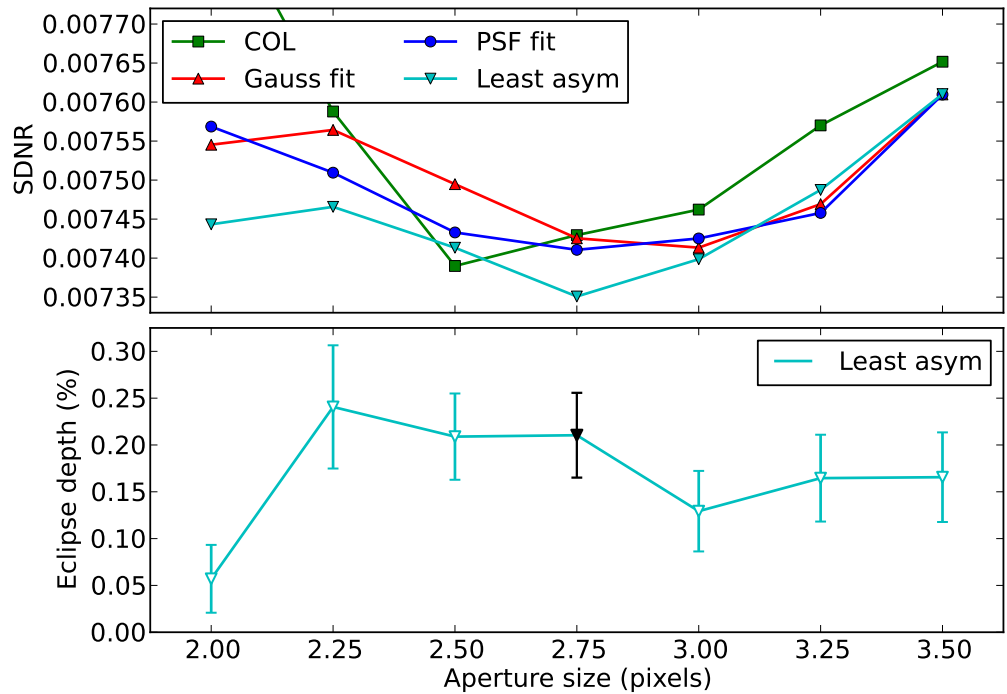}
\caption[Light curve's SNDR and eclipse depth at 8.0 {\microns}]{{\bf
    Top:} 8.0 {\micron} eclipse light-curve SDNR {\vs} aperture.  The
  legend indicates the centering method used.  All curves used the
  best ramp model from Table \ref{table:tr001bs41ramps}. {\bf Bottom:}
  Eclipse depth {\vs} aperture for least-asymmetry centering, with the
  best aperture (2.75 pixels) in black.  This data set had the
  greatest eclipse-depth variations per aperture.}
\label{fig:sdnrdepth41}
\end{figure}

This data set had eight AOR blocks.  We tested aperture photometry
from 1.75 to 3.5 pixels.  Again, least-asymmetry centering minimized
the SDNR for the 2.75-pixel aperture (Figure \ref{fig:sdnrdepth41}).
We attempted fitting with the per-AOR adjustment $A(a)$, but the seven
additional free parameters introduced a large BIC penalty, and the
many parameters certainly alias with the eclipse.  The linear ramp
provided the lowest BIC (Table \ref{table:tr001bs41ramps}).  Figure
\ref{fig:c3lightcurves} shows the data and best-fitting light curves and
rms-{\vs}-bin size plot.

\begin{table}[ht]
\centering
\caption[8.0-{\microns} Eclipse - Ramp Model Fits]
        {8.0 {\microns} Eclipse - Ramp Model Fits\sp{a}}
\label{table:tr001bs41ramps}
\begin{tabular}{cccccc}
\hline
\hline
$R(t)\,A({a})$  & Ecl. Depth (\%) & SDNR     & $\Delta$BIC & $p({\cal M}\sb{2}|D)$ \\
\hline
linramp         & 0.208(45)      & 0.0073506 & \n0.0       & $\cdots$      \\
quadramp        & 0.267(62)      & 0.0073388 & \n3.3       & 0.16          \\
risingexp       & 0.278(53)      & 0.0073389 & \n3.4       & 0.15          \\
logramp         & 0.304(45)      & 0.0073471 & \n7.0       & 0.03          \\
linramp--$A(a)$ & 0.759(185)     & 0.0073112 &  41.8       & $8\tttt{-10}$ \\
\hline
\multicolumn{6}{l}{\footnotesize {\bf Note.} \sp{a} Fits for least-asymmetry
                    centering and 2.75-pixel aperture photometry.}
\end{tabular}
\end{table}

\subsubsection[IRS-16 {\micron} Eclipses]{IRS-16 $\mu$m Eclipses}
\label{sec:tr001bs5}

\begin{figure*}[thb]
\centering
\includegraphics[width=\linewidth, clip]{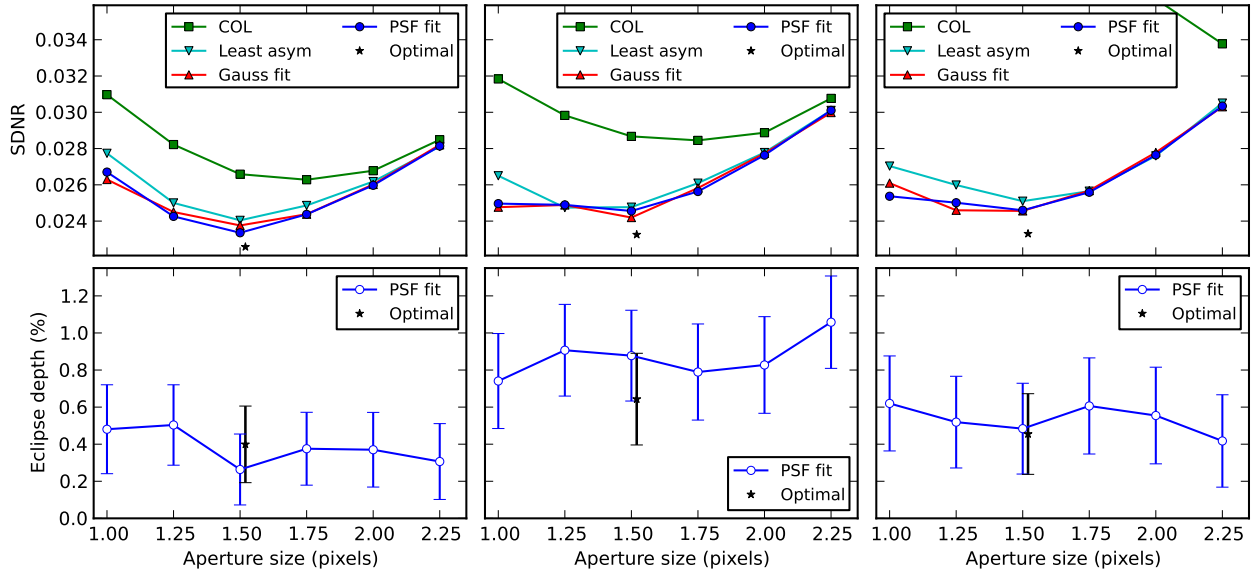}
\caption[Light curve's SNDR and eclipse depth at 16.0 {\microns}]{{\bf
    Top:} 16-{\micron} eclipse light-curves SDNR {\vs} aperture (from
  left to right, the first, second, and third visits, respectively).
  The legend indicates the centering method used; additionally, the
  optimal-photometry calculation uses the PSF-fit centering positions
  but does not involve an aperture.  We plotted the optimal-photometry
  results next to the best-aperture location (for ease of comparison).
  Each curve used the best ramp model from Tables
  \ref{table:tr001bs52ramps}, \ref{table:tr001bs53ramps}, and
  \ref{table:tr001bs51ramps}, respectively. {\bf Bottom:} Eclipse
  depth {\vs} aperture for PSF-fit centering, with the best one
  (optimal photometry) in black.}
\label{fig:sdnrdepthIRS}
\end{figure*}

These data come from three consecutive eclipses and present similar
systematics.  The telescope 
cycled among four nodding positions every five acquisitions.  As a
result, each position presented a small flux offset ($\lesssim2\%$).
Since the four nod positions are equally sampled throughout the entire
observation, they should each have the same mean level.
We corrected the flux offset by dividing each frame's flux by the
nodding-position mean flux and multiplying by the overal mean flux,
improving SDNR by $\sim6\%$.
We tested aperture photometry from 1.0 to 5.0 pixels.  In all visits
the SDNR minimum was at an aperture of 1.5 pixels; however,
optimal photometry outperformed aperture photometry
(Figure~\ref{fig:sdnrdepthIRS}).  The second visit provided the
clearest model determination (Table~\ref{table:tr001bs52ramps}).

\begin{table}[th]
\centering
\caption[16-{\microns} Eclipse, Visit 2---Individual Ramp Model Fits]
        {16 {\microns} Eclipse, Visit 2---Individual Ramp Model Fits\sp{a}}
\label{table:tr001bs52ramps}
\begin{tabular}{ccccc}
\hline
\hline
$R(t)$     & Ecl. Depth (\%) & SDNR      & $\Delta$BIC & $p({\cal M}\sb{2}|D)$\\
\hline
linramp    & 0.50(24)        & 0.0233022 & 0.0         & $\cdots$  \\
no-ramp    & 0.40(19)        & 0.0235462 & 4.1         & 0.11 \\
quadramp   & 0.74(28)        & 0.0232539 & 5.3         & 0.06 \\
risingexp  & 0.68(22)        & 0.0232595 & 5.3         & 0.06 \\
\hline
\multicolumn{5}{l}{\footnotesize {\bf Note.} \sp{a} Fits for PSF-fit centering
  and optimal photometry.}
\end{tabular}
\end{table}

At the beginning of the third visit (40 frames, $\sim$ 28 min),
the target position departs from the rest by half a pixel; omitting
the first 40 frames did not improve SDNR.
The linear ramp model minimized BIC (Table
\ref{table:tr001bs53ramps}).  Even though $\Delta$BIC between the
linear and the no-ramp models was small, the no-ramp residuals showed
a linear trend, thus we are confident on having selected the best
model.  The eclipse light curve in this visit is consistent with that
of the second visit.

\begin{table}[th]
\centering
\caption[16-{\microns} Eclipse, Visit 3---Individual Ramp Model Fits]
        {16 {\microns} Eclipse, Visit 3---Individual Ramp Model Fits\sp{a}}
\label{table:tr001bs53ramps}
\begin{tabular}{ccccc}
\hline
\hline
$R(t)$     & Ecl. Depth (\%) & SDNR      & $\Delta$BIC & $p({\cal M}\sb{2}|D)$\\
\hline
linramp    & 0.48(21)        & 0.0233010 & 0.0   & $\cdots$  \\
no-ramp    & 0.24(18)        & 0.0234888 & 1.1   & 0.37 \\
quadramp   & 0.38(22)        & 0.0233004 & 5.8   & 0.05 \\
risingexp  & 0.48(20)        & 0.0233011 & 6.2   & 0.04 \\
\hline
\multicolumn{5}{l}{\footnotesize {\bf Note.} \sp{a} Fits for PSF-fit centering
  and optimal photometry.}
\end{tabular}
\end{table}

The eclipse of the first visit had the lowest S/N of
all. The free parameters in both minimizer and MCMC
easily ran out of bounds towards implausible solutions.
For this reason we determined the best model in a joint fit combining
all three visits.  The events shared the eclipse midpoint,
duration, depth, and ingress/egress times.
We used the best data sets and models from the second and third visits
and tested different ramp models for the first visit.
With this configuration, the linear ramp model minimized the BIC of the
joint fit (Table \ref{table:tr001bs51ramps}).
Here, the target locations in the first two nodding cycles also were
shifted with respect to the rest of the frames.  
Clipping them out improved the SDNR.  Figure
\ref{fig:c3lightcurves} shows the data and best-fitting light curves and
rms-{\vs}-bin size plot.

\begin{table}[th]
\centering
\caption[16-{\micron} Eclipse, Visit 1---Ramp Model Fits]
        {16 {\micron} Eclipse, Visit 1---Ramp Model Fits\sp{a}}
\label{table:tr001bs51ramps}
\begin{tabular}{ccccc}
\hline
\hline
$R(t)$     &Ecl. Depth (\%) & SDNR      & $\Delta$BIC      & $p({\cal M}\sb{2}|D)$ \\
tr001bs51  & Joint          & Joint     & Joint            & tr001bs51    \\
\hline
linramp    & 0.35(14)       & 0.0230156 & \phantom{0}0.00  & $\cdots$     \\
quadramp   & 0.32(14)       & 0.0230172 & \phantom{0}7.10  & 0.03         \\
risingexp  & 0.36(11)       & 0.0230152 & \phantom{0}7.31  & 0.02         \\
no-ramp    & 0.33(13)       & 0.0231502 & \phantom{}10.16  & $6\tttt{-3}$ \\
\hline
\multicolumn{5}{l}{\footnotesize {\bf Note.} \sp{a} Fits for PSF-fit centering
  and optimal photometry.}
\end{tabular}
\end{table}

\subsubsection[IRS-16 {\micron} Transit]{IRS-16 $\mu$m Transit}
\label{sec:tr001bp5}

To fit this light curve we used the \citet{MandelAgol2002ApJtransits}
small-planet transit model with a quadratic limb-darkening law.  We
included priors on the model parameters that were poorly constrained
by our data.  We adopted $\cos(i)=0.0\sp{+0.019}\sb{-0.0}$ and
$a/R\sb{\star}=10.52\sp{+0.02}\sb{-0.18}$ from
\citet{TorresEtal2011StellarReanalysis} and the quadratic-limb darkening
coefficients $u\sb{1}=0.284\pm0.061$ and $u\sb{2}=0.21\pm0.12$, which
translate into our model parameters as
$c\sb{2}=u\sb{1}+2u\sb{2}=-0.7\pm0.25$ and
$c\sb{4}=-u\sb{2}=-0.21\pm0.12$ (with $c\sb{1} = c\sb{3} = 0$) from
\citet{WinnEtal2007apjTres1}.  The
midpoint and planet-to-star radius ratio completed the list of
free parameters for the transit model.

\begin{figure}[th]
\centering
\includegraphics[width=0.8\linewidth, clip]{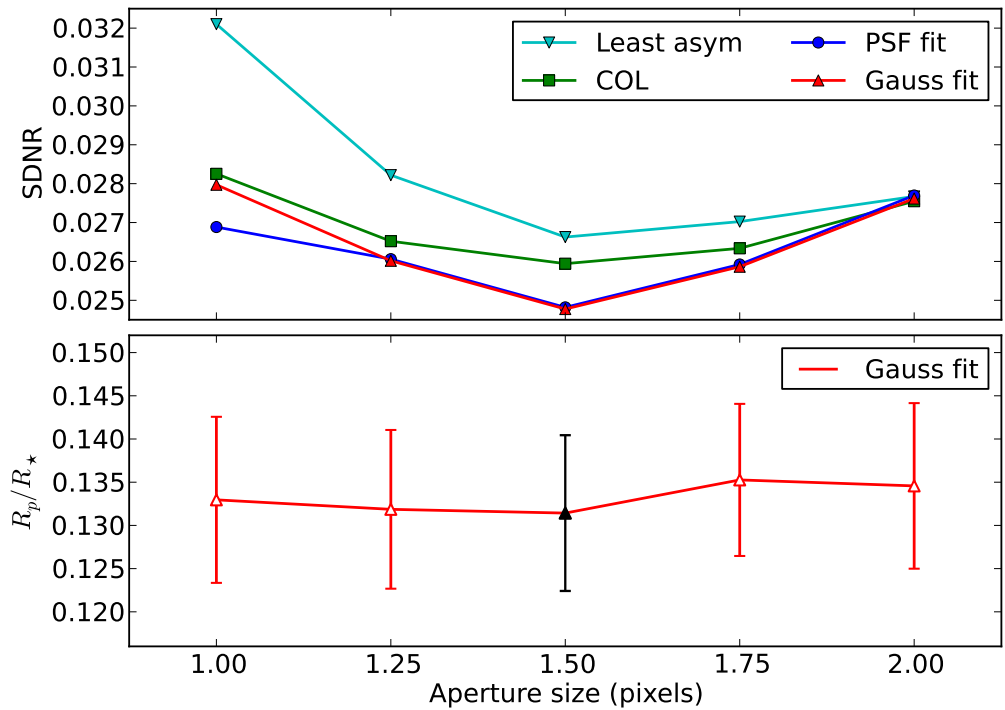}
\caption[Light curve's SNDR and planet-to-star radius ratio
 at 16.0
{\microns}]{{\bf Top:} 16-{\micron} transit light-curve SDNR {\vs}
  aperture.  All curves used the best ramp model from Table
  \ref{table:tr001bp51ramps}. {\bf Bottom:} Planet-to-star radius
  ratio {\vs} aperture for least asymmetry centering, with the best
  aperture (2.75 pixels) in black.}
\label{fig:sdnrdepth31transit}
\end{figure}

\begin{figure}[thb]
\centering
\includegraphics[width=0.5\linewidth, clip]{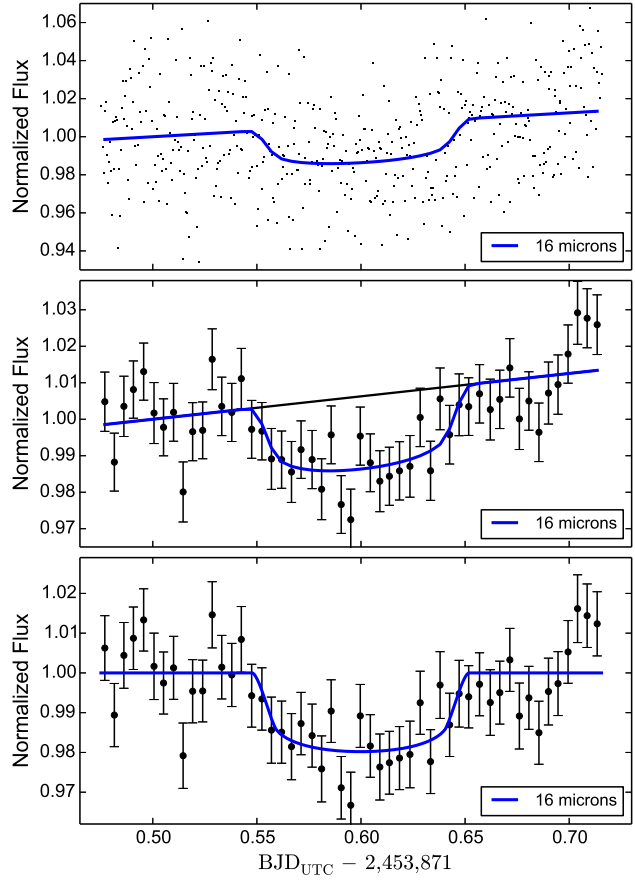}
\caption[Raw, binned, and systematics-corrected normalized TrES-1
transit light curves at 16 {\microns}]{Raw (top), binned (middle), and
  systematics-corrected (bottom) normalized TrES-1 transit light
  curves at 16 {\microns}.  The colored curves are the best-fit
  models.  The black curve is the best-fit model excluding the transit
  component.  The error bars are $1\sigma$ uncertainties.}
\label{fig:transitfit}
\end{figure}

We tested aperture photometry between 1 and 2 pixels, finding the SDNR
minimum at 1.5 pixels for the Gaussian-fit centering method
(Figure \ref{fig:sdnrdepth31transit}).  Table
\ref{table:tr001bp51ramps} shows the ramp-model fitting results.  The
linear ramp minimized BIC followed by the quadratic ramp with a 0.33
fractional probability; however, the quadratic fit shows an
unrealistic upward curvature due to high points at the end of the
observation.  Figures \ref{fig:transitfit} and
\ref{fig:c3lightcurves} show the best fit to the light curve and the
rms-{\vs}-bin size plot, respectively.

\begin{table}[th]
\centering
\caption[16-{\microns} Transit---Ramp Model Fits]
        {16 {\microns} Transit---Ramp Model Fits\sp{a}}
\label{table:tr001bp51ramps}
\begin{tabular}{ccccc}
\hline
\hline
$R(t)$     & $R\sb{p}/R\sb{\star}$ & SDNR      & $\Delta$BIC & $p({\cal M}\sb{2}|D)$ \\
\hline
linramp    &  0.1314(86)        & 0.0247755  & 0.0         & $\cdots$       \\
quadramp   &  0.1069(224)       & 0.0247118  & 1.4         & 0.33           \\
risingexp  &  0.1314(92)        & 0.0247757  & 6.2         & 0.04           \\
logramp    &  0.1316(81)        & 0.0247768  & 6.3         & 0.04           \\
no-ramp    &  0.1306(89)        & 0.0250938  & 6.9         & 0.03           \\
\hline
\multicolumn{5}{l}{\footnotesize {\bf Note.} \sp{a} Fits for Gaussian-fit
                   centering and 1.5-pixel aperture photometry.}
\end{tabular}
\end{table}

\subsubsection{Joint-fit Analysis} 
\label{sec:joint}

We used the information from all eclipse light curves combined to
perform a final joint-fit analysis.  The simultaneous fit shared a
common eclipse duration, eclipse midpoint and eclipse ingress/egress
time among all light curves.  Additionally, the three IRS eclipses
shared the eclipse-depth parameter.  We further released the duration
prior (which assumed a circular orbit).  We also performed experiments
related to the 3.6 and 4.5 {\micron} datasets.

First, to corroborate our selection of the 3.6 {\micron} model, we
compared the different 3.6 {\micron} models in the joint-fit
configuration both with the shared-midpoint constraint and with
independently-fit midpoints per waveband (Tables
\ref{table:jointramps} and \ref{table:jointmidpoint}).

\begin{table}[ht]
\centering
\caption[3.6-{\micron} Eclipse Models---Eclipse-joint Fits]
        {3.6 {\micron} Eclipse Models---Eclipse-joint Fits}
\label{table:jointramps}
\begin{tabular}{lcccc}
\hline
\hline
$R(t)A(a)$ & $\Delta$BIC   & 3.6 {\micron} Ecl. & Midpoint & Duration  \\
           & 3.6 {\micron} & Depth (\%)         & (phase)  & (phase)   \\  
\hline
\multicolumn{5}{c}{Independently fit midpoints\sp{a}\hspace{0.2cm}:} \\
\hline
$A(a)$     &   0.0         & 0.09(2)    & $\cdots$  & 0.032(1)  \\
quadramp   &   2.3         & 0.16(2)    & $\cdots$  & 0.032(1)  \\
risingexp  &   2.9         & 0.15(2)    & $\cdots$  & 0.032(1)  \\
linramp    &   6.9         & 0.10(2)    & $\cdots$  & 0.032(1)  \\
\hline
\multicolumn{5}{c}{Shared midpoint:}                                   \\
\hline
$A(a)$     &    0.0        & 0.08(2)    & 0.5015(6)        & 0.0328(9) \\
quadramp   &   13.3        & 0.14(3)    & 0.5013(5)        & 0.0331(9) \\
linramp    &   14.2        & 0.08(2)    & 0.5015(6)        & 0.0328(9) \\
risingexp  &   15.0        & 0.12(2)    & 0.5013(5)        & 0.0330(9) \\
\hline
\multicolumn{5}{l}{\footnotesize {\bf Note.} \sp{a} Midpoint values in
 Table \ref{table:jointmidpoint}.}
\end{tabular}
\end{table}

\begin{table}[ht]
\centering
\caption[Midpoint per Waveband---Eclipse-joint Fit]
        {Midpoint per Waveband---Eclipse-joint Fit}
\label{table:jointmidpoint}
\begin{tabular}{lccccc}
\hline
\hline
$R(t)A(a)$ & 3.6 {\micron} & 4.5 {\micron} & 5.8 {\micron} & 8.0 {\micron} & 16 {\micron} \\
          & (phase)  & (phase)  & (phase)  & (phase)  & (phase)  \\
\hline
$A(a)$    & 0.500(3) & 0.503(1) & 0.502(4) & 0.501(1) & 0.499(3) \\
quadramp  & 0.493(2) & 0.503(1) & 0.502(4) & 0.501(1) & 0.500(4) \\
risingexp & 0.493(1) & 0.503(1) & 0.502(4) & 0.501(1) & 0.500(3) \\
linramp   & 0.491(1) & 0.503(1) & 0.507(4) & 0.501(1) & 0.499(3) \\
\hline
\end{tabular}
\end{table}

\begin{figure}[thb]
\centering
\includegraphics[width=0.7\linewidth, clip]{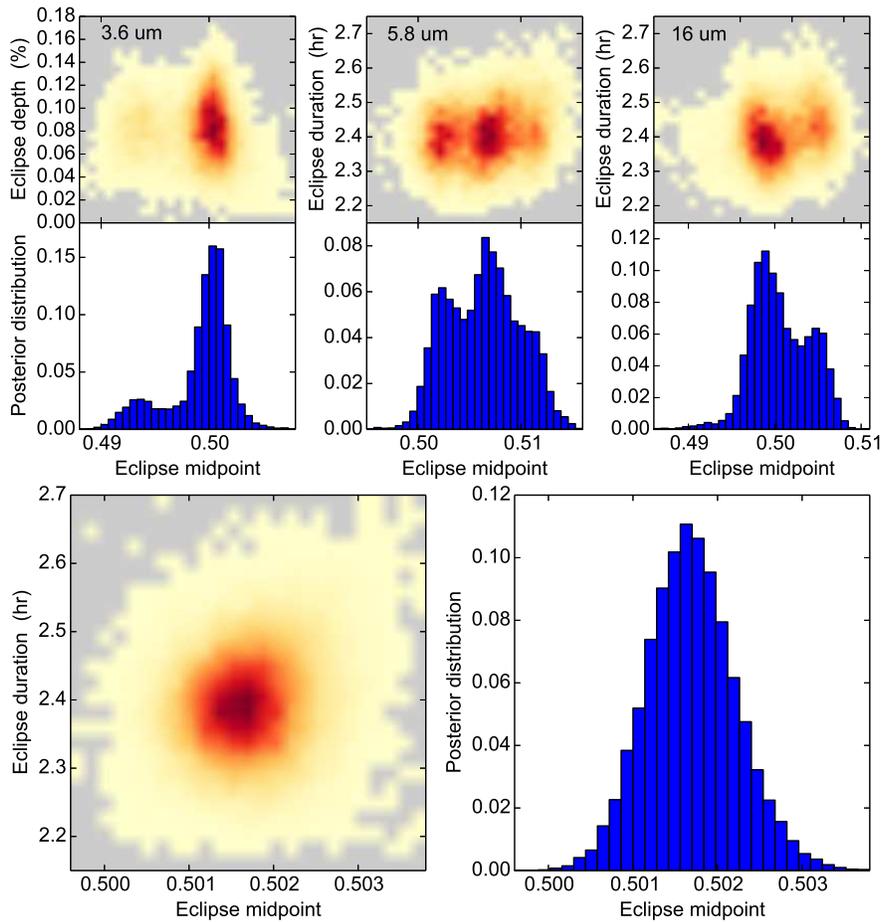}
\caption[Eclipse-midpoint pairwise and marginal
posteriors]{Eclipse-midpoint pairwise and marginal posteriors.  {\bf
    Top:} Independently-fit posterior eclipse depth or duration {\em
    vs.} midpoint for (left to right) 3.6, 5.8, and 16.0 {\microns}.
  The multi-modality did not replicate for the eclipse depth (same
  eclipse depth for each of the posterior modes).  {\bf Bottom:}
  Eclipse-duration {\em vs.} midpoint pairwise (left) and midpoint
  marginal (right) posterior distributions for the fit with shared
  midpoint.}
\label{fig:bimodalphase}
\end{figure}

All wavebands other than 3.6 {\microns} agreed with an eclipse
midpoint slightly larger than 0.5.  When we fit the midpoint
separately for each waveband, only the AOR-scale model at 3.6
{\microns} agreed with the other bands' midpoint (note that the 5.8
{\micron} data were obtained simultaneously with the 3.6 {\micron}
data, and should have the same midpoint).  The posterior distributions
also showed midpoint multimodality between these two solutions (Figure
\ref{fig:bimodalphase}).  On the other hand, with a shared midpoint,
the 3.6 {\micron} band assumed the value of the other bands for all
models, with no multimodality.  All but the AOR-scale model showed
time-correlated noise, further supporting it as the best choice.

Second, we investigated the impact of the (potentially corrupted)
4.5~{\micron} data set on the joint-fit values.  Excluding the
4.5~{\micron} event from the joint fit does not significantly alter
the midpoint (phase $0.5011~\pm~0.0006$) nor the duration
($0.0326~\pm~0.013$).  Our final joint fit configuration uses
the AOR-scaling model for the 3.6~{\micron} band, includes the
4.5~{\micron} light curve, and shares the eclipse midpoint (Table
\ref{table:jointfits}).

\subsubsection[4.5 and 8.0 {\micron} Eclipse Reanalyses]
              {4.5  and 8.0 $\mu$m Eclipse Reanalyses}
\label{sec:comparison}

Our current analysis methods differ considerably from those of nearly
a decade ago, with better centering, subpixel aperture photometry,
BLISS mapping, simultaneous fits across multiple data sets, and
evaluation of multiple models using BIC.
Furthermore, MCMC techniques were not yet prominent in most
exoplanet analyses, among other improvements.
\citet{CharbonneauEtal2005apjTrES1} used two field stars (with similar
magnitudes to TrES-1) as flux calibrators.  They extracted light
curves using aperture photometry with an optimal aperture of 4.0
pixels, based on the rms of the calibrators' flux.  At 4.5 {\microns},
they decorrelated the flux from the telescope pointing, but gave no details.
At 8.0 {\microns}, they fit a third-order
polynomial to the calibrators to estimate the ramp.
Their eclipse model had two free parameters (depth and midpoint),
which they fit by mapping {\chisq} over a phase-space grid.  Table
\ref{table:C05comparison} compares their eclipse depths with ours,
showing a marginal 1$\sigma$ difference at 4.5 {\microns}.  In
both channels our MCMC found larger eclipse-depth uncertainties
compared to those of \citet{CharbonneauEtal2005apjTrES1}, who 
calculated them
from the $\chi\sp{2}$ contour in the phase-space grid.
The introduction of MCMC techniques and the further use of more
efficient algorithms (e.g., differential-evolution MCMC) that converge
faster enabled better error estimates.  In the past, for example, a
highly-correlated posterior prevented the MCMC convergence of some
nuisance (systematics) parameters.  The non-convergence forced one to
fix these parameters to their best-fitting values.  In current
analyses, however, marginalization over nuisance parameters often
leads to larger but more realistic error estimates.

\begin{table}[th]
\centering
\caption[Eclipse-depth Reanalysis]
        {Eclipse-depth Reanalysis}
\label{table:C05comparison}
\begin{tabular}{lcc}
\hline
\hline
Eclipse depth                        & 4.5 {\microns} & 8.0 {\microns} \\
 (\%)                                &                &                \\
\hline
\citet{CharbonneauEtal2005apjTrES1}  & 0.066(13)      & 0.225(36)      \\
This work                            & 0.094(24)      & 0.213(42)      \\
\hline
\end{tabular}
\end{table}

The muxbleed correction was likely less accurately made than required
for atmospheric characterization, given the presence of a visible
muxbleed trail in the background near the star.  We cannot easily
assess either the uncertainty or the systematic offset added by the
muxbleed and its correction, given, e.g., that the peak pixel flux
varies significantly with small image motions.  Our stated 4.5
{\micron} uncertainty contains no additional adjustment for this
unquantified noise source, which makes further use of the 4.5
{\micron} eclipse depth difficult.  However, our minimizer and the
{\chisq} map of \citeauthor{CharbonneauEtal2005apjTrES1} clearly find
the eclipse, so the timing and duration appear less affected than the
depth.  In the analyses below, we include fits both with and without
this dataset.  The large uncertainty found by MCMC limits the 4.5
{\micron} point's influence in the atmospheric fit.

\section{Orbital Dynamics}
\label{sec:c3orbit}

As a preliminary analysis, we derived $e\cos(\omega)$ from the eclipse
data alone.
Our seven eclipse midpoint times straddle
phase 0.5.  After subtracting a light-time correction of $2a/c = 39$
seconds, where $a$ is the semimajor axis and $c$ is the speed of
 light, we found an eclipse phase of $0.5015 \pm 0.0006$.  This
implies a marginal non-zero value for $e\cos(\omega)$ of $0.0023 \pm
0.0009$ \citep[under the small-eccentricity approximation,][]{CharbonneauEtal2005apjTrES1}.

It is possible that a non-uniform brightness emission from the planet
can lead to non-zero measured eccentricity \citep{WilliamsEtal2006apj}.  For
example, a hotspot eastward from the substellar point can simulate a
late occultation ingress and egress compared to the uniform-brightness
case.  However, as pointed out by \citep{deWitEtal2012aapFacemap}, to constrain
the planetary brightness distribution requires a higher photometric
precision than what TrES-1 can provide.

Further, using the MCMC routine described by
\citet{CampoEtal2011apjWASP12b}, we fit a Keplerian-orbit model to our
secondary-eclipse midpoints simultaneously with 33 radial-velocity
(Table \ref{table:rv})
and 84 transit data points (Table \ref{table:transit}).
We discarded nine radial-velocity points that were affected by the
Rossiter-McLauglin effect.  We 
were able to constrain $e\cos(\omega)$ to $0.0017 \pm 0.0003$.
Although this 3$\sigma$ result may suggest a non-circular orbit,
when combined with the fit to $e\sin(\omega)$ of $-0.033 \pm 0.025$,
the posterior distribution for the eccentricity only indicates a
marginally eccentric orbit with $e = 0.033\sp{+0.015}\sb{-0.031}$.
Table \ref{table:orbit} summarizes our orbital MCMC results.

\begin{table}[ht]
\vspace{-10pt}
\centering
\caption[MCMC Eccentric Orbital Model]{MCMC Eccentric Orbital Model}
\label{table:orbit}
\begin{tabular}{lc}
\hline
\hline
Parameter                              & Best-fitting Value                \\
\hline
\math{e\sin\omega}                     & $-0.033$    {\pm}  0.025          \\
\math{e\cos\omega}                     &   0.0017    {\pm}  0.0003         \\
\math{e}                               &   0.033 $\sp{+0.015}\sb{-0.031}$  \\
\math{\omega} (\degree)                & 273      $\sp{+1.4}\sb{-2.8}$     \\
Orbital period (days)                  & 3.0300699   {\pm}  $1\tttt{-7}$   \\
Transit time, $T\sb{0}$ (MJD)\sp{a}   & 3186.80692  {\pm}  0.00005        \\
RV semiamplitude, $K$ ({\ms})          & 115.5       {\pm}  3.6            \\
system RV, $\gamma$ ({\ms})            & $-3.9$      {\pm}  1.3            \\
Reduced \math{\chi\sp{2}}              & 6.2                               \\
\hline
\multicolumn{2}{l}{\footnotesize {\bf Note.} \sp{a} MJD = BJD\sb{TDB}$-2,450,000$.}
\end{tabular}
\end{table}

\section{Atmosphere}
\label{sec:c3atmosphere}

We modeled the day-side emergent spectrum of TrES-1 with the retrieval
method of \citet{MadhusudanSeager2009apjRetrieval} to constrain the
atmospheric properties of the planet.  The code solves the
plane-parallel, line-by-line, radiative transfer equations subjected
to hydrostatic equilibrium, local thermodynamic equilibrium, and
global energy balance.  The code includes the main sources of opacity
for hot Jupiters: molecular absorption from H$\sb{2}$O, CH$\sb{4}$,
CO, and CO$\sb{2}$ \citep[Freedman, personal communication
2009]{FreedmanEtal2008apjsOpacities}, and H$_2$-H$_2$ collision
induced absorption \citep{Borysow2002H2H2}.  We assumed a Kurucz
stellar spectral model \citep{CastelliKurucz2004}.

The model's atmospheric temperature profile and molecular abundances
of H$\sb{2}$O, CO, CH$\sb{4}$, and CO$\sb{2}$ are free parameters,
with the abundance parameters scaling initial profiles that are in
thermochemical equilibrium.  The output spectrum is integrated over
the {\Spitzer} bands and compared to the observed eclipse depths by
means of $\chi\sp{2}$.  An MCMC module supplies millions of parameter
sets to the radiative transfer code to explore the phase space
\citep{MadhusudhanSeager2010apj, MadhusudhanSeager2011apjGJ436b}.

Even though the features of each molecule are specific to certain
wavelengths \citep{MadhusudhanSeager2010apj}, our independent
observations (4 or 5) are less than the number of free parameters
(10), and thus the model fitting is a degenerate problem.  Thus, we
stress that our goal is not to reach a unique solution, but to discard
and/or constrain regions of the parameter phase space given the
observations, as has been done in the past
\citep[e.g.,][]{BarmanEtal2005, Burrows2007HD209,
  KnutsonEtal2008apjHD209, MadhusudanSeager2009apjRetrieval,
  StevensonEtal2010natGJ436b, MadhusudhanEtal2011natWASP12batm}.

\begin{figure*}[th]
\centering
\includegraphics[width=0.67\linewidth,  clip]{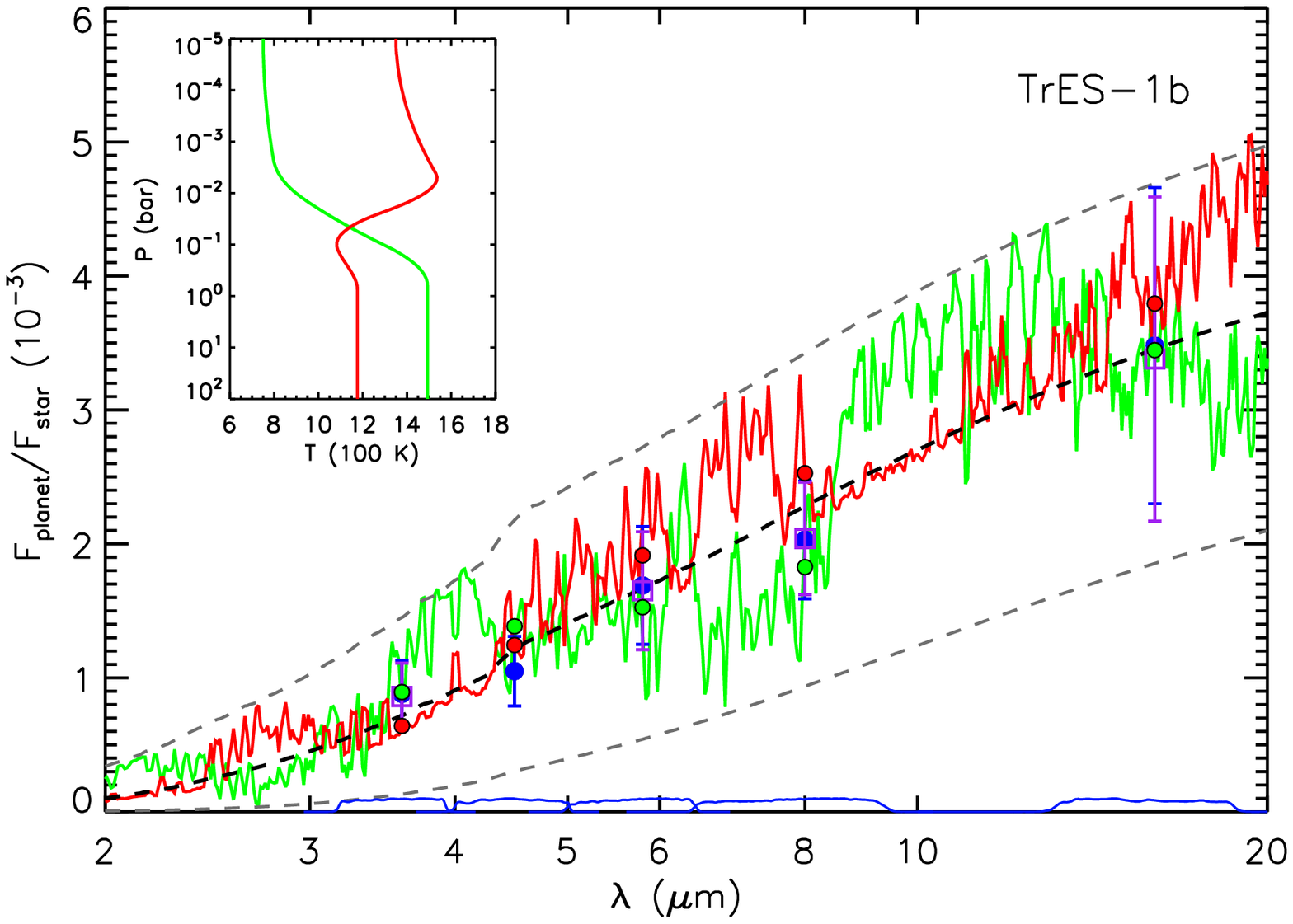}\hfill
\includegraphics[width=0.315\textwidth, clip]{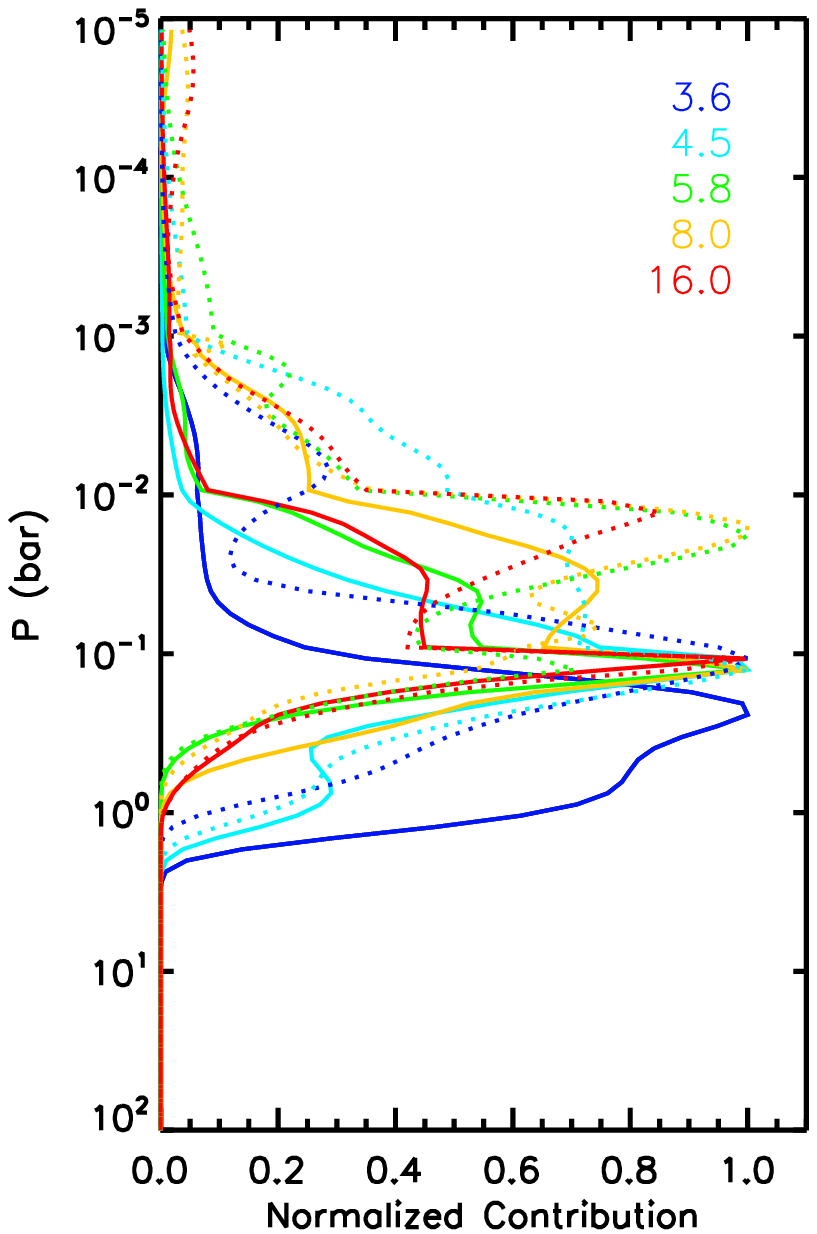}
\caption[Dayside atmospheric spectral emission and {Spitzer}-band
contribution functions of TrES-1]{{\bf Left:} Dayside atmospheric
  spectral emission of TrES-1.  The blue circles and purple squares
  with error bars are the measured eclipse depths (including and
  excluding the 4.5 {\microns} data point, respectively).  The red and
  green curves show representative model spectra with and without
  thermal inversion (see inset), based on the data including the 4.5
  {\micron} point.  Results omitting this point are similar.  Both
  models have a solar abundance atomic composition and are in chemical
  equilibrium for the corresponding temperature profiles.  The red and
  green circles give the band-integrated (bottom curves) fluxes of the
  corresponding models, for comparison to data.  The dashed lines
  represent planetary blackbody spectra with $T$ = 800, 1200, and 1500
  K.  {\bf Right:} Normalized contribution functions of the models
  over each {\Spitzer} band (see legend).  The dotted and solid lines
  are for the models with and without thermal inversion,
  respectively.}
\label{fig:c3atm}
\end{figure*}

Figure \ref{fig:c3atm} shows the TrES-1 data points and model spectra of
its day-side emission.  An isothermal model can fit the observations
reasonably well, as shown by the black dashed line (blackbody
spectrum with a temperature of 1200 K).  However, given the low S/N of
the data, we cannot rule out non-inverted nor strong thermal-inversion
models (with solar abundance composition in chemical equilibrium), as
both can fit the data equally well (green and red models).
Generally speaking, the data allow for efficient day-night
heat redistribution;  the models shown have maximum possible
heat redistributions of 60\% (non-inversion model) and 40\% (inversion
model).

As shown in Fig.\ \ref{fig:c3atm}, the data sets with and without the
4.5~{\micron} point are nearly identical.  Combined with the large
error bars (especially at 16 {\microns}), there is no significant difference
between the atmospheric model  results of the two cases. 
Both CO and CO$\sb{2}$ are dominant absorbers at 4.5 {\microns}.
Combined with the 16-{\micron} detection, which is mainly sensitive to
CO$\sb{2}$, the data could constrain the abundances of CO and CO$\sb{2}$.
Unfortunately, the error bar on the 16 {\micron} band is too large to
derive any meaningful constraint.

\section{Conclusions}
\label{sec:c3conclusions}

We have analyzed all the {\Spitzer} archival data for {\mbox{TrES-1}},
comprising eclipses in five different bands (IRAC and IRS blue
peak-up) and one IRS transit.
There has been tremendous improvement in data-analysis
techniques for {\Spitzer}, and exoplanet light curves in general,
since \citet{CharbonneauEtal2005apjTrES1}, one of the first two
reported exoplanet secondary eclipses.
A careful look at the 4.5 {\micron} data frames revealed pixels
affected by muxbleed that, although corrected by the {\Spitzer}
pipeline, still showed a clear offset output level.  Unable to know
the effect on the eclipse depth and uncertainty, we conducted
subsequent modeling both with and without the 4.5 {\micron} point.
The already-large uncertainty resulted in similar conclusions either
way.  Without adjusting our point for either the systematic or random
effects of the muxbleed correction, the depth and uncertainty at 4.5
{\micron} are both substantially larger than the original analysis.
However, at 8.0 {\microns} (which does not have similar problems) the
eclipse depths are consistent, with our MCMC giving a larger
uncertainty.

Our measured eclipse depths from our joint light-curve fitting (with and without the 4.5 {\micron} point) are
consistent with a nearly-isothermal atmospheric dayside emission at
$\sim1200$ K.  This is consistent with
the expected equilibrium temperature of 1150 K (assuming zero albedo
and efficient energy redistribution).  Furthermore, neither inverted
nor non-inverted atmospheric models can be ruled out, given the low
S/N of the data. Our transit analysis unfortunately does not improve
the estimate of the planet-to-star radius ratio
($R\sb{p}/R\sb{\star}=0.119{\pm}0.009$).  Our comprehensive orbital
analysis of the available eclipse, transit, and radial-velocity data
indicates an eccentricity of $e = 0.034\sp{+0.014}\sb{-0.032}$,
consistent with a circular orbit at the 1$\sigma$-level.  Longitudinal
variations in the planet's emission can induce time offsets in eclipse
light curves, and could mimic non-zero eccentricities
\citep[e.g.,][]{WilliamsEtal2006apj, deWitEtal2012aapFacemap}.
However, the S/N required to lay such constraints are much higher than
that of the TrES-1 eclipse data.

We also described the latest improvements of our POET pipeline.
Optimal photometry provides an alternative to aperture photometry.
We first applied optimal photometry in
\citet{StevensonEtal2010natGJ436b}, but describe it in more detail here.
Furthermore, the Differential-Evolution Markov-chain algorithm poses
an advantage over a Metropolis Random Walk MCMC, since it
automatically tunes the scale and orientation of the proposal
distribution jumps.  This dramatically increases the algorithm's
efficiency, converging nearly ten times faster.  We also now
avoid the need to orthogonalize highly correlated posterior
distributions.  

\section{Acknowledgements}

We thank D.\ Charbonneau for sharing the original {\Spitzer} pipeline
data for the 4.5 and 8.0 {\microns} bands and for helpful
discussions. We thank the amateur observers from ETD, including
Alfonso Carre\~no Garcer\'an, Zonalunar Observatory; Ferran Grau, Ca
l'Ou observatory, Sant Mart'i Sesgueioles;
 Hana Ku$\check{\rm c}$\'akov\'a,
Altan Observatory, Czech Republic and Johann Palisa
Observatory and Planetarium, Technical University Ostrava, Czech
Republic; Prof.\ Dr.\ Johannes\ M.\ Ohlert; Christopher De Pree, Agnes
Scott College and SARA; Peter Roomian, College of San Mateo
Observatory; Stan Shadick, Physics and Engineering Physics Dept.,
University of Saskatchewan; Bradley Walter, Meyer Observatory, Central
Texas Astronomical Society.  Thanks Colo Colo for its Camp30n
campaign.  We thank contributors to SciPy, Matplotlib, and the Python
Programming Language, the free and open-source community, the NASA
Astrophysics Data System, and the JPL Solar System Dynamics group for
software and services.  PC is supported by the Fulbright Program for
Foreign Students.  This work is based on observations made with the
{\em Spitzer Space Telescope}, which is operated by the Jet Propulsion
Laboratory, California Institute of Technology under a contract with
NASA.  Support for this work was provided by NASA through an award
issued by JPL/Caltech and through the NASA Science Mission
Directorate's Astrophysics Data Analysis Program, grant NNH12ZDA001N.

\pagebreak

\section{Joint Best Fit}
\label{sec:app}

Table \ref{table:jointfits} summarizes the model setting and results
of the light-curve joint fit.  The midpoint phase parameter was
shared among the IRS eclipse observations.

Table \ref{table:rv} lists the aggregate TrES1 radial-velocity
measurements.

Table \ref{table:transit} lists the aggregate TrES1 transit-midpoint
measurements.



\begin{landscape}
\begin{table*}[ht]
\vspace{0.3cm}
\scriptsize
\centering
\caption{\label{table:jointfits} Best-Fit Eclipse Light Curve Parameters}
\begin{tabular}{rcccccccc}
\hline
\hline
Parameter                                       & tr001bs11       & tr001bs21\sp{a} & tr001bs31       & tr001bs41       & tr001bs51      & tr001bs52      & tr001bs53       & tr001bp51 \\
\hline
Centering algorithm                             & Gauss fit       & Center of Light & Least Asymmetry & Least Asymmetry & PSF fit        & PSF fit        & PSF fit         & Gauss fit   \\
Mean $x$ position (pix)                         & 119.95          & 169.02          & 113.74          & 167.92          & $\cdots$       & $\cdots$       & $\cdots$        & $\cdots$  \\
Mean $y$ position (pix)                         &  82.58          & 118.63          & 83.29           & 117.62          & $\cdots$       & $\cdots$       & $\cdots$        & $\cdots$  \\
$x$-position consistency\sp{b} (pix)           & 0.014           & 0.019           & 0.021           & 0.019           & 0.038          & 0.036          & 0.040           & 0.045     \\
$y$-position consistency\sp{b} (pix)           & 0.026           & 0.025           & 0.024           & 0.030           & 0.044          & 0.036          & 0.043           & 0.037     \\
Optimal/Aperture  photometry size (pix)         & 2.50            & 3.75            & 2.75            & 2.75            & optimal        & optimal        & optimal         &  1.5      \\
Inner sky annulus (pix)                         & 7.0             & 7.0             & 7.0             & 7.0             & 5.0            & 5.0            & 5.0             &  5.0      \\
Outer sky annulus (pix)                         & 15.0            & 15.0            & 15.0            & 15.0            & 10.0           & 10.0           & 10.0            & 10.0      \\
BLISS mapping                                   &  Yes            & Yes             & No              & No              &  No            &  No            &  No             & No        \\
Minimum Points Per Bin                          &   4             &  4              &  $\cdots$       & $\cdots$        &  $\cdots$      &  $\cdots$      &  $\cdots$       & $\cdots$  \\
System flux \math{F\sb{s}} (\micro Jy)          & 33191.4(5.9)    & 21787.0(2.3)    & 14184.5(3.3)    & 8440.7(2.3)     & 1792.3(2.1)    & 1797.2(2.3)    & 1796.6(2.3)     & 857(1.8)  \\
Eclipse depth (\%)                              & 0.083(24)       & 0.094(24)       & 0.162(42)       & 0.213(42)       & 0.33(12)       & 0.33(12)       & 0.33(12)        & $\cdots$   \\
Brightness temperature (K)                      & 1270(110)       & 1126(90)        & 1205(130)       & 1190(130)       & 1270(310)      & 1270(310)      & 1270(310)       & $\cdots$   \\
Eclipse midpoint (orbital phase)                & 0.5015(5)       & 0.5015(5)       & 0.5015(5)       & 0.5015(5)       & 0.5015(5)      & 0.5015(5)      & 0.5015(5)       & $\cdots$   \\
Eclipse/Transit midpoint (MJD\sb{UTC})\sp{c}   & 3630.7152(16)   & 3309.5283(16)   & 3630.7152(16)   & 3309.5283(16)   & 3873.1204(16)  & 3876.1504(16)  & 3879.1805(16)   & $3871.5998(38)$ \\
Eclipse/Transit midpoint (MJD\sb{TDB})\sp{c}   & 3630.7159(16)   & 3309.5290(16)   & 3630.7159(16)   & 3309.5290(16)   & 3873.1211(16)  & 3876.1512(16)  & 3879.1812(16)   & $3871.6005(38)$ \\
Eclipse/Transit duration (\math{t\sb{\rm 4-1}}, hrs) & 2.39(7)    & 2.39(7)         & 2.39(7)         & 2.39(7)         & 2.39(7)        & 2.39(7)        & 2.39(7)         & 2.496(33)  \\
Ingress/egress time (\math{t\sb{\rm 2-1}}, hrs) & 0.31(1)         & 0.31(1)         & 0.31(1)         & 0.31(1)         & 0.31(1)        & 0.31(1)        & 0.31(1)         & 0.28(2)    \\
$R\sb{p}/R\sb{\star}$                           & $\cdots$        & $\cdots$        & $\cdots$        & $\cdots$       & $\cdots$       & $\cdots$        & $\cdots$        & 0.1295(95) \\
$\cos(i)$                                       & $\cdots$        & $\cdots$        & $\cdots$        & $\cdots$       & $\cdots$       & $\cdots$        & $\cdots$        & $0.0\sp{+0.000008}\sb{-0.0}$   \\
$a/R\sb{\star}$                                 & $\cdots$        & $\cdots$        & $\cdots$        & $\cdots$       & $\cdots$       & $\cdots$        & $\cdots$        & $10.494\sp{+0.092}\sb{-0.135}$ \\
Limb darkening coefficient, $c2$                & $\cdots$        & $\cdots$        & $\cdots$        & $\cdots$       & $\cdots$       & $\cdots$        & $\cdots$        &   0.75(22)  \\
Limb darkening coefficient, $c2$                & $\cdots$        & $\cdots$        & $\cdots$        & $\cdots$       & $\cdots$       & $\cdots$        & $\cdots$        & $-0.19(11)$ \\
Ramp equation ($R(t)$)                          & $A(a)$          & None            & None            & linramp         & linramp        & linramp        & linramp         & linramp    \\
Ramp, linear term ($r\sb{1}$)                   & $\cdots$        & $\cdots$        & $\cdots$        & 0.2455(82)      & 0.182(49)      & 0.151(42)      & 0.118(47)       & 0.063(17)  \\
AOR scaling factor ($A(a\sb{2})$)               & 1.00234(33)     & $\cdots$        & $\cdots$        & $\cdots$        & $\cdots$       & $\cdots$       & $\cdots$        & $\cdots$   \\
Number of free parameters\sp{d}                & 6               & 5               & 5               & 6               & 6              & 6              & 6               & 8          \\
Total number of frames                          & 3904            & 1518            & 1952            & 1518            & 500            & 500            & 500             & 500        \\
Frames used\sp{e}                              & 3827            & 1407            & 1763            & 1482            & 460            & 500            & 500             & 492        \\
Rejected frames (\%)                            & 1.97            & 7.31            & 9.68            & 2.37            & 8.0            & 0.0            & 0.0             & 1.6        \\
BIC value                                       & 10103.0         & 10103.0         & 10103.0         & 10103.0         & 10103.0        & 10103.0        & 10103.0         & 533.4      \\
SDNR                                            & 0.0053766       & 0.0026650       & 0.0083273       & 0.0074324       & 0.0223287      & 0.0233603      & 0.0233306       & 0.0248263  \\
Uncertainty scaling factor                      & 0.946           & 1.065           & 1.186           & 0.962           & 0.543          & 0.574          & 0.590           & 0.489      \\
Photon-limited S/N (\%)                         & 99.34           & 89.67           & 74.04           & 63.01           & 8.34           & 7.98           & 7.99            & 10.7       \\
\hline
\multicolumn{9}{l}{{\bf Notes.}} \\
\multicolumn{9}{l}{\sp{a} Data corrupted by muxbleed.} \\
\multicolumn{9}{l}{\sp{b} rms frame-to-frame position difference.} \\
\multicolumn{9}{l}{\sp{c} MJD = BJD $-$ 2,450,000.} \\
\multicolumn{9}{l}{\sp{d} In the individual fits.} \\
\multicolumn{9}{l}{\sp{e} We exclude frames during instrument/telescope
  settling, for insufficient points at a given BLISS bin, and for bad
  pixels in the photometry aperture.}
\end{tabular}
\thispagestyle{empty}

\begin{minipage}[t]{\linewidth}
\vspace{1.0cm}
\centerline {\normalsize \phantom{aaaa} 91}
\end{minipage}
\end{table*}
\end{landscape}

\pagebreak

\begin{table}[ht]
\footnotesize
\centering
\caption[TrES-1 Radial-velocity Data]{TrES-1 Radial-velocity Data}
\label{table:rv}
\begin{tabular}{cr@{\,{\pm}\,}lc}
\hline
\hline
Date                   & \mctc{RV}               & Reference \\
BJD(TDB) $-$ 2450000.0 & \mctc{(m\,s$\sp{-1}$)}  &           \\
\hline
3191.77001             &  $  60.4 $    &  12.8   & 1   \\
3192.01201             &  $ 115.1 $    &   8.3   & 1   \\
3206.89101             &  $  87.1 $    &  16.0   & 1   \\
3207.92601             &  $  15.8 $    &  10.4   & 1   \\
3208.73001             &  $-113.3 $    &  15.0   & 1   \\
3208.91701             &  $ -98.1 $    &  19.8   & 1   \\
3209.01801             &  $-118.4 $    &  15.3   & 1   \\
3209.73101             &  $  49.8 $    &  15.7   & 1   \\
3237.97926             &  $  68.32$    &  3.66   & 2   \\
3238.83934             &  $-102.23$    &  3.27   & 2   \\
3239.77361             &  $ -24.53$    &  3.25   & 2   \\
3239.88499             &  $  10.00$    &  3.11   & 2   \\
3240.97686             &  $  70.68$    &  3.73   & 2   \\
3907.87017             &  $  18.7 $    &  14.0   & 3   \\
3907.88138             &  $  30.5 $    &  12.5   & 3   \\
3907.89261             &  $  54.6 $    &  12.0   & 3   \\
3907.90383             &  $  24.3 $    &  10.4   & 3   \\
\n3907.91505\sp{a}     &  $  26.4 $    &  11.4   & 3   \\
\n3907.92627\sp{a}     &  $  30.4 $    &  10.9   & 3   \\
\n3907.93749\sp{a}     &  $  22.4 $    &  14.3   & 3   \\
\n3907.94872\sp{a}     &  $   2.9 $    &  11.0   & 3   \\
\n3907.95995\sp{a}     &  $  -7.1 $    &  12.1   & 3   \\
\n3907.97118\sp{a}     &  $ -22.3 $    &  13.3   & 3   \\
\n3907.98240\sp{a}     &  $ -40.5 $    &  13.3   & 3   \\
\n3907.99363\sp{a}     &  $ -39.2 $    &  13.0   & 3   \\
\n3908.00487\sp{a}     &  $ -9.8  $    &  12.2   & 3   \\
\n3908.01609\sp{a}     &  $ -30.5 $    &  13.8   & 3   \\
3908.02731             &  $ -17.7 $    &  13.6   & 3   \\
3908.03853             &  $ -24.7 $    &  12.2   & 3   \\
3908.04977             &  $ -27.5 $    &  11.1   & 3   \\
3908.06099             &  $ -38.2 $    &  13.3   & 3   \\
3908.07222             &  $ -23.7 $    &  11.2   & 3   \\
3908.08344             &  $ -23.0 $    &   9.6   & 3   \\
\hline
\multicolumn{4}{l}{{\bf Note.}} \\
\multicolumn{4}{l}{\sp{a} Discarded due to Rossiter-McLaughlin effect.}  \\
\multicolumn{4}{l}{{\bf References.} (1) \citealp{AlonsoEtal2004apjTrES1disc};
                   (2) \citealp{LaughlinEtal2005TrES1followup};} \\
\multicolumn{4}{l}{(3) \citealp{NaritaEtal2007TrES1RLmeasurements}.}  \\
\end{tabular}
\end{table}

\pagebreak

\begin{table}[ht]
\scriptsize
\centering
\caption[TrES-1 Transit Midpoint Data]{TrES-1 Transit Midpoint Data}
\label{table:transit}
\begin{tabular}{p{4cm}p{3cm}l}
\hline
\hline
Midtransit Date         &  Error    & Source\sp{a} \\
BJD(TDB) $-$ 2450000.0  &           &        \\
\hline
6253.23986 & 0.00105 & ETD: Sokov E. N.                                    \\
6198.69642 & 0.00119 & ETD: Roomian P.                                     \\
6198.69600 & 0.00056 & ETD: Shadic S.                                      \\
6177.47937 & 0.00099 & ETD: Emering F.                                     \\
6168.39577 & 0.00042 & ETD: Mravik J., Grnja J.                            \\
6107.79376 & 0.00032 & ETD: Shadic S.                                      \\
6074.46334 & 0.00117 & ETD: Bachschmidt M.                                 \\
6074.46253 & 0.00112 & ETD: Emering F.                                     \\
6071.43377 & 0.00055 & ETD: Carre\~no                                      \\
6071.43165 & 0.00072 & ETD: Gaitan J.                                      \\
6071.43099 & 0.0007  & ETD: Horta F. G.                                    \\
5886.59953 & 0.00048 & ETD: Shadic S.                                      \\
5801.75506 & 0.0004  & ETD: Shadic S.                                      \\
5798.73056 & 0.00049 & ETD: Shadic S.                                      \\
5795.69991 & 0.00053 & ETD: Walter B., Strickland W., Soriano R.           \\
5795.69903 & 0.00064 & ETD: Walter B., Strickland W., Soriano R.           \\
5795.69797 & 0.00055 & ETD: Walter B., Strickland W., Soriano R.           \\
5777.51807 & 0.00056 & ETD: Centenera F.                                   \\
5768.42617 & 0.00042 & ETD: V. Krushevska, Yu. Kuznietsova, M. Andreev     \\
5765.39585 & 0.0004  & ETD: V. Krushevska, Yu. Kuznietsova, M. Andreev     \\
5762.36407 & 0.00037 & ETD: V. Krushevska, Yu. Kuznietsova, M. Andreev     \\
5759.33530 & 0.00049 & ETD: V. Krushevska, Yu. Kuznietsova, M. Andreev     \\
5707.81338 & 0.00093 & ETD: Marlowe H., Makely N., Hutcheson M., DePree C. \\
5680.55402 & 0.00064 & ETD: Sergison D.                                    \\
5671.46700 & 0.00114 & ETD: Ku\v{c}\'akov\'a H.                            \\
5671.46384 & 0.00088 & ETD: Vra\v{s}t\'ak M.                               \\
5671.46382 & 0.0009  & ETD: Br\'at L.                                      \\
5371.48766 & 0.00074 & ETD: Mihel\v{c}i\v{c} M.                            \\
5304.82572 & 0.00084 & ETD: Shadick S.                                     \\
5095.75034 & 0.00075 & ETD: Rozema G.                                      \\
5089.69043 & 0.00109 & ETD: Vander Haagen G.                               \\
5068.48006 & 0.00062 & ETD: Trnka J.                                       \\
5062.42088 & 0.00053 & ETD: Sauer T.                                       \\
5062.42078 & 0.00046 & ETD: Trnka J., Klos M.                              \\
5062.42012 & 0.00046 & ETD: D\v{r}ev\v{e}n\'y R., Kalisch T.               \\
5062.41959 & 0.0006  & ETD: Br\'at L.                                      \\
5062.41797 & 0.00102 & ETD: Ku\v{c}\'akov\'a H., Speil J.                  \\
4998.79649 & 0.0016  & ETD: Garlitz                                        \\
4971.51779 & 0.001   & ETD: Gregorio                                       \\
4968.48904 & 0.00192 & ETD: P\v{r}ib'ik V.                                 \\
4968.48811 & 0.00053 & ETD: Trnka J.                                       \\
4968.48753 & 0.00028 & ETD: Andreev M., Kuznietsova Y., Krushevska V.      \\
4671.54149 & 0.0021  & ETD: Mendez                                         \\
4662.44989 & 0.001   & ETD: Forne                                          \\
4383.68459 & 0.0019  & ETD: Sheridan                                       \\
4380.65579 & 0.0014  & ETD: Sheridan                                       \\
\end{tabular}
\end{table}

\pagebreak

\begin{table}[ht]
\scriptsize
\renewcommand\thetable{3.16}
\centering
\caption{TrES-1 Transit Midpoint Data --- (Continued)}
\begin{tabular}{p{4cm}p{3cm}l}
\hline
\hline
Midtransit Date         &  Error    & Source\sp{a} \\
BJD(TDB) $-$ 2450000.0  &           &               \\
\hline
4362.47423 & 0.0002  & \citet{HrudkovaEtal2009TTVsearch}                   \\
4359.44430 & 0.00015 & \citet{HrudkovaEtal2009TTVsearch}                   \\
4356.41416 & 0.0001  & \citet{HrudkovaEtal2009TTVsearch}                   \\
4356.41324 & 0.00096 & ETD: Andreev M., Kuznietsova Y., Krushevska V.      \\
4350.35296 & 0.00036 & ETD: Andreev M., Kuznietsova Y., Krushevska V.      \\
4347.32322 & 0.00028 & ETD: Andreev M., Kuznietsova Y., Krushevska V.      \\
3907.96406 & 0.00034 & \citet{NaritaEtal2007TrES1RLmeasurements}           \\
3901.90372 & 0.00019 & \citet{WinnEtal2007apjTres1}                        \\
3901.90371 & 0.0016  & \citet{NaritaEtal2007TrES1RLmeasurements}           \\
3898.87342 & 0.00014 & \citet{NaritaEtal2007TrES1RLmeasurements}           \\
3898.87341 & 0.00014 & \citet{WinnEtal2007apjTres1}              \\
3898.87336 & 0.00008 & \citet{WinnEtal2007apjTres1}              \\
3895.84298 & 0.00015 & \citet{NaritaEtal2007TrES1RLmeasurements} \\
3895.84297 & 0.00018 & \citet{WinnEtal2007apjTres1}              \\
3856.45180 & 0.0005  & ETD: Hentunen                             \\
3650.40752 & 0.00045 & ETD: NYX                                  \\
3550.41568 & 0.0003  & ETD: NYX                                  \\
3547.38470 & 0.0012  & ETD: NYX                                  \\
3256.49887 & 0.00044 & ETD: Ohlert J.                            \\
3253.46852 & 0.00057 & ETD: Pejcha                               \\
3253.46812 & 0.00038 & ETD: Ohlert J.                            \\
3247.40751 & 0.0004  & \citet{CharbonneauEtal2005apjTrES1}       \\
3189.83541 & 0.0019  & \citet{CharbonneauEtal2005apjTrES1}       \\
3186.80626 & 0.00054 & \citet{AlonsoEtal2004apjTrES1disc}        \\
3186.80611 & 0.0003  & \citet{CharbonneauEtal2005apjTrES1}       \\
3183.77521 & 0.0005  & \citet{CharbonneauEtal2005apjTrES1}       \\
3174.68641 & 0.0004  & \citet{CharbonneauEtal2005apjTrES1}       \\
2868.65031 & 0.0022  & \citet{CharbonneauEtal2005apjTrES1}       \\
2856.52861 & 0.0015  & \citet{CharbonneauEtal2005apjTrES1}       \\
2847.43631 & 0.0015  & \citet{CharbonneauEtal2005apjTrES1}       \\
2850.47091 & 0.0016  & \citet{CharbonneauEtal2005apjTrES1}       \\
3171.65231 & 0.0019  & \citet{CharbonneauEtal2005apjTrES1}       \\
3192.86941 & 0.0015  & \citet{CharbonneauEtal2005apjTrES1}       \\
3180.75291 & 0.0010  & \citet{CharbonneauEtal2005apjTrES1}       \\
4356.41492 & 0.00010 & \citet{HrudkovaEtal2009TTVsearch}         \\
4359.44506 & 0.00015 & \citet{HrudkovaEtal2009TTVsearch}         \\
4362.47499 & 0.00020 & \citet{HrudkovaEtal2009TTVsearch}         \\
\hline
\multicolumn{3}{l}{\footnotesize {\bf Note.} \sp{a} ETD: amateur transits from the Exoplanet
 Transit Database (\href{http://var2.astro.cz/ETD/index.php}{http://var2.astro.cz/ETD/index.php})} \\
\multicolumn{3}{l}{\footnotesize with reported error bars and quality indicator of 3 or better.}\end{tabular}
\end{table}

\pagebreak

\pagebreak

\section{Light Curve Data Sets}
\label{sec:app2}
All the light curve data sets are available in Flexible Image
Transport System (FITS) format in a tar.gz package in the electronic
edition.


\bibliographystyle{apj}
\bibliography{chap3-TrES1}

\chapter{ON CORRELATED-NOISE ANALYSES APPLIED TO EXOPLANET LIGHT CURVES}
\label{chap:rednoise}

{\singlespacing
\noindent{\bf 
Patricio Cubillos\sp{1},
Joseph Harrington\sp{1},
Thomas J. Loredo \sp{2},
Nate~B.~Lust\sp{1},
Jasmina Blecic\sp{1}, and
Madison~M.~Stemm\sp{1}
}

\vspace{1cm}

\noindent{\em
\sp{1}Planetary Sciences Group, Department of Physics, University of Central Florida, Orlando, FL 32816-2385, USA \\
\sp{2}Center for Radiophysics and Space Research, Space Sciences Building, Cornell University,
    Ithaca, NY 14853-6801, USA
}

\vspace{1cm}

\centerline{In preparation for {\em The Astrophisical Journal}.}


}

\clearpage
\setcitestyle{authoryear, round}

\section{Abstract}

  Time-correlated noise is a significant source of uncertainty when
  modeling exoplanet light-curve data.  A correct assessment of the
  correlated noise is fundamental to determining the true statistical
  significance of our findings.  In the field, the time-averaging,
  residual-permutation, and wavelet-based likelihood methods are three
  of the most widely used time-correlated noise estimators.  Yet,
  there are few studies that investigate these methods quantitatively.
  We have reviewed these three techniques and evaluated their
  performance on the eclipse-depth estimation for synthetic exoplanet
  light-curve data.  For the time-averaging method we found artifacts
  in the rms-vs.-bin-size curves that led to underestimated
  uncertainties.  This is most noticeable when the time scale of the
  feature under study is a large fraction of the total observation
  time.  We show that the residual-permutation method is unsound as a
  tool for estimating uncertainty in parameter estimates.  For the
  wavelet-likelihood method, we noted errors in the published
  equations and provide a list of corrections.  We further tested
  these techniques by injecting eclipse signals into a synthetic
  dataset and into a real Spitzer dataset, allowing us to assess the
  competence of the methods for data with controlled and real correlated noise,
  respectively.  Both the time-averaging and wavelet-likelihood
  methods improved the estimate of the eclipse-depth uncertainty with
  respect to a white-noise analysis (a Markov-chain Monte Carlo
  exploration that assumes uncorrelated data errors).  The
  time-averaging and wavelet-analysis methods estimated eclipse-depth
  errors within $\sim20-50$\% of the expected value.  Lastly, we
  present our open-source model-fitting tool, Multi-Core Markov-Chain
  Monte Carlo ({\mcc}).  This module uses Bayesian statistics to
  estimate the best-fitting values and the credible region of the
  (user-provided) model fitting parameters.  {\mcc} is a Python and C
  code, available to the community at
  https://github.com/pcubillos/MCcubed.

\section{Introduction}
\label{sec:c4introduction}

Whether our goal is the detection or the characterization of
exoplanets through transit or eclipse observations, the large
contrast between the stellar and planetary emission (about a thousand
times in the infrared) make the data analysis an intrinsically
challenging task.  For example, for the Spitzer Space Telescope, most
planetary signals \citep[e.g.,][]{StevensonEtal2010natGJ436b,
 DemoryEtal2012apjl55cnceEclipse} lie below the instrument's design
 criteria for
photometric stability \citep{FazioEtal2004apjsIRAC}.
Extracting planetary signals at this precision requires meticulous
data reduction.  Despite our best attempts to account for all known
systematics, time-correlated residuals (or red noise)
between the data and models often remain.  These systematics may originate
from instrumental or astrophysical sources, for example: stellar flux
variations from flares or granulation; imperfect flat fielding; or
telluric variations from 
changing weather conditions, differential extinction, or imperfect
telescope systematics corrections from changing telescope pointing.
Many authors have acknowledged correlated
noise as an important source of noise in time-series data sets
\citep[e.g.,][]{PontEtal2006mnrasRednoise, WinnEtal2007ajHATP1b,
 AgolEtal2010apjHD189, CubillosEtal2013apjWASP8b}.

Correlated noise affects both the accuracy and the precision of 
model parameters.  The typical statistical analyses neglect the
correlation between data points (e.g., likelihood functions based on
uncorrelated noise).  Hence, their estimated best-fitting values may
be biased, whereas their credible regions (Appendix
\ref{sec:CredRegion}) are too small, because they do not account for
correlated noise.
This paper examines three correlated-noise estimators found in
the exoplanet literature.
First, the time-averaging method \citep{PontEtal2006mnrasRednoise,
 WinnEtal2007ajHATP1b} compares the standard deviation of the data to
the (expected) uncorrelated-noise standard deviation, scaling the
uncertainties accordingly.
Next, the residual-permutation (or ``prayer bead'') method
\citep{BouchyEtal2005CorrNoise} uses a bootstrap algorithm that
preserves the structure of the residuals.
Lastly, \citet{CarterWinn2009apjWavelets} calculated the likelihood
function in a wavelet basis, where the correlation between the wavelet
coefficients is negligible.
Qualitatively speaking, these methods do return larger parameter
uncertainties for stronger correlated noise.  However, besides
\citet{CarterWinn2009apjWavelets}, there are few efforts to validate
their quantitative accuracy.

We have implemented these methods, testing them
with real and synthetic exoplanet eclipse data.  We did
not consider other correlated-noise modeling frameworks.  We
explicitly excluded Gaussian processes
\citep[][]{GibsonEtal2012mnrasGaussProc} since this technique becomes
computationally prohibitive for large datasets like the ones explored
in this work \citep[over $\sim 1000$ data
points,][]{Gibson2014mnrasModelSelecGP}.  With a focus on
atmospheric characterization, we concentrated on estimating the depth
of exoplanet light curves.  The eclipse depth constrains the
planet-to-star flux ratio, whereas the transit depth constrains the
planet-to-star radius ratio \citep[][]{SeagerDeming2010AnnualRev}.
Moreover, a multi-wavelength analysis of the depth samples the
planetary spectrum \citep[e.g.,][]{MadhusudhanSeager2009apjRetrieval},
constraining the atmospheric properties of the planet.

We focused on Spitzer observations, since they
represent the largest and best-quality sample of exoplanet data beyond
2 {\microns}.  Additionally, many of the Spitzer light curves show
time-correlated noise.  We simulated realistic Spitzer
secondary-eclipse observations in terms of known systematics,
signal-to-noise ratio (S/N), cadence, duration, and
eclipse shape.  Spitzer data are affected by two well-known
systematics: time-varying sensitivity (ramp) and intra-pixel sensitivity
variations \citep{KnutsonEtal2009apjHD149026bphase,
 CharbonneauEtal2005apjTrES1}.  Although several models have been
proposed to correct for these systematics
\citep{HarringtonEtal2007natHD149026b, KnutsonEtal2008apjHD209, BallardEtal2010paspIntraPixel,
 AgolEtal2010apjHD189, StevensonEtal2012apjHD149026b,
 LewisEtal2013apjHAT2bPhase, DemingEtal2014IntraPixelModel}, the
corrections are not always perfect, and thus many light-curve fits
exhibit time-correlated residuals.

In Section \ref{sec:estimators}, we summarize the model-fitting
estimation problem and review the correlated-noise estimators.  In
Section \ref{sec:simulations}, we test and compare the
correlated-noise methods by retrieving injected synthetic eclipse curves into synthetic and real light-curve data.  In Section \ref{sec:mc3} we present our open-source
module, Multi-Core Markov-Chain Monte Carlo (MC\sp{3}), to calculate
the model-parameters' credible regions.  Finally, in Section
\ref{sec:c4conclusions} we present our conclusions.

\section{Model-Fitting Estimation}
\label{sec:estimators}

A meaningful model-fitting analysis consists of two fundamental tasks,
estimating the model parameters' best-fitting values and estimating the
credible
region.  In this section we provide a brief description of the
model-fitting problem under the Bayesian-statistics perspective.  For
a more detailed discussion see, for example,
\citet{Gregory2005BayesianBook} and \citet{SiviaSkilling2006Bayesian}.
The following subsections then describe three of the most common
methods to account for correlated noise.

\subsection{Overview}
\label{sec:overview}

To determine the set of best-fitting values, {\bf x}, of a model, $M$,
given a data set, ${\bf D}$, one maximizes the posterior probability
distribution:
\begin{equation}
P({\bf x}|{\bf D},M) = P({\bf x}|M) P({\bf D}|{\bf x}, M),
\end{equation}
where $P({\bf x}|M)$ is the prior probability distribution and
$P({\bf D}|{\bf x}, M)$ is the likelihood function,
$\mathcal{L}({\bf x})$.  In the most general case, the likelihood
function is given by
\begin{equation}
\label{eq:generallike}
\mathcal{L}({\bf x}) =
   \frac{1}{(2\pi)\sp{N/2}|\Sigma|\sp{\frac{1}{2}}}
   \exp{\left(-\frac{1}{2} {\bf r}\sp{T}\Sigma\sp{-1}{\bf r} \right)},
\end{equation}
where $\Sigma$ is the data covariance matrix and
${\bf r}={\bf r}({\bf x})$ are the residuals between the data points
and the model.  If the data values are uncorrelated, the off-diagonal
terms in the covariance matrix become negligible, and the likelihood
function simplifies to:
\begin{equation}
\label{eq:whitelike}
\mathcal{L}({\bf x}) = \prod\sb{i}\frac{1}{\sqrt{2\pi\sigma\sb{i}\sp{2}}}
   \exp{\left(-\frac{r\sb{i}\sp{2}({\bf x})} {2\sigma\sb{i}\sp{2}}\right)},
\end{equation}
with $\sigma\sb{i}\sp{2} \equiv \Sigma\sb{ii}$ the variance of the
data point $i$.  When the data-point uncertainties are known,
maximizing the likelihood function translates into minimizing chi
squared, $\chisq = \sum (r\sb{i}/\sigma\sb{i})\sp{2}$.  The reduced chi
squared, {\redchisq} ({\chisq} divided by the number of degrees of
freedom), is the ratio of the variance of the fit to the variance of
the data \citep{BevingtonRobinson2003DataReduction}.  Then,
{\redchisq} serves as a measure of the goodness of fit; if the model
is a good approximation of the observations, we expect
$\redchisq \approx 1$.

In the Bayesian framework, the credible region of the parameters can be
estimated via the Markov-chain Monte Carlo (MCMC) algorithm.  The MCMC
method generates random samples from
the parameter phase space with a probability density proportional to
the posterior probability distribution, $P({\bf x}|{\bf D},M)$.  The
credible region for each parameter is then obtained from the interval
that contains a certain fraction of the highest posterior density
(typically 68\%, 95\%, or 99\%) of the marginalized posterior (see Appendix \ref{sec:CredRegion}).  For example, when the posterior
follows a normal distribution, the 68.3\% credible interval
corresponds to the interval contained within 1 standard deviation from
the mean.

The data uncertainties, $\sigma\sb{i}$, determine the span of the
sampled distribution: the smaller the uncertainties, the more
concentrated are the marginal posterior distributions.  
If there is no information on the data uncertainties, one might adopt a
uniform uncertainty value for every data point.  If the uncertainties
of a time-series data set are known to be under- (or over-) estimated
($\redchisq \ne 1$), one often scales them:
\begin{equation}
\sigma\sb{i}' = \sqrt{\redchisq}\sigma\sb{i}.
\end{equation}

An MCMC guided by Equation (\ref{eq:whitelike}) works well when the
data points are independent and identically distributed; however, it
does not account for time-correlated noise.  Alternatively, an MCMC
that uses the full covariance matrix of Equation (\ref{eq:generallike})
should account for correlated noise, although its calculation often
becomes computationally prohibitive.

\subsection{Time Averaging}
\label{sec:timeavg}

\citet{PontEtal2006mnrasRednoise} developed a method to estimate the
uncertainty of a transit or eclipse-depth calculation using the
light-curve data points themselves.  They considered the noise as the
sum in quadrature of two components, a purely white (uncorrelated)
source (characterized by a standard deviation per data point
$\sigma\sb{w}$), and a purely time-correlated source (characterized by
$\sigma\sb{r}$).  Considering this, the white-noise component of
the transit-depth uncertainty scales as:
\begin{equation}
\sigma\sb{d} = \frac{\sigma\sb{w}}{\sqrt{n}},
\label{eq:deptherrorwhite}
\end{equation}
with $n$ the number of data points in the transit.  On the other hand,
the time-correlated standard deviation, $\sigma\sb{r}$, is independent
of the number of data points.  Hence, for any given signal, the
uncertainty of a measurement should scale as:
\begin{equation}
\sigma\sb{d} = \sqrt{\frac{\sigma\sb{w}\sp{2}}{n} + \sigma\sb{r}\sp{2}}.
\label{eq:deptherror}
\end{equation}

Note that for small $n$, $\sigma\sb{d}$ is dominated by
$\sigma\sb{w}/\sqrt{n}$, whereas as $n$ increases, $\sigma\sb{d}$
approaches $\sigma\sb{r}$.  The time-averaging method uses this fact
to estimate the contribution from the correlated noise.  We followed
the procedure described by \citet{WinnEtal2007ajHATP1b}.  First, we
calculated the residuals between the data points and the best-fitting
model.  Then, we grouped the residuals in bins of $N$ elements each,
and calculated their mean values.  Lastly, we calculated the
standard deviation (or root mean squared, rms) of the binned
residuals, rms$\sb{N}$.  We repeated the process for a range of bin
sizes from one to half the data size.

Now, let $\sigma\sb{1}$ be the rms value of the non-binned residuals
(which is dominated by white noise).  In the absence of correlated
noise, the rms for the set of $M$ bins, each containing $N$ points, is
given by the extrapolation of $\sigma\sb{1}$
\citep{WinnEtal2008apjXO3bRedNoise}:
\begin{equation}
\sigma\sb{N} = \frac{\sigma\sb{1}}{\sqrt{N}} \sqrt{\frac{M}{M-1}}.
\label{eq:rmsvsbin}
\end{equation}

\begin{figure}[tb]
\centering
\includegraphics[width=0.7\linewidth, clip]{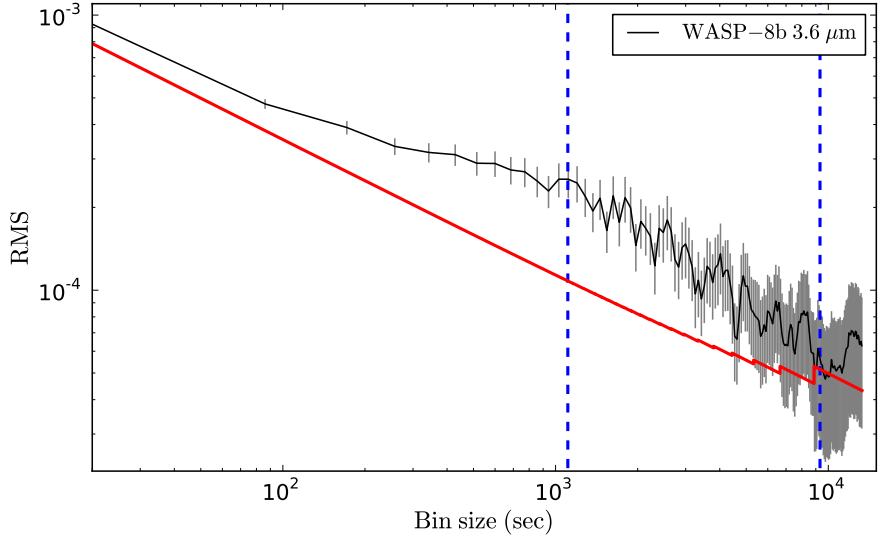}
\caption[Binned residuals rms vs.\ bin size of WASP-8b Spitzer eclipse]
        {Binned residuals rms vs.\ bin size (black curve with gray
 error bars) of WASP-8b Spitzer eclipse at 3.6 {\microns} (PI J. Harrington, Program ID 60003, see \citealp{CubillosEtal2013apjWASP8b}).  The red curve
 corresponds to the expected rms for white noise (Equation
 \ref{eq:rmsvsbin}).  The vertical dashed lines mark
 the durations of ingress/egress
 (left) and eclipse (right).  The $1\sigma$
 uncertainties of the rms curve correspond to
 $\sigma\sb{N}/\sqrt{2N}$ \citep{StevensonEtal2012apjHD149026b}; see
 Appendix \ref{sec:StdUncert} for the derivation.}
\label{fig:wasp8bRMS}
\end{figure}

The rms$\sb{N}$ and $\sigma\sb{N}$ curves are analogous to Equations
(\ref{eq:deptherror}) and (\ref{eq:deptherrorwhite}), respectively.
Their ratio, $\beta\sb{N} = {\rm rms}\sb{N} / \sigma\sb{N}$, serves as
a scaling factor to correct the uncertainties for time-correlated
noise.  One typically visualizes both curves in an rms vs.\ bin size
plot (Figure \ref{fig:wasp8bRMS}).

\subsubsection{Gaussian-noise Test}
\label{sec:TAzerotest}

We have noted in the literature that some exoplanet-fit rms curves
deviate below the $\sigma\sb{N}$ curve
\citep[e.g.,][]{StevensonEtal2012apjHD149026b,
 CubillosEtal2013apjWASP8b,
 BlecicEtal2013apjWASP14b}.  We
studied that behavior further by analyzing the rms curves for
zero-mean, random, normally-distributed signals.  Such signals represent
the residuals of an ideal fit to a data set affected only by
uncorrelated noise.  In this case, we expect the $\sigma\sb{N}$ curve
to be a good estimator of rms$\sb{N}$.

We generated four zero-mean normal-distribution sets of 1000 realization each, using the Python
routine \texttt{numpy.random.normal} with a standard deviation of 1.
We determined the number of datapoints per trial by simulating
cadences of 0.4 and 2 seconds and observation durations of 6 and 10
hours.  These configurations correspond to the typical values for
Spitzer transit or eclipse observations.

\begin{figure}[tb]
\centering
\includegraphics[width=0.7\linewidth, clip]{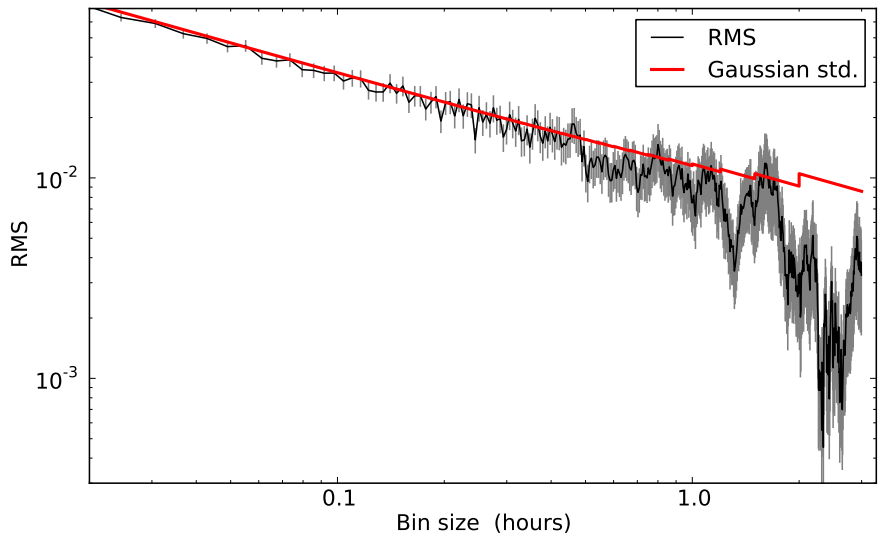}
\caption[rms vs.\ bin size for a zero-mean normal-distribution]
        {rms vs.\ bin size for a zero-mean normal-distribution.
 The error bars denote the $1\sigma$ uncertainties.  The red curve
 corresponds to $\sigma\sb{N}$.  There are 54,000 points,
 simulating a cadence of 0.4 seconds for a 6 hour observation.}
\label{fig:RMSzeroTrial}
\end{figure}

As expected, the rms curves deviated above $\sigma\sb{N}$ (by more
than one standard deviation) in less than 1\% of the trials.  On the
other hand, more than a third of the trials showed large deviations of
the rms curve ($> 3\sigma$) below $\sigma\sb{N}$ (Figure
\ref{fig:RMSzeroTrial}), particularly when the bin size approached the
observation time span.  All four sets returned similar fractions of
trials above and below $\sigma\sb{N}$.  These not uncommon artifacts
can partially or totally obstruct our ability to estimate correlated
noise.

\subsection{Residual Permutation}
\label{sec:prayer}

Residual permutation (also called prayer bead) is a nonparametric
bootstrapping method from frequentist statistics.  Nonparametric
bootstrapping methods directly use the sampled data (typically via
resampling) to generate a distribution that approximates the
probability distribution of the data.
The motivating idea is to shift the data while preserving the time
ordering and, thus, preserving the correlation structure.
While this is true, when there is
correlated noise, the shifted datasets do not correspond to
independent replications from any distribution, and thus do not
exhibit the variability necessary for uncertainty quantification
(e.g., computing confidence levels or estimator bias).

In the exoplanet field, the residual-permutation technique has been
repeatedly used to estimate parameter uncertainties.  However, the
name of the technique has been loosely used to describe similar, but
not equivalent procedures over the past decade.
\citet{BouchyEtal2005CorrNoise}, \citet{GillonEtal2007aaGJ436bspitz},
and \citet{Southworth2008HomogeneousStudyI} all describe different
methods, when referring to residual permutations.  Some authors
reference \citet{JenkinsEtal2002apjDetectionConfidence}, who actually
use a ``segmented bootstrap'', applying the method for
detection instead of parameter estimation.  Furthermore, several
authors have wrongly attributed the method to
\citet{MoutouEtal2004aaOgle132b}.

Currently, the most widely-used version of residual permutation is the
one described by \citet{Southworth2008HomogeneousStudyI} or
\citet{WinnEtal2008apjXO3bRedNoise}.  This implementation computes the
residuals between the light curve and the best-fitting model,
cyclically shifts the residuals (preserving the point-to-point
structure and thus the ``redness'' of the noise) by a given number of
data points, adds the residuals back to the model, and finds a new set
of best-fitting parameters.  Usually, either one repeats the
shift--fit process for a large number of iterations with random
shifts, or one sequentially shifts the residuals by one data point at
a time, fitting all possible shifts.  Each parameter uncertainty is
then given by the respective standard deviation of the distribution of
best-fitting values.  Section \ref{sec:synthetic} tests this
implementation for an example with synthetic data.  As expected, the
method fails to estimate the parameter uncertainties.

\subsection{Wavelet Analysis}
\label{sec:wavelet}

\citet{CarterWinn2009apjWavelets} introduced to the exoplanet field a
technique where the time-correlated noise is modeled using wavelet
transforms \citep{DericheTewfik1993tspLikeFractalNoise,
 Wornell1993ieeeWavelet,
 WornellOppenheim1992ieeeFractalSignalEstimation,
 WornellOppenheim1992ieeeWaveletFractalModulation}.  This method
projects the time-series residuals into an orthonormal wavelet basis,
where the off-diagonal terms of the covariance matrix become
negligible, thus simplifying the likelihood function calculation.
Furthermore, they assumed noise that has a power spectral density with frequency $f$, varying as $1/f\sp{\gamma}$. 
They parameterized
the noise with three
parameters, $\gamma$, $\sigma\sb{\omega}$, and $\sigma\sb{r}$, as described in Equations (41)--(43) of \citet{CarterWinn2009apjWavelets}.

A thorough review of wavelets is beyond the scope of this work;
see \citet{Mallat2008WaveletTour} and \citet{Wornell1996Wavelet} for 
more comprehensive discussions.  Briefly, a wavelet transform
projects a time-series signal onto a basis of functions that are
dilations and translations of a compact parent (``wavelet'') function.
The resulting transform has two dimensions, scale and time.
The discrete wavelet transform (DWT) consists of the hierarchical
application over $M$ dilation scales of an orthonormal wavelet
transform on a discrete time-series signal.  For a signal consisting
of $N=N\sb{0}2\sp{M}$ uniformly-spaced samples (with $N\sb{0}$
integer), and a wavelet function with $2N\sb{0}$ coefficients, the DWT
produces $N\sb{0}$ scaling coefficients and $N\sb{0}2\sp{m-1}$ wavelet
coefficients at each scale $m$, totaling $N\sb{0}(2\sp{M}-1)$ wavelet
coefficients.

\subsubsection{Wavelet-based Likelihood}
\label{sec:waveletlike}

The likelihood function in the wavelet analysis is calculated in the
following way.  Let $\epsilon(t)$ be the fitting residuals of a
time-series signal.  Considering $\epsilon(t)$ as the contribution of
a time-correlated ($\gamma \neq 0$) and an uncorrelated ($\gamma = 0$)
component:
\begin{equation}
\epsilon(t) = \epsilon\sb{\gamma}(t) + \epsilon\sb{0}(t),
\end{equation}
this method calculates the DWT of $\epsilon(t)$ to produce the
wavelet, $r\sb{n}\sp{m}$, and scaling, $\bar{r}\sb{n}\sp{1}$,
coefficients of the signal.  The variances of these coefficients are
estimated, respectively, as:
\begin{eqnarray}
\label{eq:varwavelet}
\sigma\sb{W}\sp{2} & = & \sigma\sb{r}\sp{2}2\sp{-\gamma m}          +
                        \sigma\sb{\omega}\sp{2} \\ 
\label{eq:varscale}
\sigma\sb{S}\sp{2} & = & \sigma\sb{r}\sp{2}2\sp{-\gamma}  g(\gamma) +
                        \sigma\sb{\omega}\sp{2},
\end{eqnarray}
where $\sigma\sb{\omega}$ and $\sigma\sb{r}$ parameterize the standard
deviation of the uncorrelated and the correlated-noise signals,
respectively.  See Appendix \ref{sec:WaveletVariance} for the
derivation of $g(\gamma)$.
Therefore, the wavelet-based likelihood function is given by
\begin{eqnarray}
\mathcal{L}({\bf x}, \sigma\sb{\omega}, \sigma\sb{r}) 
       & = & \left\{ \prod\sb{m=1}\sp{M} \prod\sb{n=1}\sp{N\sb{0}2\sp{m-1}}
             \frac{1}{\sqrt{2\pi\sigma\sb{W}\sp{2}}}
             \exp\left[{-\frac{(r\sb{n}\sp{m})\sp{2}}{2\sigma\sb{W}\sp{2}}} \right] \right\} \times \nonumber \\
         &  & \left\{ \prod\sb{n=1}\sp{n\sb{0}}
              \frac{1}{\sqrt{2\pi\sigma\sb{S}\sp{2}}}
              \exp\left[{-\frac{(\bar{r}\sb{n}\sp{1})\sp{2}}{2\sigma\sb{S}\sp{2}}} \right] \right\}.
\label{eq:wavelike}
\end{eqnarray}

Equation (\ref{eq:wavelike}) allows one to fit a model, sample its
parameter's posterior distribution, and determine the credible region,
while taking into account the effects of time-correlated noise.

\subsubsection{Wavelet-likelihood Errata}
\label{sec:waverrata}

During our review and implementation of the wavelet-likelihood
technique from \citet{CarterWinn2009apjWavelets}, we found a few
oversights in their equations and code (available in the
Astronomical Source Code Library, ASCL\footnote{\href{http://asterisk.apod.com/viewtopic.php?f=35\&t=21675}
 {http://asterisk.apod.com/viewtopic.php?f=35\&t=21675}}).  First,
the index for the scale, $m$, of the likelihood function in their Equations
(32) and (41) should start from 1 (however, the online code has the
correct index value).

Next, the variance of the scaling coefficient in the ASCL code, Equation
(34) of the paper, for $\gamma=1$, is missing the factor
$2\sp{-\gamma}=2\sp{-1}$.  The corrected equation should read:
\begin{equation}
\sigma\sb{S}\sp{2} = \frac{\sigma\sb{r}\sp{2}}{4\ln{2}} +
                    \sigma\sb{\omega}\sp{2}.
\end{equation}

Lastly, Section 4.1 of \citet{CarterWinn2009apjWavelets} mentions
that they used a data set of 1024 elements, and that their DWT
produced 1023 wavelet coefficients and 1 scaling coefficient (implying
$N\sb{0}=1$).  This is inconsistent with the wavelet used (a
second-order Daubechies wavelet), for which $N\sb{0}=2$.  This
wavelet's DWT returns 2 scaling coefficients and 1022 wavelet
coefficients (for the given dataset).  The ASCL code is also suited to
perform a likelihood calculation assuming $N\sb{0}=1$, resulting in
each likelihood term having an $m$ value offset by 1.

\section{Correlated-noise Tests for Exoplanet Eclipse Data}
\label{sec:simulations}

We carried out simulations to assess the performance
of the correlated-noise estimators described in Section
\ref{sec:estimators}.  We focused on estimating the
secondary-eclipse depth in a light curve observation,
creating synthetic light curves that
represent Spitzer InfraRed Array Camera (IRAC)
observations in terms of the S/N, known systematics, cadence,
observation duration, and eclipse shape.

In our first experiment we tested the estimators' performances when the
time-correlated noise is described by a signal with a $1/f$ power
spectral distribution \citep[similar to the experiment
of][]{CarterWinn2009apjWavelets}.  We tested the case when the
observation time span is similar to the eclipse-event duration
(Section \ref{sec:synthetic}, typical of secondary eclipse
observations) and for the hypothetical case when the time span lasted
an order of magnitude longer than the eclipse event (Section
\ref{sec:synthetic2}).  In a second experiment (Section
\ref{sec:semisynth3}) we tested the estimators on a more realistic
case by injecting a synthetic eclipse feature into a Spitzer
phase-curve dataset.

\subsection{Synthetic-noise Simulation}
\label{sec:synthetic}

For this simulation we generated synthetic light curves by combining a
\citet{MandelAgol2002ApJtransits} eclipse model, a linear ramp model,
and a signal with both correlated and uncorrelated noise.  The light-curve parameters
closely follow those of a Spitzer observation of
the WASP-12 system (Table \ref{table:parameters}).  The signal
consisted of 1700 data points, with a cadence of $\sim$12 seconds
between data points, spanning an orbital-phase range from 0.39 to 0.63,
about twice the eclipse duration.

\begin{table}[ht]
\centering
\caption[Synthetic Light-curve Parameters]
        {Synthetic Light-curve Parameters}
\begin{tabular}{cc}
\hline
\hline
Parameter                           & Value           \\
\hline
Eclipse depth (counts)              & 98.1            \\
Eclipse duration (phase)            & 0.1119          \\
Eclipse mid point (phase)           & 0.5015          \\
Eclipse ingress/egress time (phase) & 0.013           \\
Ramp slope (counts/phase)           & 0.006           \\
System flux (counts)                & 25815           \\
$\sigma\sb{\omega}$ (counts)        & 64.5            \\
$\sigma\sb{r}$ (counts)             & 0, 230, and 459 \\
\hline
\end{tabular}
\label{table:parameters}
\end{table}

We created three sets of 5000 light-curve realizations each.  For each
realization, we
generated a zero-mean random normal
distribution, which we added to the light curve as the uncorrelated
noise.  We adjusted the variance of this signal
($\sigma\sb{\omega}\sp{2}$) to yield an eclipse-depth signal-to-noise ratio
of 30.  Additionally, we generated purely-correlated $1/f$ signals
($\sigma\sb{\omega}=0$) using a Gaussian random number generator with variances given
by Equations (\ref{eq:varwavelet}) and (\ref{eq:varscale}).  Then, we
applied the inverse DWT to transform the signal from the wavelet basis
to the time domain.  Following the notation of
\citet{CarterWinn2009apjWavelets}, $\alpha$ denotes the ratio
between the rms of the uncorrelated and correlated noise signals.

\begin{figure*}[htb]
\centering
\includegraphics[width=\linewidth, clip]{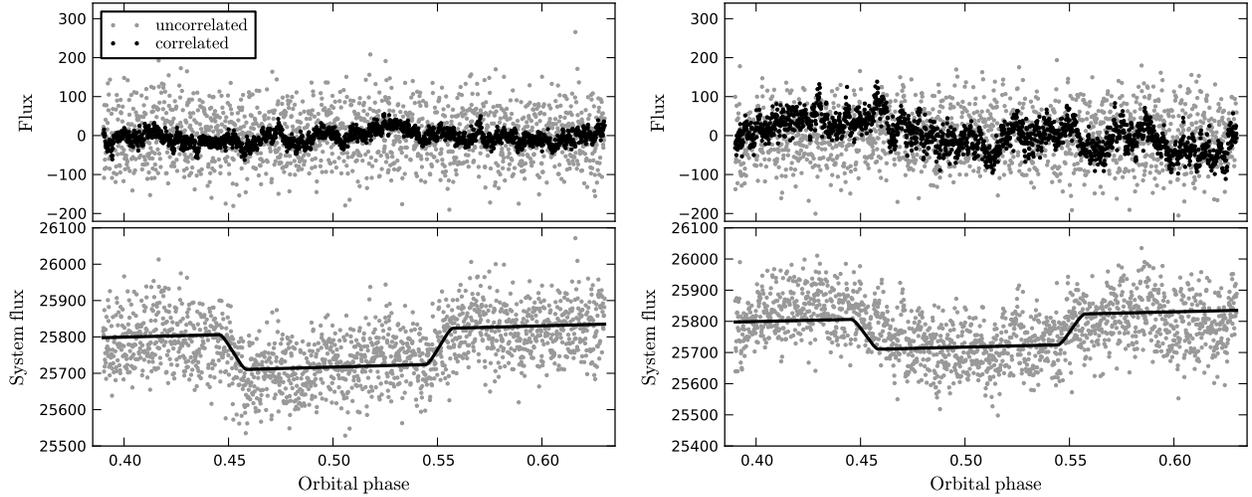}
\caption[Simulated Spitzer time-series data sets]
        {Simulated Spitzer time-series data sets.  The top panels
 overplot the correlated (black) and uncorrelated (grey) noise
 components of the light curves vs.\ orbital phase.  The bottom panels
 show the synthetic light curves (eclipse, ramp, and noise)
 vs.\ orbital phase in gray.  The black solid line shows the noiseless
 model.  The noise rms ratios are $\alpha=0.25$ (left panels) and
 $\alpha=0.5$ (right panels).}
\label{fig:synthetic}
\end{figure*}

We left the first set with pure uncorrelated noise ($\alpha=0$), we
added to the second set a weak time-correlated signal
($\alpha=0.25$), and we added to the third set a strong correlated
signal ($\alpha=0.5$).  Figure \ref{fig:synthetic} shows two synthetic
light curves for $\alpha=0.25$ and $\alpha=0.5$.  Note that our
designations of ``weak'' and ``strong'' are, to some extent,
arbitrary.  We selected these limits based on our experience and
tests: we observed that for $\alpha \lesssim 0.20$, the
time-correlated signal becomes negligible compared to the
uncorrelated-noise signal.  On the other hand, ratios of
$\alpha \sim 0.5$ are on the level of what we have observed in some
cases \citep[e.g., WASP-8b,][]{CubillosEtal2013apjWASP8b}.

For each realization, we estimated the parameter posteriors using the
methods described in Section \ref{sec:estimators}.  
Our model-fitting routines only fixed the eclipse ingress/egress-time
parameter (usually poorly-constrained by eclipse data), leaving the system flux,
eclipse depth, eclipse midpoint, eclipse duration, and ramp slope
free.  First, we carried
out a ``white analysis'' by using Equation (\ref{eq:whitelike}) to
estimate the model-parameter best-fitting values (using the
Levenberg-Marquardt algorithm) and their posterior distributions (using
MCMC).

\begin{figure}[htb]
\centering
\includegraphics[width=0.7\linewidth, clip]{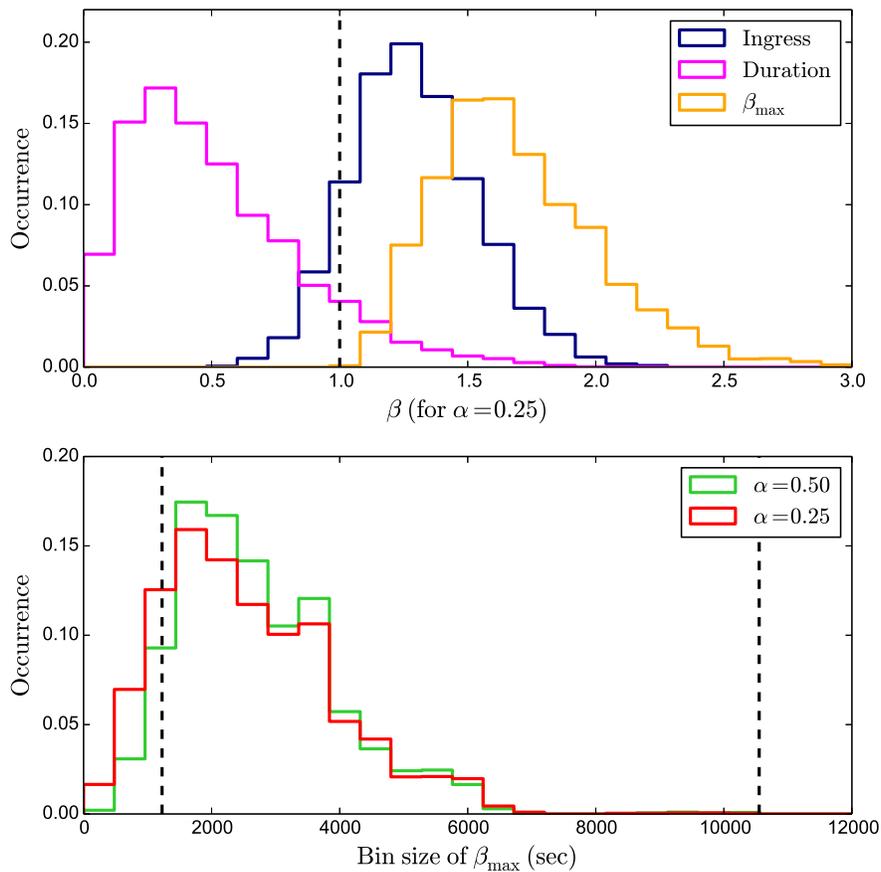}
\caption[$\beta$ distributions]
        {{\bf Top:} Normalized distribution of $\beta$ for the
 $\alpha=0.25$ set.  The histograms represent $\beta$ measured at
 the ingress-time and eclipse-duration timescales, and at the maximum value
 of $\beta$.  The distributions for the other two sets ($\alpha=0$
  and 0.5) were similar.  {\bf Bottom:} Normalized distribution of the bin
 sizes for $\beta\sb{\rm max}$.  The vertical dashed lines indicate
 the ingress time and eclipse duration.}
\label{fig:beta}
\end{figure}

Next, we used the MCMC results to calculate the time-averaging
rms-vs.-bin-size curves.  We retrieved the $\beta$ factor at three
timescales: at the ingress time, at the eclipse duration, and at the
time of maximum $\beta$ ($\beta\sb{\rm max}$, Figure \ref{fig:beta}).
In accordance with the discussion in Section \ref{sec:TAzerotest}, most
$\beta$ values at the eclipse-duration timescale (similar to the total
observation duration) were underestimated ($\beta<1$).  We
adopted $\beta\sb{\rm max}$ as the scaling factor to calculate the
time-averaging method uncertainties.
We also ran the residual-permutation method by applying Equation
(\ref{eq:whitelike}) with the Levenberg-Marquardt algorithm.  Finally,
we applied the wavelet-based likelihood method in an MCMC guided by
Equation (\ref{eq:wavelike}).  We simultaneously fit the noise
parameters ($\sigma\sb{\omega}$ and $\sigma\sb{r}$) and the model
parameters, while keeping $\gamma$ fixed at 1.  We found that a
Jeffrey's non-informative prior on $\sigma\sb{r}$ handled the
case with no correlated noise better.
A Jeffrey's prior is a scale-invariant prior that has an equal
probability per order of magnitude.  It is a more convenient prior
when the parameter may range over several orders of magnitude
\citep{Gregory2005BayesianBook}.  The only requirement is that the
parameter value must be positive.

\subsubsection{Results}
\label{sec:simresults}

To analyze the results, we proceded in a similar way as
\citet{CarterWinn2009apjWavelets}: for the different sets and
methods, we calculated the accuracy-to-uncertainty ratio
(${\cal N}\sb{p}$, the difference between the estimated, $\hat{p}$,
and the true, $p$, value of a parameter, divided by the estimated
uncertainty, $\hat{\sigma}\sb{p}$):
\begin{eqnarray}
{\cal N}\sb{p} = \frac{{\hat{p} - p}}{\hat{\sigma}\sb{p}}.
\label{eq:sigmastats}
\end{eqnarray}
A higher noise level in the light curve should worsen the model
accuracy and increase the parameter uncertainties.  In the ideal case,
we expect the variation of these two quantities to scale such that the
set's standard deviation of the accuracy-to-uncertainty ratio,
$\sigma\sb{\cal N}$, approaches unity for large $N$.  If
$\sigma\sb{\cal N} > 1$, the method underestimates the uncertainties,
and vice versa.

\begin{figure}[htb]
\centering
\includegraphics[width=0.7\linewidth, clip]{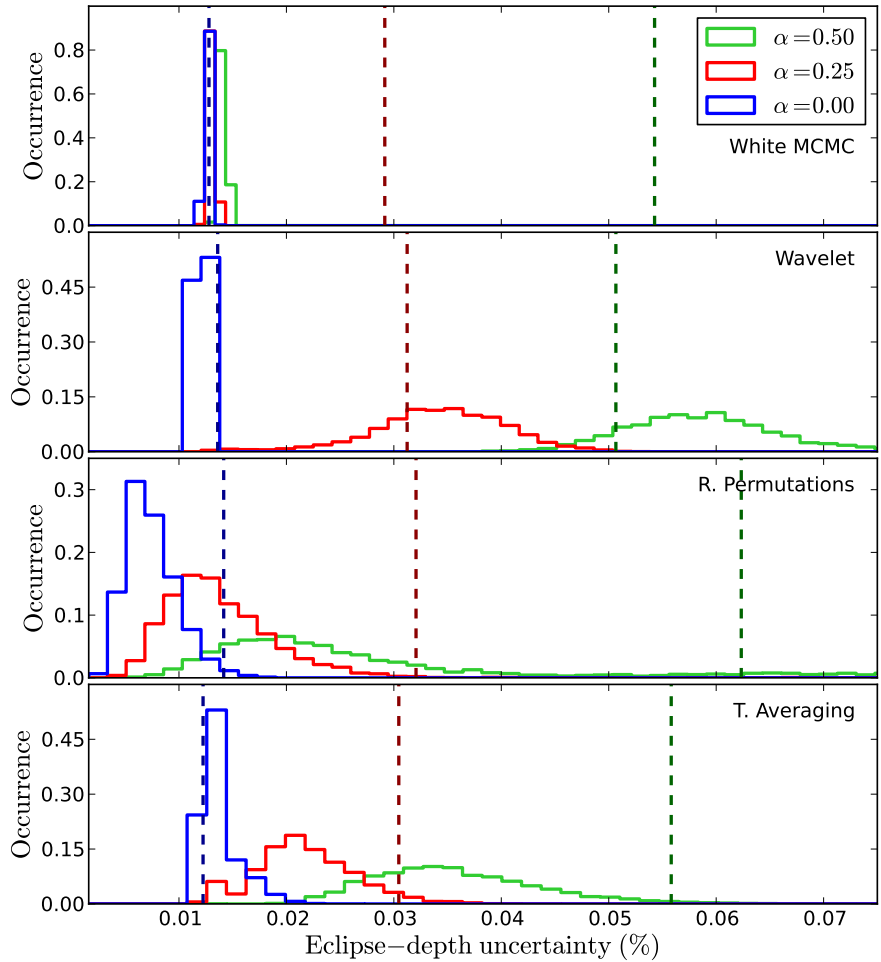}
\caption[Distribution of estimated eclipse-depth uncertainties]
        {Distribution of the estimated eclipse-depth uncertainties for
 the MCMC (top panel), wavelet (second panel), residual-permutation
 (third panel), and time-averaging analysis (bottom panel).  The
 vertical dashed lines indicate the expected mean uncertainty that
 would have yielded $\sigma\sb{\cal N}=1$.}
\label{fig:syntheticResults}
\end{figure}

Figure \ref{fig:syntheticResults} shows the distribution of
estimated eclipse-depth uncertainties for each method and set.  To
find the ideal uncertainty level, we scaled the uncertainties of each
set and method by a constant such that $\sigma\sb{\cal N}=1$.  The
vertical dashed lines indicate the mean value of the scaled uncertainties.
Note that $\sigma\sb{\cal N}$ depends on the particular distribution
of $\{\hat{p},\hat{\sigma}\sb{p}\}$ pairs, thus, the expected
uncertainties differ for each method.  Table \ref{table:results}
shows the standard deviation of the accuracy-to-uncertainty ratio for
each set.

\begin{table}[ht]
\centering
\caption[$\sigma_{\cal N}$   for Synthetic-data Sets]
        {$\sigma\sb{\cal N}$ for Synthetic-data Sets}
\label{table:results}
\strut\hfill
\begin{tabular}{ccccc}
\hline
\hline
$\alpha$  & White MCMC & R. Permutation & T. Averaging  & Wavelet  \\
\hline
0.00      &  1.01      &  2.00          &  0.95         &  1.13    \\
0.25      &  2.23      &  2.47          &  1.42         &  0.90    \\
0.50      &  3.83      &  2.27          &  1.58         &  0.85    \\
\hline
\end{tabular}
\hfill\strut
\end{table}

As expected, the white-MCMC analysis correctly estimated the
uncertainties for the uncorrelated noise set, but it is insensitive
to time-correlated noise.  This is reflected in the clustering of the
uncertainties at the same value, regardless of the amount of
time-correlated noise (Fig.\ \ref{fig:syntheticResults}, top panel) and
the increase of $\sigma\sb{\cal N}$ with increasing $\alpha$ (Table
\ref{table:results}).  The uncertainty distributions for the three
other methods do correlate with $\alpha$, but only at a qualitative
level.  The residual-permutation method produced the least-accurate
results, underestimating the uncertainties, including the case \linebreak[4]

\noindent without
correlated noise.  The time-averaging method underestimated the
uncertainties, whereas the wavelet analysis slightly overestimated the
uncertainties (Fig.\ \ref{fig:syntheticResults} and Table
\ref{table:results}).


\subsection{Long-duration Synthetic-noise Simulation}
\label{sec:synthetic2}

We tested the correlated-noise estimators in the
regime where the time-averaging method is not affected by the
artifacts described in Section \ref{sec:TAzerotest}.
Here we replicated the previous experiment, but for a total
observation time $\sim 20$ times longer than the eclipse duration
(akin to a phase-curve observation).  We generated the light curve
with the same eclipse configuration as in Section
\ref{sec:synthetic2}, keeping the cadence (17,000 data points total)
and the value of $\sigma\sb{\omega}$ at 64.5.  To conserve the noise
rms ratios at $\alpha=0.25$ and 0.5, we set $\sigma\sb{r}=774$ and
1549, respectively.

\begin{figure}[htb]
\centering
\includegraphics[width=0.7\linewidth, clip]{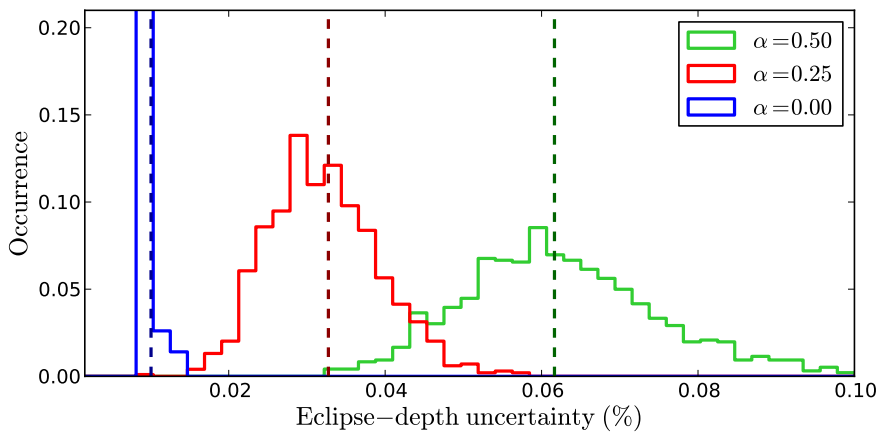}
\caption[Distribution of estimated eclipse-depth uncertainties for
 the time-averaging analysis]
        {Distribution of estimated eclipse-depth uncertainties for
 the time-averaging analysis (using $\beta$ of the eclipse duration).
 For these simulations, the length of the light curves was 20 times
 that of the eclipse duration.  The vertical dashed lines indicate
 the expected mean uncertainty that would have yielded
 $\sigma\sb{\cal N}=1$.}
\label{fig:longrun}
\end{figure}

Upon applying the time-averaging analysis we confirmed that the
$\beta$ factor corresponding to the eclipse duration is less affected
by the method's artifacts.  Furthermore, we found that this $\beta$
factor accurately corrects the uncertainties for the time-correlated
noise (Figure \ref{fig:longrun}).

\subsection{Simulation with Spitzer-IRAC Noise}
\label{sec:semisynth3}

In this section we tested the correlated-noise estimators for a real
exoplanet signal known to be affected by correlated noise.  Hence, we
do not assume any functional form for the noise, we rather use the
true instrumental noise as detected by the telescope.  We selected the
4.5 {\micron} Spitzer IRAC phase curve of the extrasolar planet
HD~209458b (PI H. Knutson, Program ID 60021, AOR 38703872) published
by \citet{ZellemEtal2014apjHD209Phase}.  We generated a set of
trials by injecting a synthetic eclipse signal, which we then
recovered with the time-averaging and wavelet-likelihood techniques.

First we constructed a flat baseline light curve by removing the
eclipse, transit, and phase-curve sinusoidal signals from the data set
(but conserving the systematics).  The data consisted of a 24 hour-long
continuous observation of the HD~209458 system, starting shortly
before an eclipse event and ending shortly after a transit event,
covering the orbital phase from 0.42 to 1.05 (orbital period 3.52
days), with a mean cadence of 0.4 seconds (1.31$\tttt{-6}$ orbital
phase span).  We processed the Spitzer BCD data to obtain a raw light
curve using the Photometry for Orbits, Eclipses, and Transits (POET)
pipeline \citep{StevensonEtal2010natGJ436b,
 StevensonEtal2012apjHD149026b, StevensonEtal2012apjGJ436c,
 CampoEtal2011apjWASP12b, NymeyerEtal2011apjWASP18b,
 CubillosEtal2013apjWASP8b, CubillosEtal2014apjTrES1}.  We trimmed
the data to the time span contained after the eclipse and before
the transit (orbital phase from 0.52 to 0.98).  Since the ramp
variation is negligible after a few hours of observation, the
resulting section of the dataset contained only the astrophysical
signal (the stellar background and the planetary phase-curve
variation) and the instrument's intrapixel systematics
\citep{CharbonneauEtal2005apjTrES1}.

We then fit the remaining light curve with a BLISS map model
\citep[the intrapixel effect,][]{StevensonEtal2012apjHD149026b}, and a
sinusoidal function \citep[for the phase-curve variation,
following][]{ZellemEtal2014apjHD209Phase}:
\begin{equation}
F(t) = 1 + c\sb{0} + c\sb{1} \cos(2\pi t) + c\sb{2} \sin(2\pi t),
\end{equation}
where $c\sb{0}$, $c\sb{1}$, and $c\sb{2}$ are the model fitting
parameters, and $t$ is the time of the observation (measured in
orbital phase).  To avoid degeneracy with the other fitting
parameters, we constrained $c\sb{0}$ by requiring
$F(t\sb{0}) \equiv 1$, with $t\sb{0}$ the eclipse midpoint time.  We
finally subtracted the best-fitting sinusoidal model to remove the
phase-curve variation.

To construct the trial samples, we injected a Mandel-Agol eclipse
curve at random midpoints into the baseline.  We drew the midpoints
from a uniform distribution between orbital phases 0.65 and 0.85.  We
set the eclipse ingress/egress time to 19.8 min (0.0039 orbital phase
span) and the eclipse depth to 0.125\% of the system flux.  To test
how the time-averaging method performs as a function of the observing
length-to-eclipse time span ratio, we generated two sets with
eclipse durations of 2.9 hr (0.034 orbital phase span) and 1.5 hr
(0.018 orbital phase span), with 3000 realizations each.

\begin{figure}[htb]
\centering
\includegraphics[width=0.7\linewidth, clip]{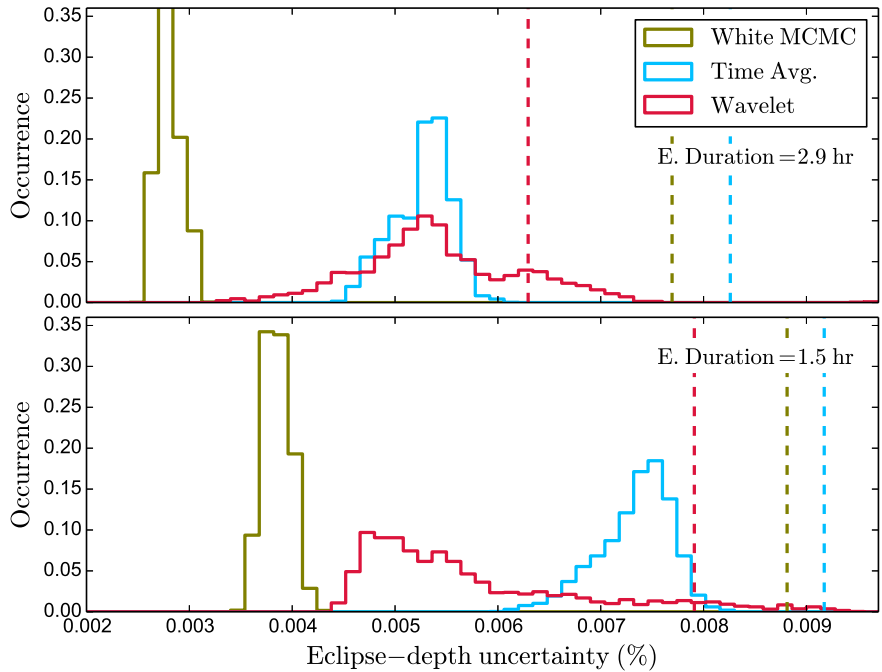}
\caption[Distribution of estimated eclipse-depth uncertainties]
        {Distribution of estimated eclipse-depth uncertainties
 for long (top panel) and short (bottom panel) eclipse duration simulations.
 The vertical dashed lines
 indicate the expected mean uncertainty that would have yielded
 $\sigma\sb{\cal N}=1$ for each method.}
\label{fig:HD209simulation}
\end{figure}

Our fitting model included the eclipse (Mandel and Agol) and the BLISS-map
model.  For the wavelet-likelihood analysis we trimmed the dataset to
a time span of 7.2 hr (0.085 orbital phase span) around the eclipse
midpoint, since the MCMC often failed to converge
for larger datatets.  We analyzed the results in the same manner as in
Section \ref{sec:simresults}.  Figure \ref{fig:HD209simulation} shows
the distribution of the eclipse-depth uncertainties for each method
and set.  Table \ref{table:HD209results} presents the
standard deviation of the uncertainty-to-accuracy ratios.

\begin{table}[ht]
\centering
\caption[$\sigma_{\cal N}$   for the HD~209458b Sets]
        {$\sigma\sb{\cal N}$ for the HD~209458b Sets}
\label{table:HD209results}
\begin{tabular}{ccccc}
\hline
\hline
Eclipse Duration (hr) & White MCMC & T. Averaging & Wavelet \\
\hline
2.9                   & 2.772      & 1.553        & 1.176   \\
1.5                   & 2.293      & 1.241        & 1.462   \\
\hline
\end{tabular}
\end{table}

Figure \ref{fig:HD209simulation} shows that, in general, the
longer-eclipse set returns smaller uncertainties than the
shorter-eclipse set.  This is expected given the lower number of
data points contributing in a shorter eclipse.  Table
\ref{table:HD209results} shows that both methods improved the
estimation of the eclipse depth uncertainty, underestimating it from
$\sim20$\% to 50\%.  The time-averaging method, as expected, performed
better for the shorter-eclipse set (observation span-to-eclipse
duration ratio of 25.5, compared to 15.5 for the longer-eclipse
set).  The wavelet-likelihood method, on the other hand, performed
better for the longer-eclipse set.

\section{Multi-Core Markov-Chain Monte Carlo (MC\sp{3}) Code}
\label{sec:mc3}

We implemented all of the discussed statistical methods into the
open-source Python module Multi-Core Markov-Chain Monte Carlo
(\href{https://github.com/pcubillos/MCcubed}{{\mcc}},
\href{https://github.com/pcubillos/MCcubed}{github.com/pcubillos/MCcubed}).
Unlike other exoplanet model-fitting tools that are tailored to specific
tasks,
{\mcc} allows the user to define the modeling function and,
thus, it is a general-purpose statistical package.  We
developed the main bulk of the code in Python, with several extensions
written in C, combining simplicity and high performance.  The code
runs in a single CPU or in multiple parallel processors (through
Message Passing Interface, MPI).  {\mcc} provides statistically-robust
model optimization (via Levenberg-Marquardt minimization) and
credible-region estimation (via MCMC sampling) routines.

The MCMC random sampling is done via the Metropolis Random Walk (MRW,
using multivariate Gaussian proposals) or the Differential-Evolution
Markov-chain Monte Carlo algorithm \citep[DEMC,
][]{Braak2006DifferentialEvolution}.  While the proposal step sizes of
the MRW are predetermined by the user and have to be manually adjusted
before each run, DEMC automatically adjusts the scale and orientation
of the proposal distribution.  To do so, DEMC runs several chains in
parallel, computing the proposed jump for a given chain from the
difference between the parameter states of two other randomly selected
chains.  As the chains converge toward the posterior distribution, the
proposal jumps will be mainly oriented along the desired distribution
and will have adequate scales.  Therefore, DEMC improves the MCMC
efficiency in two ways: (1) it increases the acceptance rate to
optimal levels \citep[$\gtrsim 25$\%,][]{RobertsEtal1997} by better
sampling the phase space, and (2) it eliminates the heuristic need for
the user to adjust the proposal jump scales.

The Metropolis-Hastings acceptance rule implements both the regular
likelihood function (Eq.\ \ref{eq:whitelike}) and the wavelet-based
likelihood (Eq.\ \ref{eq:wavelike}).  The priors can be bounded or
unbounded uniform, log-scale uniform (Jeffrey's), or Gaussian.
To assess that the MCMC is working properly, the code performs a
chain-convergence test using the
\citet{GelmanRubin1992} statistics.  The code also produces several
plots to help visualize the results: trace, rms-vs.-bin-size,
marginal-posterior, and pairwise-posterior plots can indicate
non-convergence, multi-modal posteriors, parameter correlations,
correlated noise, or incorrect priors.  At the end of the MCMC run the
code returns the sampled posterior distribution of the parameters,
their best-fitting values, their 68\% credible region, and the
acceptance rate of the MCMC.
The majority of the routines of this module derive from our POET
pipeline and, thus, have been thoroughly tested for years.

The core structure of {\mcc} consists of a central hub, which drives
the MCMC exploration, and the workers, which evaluate the model for
the given free parameters.
When run in parallel, the hub communicates with the workers through
MPI, sending the free parameters and receiving the evaluated models
back (transmitting the data as one-dimensional arrays).  {\mcc}
assigns one CPU to each worker (i.e., one for each chain).  When run
in a single CPU, the hub evaluates the models {\it in situ}).
Each cycle (iteration) of the MCMC comprises the following steps: (1)
the hub generates the proposal state (the set of free parameters) for
each chain,  (2) the workers (or the hub) evaluate the model for the
proposed state,  (3) the hub computes the Metropolis ratio and
accepts/rejects the proposal state for each chain.

The {\mcc} code runs from both the shell prompt and the Python
interactive interpreter.  The user can configure the MCMC run either
through a configuration file, command line arguments (prompt), and/or
function arguments (Python interpreter).  The minimum required inputs
are the modeling function, the data being fitted, and starting
estimate values for the free parameters.  As optional arguments, the
user can supply the data uncertainties, priors, and any extra
arguments of the modeling function (in a manner much like the
\texttt{scipy.optimize.leastq} routine).  Additionally, the package
allows the user to configure multiple features of the MCMC, e.g.:
number of chains, number of iterations, burn-in length, thinning
factor, etc.  The repository of the code includes a user manual and
guided examples.

\section{Conclusions}
\label{sec:c4conclusions}

Time-correlated noise is an important source of uncertainty for faint
signals such as exoplanet light curves.  Unless all systematics of the
data are well understood, the correlated noise must be taken into
account to obtain a reliable estimation of the parameter
uncertainties.  We have reviewed three of the most widely used
methods to assess time-correlated noise in exoplanet time-series
data: time averaging, residual permutation, and wavelet-based
likelihood.  We expanded the limited literature of tests to assess the
quantitative results of these techniques, focusing specifically on the
case of Spitzer secondary-eclipse time-series data.

We found that the rms-vs.-bin-size curve of the time-averaging method
can often present artifacts that partially or totally obstruct our
assessment of the correlated noise with this method for typical eclipse or transit data.  These artifacts are particularly
prominent at timescales with the same order of magnitude as the
total observation duration.
The residual-permutation method is unsound as a tool for quantifying
uncertainty in parameter estimates, because it does not produce
ensembles that mimic the behavior of independent draws from a
probability distribution.
In our review and implementation of the wavelet method by
\citet{CarterWinn2009apjWavelets}, we found and corrected errors in
the equations of the publication and the online code (Section
\ref{sec:waverrata}).

We developed the Multi-Core Markov-Chain
Monte Carlo open-source Python module (\href{https://github.com/pcubillos/MCcubed}{{\mcc}},
\href{https://github.com/pcubillos/MCcubed}
{https://github.com/pcubillos/MCcubed}).  {\mcc} implements all of the
statistical routines described in this paper, allowing the user to
estimate best-fitting model parameters and their credible region,
while letting the user provide the modeling function.  Given the
challenging nature of exoplanet observations, most analyses require a
great deal of fine-tuning and use of advanced reduction and
statistical techniques.  By releasing our code to the community, we
hope not only to provide access to the routines discussed here, but
also to encourage researchers to consider open development and
cross-validation of the software tools used in the field.

We tested the time-correlated analysis methods on synthetic exoplanet
eclipse light curves, focusing on eclipse-depth estimation.  In our
first simulation, we tested the case when the time-correlated noise
has a power spectral density of the form $1/f$.  We generated
synthetic light curves consistent with the S/N level, cadence, known
telescope systematics, and observation duration of a Spitzer IRAC
eclipse observation.  The light curves are composed of a Mandel and Agol
eclipse model, a linear ramp model, uncorrelated noise (random normal
distribution), and $1/f$ time-correlated noise.  The
residual-permutation method returned the poorest results, largely
underestimating the uncertainties.  Both the time-averaging and the
wavelet-likelihood methods provided better estimations of the
uncertainties than a white MCMC analysis, although with some
limitations: the wavelet-likelihood method overestimated the
eclipse-depth uncertainty by about 10\%, whereas the time-averaging
method underestimated the uncertainties by about 50\%.  When the
eclipse duration is short compared to the total observation duration
(e.g., a phase-curve observation), the time-averaging correction
accurately estimates the eclipse-depth uncertainty for uncorrelated
and $1/f$-correlated noise.

In a further simulation, we generated eclipse light-curve samples by
injecting an eclipse signal into a real Spitzer IRAC (4.5 {\micron})
time-series dataset as a baseline.  This experiment allowed us to
assess the performance of the time-correlated estimators without
assuming a specific shape of the time-correlated signal.  We studied
the results for two sets with a different eclipse duration each.
Both the time-averaging and the wavelet-likelihood methods
significantly improved the uncertainty estimations compared to a white
MCMC analysis.  However, they are not perfect, as in both cases the
uncertainties are overestimated by $\sim20-50$\%.  Similar to the
previous simulation, the time-averaging method performed better when
the eclipse duration is shorter compared to the total observation
time.  In contrast, the wavelet-likelihood analysis performed better
for the simulation with a longer eclipse duration.
If the correlated noise present in other Spitzer data sets behaves in
a similar manner as the one treated here, there is not yet a
universally best method for assessing time-correlated noise.

\section{Acknowledgements}
We thank contributors to SciPy, Matplotlib, and the Python Programming
Language, the free and open-source community.  PC is supported by the
Fulbright Program for Foreign Students.  Part of this work is based on
observations made with the {\em Spitzer Space Telescope}, which is
operated by the Jet Propulsion Laboratory, California Institute of
Technology under a contract with NASA.  Support for this work was
provided by NASA through an award issued by JPL/Caltech and through
the NASA Science Mission Directorate's Astrophysics Data Analysis
Program, grant NNH12ZDA001N.

\section{Bayesian Credible Region}
\label{sec:CredRegion}


In the Bayesian context, given the posterior probability density,
$p(\theta|{\bf D})$, of a parameter, $\theta$, given the dataset,
${\bf D}$, the highest posterior density region (or credible
region), $R$, is defined by
\begin{equation}
C = \int\sb{R} {\rm d}\theta\;p(\theta|{\bf D})
\end{equation}
where C is the probability contained in the credible region.  The
region $R$ is selected such that the posterior probability of any
point inside $R$ is larger than that of any point outside.

In the practice, to calculate the credible region, one constructs a
histogram of the sampled posterior distribution (normalized such that
the sum equals one) and sorts the bins in descending order.  Then one
sequentially adds the values of $p$ until reaching $C$.  The
credible-region boundaries are given by the smallest and largest
values of $\theta$ for the samples considered in the sum, if the
region is contiguous.

\section{Standard Deviation Uncertainty}
\label{sec:StdUncert}

The uncertainty of a parameter estimate in a problem with a fixed model
dimension (number of parameters) and growing sample size typically decreases asymptotically
at the $\sqrt{N}$ rate.  That is, for estimating a Gaussian mean from $N$
samples with the standard deviation $\sigma$, which is
known, the uncertainty is $\sigma/\sqrt{N}$.  However, this result says
nothing about the actual {\em size} of the uncertainty at any
particular sample size.  When $\sigma$ is unknown, it becomes the
target of estimation, instead of (or in addition to) the mean.  
Here, we elaborate on the derivation of the uncertainty for the
standard deviation of a Gaussian.  The derivation uses the Laplace approximation for
a normal standard deviation and its uncertainty, i.e., it finds a
Gaussian distribution with a peak and curvature matching the marginal
probability density function.

Given a normal distribution of values with unknown mean $\mu$ and
standard deviation $\sigma$, let $b\sb{i}$ be the means for a sample
of $N$ groups of samples (``bins'') drawn from this distribution.
The sample mean, $\bar{b}$, and the sample variance, $s\sp{2}$,
are defined as usual:
\begin{equation}
\label{Aeq:meanstd}
\bar{b} = \frac{1}{N}\sum\sb{i} b\sb{i}, \qquad
s\sp{2} = \frac{1}{N}\sum\sb{i} (b\sb{i} - \bar{b})\sp{2}.
\end{equation}

If the residual $r\sb{i} = b\sb{i} - \bar{b}$ and $r\sp{2} = \sum\sb{i} r\sb{i}\sp{2}$, then the sample
variance becomes $s\sp{2} = r\sp{2}/N$.

The likelihood function for our normal distribution with ($\mu$,
$\sigma$) is:
\begin{eqnarray}
\label{Aeq:likelihood}
\cal{L}(\mu, \sigma) & = & \prod\sb{i} p\,(b\sb{i}|\mu, \sigma) \nonumber \\
                    & = & \prod\sb{i} \frac{1}{\sigma\sqrt{2\pi}}
                 \exp\left[-\frac{(b\sb{i}-\mu)\sp{2}}{2\sigma\sp{2}}\right],
\end{eqnarray}

\noindent so the likelihood can be written is terms of $\bar{b}$ and $r$ as:
\begin{eqnarray}
\label{Aeq:likelihood2}
\cal{L}(\mu,\sigma)
           & = & \frac{1}{\sigma\sp{N} (2\pi)\sp{N/2}}
                 \exp\left(-\frac{r\sp{2}}{2\sigma\sp{2}}             \right) 
                 \exp\left(-\frac{N(\mu-\bar{b})\sp{2}}{2\sigma\sp{2}}\right).
\end{eqnarray}

To estimate $\mu$ and $\sigma$, we will adopt a flat prior for $\mu$
and a log-flat prior for $\sigma$, corresponding to $p(\sigma) \propto
1/\sigma$. Then, the joint posterior probability $p(\mu,\sigma|D)$ for
$\mu$ and $\sigma$, given the data $D$ is:

\begin{equation}
\label{eq:posterior1}
p(\mu,\sigma|D) \propto p(\sigma) \times \cal{L}(\mu,\sigma),
\end{equation}
with $p(\sigma)$ the prior probability on $\sigma$.

\begin{eqnarray}
\label{eq:posterior2}
p(\mu,\sigma|D) 
 & \propto & \frac{1}{\sigma\sp{N+1}}  
             \exp\left(-\frac{r\sp{2}}{2\sigma\sp{2}}\right)
             \exp\left(-\frac{N(\mu-\bar{b})\sp{2}}{2\sigma\sp{2}}\right).
\end{eqnarray}

Calculate the marginal posterior density for $\sigma$ by
integrating over $\mu$:
\begin{eqnarray}
\label{eq:margsigma}
p(\sigma|D)
     & \propto & \int  \frac{1}{\sigma\sp{N+1}}
                 \exp\left( -\frac{r\sp{2}}{2\sigma\sp{2}} \right) 
                 \exp\left( -\frac{N(\mu-\bar{b})\sp{2}}{2\sigma\sp{2}}\right)
                 {\rm d}\mu.
\end{eqnarray}

The $\mu$ dependence is in the last exponential factor, a Gaussian that
integrates to $\sigma\sqrt{2\pi}$.  We denote the result as $f(\sigma)$:
\begin{eqnarray}
\label{eq:margsigma2}
\label{eq:fSigma}
p(\sigma|D) \propto \frac{1}{\sigma\sp{N}}
                   \exp\left( -\frac{r\sp{2}}{2\sigma\sp{2}} \right)
                   = f(\sigma).
\end{eqnarray}

We estimate $\sigma$ with its mode, $\hat\sigma$, which maximizes
$f(\sigma)$. The first derivative of $f(\sigma)$ is:
\begin{eqnarray}
\label{eq:Tom15}
f'(\sigma) = f(\sigma)\left(\frac{r\sp{2}}{\sigma\sp{3}} - 
                           \frac{N}{\sigma}             \right),
\end{eqnarray}
so that setting $f'(\hat\sigma) = 0$ gives $\hat\sigma = r/\sqrt{N} = s$,
as one might expect.

For a simple estimate of the uncertainty, let's consider a Gaussian
approximation with mean at $\hat\sigma$.  The curvature (second
derivative) of $\sigma$ at $\hat\sigma$:
\begin{eqnarray}
\label{eq:Tom16}
f''(\sigma) =
f'(\sigma) \left(\frac{r\sp{2}}{\sigma\sp{3}} - \frac{N}{\sigma}\right) +
 f(\sigma) \left(\frac{N}{\sigma\sp{2}} - \frac{3r\sp{2}}{\sigma\sp{4}}\right),
\label{fpp}
\end{eqnarray}
determines the standard deviation.  When $f(x)$ is of the form of a
normal distribution with mean $m$ and standard deviation $w$, it is
easy to show that $f''(m) = -f(m)/w\sp{2}$.
So, if $\delta$ is the standard deviation for $\sigma$, in the normal
approximation, that matches the curvature at the peak, we have
$\delta\sp{2} \approx - f(\hat\sigma)/f''(\hat\sigma)$.  Evaluating
Equation (\ref{eq:Tom16}) at $\hat\sigma$, the first term vanishes
(since $f'(\hat\sigma)=0$), and the remaining term gives an
approximate standard deviation of:
\begin{eqnarray}
\label{eq:Tom17}
\delta \approx \frac{r}{N\sqrt{2}} = \frac{s}{\sqrt{2N}}.
\label{delta-approx}
\end{eqnarray}

So, the mean and standard deviation sum for $\sigma$ for large $N$ is:
\begin{eqnarray}
\label{eq:Tom18}
\sigma = s \pm \frac{s}{\sqrt{2N}}.
\label{sigma-est}
\end{eqnarray}

\section{Wavelet Coefficients Variance}
\label{sec:WaveletVariance}

Starting from the Equation (37) of \citet{Wornell1993ieeeWavelet}, we
calculate the variance of the wavelet coefficients as:

\begin{equation}
{\rm var}\ x^{m}_{n} = E[x^{m}_{n}x^{m}_{n}] = \frac{2^{-m}}{2\pi}\int^{\infty}_{-\infty}
\frac{\sigma^{2}_{x}}{|\omega|^\gamma}|\Psi(2^{-m}\omega)|^{2}{\rm d}\omega,
\end{equation}

with a change of variable, $u = 2^{-m}\omega$, we have:

\begin{equation}
\label{eq:varw}
{\rm var}\ x^{m}_{n} = 2^{-\gamma m} \frac{1}{2\pi}\int^{\infty}_{-\infty}
\frac{\sigma^{2}_{x}}{|u|^\gamma}|\Psi(u)|^{2}{\rm d}u.
\end{equation}

Now, let
\begin{equation}
\sigma^2 = \frac{1}{2\pi} \int^{\infty}_{-\infty}
\frac{\sigma^{2}_{x}}{|\omega|^\gamma}|\Psi(\omega)|^{2}{\rm d}\omega,
\end{equation}
we then obtain the equation that follows Equation (37) of
\citet{Wornell1993ieeeWavelet}:
\begin{equation}
{\rm var}\ x^{m}_{n} = 2^{-\gamma m} \sigma^2.
\end{equation}

This variance corresponds to Equation (24) of
\citet{CarterWinn2009apjWavelets}. In their notation:
$\langle \epsilon^{m}_{n}\epsilon^{m}_{n} \rangle = {\rm var}\
x^{m}_{n}$, and $\sigma_r^2 = \sigma^2$. \newline

If we assume an ideal bandpass ---i.e., Eq.\ (3) of
\citet{Wornell1993ieeeWavelet}--- we can work our Eq.\ (\ref{eq:varw})
to obtain Eq.\ (45) of \citet{Wornell1993ieeeWavelet}.  Note that
Eq.\ (3) of \citet{Wornell1993ieeeWavelet} should read $\Psi(\omega)$
instead of $\psi(\omega)$.  Then:
\begin{eqnarray}
{\rm var}\ x^{m}_{n} & = & 2^{-\gamma m} \frac{2}{2\pi}\int^{2\pi}_{\pi}
\frac{\sigma^{2}_{x}}{|\omega|^\gamma}d\omega.  \\
                   & = & 2^{-\gamma m} \frac{\sigma^{2}_{x}}{\pi^\gamma}
                         \frac{[2^{1-\gamma}-1]}{1-\gamma},
\end{eqnarray}
For $\gamma \ne 1$, in the notation of
\citet{CarterWinn2009apjWavelets}, we have:
\begin{equation}
\label{eq:sigmar}
\sigma_r^2 = \frac{\sigma^{2}_{x}}{\pi^\gamma}
                     \frac{[2^{1-\gamma}-1]}{1-\gamma}.
\end{equation}

In analogy to Eq.\ (\ref{eq:varw}), we can repeat the process for the scaling
coefficient:
\begin{equation}
{\rm var}\ a^{m}_{n} = 2^{-\gamma m} \frac{1}{2\pi}\int^{\infty}_{-\infty}
                \frac{\sigma^{2}_{x}}{|\omega|^\gamma}|\Phi(\omega)|^{2}d\omega.
\end{equation}
Assuming again an ideal bandpass ---i.e., Eq.\ (10) of
\citet{Wornell1993ieeeWavelet}--- we obtain for $\gamma \ne 1$:
\begin{eqnarray}
\label{eq:vars}
{\rm var}\ a^{m}_{n} & = & 2^{-\gamma m} \frac{2}{2\pi}\int^{\pi}_{0}
                         \frac{\sigma^{2}_{x}}{|\omega|^\gamma}d\omega. \\
                   & = & 2^{-\gamma m} \frac{\sigma^{2}_{x}}{\pi^\gamma}
                         \frac{1}{1-\gamma}.
\end{eqnarray}

If we plug in $\sigma_r^2$ from Eq.\ (\ref{eq:sigmar}), we have:

\begin{equation}
\label{eq:vars2}
{\rm var}\ a^{m}_{n} = 2^{-\gamma m} \sigma^{2}_{r}
                     \frac{1}{2^{1-\gamma}-1},
\end{equation}
where we can recognize $g(\gamma)$ from
\citet{CarterWinn2009apjWavelets}:
\begin{equation}
 g(\gamma) = \frac{1}{2^{1-\gamma}-1}.
\end{equation}

This is consistent with \citet{CarterWinn2009apjWavelets} code for
$\gamma \ne 1$, where he has:
\begin{eqnarray}
{\rm sm2} & = & \sigma^{2}_{r} 2^{-\gamma m} \frac{1}{2^{1-\gamma}-1}, \\
         & = & \sigma^{2}_{r} \frac{1}{2-2^{\gamma}}.  \qquad ({\rm for }\ m=1)
\end{eqnarray}


\bibliographystyle{apj}
\bibliography{chap4-rednoise}

\chapter{
THE BAYESIAN ATMOSPHERIC RADIATIVE TRANSFER CODE FOR EXOPLANET MODELING AND APPLICATION TO THE EXOPLANET HAT-P-11b}
\label{chap:BART}

{\singlespacing
\noindent{\bf 
Patricio~Cubillos\sp{1},
Joseph~Harrington\sp{1},
Jasmina~Blecic\sp{1},
Patricio~M.~Rojo\sp{2},
Nate~B.~Lust\sp{1},
Ryan~C.~Challener\sp{1},
M.~Oliver~Bowman\sp{1},
Madison~M.~Stemm\sp{1},
Austin~J.~Foster\sp{1},
Sarah D. Blumenthal\sp{1},
Andrew~S.~D.~Foster\sp{1},
Dylan~K.~Bruce\sp{1},
Emerson~DeLarme\sp{1},
Justin~Garland\sp{1}
}

\vspace{1cm}

\noindent{\em
\sp{1}Planetary Sciences Group, Department of Physics, University of Central Florida, Orlando, FL 32816-2385, USA \\
\sp{2}Department of Astronomy, Universidad de Chile, Santiago, Chile
}

\vspace{1cm}

\centerline{In preparation for {\em The Astrophisical Journal}.}


}

\clearpage
\setcitestyle{authoryear,round}
\defcitealias{FraineEtal2014natHATP11bH2O}{F14}

\section{Abstract}

This and companion papers \citep[][]{HarringtonEtal2015apjBART,
  BlecicEtal2015apjBART} present the Bayesian Atmospheric Radiative
Transfer (BART) code, an open-source, open-development package to
characterize extrasolar-planet atmospheres.  BART combines a
thermochemical\-/equilibrium abundances (TEA), a radiative-transfer
(Transit), and a Bayesian statistical ({\mcc}) module to constrain
atmospheric temperatures and molecular abundances for given
spectroscopic observations.  The TEA Python code calculates
thermochemical-equilibrium mixing ratios for gaseous molecular species
\citep{BlecicEtal2015apsjTEA}.  Transit is an efficient
one-dimensional line-by-line radiative-transfer C code, developed by
P. Rojo and further modified by the UCF exoplanet group.  This
radiative code produces transmission and hemisphere-integrated
emission spectra.  Transit handles the  HITRAN,
\citeauthor{PartridgeSchwenke1997jcpH2O}'s {\water}, \citeauthor{Schwenke1998TiO}'s TiO,
and \citeauthor{Plez1998aaTiOLineList}'s VO line-by-line opacity data; and cross-section opacities
from HITRAN, ExoMol, and \citeauthor{BorysowEtal2001jqsrtH2H2highT}.
Transit emission-spectra models agree with models from C. Morley
(priv.\ comm.) within a few percent.  The statistical package, {\mcc},
is a general-purpose, model-fitting Python code that implements the
classical and differential-evolution Markov-chain Monte Carlo
algorithms in a multiprocessor environment.  We applied BART to the
Spitzer and Hubble transit observations of the Neptune-sized planet
HAT-P-11b.  We reproduced the conclusions of
\citet{FraineEtal2014natHATP11bH2O}, constraining the {\water}
abundance and finding an atmosphere enhanced in heavy elements.  The
BART source code and documentation is available at
\href{https://github.com/exosports/BART} {https://github.com/exosports/BART}.

\section{Introduction}
\label{sec:c5introduction}

Transiting exoplanets offer the most favorable scenario to
characterize exoplanet atmospheres; as planets pass in front of or
behind their host stars, the observed flux reveals the planetary size
and emission, respectively.  Furthermore, the planned survey missions
will use the transit method to find exoplanets around the brightest
stars in the solar neighborhood \citep[][]{RickerEtal2014spieTESS,
  WheatleyEtal2013NGTS, BroegEtal2013CHEOPS}.  Based on the current estimates of the exoplanet 
occurrence rate in our galaxy \citep[e.g.,][]{BonfilsEtal2011PlanetOccurrenceMdwarf,
  FressinEtal2013KeplerRate,
  DressingCharbonneau2015apjOcurrenceHabitableMdwarfs}, we expect to find thousands of planets.

Furthermore, we are expecting a profound
transformation in the state of exoplanet characterization with the arrival of the James Webb Space Telescope
(JWST).
Thanks to its superior collecting area (25
m$\sp{2}$), spectral coverage ($\sim 0.6 - 28.0$ {\microns}), and
resolving power ($R=4-3000$), JWST will be able to characterize a broad range of exoplanet atmospheres,
with unprecedented detail.
Hopefully,
future ground- and space-based telescopes will help to overcome 
the current limitations, and reveal the properties of exoplanets, from
rocky Earth-like worlds
\citep[e.g.,][]{JenkinsEtal2015apjKepler452b} to hot gas giants.

\subsection{Atmospheric Modeling and Retrieval}

There are numerous physical processes that shape planetary atmospheres:
radiative processes, scattering
\citep[e.g.,][]{BiddleEtal2014mnrasGJ3470b}, circulation dynamics
\citep[e.g.,][]{ShowmanEtal2012DopplerSignatures}, chemical kinetics
\citep[e.g.,][]{AgundezEtal2012aaChem}, photochemistry
\citep[e.g.,][]{MosesEtal2011apjDisequilibrium}, cloud physics
\citep[e.g.,][]{Fortney2005mnrasClouds}, etc.
In addition, exoplanet signals are intrinsically faint, and the current
data are sparse and of low signal-to-noise ratios.  Thus, to properly extract
and interpret the data requires advanced and robust techniques.

The Bayesian retrieval approach provides an ideal framework to fit
poorly-constrained models in a statistically-robust manner.  A
retrieval that uses Markov-chain Monte Carlo (MCMC) algorithms can
determine the best-fitting model solution and the parameters' credible
intervals.  An MCMC draws a large number of samples from the
parameters' phase space, generating a posterior distribution
proportional to the likelihood of the model.  In this process, the
data drive the exploration towards the most-probable solutions; the
data quality (i.e., the error bars) determines the credible region for
the model parameters ($\sim$ the span of the distribution).  Modern
MCMC algorithms enable more efficient and automated exploration than
in the past 
\citep{Braak2006DifferentialEvolution, Braak2008SnookerDEMC}.

In the exoplanet field, the retrieval approach was pioneered by
\citet{MadhusudhanSeager2010apjRetrieval}.  They parameterized the
temperature profile and scaled thermochemical-equilibrium molecular
abundances of {\water}, {\methane}, CO, and {\carbdiox}.  Then, other
groups followed, implementing different modeling schemes.  For example,
\citet{BennekeSeager2012apjRetrieval} implemented a self-consistent
chemistry and radiative-transfer model, parameterizing the
metallicity, carbon-to-oxygen ratio, internal heat, albedo, heat
redistribution, diffusion, and cloud properties.
\citet{LeeEtal2012mnrasRetrieval} introduced the optimal-estimation
algorithm for exoplanet atmospheric modeling.
\citet{LineEtal2013apjRetrievalI} parameterized the thermal profile
and constant-with-height molecular abundances.
\citet{WaldmannEtal2014TauRexI} presented a retrieval code using
pattern recognition and nested sampling.

The radiative-transfer equation links the observed spectra to the
atmospheric properties.  The temperature, pressure, and composition of
the atmosphere determine the shape of the observed spectrum.  We can
distinguish particular atmospheric species, since each one imprints a
characteristic absorption pattern in the spectrum.  The atmospheric
temperature, gravity, and mean molecular mass determine the
atmospheric scale height, and thus modulate the amplitude of the
spectral features.  High-altitude haze layers can flatten
a spectrum \citep[e.g.,][]{KreidbergEtal2014natCloudsGJ1214b,
  KnutsonEtal2014natGJ436b}.

Although the physics behind the radiative processes is well
understood, we lack laboratory data for absorption at temperatures
above $\sim1200$ K ---most available opacity databases were conceived
for Earth-like temperatures.  Besides, it is still little understood
how the different processes interact under each particular circumstance
to determine the atmospheric
composition.  As a consequence, the atmospheric modeling
parameterization widely differs from group to group.  Ideally, as
better-quality data permit more robust constraints, we will
incorporate this information into the models, and will better
understand the physics.

This work is part of a series of three papers introducing the
open-source Bayesian Atmospheric Radiative Transfer (BART) package for
exoplanet characterization.  Here, we focus on the radiative-transfer
treatment and the description of the statistical module. For
additional details about the project, see the companion papers
\citet{HarringtonEtal2015apjBART} and \citet{BlecicEtal2015apjBART}.
Section \ref{sec:c5code} introduces the BART modeling package.  This
section describes the radiative-transfer and the Bayesian statistical
modules, and presents validation examples.  Section
\ref{sec:c5analysis} shows our retrieval analysis of the extrasolar
planet HAT-P-11b using BART.  Finally, Section \ref{sec:discusison}
presents our prospects for the BART project and summarizes our
conclusions.

\section{The Bayesian Atmospheric Radiative Transfer Package}
\label{sec:c5code}

The BART package is a Python and C code that characterizes the
atmospheres of astrophysical bodies for given spectroscopic data.
BART uses a Bayesian approach to constrain the atmospheric properties
in a statistically-robust manner.  Following the `Reproducible
Research' principles \citep[see further details
in][]{HarringtonEtal2015apjBART}, this is an open-source,
open-development project, available under version control at
\href{https://github.com/exosports/BART} {https://github.com/exosports/BART}.
In its conceived design, the code retrieves the temperature and
abundance profiles of spectroscopically-active species \citep[as
in][]{MadhusudhanSeager2010apjRetrieval}.  Note that, although we
describe BART in the context of exoplanet atmospheric retrieval, the
code can be extended for other applications.

The BART package incudes three self-sufficient modules: the
one-dimensional radiative-transfer code
(\href{https://github.com/exosports/transit} {Transit}), the
Thermochemical-Equilibrium Abundances code
\citep[\href{https://github.com/dzesmin/TEA}
{TEA},][]{BlecicEtal2015apsjTEA}, and the
Multi-Core Markov-Chain Monte Carlo code
\citep[\href{https://github.com/pcubillos/MCcubed}
{\mcc},][]{CubillosEtal2015apjRednoise}.  The repository includes a
user manual that explains the features of the code,
describes the code inputs and outputs, and provides sample runs.  A
second document aimed at developers, the code manual, details in depth
the data structures, the file formats, and the code workflow.

\begin{figure}[htb]
\centering
\includegraphics[width=0.5\linewidth, clip]{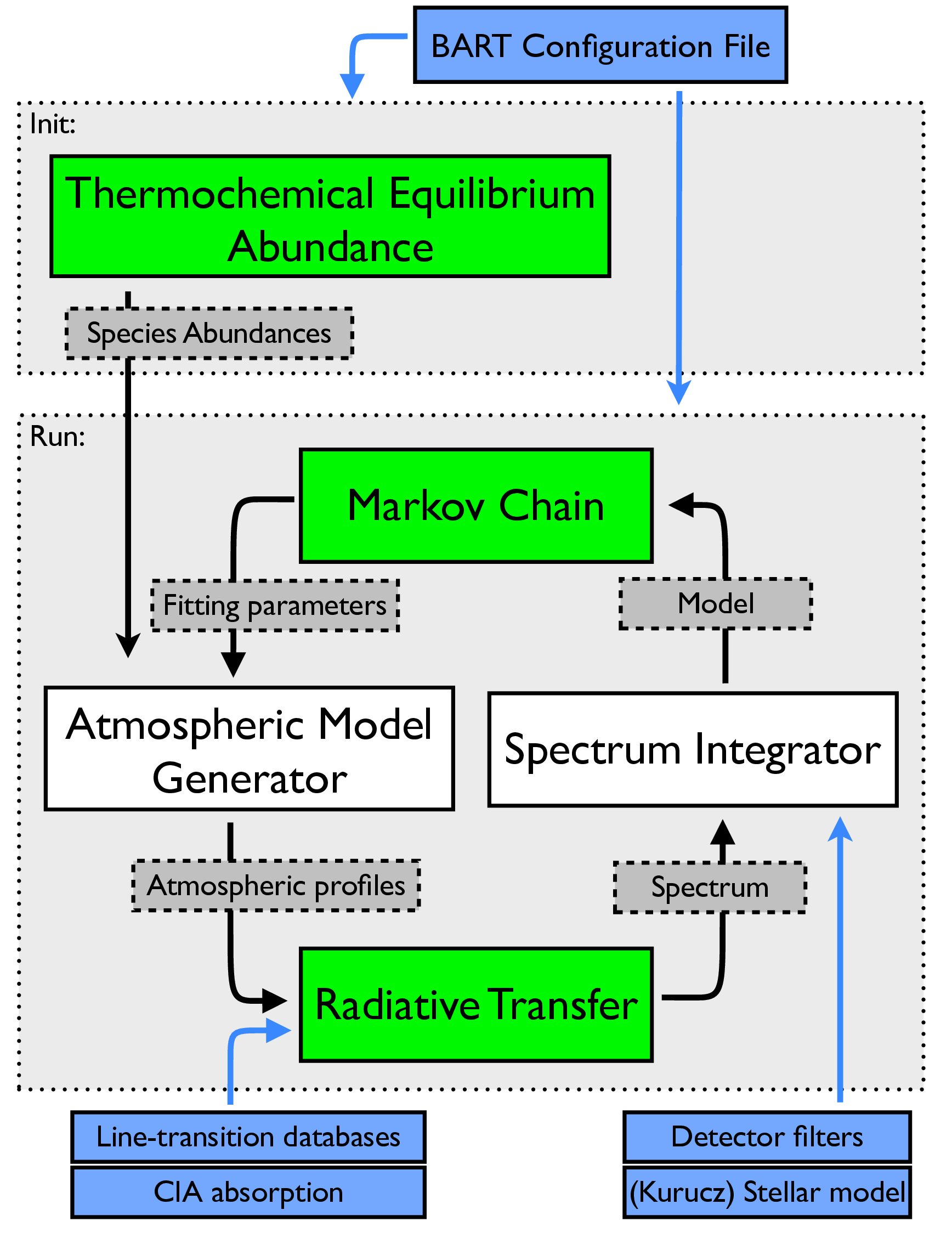}
\caption[BART flow chart]{\small BART flow chart.  The green boxes are
  the project's submodules, the white boxes are additional BART
  routines, the blue boxes are input files, and the gray boxes are the
  data passed around.  The BART configuration file sets up the
  parameters and inputs for all modules.  The TEA module provides
  initial species' abundance profiles for the atmospheric model.  The
  statistical module drives the parameter exploration by drawing
  samples in a Markov-chain loop.  The MCMC feeds the fitting
  parameters to the atmospheric model generator, which then computes
  the temperature, altitude, and abundance profiles.  The
  radiative-transfer module calculates the planetary spectrum, which
  is later integrated over the bandpasses of the observations.  The results
  are sent back to the MCMC, which compares them to the data and generates
  the set of fitting parameters for the next iteration.}
\label{fig:BARTflow}
\end{figure}

Figure \ref{fig:BARTflow} illustrates the interaction between BART and
its submodules.  The BART program is divided into two main sections.
The initialization section constructs a one-dimensional atmospheric
model of the pressure, temperature, and species' abundances.  Adopting
the pressure as the independent variable, BART calculates the
temperature profiles implementing the three-stream Eddington
approximation model \citep[][]{LineEtal2013apjRetrievalI}.  For the
abundances, the user can opt for thermochemical-equilibrium
\citep[through the TEA module,][]{BlecicEtal2015apsjTEA}, or set
vertically uniform values.

After the initialization, BART samples the parameter phase space via a
MCMC.  First, {\mcc} draws a set of parameters that will define the
atmospheric-model properties.  Next, an atmospheric generator computes
the temperature, abundance, and altitude of the atmospheric layers.
Then, Transit calculates the spectrum for the given atmospheric model.
After that, BART integrates the spectrum over the detector
spectral-response filters.  Finally, {\mcc} compares the
filter-integrated values to the data points (through $\chi\sp{2}$
statistics) and generates a new set of fitting parameters.  The data
drives the parameter exploration though the Metropolis ratio.  To
estimate the parameters' credible region, the MCMC iterates this cycle
thousands of times, populating the phase-space.  The resulting
posterior distribution of samples has a density proportional to the
model probability.  The following sections further detail
the components of the BART code.

\subsection{Thermochemical Equilibrium Abundance Module}
\label{sec:TEA}

The Python TEA code \citep{BlecicEtal2015apsjTEA} calculates the mole
mixing ratios (abundances) of atomic and molecular gaseous species
under thermochemical equilibrium.  TEA implements the methodology of
\citet{WhiteEtal1958jcpChemicalEquilibrium} and \citet{Eriksson1971};
given the elemental abundances, temperature, and pressure values, TEA
minimizes the Gibbs free-energy of the system in an iterative,
Lagrangian-optimization scheme.

Using TEA, \citet{BlecicEtal2015apsjTEA} successfully reproduced results
from \citet{BurrowsSharp1999apjChemicalEquilibrium}, the Chemical
Equilibrium with Applications (CEA) code, and
\citet{WhiteEtal1958jcpChemicalEquilibrium}.  The TEA code and
documentation are available at \href{https://github.com/dzesmin/TEA}
{https://github.com/dzesmin/TEA} under an open-source license.

BART uses the TEA module to calculate initial atmospheric abundances, for an initial, user-defined temperature profile.  These initial profiles are used as the base to scale and explore different atmospheric abundances (see Sec. \ref{sec:AbundProfiles}).

\subsection{Atmospheric Model Generator}
\label{sec:atmgen}

The model generator connects the statistical-driver with the
radiative-transfer module.  This module receives the free
parameters, computes the corresponding atmospheric profiles
(temperature, altitude, and abundances), and feeds the profiles to
Transit.
The temperature and altitude profiles are calculated from the given
physical properties and model parameters, whereas the abundance
profiles are calculated by scaling the input initial abundance
profiles with the model free parameters (scaling factors).

\subsubsection{Temperature Profile}
\label{sec:TPprofile}

The temperature profiles are calculated by applying the three-stream
Eddington-approximation model of \citet{LineEtal2013apjRetrievalI},
which is a variation of the model of
\citet{ParmentierGuillot2014aapTmodel}.  This model considers two heat
sources, the internal planetary energy (infrared), and the incident
stellar irradiation (optical).  The energy transfer is modulated by an
infrared and two optical streams, each one parameterized by a Plank
mean opacity; in practice, the thermal stream is parameterized by the
Plank mean thermal opacity, $\kappa$, whereas the optical streams are
parameterized by the optical-to-infrared ratio of the mean opacities:
$\gamma\sb{1}$ and $\gamma\sb{2}$.

A parameter $\beta$ modulates the absorbed stellar energy, working as
a proxy for the albedo and day-to-night energy-redistribution factor.
The input stellar irradiation is characterized by the temperature:
\begin{equation}
T\sb{\rm irr} = \beta \left( \frac{R\sb{s}}{2a}\right)\sp{1/2} T\sb{s},
\end{equation}
where $T\sb{s}$ and $R\sb{s}$ are the stellar temperature and radius,
respectively, and $a$ is the orbital semi-major axis.  A
$\beta$ value of unity corresponds to a planet
 with zero albedo and efficient energy redistribution.
The internal heat is characterized by an internal temperature,
$T\sb{\rm int}$.

The temperature at each pressure layer, $p$, is then given by:
\begin{equation}
T\sp{4}(p) = \frac{3 T\sb{\rm int}\sp{4}}{4} \left( \frac{2}{3} + \tau \right) + 
             \frac{3 T\sb{\rm irr}\sp{4}}{4} (1-\alpha) \xi\sb{1}(\tau)        +
             \frac{3 T\sb{\rm irr}\sp{4}}{4}    \alpha  \xi\sb{2}(\tau),
\end{equation}
with
\begin{equation}
\xi\sb{i}(\tau) = \frac{2}{3} + \frac{2}{3\gamma\sb{i}} \left[1 + \left(\frac{\gamma\sb{i}\tau}{2}-1\right)\exp(-\gamma\sb{i}\tau)\right] +
   \frac{2\gamma\sb{i}}{3} \left(1-\frac{\tau\sp{2}}{2}\right)E\sb{2}(\gamma\sb{i}\tau),
\end{equation}
where $\tau(p) = \kappa p/g$ is the thermal optical depth for the
given atmospheric gravity, $g$; $E\sb{2}(\gamma\sb{i}\tau)$ is the
second-order exponential integral; and the parameter $\alpha$
partitions the flux between two optical streams.
Since the internal flux has little impact on the spectra,
$T\sb{\rm int}$ is fixed during the MCMC for our runs; the free
parameters are then: $\kappa$, $\gamma\sb{1}$, $\gamma\sb{2}$,
$\alpha$, and $\beta$.

\subsubsection{Molecular-Abundance Profiles}
\label{sec:AbundProfiles}

The code modifies the atmospheric composition by scaling the
entire initial abundance profile with the abundance free parameter
($f\sb{X}$) for selected species $X$.  Let $q\sp{0}\sb{X}(p)$ be the
initial abundance profile at pressure $p$, then the modified
abundances are calculated as:
\begin{equation}
  q\sb{X}(p) = q\sp{0}\sb{X}(p) \times 10\sp{f\sb{X}}.
\end{equation}
Since the abundances may vary over several orders of magnitude, the
abundance free parameters modify the log-scale abundances.
To preserve the total mixing ratio at 1.0, the code adjusts the
abundances of {\molhyd} and He at each layer (keeping the {\molhyd}/He
abundance ratio constant).

By randomly scaling the abundances, the code can explore
disequilibrium compositions.  The user can define which atmospheric
species vary their abundances.  {\water}, {\methane}, CO, and
{\carbdiox} are the most-abundant spectroscopically-active species
that shape the infrared spectrum.  These are the standard species
included in most exoplanet retrievals
\citep[e.g.,][]{MadhusudhanSeager2009apjRetrieval,
  LineEtal2013apjRetrievalI, WaldmannEtal2014TauRexI}.

\subsubsection{Altitude -- Pressure}
\label{sec:radpress}

BART calculates the altitude (the radius) for each layer
using the hydrostatic\-/equilibrium equation:
\begin{equation}
\frac{\der p}{\der z} = -\rho g,
\end{equation}
where $p$, $z$, and $\rho$ are the pressure, altitude, and mass
density of the layers, respectively, and $g$ is the atmospheric
gravity.

For eclipse geometry, the emission spectrum depends on the relative
altitude between the layers ($\Delta z$); however, for transit
geometry, the transmission spectrum depends on the absolute altitude of
the layers.  Thus, in this case, the code fits for the radius
at a fiducial pressure level (0.1 bar by default, but adjustable by
the user).

\subsection{Radiative Transfer}
\label{sec:transit}

The radiative-transfer equation describes how light propagates as it
travels through a medium.  Let the specific intensity, $I\sb{\nu}$,
denote the power carried by rays per unit area, $\der A$, per unit
wavenumber, $\nu$, in the interval $\der \nu$ within a solid angle
$\der\Omega$.  Then, the radiative-transfer equation for the specific
intensity is given by:
\begin{equation}
  \frac{{\rm d}I\sb{\nu}}{{\rm d}s} = -e\sb{\nu}(I\sb{\nu} - S\sb{\nu}),
\label{eq:RadTran}
\end{equation}
where $s$ is the path traveled by the light ray, $e\sb{\nu}$ is
the atmospheric opacity, and $S\sb{\nu}$ is the source function (the
intensity contributed by the atmosphere into the beam).  The opacity
depends on the atmospheric composition, pressure, and temperature.

The observed flux is the integral of the intensity over the solid
angle.  Therefore, by solving the radiative-transfer equation for the
given observing geometry, we can model the observed transit and
eclipse depths as a function of wavelength.

The Transit module solves the one-dimensional radiative-transfer
equation for two relevant cases of exoplanet observations:
the transmission spectrum for transit observations, and
the hemisphere-integrated emission spectrum for eclipse observations.
The model assumes hydrostatic balance, local thermodynamic
equilibrium, and ideal gas law.
The opacity comes from electronic, rotational, and vibrational
line-transition absorptions (hereafter, simply called
``line transitions'') and collision-induced absorption (CIA).
Transit requires as inputs: (1) a configuration file that indicates
the wavenumber sampling, observing geometry, input files, etc.;
(2) a one-dimensional atmospheric model that specifies the atmospheric
composition and the pressure, altitude, temperature, and species
abundances of each layer; and
(3) line-by-line and/or cross-section opacity files.
The code was originally developed at Cornell University by Patricio
Rojo, as part of his dissertation project with Joseph Harrington \citep{RojoPhDT2006}.
Transit is a C, modular, object-oriented code, wrapped with
SWIG\footnote{\href{http://swig.org/}{swig.org/}} for use in Python.
Our version of Transit is an open-source project hosted at
\href{https://github.com/exosports/transit}
{https://github.com/exosports/transit}.

\begin{figure}[htb]
\centering
\includegraphics[width=0.5\linewidth, clip]{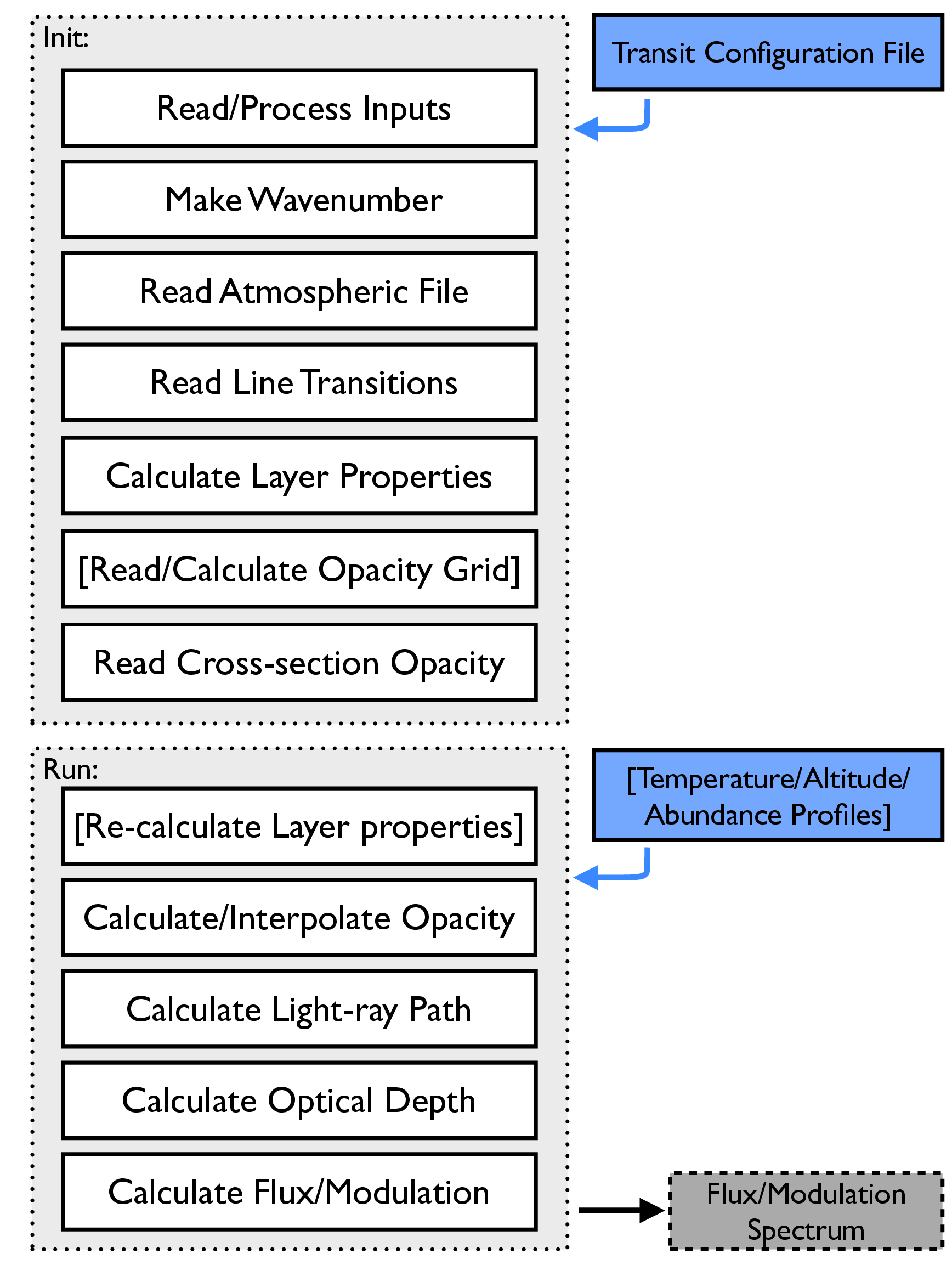}
\caption[Transit flow chart]{\small Transit flow chart.  The white
  boxes denote the main routines, the blue boxes the code inputs, and
  the gray box the output spectrum file.  The items in brackets are
  optional.  Transit consists of two main blocks of code,
  initialization (read input files) and run (evaluate spectrum).
  The code sequentially executes the
  routines from the top down.  When used for BART,
  the initialization block is executed once at the beginning of the
  run.  Then, BART executes the spectrum-calculation block for each
  iteration of the Markov chain.}
\label{fig:TransitFlow}
\end{figure}

The Transit program divides the spectrum calculation into two main
sections: initialization and run (Figure \ref{fig:TransitFlow}).  Once
Transit executes the initialization, it can produce multiple
spectrum models, updating the atmospheric model each time.

The first section of Transit reads and processes the input files.
The routines in this section (1)
read the configuration file, (2) create the (equi-spaced)
wavenumber array, (3) read the atmospheric-model and line-transition
opacity files, (4) calculate the density and altitude of the
atmospheric layers, along with the partition function of the species,
(5) compute or read a tabulated opacity grid (optional), and (6) read
cross-section opacity files.

The opacity grid is a four-dimensional table of precomputed opacities.
This table contains the opacities (in cm$\sp{2}$ gr$\sp{-1}$)
evaluated over the wavenumber array, at each atmospheric pressure
level, for a grid of temperatures, and for each absorbing species.
The opacity grid speeds up the spectrum evaluation allowing Transit to interpolate
the opacities from the table, instead of repeatedly computing the
line-by-line calculations.

The second section of Transit computes the emission or transmission
spectrum.  This section's routines (1) update the atmospheric
model (optional), (2) compute the opacity, either interpolating from
the opacity grid or from line-by-line calculations, (3) calculate the
light-ray path, (4) integrate the opacity over the ray paths to obtain
the optical depth, and (5) calculate the intensity and flux spectra.
The following sections further detail the spectral calculations, the
opacity calculations, and the available databases.

\subsubsection{Transit Geometry}
\label{sec:tgeom}

\begin{figure}[htb]
\centering
\includegraphics[width=0.7\linewidth, clip]{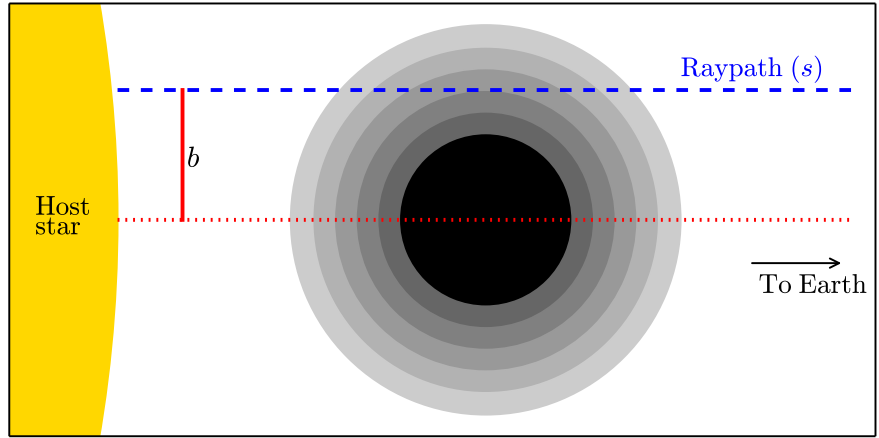}
\caption[Diagram for transit-geometry observation]{\small
  Transit-geometry observation diagram.  The atmospheric model consists
  of a set of one-dimensional spherically-symmetrical shells (gray
  gradient) around the planet center (black).  Given the geometry of
  the observation, all the observed light rays (e.g., blue dashed line)
  travel parallel to each other and, hence, the optical depth for each
  ray depends exclusively on the impact parameter ($b$) of the ray
  path.}
\label{fig:transitSketch}
\end{figure}

During a transit event (Fig.\ \ref{fig:transitSketch}) the planet
blocks a fraction of the stellar light, which is proportional to the
planet-to-star area ratio, $R\sb{\rm p}\sp{2}$/$R\sb{\rm s}\sp{2}$.
Since each species imprints a characteristic absorbing pattern as a
function of wavelength, the planetary atmosphere modulates the
transmission (or modulation) spectrum:
\begin{equation}
  M\sb{\nu}  = \frac{F\sb{\nu,O}-F\sb{\nu,T}}{F\sb{\nu,O}},
  \label{eq:modulation}
\end{equation}
where $F\sb{\nu,T}$ and $F\sb{\nu,O}$ are the observed fluxes during
transit and out of transit, at wavenumber $\nu$.

For transmission geometry, the planetary emission is small compared to
the stellar intensity.  Then, we can neglect the source-function term
from Equation (\ref{eq:RadTran}):
\begin{equation}
  \frac{{\rm d}I\sb{\nu}}{{\rm d}s} = -e\sb{\nu}I\sb{\nu}.
\end{equation}

Let $I\sb{\nu}\sp{0}$ be the stellar specific intensity.  By defining
the optical depth along the path as:
\begin{equation}
\label{eq:tau}
\tau\sb{\nu} = \int\sb{\rm path} e\sb{\nu} {\der}s,
\end{equation}
the solution for the transmission radiative-transfer equation becomes:
\begin{equation}
I\sb{\nu} = I\sb{\nu}\sp{0} e\sp{-\tau\sb{\nu}}.
\end{equation}

The specific flux, $F\sb{\nu}$, is the integral of the
specific intensity in the direction of the observer over the solid
angle, $\der\Omega$:
\begin{equation}
F\sb{\nu} = \int I\sb{\nu} cos\theta \der\Omega,
\label{eq:SpecificFlux}
\end{equation}
where $\theta$ is the angle between the ray beams and the normal vector of
the detector.  Since the distance from Earth to the system,
$d$, is much larger than the distance from the star to the planet, the
stellar radius, and the planetary radius ($d \gg a$,
$d \gg R\sb{\rm s}$, and $d \gg R\sb{\rm p}$, respectively), we can
assume that the observed rays travel in a parallel beam thorough the
planetary atmosphere.  Then, we can rewrite the solid-angle integral
as an integral over the projected disk of the system:
\begin{equation}
F\sb{\nu}  = \frac{2\pi}{d\sp{2}} \int\sb{0}\sp{R\sb{\rm s}} I\sb{\nu} r \der r.
\label{eq:tranflux}
\end{equation}

For the out-of-transit flux, Equation (\ref{eq:tranflux}) gives:
\begin{equation}
F\sb{\nu,O}  = \pi I\sb{\nu}\sp{0} \left(\frac{R\sb{\rm s}}{d}\right)\sp{2}.
\label{eq:ootFlux}
\end{equation}

For the in-transit flux, we assume that the planetary atmosphere is a set
of spherically-symmetrical homogeneous layers (Fig.\
\ref{fig:transitSketch}).  Then, the optical depth becomes a function
of the impact parameter of the light ray, $b$:
$\tau\sb{\nu} = \tau\sb{\nu}(b)$.

Consider now a planetary altitude, $R\sb{\rm top}$, high enough such that
$e\sp{-\tau\sb{\nu}(R\sb{\rm top})} \approx 1$ (in practice, the top
layer of the atmospheric model).
In addition, we neglect the variation of limb-darkening over the
projected area of the planet onto the star, and we correct the
limb-darkening factor for the measured transit depth.  Then, the
in-transit specific flux becomes:
\begin{eqnarray}
F\sb{\nu,T} & = & \frac{2\pi}{d\sp{2}} 
   \int\sb{0}\sp{R\sb{\rm s}} I\sb{\nu}\sp{0} e\sp{-\tau\sb{\nu}(b)}r{\der}r, \\
            & = & \frac{2\pi}{d\sp{2}} I\sb{\nu}\sp{0}
   \left[ \int\sb{0}\sp{R\sb{\rm top}} e\sp{-\tau\sb{\nu}} b \der b\ +\
        \frac{R\sb{\rm s}\sp{2} - R\sb{\rm top}\sp{2}}{2} \right].
\label{eq:itFlux}
\end{eqnarray}

Finally, substituting Equations (\ref{eq:ootFlux}) and (\ref{eq:itFlux})
into (\ref{eq:modulation}), we calculate the modulation spectrum as:
\begin{equation}
M\sb{\nu}  = \frac{1}{R\sb{\rm s}\sp{2}}\left[ R\sb{\rm top}\sp{2} - 
                2\int\sb{0}\sp{R\sb{\rm top}} e\sp{-\tau\sb{\nu}} b \der b
             \right].
\end{equation}

\subsubsection{Eclipse Geometry}
\label{sec:eclipse}

\begin{figure}[htb]
\centering
\includegraphics[width=0.7\linewidth, clip]{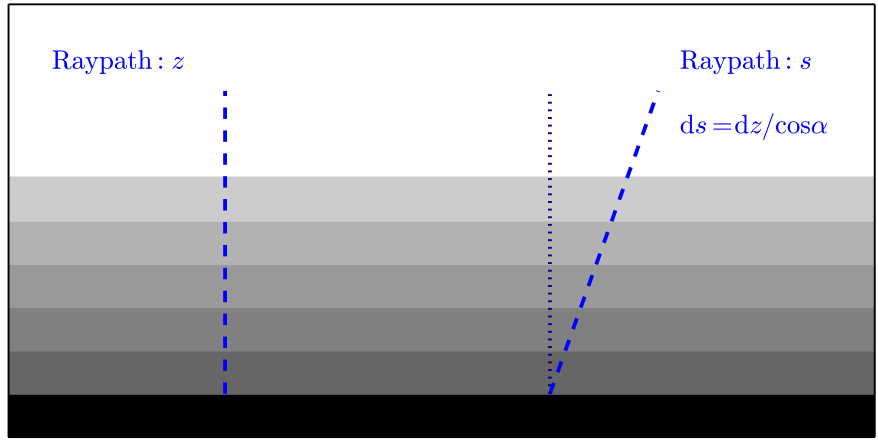}
\caption[Diagram for eclipse-geometry observation]{\small Diagram for
  eclipse-geometry observation.  The atmospheric
  model consists of a set of plane-parallel layers (gray gradient)
  above the center of the planet (black).  Transit calculates the
  emerging intensity along a set of paths (blue dashed lines) with a
  given incident inclination with respect to the normal vector of the planet,
  $\alpha$.}
\label{fig:eclipseSketch}
\end{figure}

An eclipse event reveals the planetary day-side emission.  The Transit
eclipse-geometry code (mainly implemented by J. Blecic) solves the
radiative-transfer equation to obtain the emerging intensity at the
top of the atmosphere.  For this case, we adopt the plane-parallel
approximation (Fig.\ \ref{fig:eclipseSketch}), which models the
atmosphere as a stratified set of plane horizontal homogeneous layers.
Additionally, we adopt the Local Thermodynamic Equilibrium
approximation, where the source function becomes the Planck function
$B\sb{\nu}(T)$.

Let $\der \tau\sb{\nu,z} = -e\sb{\nu} \der z$ be the vertical optical
depth (with origin $\tau=0$ at the top of the atmosphere).  The path
($\der s$) of a ray with an angle $\alpha$, with respect to the normal vector,
is related to the vertical path as: $\der s = \der z/\cos\alpha \equiv
\der z / \mu$.  Then, the radiative-transfer equation becomes:
\begin{equation}
-\mu\frac{\der I\sb{\nu}}{\der \tau\sb{\nu,z}} = - I\sb{\nu}  +  S\sb{\nu},
\end{equation}
which can be rewritten as:
\begin{equation}
-\mu\frac{\der}{\der \tau\sb{\nu,z}}\left( I\sb{\nu}e\sp{-\tau\sb{\nu,z}/\mu} \right) 
 =  S\sb{\nu}e\sp{-\tau\sb{\nu,z}/\mu}.
\label{eq:RTeclipse}
\end{equation}

Transit calculates the emergent intensity by integrating Equation
(\ref{eq:RTeclipse}) from the deep layers to the top of the
atmosphere.  At depth, the atmosphere is optically thick
($\tau=\tau\sb{b} \gg 1$), such that $\exp(-\tau\sb{b}/\mu) \to 0$.
Therefore, the intensity at the top of the atmosphere is given by:
\begin{equation}
I\sb{\nu}(\tau=0, \mu) = 
     \int\sb{0}\sp{\tau\sb{b}}B\sb{\nu}e\sp{-\tau/\mu}\der\tau/\mu.
\label{eq:EclipseIntensity}
\end{equation}

By changing variables from the angle $\theta$ to the angle on the
planet hemisphere, $\alpha$ (related by: $d \sin\theta =
R\sb{p}\sin\alpha$),  the
emergent flux (Eq.\ \ref{eq:SpecificFlux}) becomes:
\begin{eqnarray}
F\sb{\nu} & = & \left(\frac{R\sb{p}}{d}\right)\sp{2} 
                \int\sb{0}\sp{2\pi}\int\sb{0}\sp{\pi/2} 
                   I\sb{\nu} \cos(\alpha)\sin(\alpha) \der\alpha \der\phi \\
          & = & 2\pi \left(\frac{R\sb{p}}{d}\right)\sp{2}
                  \int\sb{0}\sp{1} I\sb{\nu} \mu\der\mu.
\end{eqnarray}

Transit approximates this integral by summing over a discrete set of
angles:
\begin{equation}
F\sb{\nu} \approx \pi \left(\frac{R\sb{p}}{d}\right)\sp{2}
           \sum\sb{i} <I\sb{\nu}(\mu\sb{i})> \Delta\mu\sb{i}\sp{2}.
\label{eq:pflux}
\end{equation}

Additionally, Transit approximates the average intensity in each
angle interval by the intensity of the mean angle,
$<I\sb{\nu}(\mu)>\ \approx I\sb{\nu}(<\mu>)$.  Transit computes
$I\sb{\nu}(<\mu>)$ through a Simpson numerical integration of Equation
(\ref{eq:EclipseIntensity}) and then returns the surface emergent
flux:
\begin{equation}
  F\sb{\nu}\sp{\rm surf} \approx \pi \sum\sb{i} <I\sb{\nu}(\mu\sb{i})> \Delta\mu\sb{i}\sp{2}.
\end{equation}

\subsubsection{Collision-induced Absorption}

Collision-induced absorption is one of the main sources of atmospheric opacity. Sec.\ \ref{sec:databases} details how Transit incorporate the CIA calculations.
CIA occurs when particles without an
intrinsic electric dipole moment collide.  The collisions induce a
transient dipole moment, which allows dipole transitions.  The short
interaction time of the collisions broadens the line profiles,
generating a smooth CIA spectrum.  The CIA opacity
scales with the density of the colliding species, and thus becomes
more relevant at the deeper, higher-pressure layers of the atmosphere
\citep{SharpBurrows2007apjOpacities}.  For gas-giant planets, the two
most important CIA sources are {\molhyd}--{\molhyd} and, to a lesser extent,
{\molhyd}--He collisions.

\subsubsection{Line-transition Absorption}

A species absorbs or emits photons at specific wavelengths,
corresponding to the characteristic energies between its electronic,
rotational, and vibrational transitions, giving rise to
line-transition opacities.  The atmospheric temperature and pressure
determine the strength and shape of a species' line transitions.

Transit calculates the line-transition opacity, $e\sb{\nu}$, in a line-by-line
scheme, adding the contribution from each broadened line-transition, $j$:
\begin{equation}
e\sb{\nu} = \sum\sb{j} S\sb{j} V(\nu-\nu\sb{j}),
\label{eq:opacity}
\end{equation}
where $\nu\sb{j}$ is the wavenumber of the line transition, $S\sb{j}$ is the
line strength (in cm{\tnt}), and $V$ is the line profile (Voigt).  The
line strength is given by:
\begin{equation}
S\sb{j} = \frac{\pi e\sp{2}}{m\sb{e}c\sp{2}} \frac{(gf)\sb{j}}{Z\sb{i}(T)}
          n\sb{i} \exp\left(-\frac{h c E\sb{\rm low}\sp{j}}{k\sb{B}T}\right)
          \left\{1-\exp\left(-\frac{hc\nu\sb{j}}{k\sb{B}T}\right)\right\},
\end{equation}
where $gf\sb{j}$ and $E\sp{j}\sb{\rm low}$ are the weighted oscillator
strength and lower state energy level (in cm$^{-1}$) of the line
transition, respectively; $Z\sb{i}$ and $n\sb{i}$ are the partition
function and number density of the isotope $i$, respectively; $T$ is
the atmospheric temperature; $e$ and $m\sb{e}$ are the electron's charge
and mass, respectively; $c$ is the speed of light, $h$ is Planck's
constant; and $k\sb{\rm B}$ is the Boltzmann's constant.

The Voigt profile considers the Doppler and the Michelson-Lorentz
collision broadening.  The Doppler and Lorentz half-widths at half
maximum \citep[HWHM,][]{Goody1996AtmosphericPhysics} are,
respectively:
\begin{eqnarray}
\label{eq:alphad}
\alpha\sb{d} & = & \frac{\nu\sb{j}}{c}
                   \sqrt{\frac{2k\sb{\rm B}T\ln{2}}{m\sb{i}}}, \\
\label{eq:alphal}
\alpha\sb{l} & = & \frac{1}{c} \sum\sb{a} n\sb{a}d\sb{a}\sp{2}
                        \sqrt{\frac{2k\sb{\rm B}T}{\pi}
                             \left(\frac{1}{m\sb{i}} + \frac{1}{m\sb{a}}\right)}
\end{eqnarray}
where $m\sb{i}$ is the mass of the absorbing species and $d\sb{a}$ is
the collision diameter between the interacting particles.  The sum
goes over all species, $a$, in the atmosphere.  Transit pre-computes,
during the initialization section, a set of Voigt profiles
\citep[following the algorithm of][]{Pierluissi1977jqsrtVoigt} for a
grid of Doppler and Lorentz HWHM.

The Transit code implements a series of
steps to improve the performance of the line-by-line opacity
calculation (Eq.\ \ref{eq:opacity}).
When multiple lines of a same isotope fall in the same wavenumber bin,
the code adds the line strengths before broadening the line over the
wavenumber array---note that the line-strength calculation is fast
compared to the Voigt-profile broadening. 
Additionally, a line-strength cutoff (a user-adjustable parameter)
prevents the computation of the weaker lines that do not contribute
significantly.  Our cutoff is relative to the largest line strength in
each layer instead of the fixed-temperature cutoff of previous works
\citep[e.g.,][]{SharpBurrows2007apjOpacities,
  LineEtal2012apjExospecinfo} because of the large strength variation
with temperature.
Lastly, Transit automatically adjusts the wavenumber sampling at each
layer to avoid under- or oversampling the line profiles.

\subsubsection{Opacity Databases}
\label{sec:databases}

The CIA input files of Transit (cross-section opacity files) are ASCII
tables of the opacity (in units of cm$\sp{-1}$amagat$\sp{-2}$, with $1\
{\rm amagat} = n\sb{0} = 2.68679\tttt{19}$ molecules cm$\sp{-3}$) as a
function of wavenumber and temperature.
Transit provides Python scripts to reformat the CIA data files given
by A.
Borysow\footnote{\href{http://www.astro.ku.dk/~aborysow/programs}
  {astro.ku.dk/{\sim}aborysow/programs}} and HITRAN
\citep{RichardEtal2012jqsrtCIA} into its internal format.

\begin{figure}[ht]
\centering
\includegraphics[width=0.7\linewidth, clip]{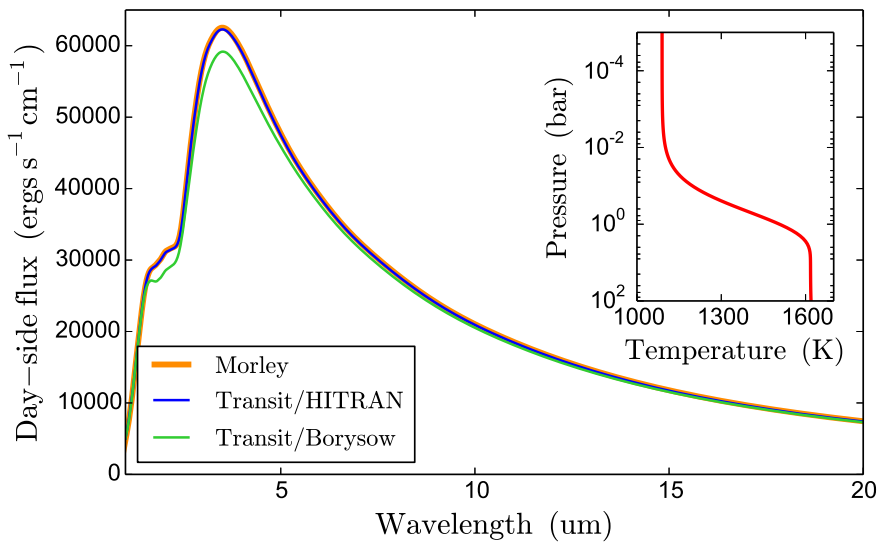}
\caption[Model emission spectra for pure collision-induced absorption]
{\small  Model emission spectra for pure collision-induced absorption.
  The blue and green curves show the Transit spectra for HITRAN and
  Borysow CIA opacities, respectively.  The orange curve shows the
  HITRAN CIA spectrum of C. Morley.  All spectra were calculated
  for an atmospheric model with uniform mole mixing ratios composed of
  85\% {\molhyd} and 15\% He, for a planet with a 1 {\rjup} radius and
  a surface gravity of 22 m\;s$\sp{-2}$.  The inset shows the
  atmospheric temperature profile as a function of pressure.}
\label{fig:CIAvalid}
\end{figure}

To evaluate the CIA opacity in the atmosphere, Transit performs a
bicubic interpolation (wavenumber and temperature) from the tabulated
CIA opacities ($e\sb{\rm cia}\sp{\rm t}$), and scales the values to
units of cm$\sp{-1}$ ($e\sb{\rm cia}$), multiplying by the number
density of the species:
\begin{equation}
e\sb{\rm cia} = e\sb{\rm cia}\sp{\rm t} \frac{n\sb{1}}{n\sb{0}}
                                        \frac{n\sb{2}}{n\sb{0}},
\end{equation}
where $n\sb{1}$ and $n\sb{2}$ are the number densities of the colliding
species (in units of molecules cm$\sp{-3}$).  Figure
\ref{fig:CIAvalid} shows Transit emission spectra for pure
{\molhyd}--{\molhyd} and {\molhyd}--He CIA opacities.  When compared
against the models of C. Morley \citetext{priv.\ comm.}, we agree to
better than 0.5\%.  We also noted that the HITRAN absorption is weaker
than the Borysow absorption, producing emission spectra 2\%--8\%
stronger for this atmospheric model.

Transit treats the cross-section line-transition opacity inputs in a
similar manner as the CIA data files, except that the opacity is given
in cm$\sp{-1}$amagat$\sp{-1}$ units.  Transit provides a routine to
re-format HITRAN and ExoMol\footnote{
  \href{http://www.exomol.com/data/data-types/xsec}
  {http://exomol.com/data/data-types/xsec}} cross-section data files into the
format required by Transit.

Transit stores the line-by-line opacity data into a Transit Line
Information (TLI) binary file.  A TLI file contains a header and the
line-transition data.  The header contains the number and names of
databases, species, and isotopes, and the partition function per
isotope as a function of temperature.  The line-transition data consist
of four arrays with the transition's wavelength, lower-state energy,
oscillator strength, and isotope ID.  The original Transit line-reading code has been rewritten in Python to maje it easier for uers to add functions to read additional line-list formats.  As of this writing, the Transit line reader can
process line-transition files from the HITEMP/HITRAN lists
\citep{RothmanEtal2010jqsrtHITEMP, Rothman2013JqsrtHITRAN}, the
{\water} list from \citet{PartridgeSchwenke1997jcpH2O}, the TiO list
from \citet{Schwenke1998TiO}, and the VO list from B. Plez
(priv.\ comm.).

Most line-by-line databases provide tabulated partition-function
files.  For the HITRAN and HI-TEMP databases, Transit provides
an adaptation\footnote{\href{https://github.com/pcubillos/ctips}
  {https://github.com/pcubillos/ctips}} of the Total Internal Partition Sums
code\footnote{\href{http://faculty.uml.edu/robert\_gamache/software/index.htm}
  {http://faculty.uml.edu/robert\_gamache/software/index.htm}}
\citep[TIPS,][]{LaraiaEtal2011icarusTIPS}.

\begin{figure}[htb]
\centering
\includegraphics[width=0.4\linewidth]{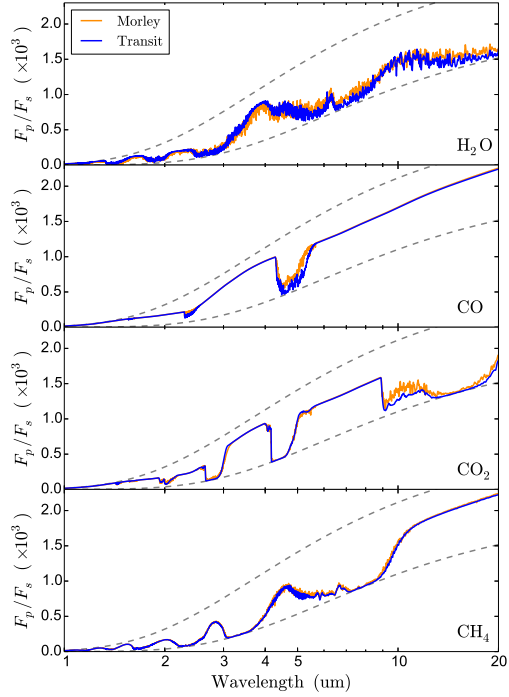}
\caption[Model planet-to-star flux ratio spectra for water,
  carbon monoxide, carbon dioxide, and methane]{
\small Model planet-to-star flux ratio spectra for water,
  carbon monoxide, carbon dioxide, and methane (top to bottom panels).  The
  blue and orange solid curves show the Transit and the Morley
  radiative-transfer spectra, respectively (Gaussian-smoothed for
  better visualization).  The planetary atmospheric model is the same as
  in Fig.\ (\ref{fig:CIAvalid}), with additional uniform mixing ratios
  of $\ttt{-4}$ for the respective species in each panel.  The
  stellar model corresponds to a blackbody spectrum of a 1 {\rsun}
  radius and 5700 K surface-temperature star.  The dashed grey lines
  indicate the flux ratio for planetary blackbody spectra at 1090 and
  1620 K (atmospheric maximum and minimum temperatures).  The CIA
  opacity comes from \citet{RichardEtal2012jqsrtCIA}.  For {\water}
  and {\methane} both models used the line lists from
  \citep{PartridgeSchwenke1997jcpH2O} and
  \citet{YurchenkoTennyson2014mnrasExomolCH4}, respectively.  For CO,
  the Morley models used the line list from
  \citet{Goorvitch1994apjsCOlinelist}, whereas Transit used the HITEMP
  line list (which is based on the
  \citeauthor{Goorvitch1994apjsCOlinelist} line list).  For
  {\carbdiox}, Morley used \citet{HuangEtal2013jqsrtCO2} and \citet{HuangEtal2014jqsrtCO2}, whereas Transit
  used HITEMP.  The largest differences correspond to CO and {\carbdiox}, the molecules that correspond to different databases.  There are also some differences in the {\water} spectrum.}
\label{fig:emission}
\end{figure}

\begin{figure}[htb]
\centering
\includegraphics[width=0.7\linewidth, clip]{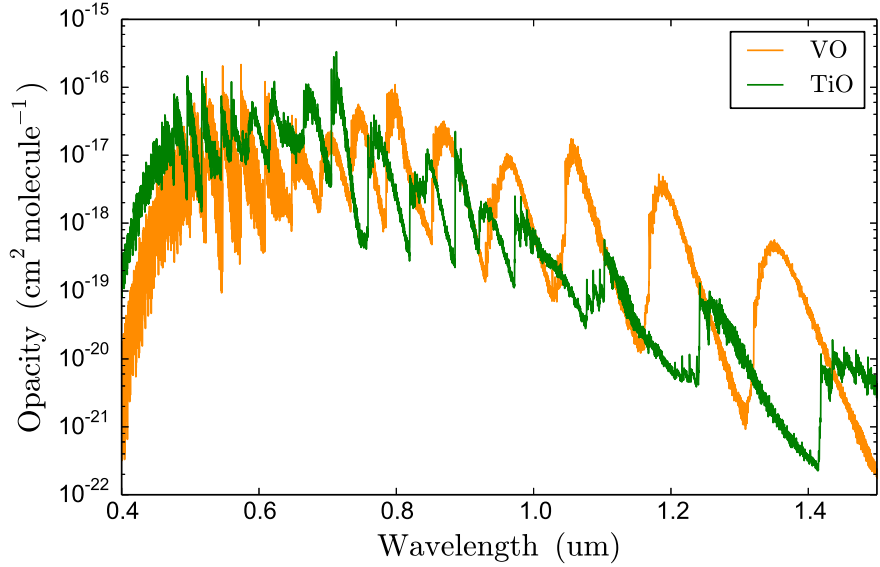}
\caption[Near-infrared titanium- and vanadium-oxide
  opacity spectra]{\small Near-infrared titanium- and vanadium-oxide
  opacity spectra.  Transit calculated the opacities
  at a temperature of 2200 K and a pressure of 10.13 bar
  (10 atm).  Our models are consistent with Figures (4) and (5) of
  \citet{SharpBurrows2007apjOpacities}.}
\label{fig:TiO-VOopacity}
\end{figure}

Figure \ref{fig:emission} shows an example of the emission spectra of
{\water}, CO, {\carbdiox}, and {\methane}.  A comparison with the
Morley models shows a good agreement for all four cases.  Figure
\ref{fig:TiO-VOopacity} shows an example of the 
TiO and VO opacity spectra.  The Transit spectra 
agree with that of \citet{SharpBurrows2007apjOpacities}.

\subsection{Spectrum Band Integration}
\label{sec:outcon}

To compare the model spectra to the data, BART integrates the spectra
over the detector transmission filters (also called filter bands).
The filters describe the spectral response curve for each observing
band.  For transit geometry, the observed transit depths are directly
compared to the band-integrated transmission spectra.  On the other
hand, for eclipse geometry, the observed eclipse depths correspond to
the planet-to-star flux ratio:
\begin{equation}
  F\sb{p}/F\sb{s} = \frac{F\sb{p}\sp{\rm surf}}{F\sb{s}\sp{\rm surf}}\left(\frac{R\sb{p}}{R\sb{s}}\right)\sp{2},
\end{equation}
where $F\sb{p}\sp{\rm surf}$ is the surface flux spectrum of the star.
BART incorporates the Kurucz models for the stellar spectra
\citep{CastelliKurucz2004}.

\subsection{Statistical MCMC Driver}
\label{sec:MC3}

To drive the MCMC exploration, BART uses the open-source module,
Multi-Core Markov-Chain Monte Carlo
\citep[{\mcc},][]{CubillosEtal2015apjRednoise}, hosted at
\href{https://github.com/pcubillos/MCcubed}
{https://github.com/pcubillos/MCcubed}.  {\mcc} is a general-purpose,
model-fitting software.
The code is written in Python with several C\-/code extensions,
and has parallel multiprocessor capacity.
To explore the parameter phase space, BART uses the Differential\-/Evolution Markov\-/chain Monte Carlo
algorithm \citep[DEMC,][]{Braak2006DifferentialEvolution} implemented in {\mcc}.

The posterior in the Metropolis-Hastings acceptance rule can take
uniform non\-/informative, Jeffrey's non\-/informative, or Gaussian
informative priors.  {\mcc} checks for the MCMC convergence via the
\citet{GelmanRubin1992} statistics.  An {\mcc} run returns the sampled
parameter posterior distribution, the best-fitting values, the limits
of the 68\% credible region, and the acceptance rate.
At the end of the MCMC run, the program produces several plots to help
visualize the results.  Trace plots show the sequence of values along the MCMC iterations for each free parameter.  Marginal-posterior plots show the posterior probability distribution for each free parameter.  Pairwise-posterior plots show the two-dimensional posterior distribution for all the combinations of free-parameter pairs.
These plots help to identify non-convergence, multi-modal
posteriors, correlations, or incorrect priors.

\subsubsection{{\mcc} Usage for BART}
\label{sec:MC3BART}

BART and {\mcc} implement the same configuration-file format, allowing the user to define all parameters in a single file.
The BART code runs best on multiple processors.  BART includes
a modified worker program that allows the code to separate the
initialization from the MCMC-loop code (see Sec.\ \ref{sec:c5code}).
This worker routine is comprised of the atmospheric model generator, the run
section of Transit (see Fig.\ \ref{fig:transitSketch}), and the
spectrum integrator.  During the MCMC loop, the data transfered
through MPI are the free-parameter values and the simulated observations obtained by band-integrating the model spectra (see Fig.\ \ref{fig:BARTflow}).
The line-by-line opacity calculation in Transit is the most
computationally demanding task (typically 95\% of the run time).  Thus, a
BART run avoids this step by using an opacity table, either loaded or
calculated during the initialization.
Transit running times vary largely depending on the CPU performance
and modeling configuration.

\subsection{Retrieval Validation Test}
\label{sec:BARTval}

\begin{figure*}[htb]
\centering
\includegraphics[width=\linewidth, clip]{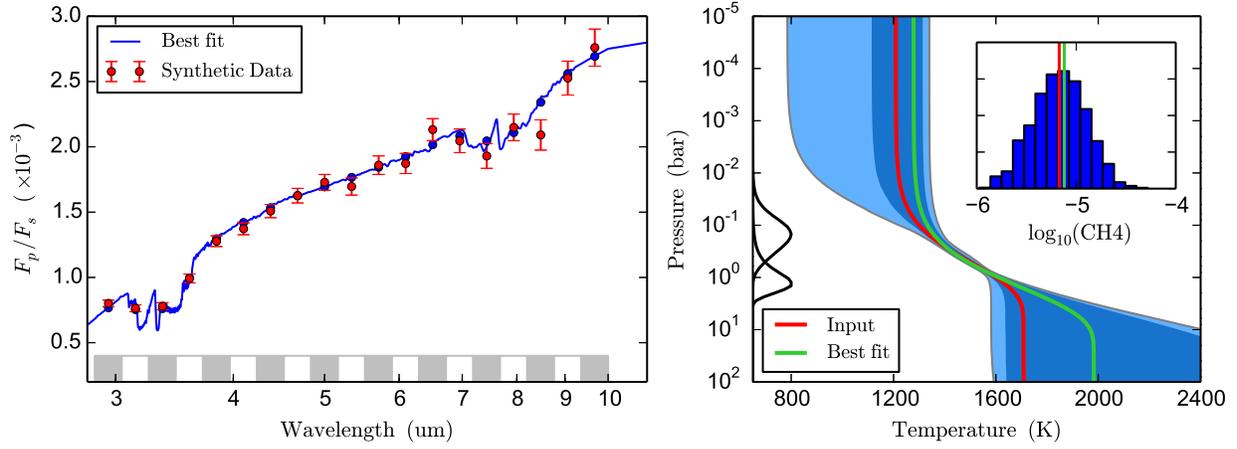}
\caption[Retrieval for a synthetic model spectrum]{\small BART retrieval for a
  synthetic model spectrum.  The left panel shows the planet-to-star
  flux ratio for a synthetic observation (red points with error bars)
  and the retrieved best-fitting model (blue curve).  The bottom
  gray/white boxes show the filter bands for each data point.  The
  right panel shows the retrieved temperature profile and 
  methane-abundance posteriors.  The dark- and light-blue shaded areas show
  the 68\% and 95\% credible regions, respectively.  The black solid
  curves on the left show two typical contribution functions for this
  simulation \citep[for further details, see][]{BlecicEtal2015apjBART}. The inset shows the posterior histogram for the methane
  abundance scale factor.}
\label{fig:retrieval}
\end{figure*}

In this section, we applied the BART analysis to synthetic
observations to test the retrieval code.
We generated an atmospheric model for a hot-Jupiter planet with the
characteristics of the system HD~209458.  The system parameters are
$R\sb{p}=1.35$ {\rjup}, $R\sb{s}=1.145$ {\rsun}, and $T\sb{s}= 6075$
K.  Figure \ref{fig:retrieval}, right panel, shows the input
temperature-pressure profile (red curve).  TEA calculated the
thermochemical-equilibrium abundances for this atmospheric model,
given solar elemental abundances
\citep{AsplundEtal2009araSolarComposition}.

The atmospheric model considered the opacities only from
{\molhyd}-{\molhyd} and {\molhyd}-He CIA
\citep{RichardEtal2012jqsrtCIA}, and {\methane} line transitions
\citep{Rothman2013JqsrtHITRAN}.  We generated a synthetic emission
spectrum, with $\sim6\times$ enhanced methane abundance, over the
2--10 {\micron} range (the range covered by the Spitzer's Infrared Array
Camera detectors).  We integrated the spectrum over 19 synthetic broad
filter bands to generate the synthetic observations.  We further added
Poisson noise to the synthetic datapoints, giving a signal-to-noise
ratio of 26 per point on average.

We configured BART to retrieve the methane abundance and a simplified
version of the three-stream Eddington-approximation temperature model.
We kept $\alpha$ fixed at a value of 0.0, which gives zero weight to
the stream controlled by $\gamma\sb{2}$.  Thus, only $\log\kappa$,
$\log\gamma\sb{1}$, and $\beta$ remained as fitting parameters.
Figure \ref{fig:retrieval} shows the retrieval results and the input
models.  The best-fitting model and the posterior distributions of the
temperature profile and the {\methane} abundace agree with the input
values to within the 68\% credible region.  The better
constrained region of the atmosphere (around 1 bar) agrees well with
the peak of the contribution curves for these observing
bands (the altitude where the atmosphere
changes from optically thin to optically thick).

\section{Application to HAT-P-11b}
\label{sec:c5analysis}

In this section we analyzed the transit observations of the
Neptune-sized planet HAT-P-11b.  We re-analyzed the Spitzer data,
retrieved the planet's atmospheric properties, and compared our
results to the similar analysis by \citet[][hereafter
F14]{FraineEtal2014natHATP11bH2O}.

The exoplanet HAT-P-11b \citep{BakosEtal2010apjHATP11b} is slightly
larger than Neptune in mass (26 $M\sb{\oplus}$) and radius (4.7
$R\sb{\oplus}$).  The planet orbits a K4 dwarf star ($R\sb{s}=0.75$
{\rsun}, $T\sb{s}= 4780$ K), at a distance of 0.053 AU, with a period
of 5 days.  Given these parameters, the planetary equilibrium
temperature (temperature at which the emission as blackbody balances
the absorbed energy, assuming zero albedo and efficient heat
redistribution) is $T\sb{\rm eq} = 878$ K.

\citetalias{FraineEtal2014natHATP11bH2O} observed transits of
HAT-P-11b with the Hubble Space Telescope's (HST) Wide Field Camera 3
(WFC3) and the Spitzer's Infrared Array Camera (IRAC) 3.6 and
4.5~{\microns} bands.  The WFC3 spectroscopic data covers the 1.1 to
1.7~{\micron} region of the spectrum with a spectral resolution of
$\sim 75$.  \citetalias{FraineEtal2014natHATP11bH2O} obtained Kepler
Space Telescope data concurrently with the Spitzer observations,
allowing them to model star spots on the surface of the star.
\citetalias{FraineEtal2014natHATP11bH2O} characterized the atmospheric
composition with the Self-Consistent Atmospheric Retrieval Framework
for Exoplanets (SCARLET) tool
\citep{Benneke2015apjSCARLETretrieval}.  They detected water
absorption features and estimated a best-fitting metallicity of 190
times the solar metallicity.

Note, however, that the activity of the host star may introduce a
variation in the observed transit depths at
different epochs.  Although \citetalias{FraineEtal2014natHATP11bH2O}
estimated a transit depth uncertainty of $51$ p.p.m., their
best-fitting offset between the WFC3 and Spizer data was $\sim 93$
p.p.m.

\subsection{Analysis of the Spitzer Data}
\label{sec:c5HATana}

We re-analyzed the HAT-P-11b Spitzer transmission light curves with
our Photometry for Orbits, Eclipses, and Transits (POET) pipeline
\citep{StevensonEtal2010natGJ436b, StevensonEtal2012apjHD149026b,
  StevensonEtal2012apjGJ436c, CampoEtal2011apjWASP12b,
  NymeyerEtal2011apjWASP18b, CubillosEtal2013apjWASP8b,
  CubillosEtal2014apjTrES1}.  The Spitzer Space Telescope
(warm-mission) obtained four transit light curves of HAT-P-11b (PI
Deming, program ID 80128) using the IRAC instrument: two
visits at 3.6 {\microns} (2011 Jul 07 and Aug 15) and two visits at
4.5 {\microns} (2011 Aug 05 and Aug 29).  The telescope observed in
sub-array mode with a cadence of 0.4 s.

The POET analysis started by reading the Spitzer basic calibrated data
(BCD) frames (Spitzer pipeline version 18.18.0).  POET discarded bad
pixels, determined the target position on the detector (fitting a
two-dimensional Gaussian function), and calculated aperture photometry
to produce a raw light curve.  For each event, we tested circular
apertures with radii ranging from 1.75 to 4.0 pixels (in 0.25 pixel
increments).

POET simultaneously modeled the out-of-transit system flux, eclipse
curve, and telescope systematics with a Levenberg-Marquardt minimizer
and a Markov-chain Monte Carlo routine.
The Spitzer IRAC systematics include temporal and intra-pixel
sensitivity variations \citep{CharbonneauEtal2005apjTrES1}.  We
modeled the temporal systematic with a set of time-dependent ``ramp''
models \citep[polynomial, exponential, and logarithmic functions, and
combinations of them;][]{CubillosEtal2013apjWASP8b,
  CubillosEtal2014apjTrES1}.  We modeled the intra-pixel systematics
with the Bi-Linearly Interpolated Sub-pixel Sensitivity (BLISS) map
model \citep{StevensonEtal2012apjHD149026b}.
We used a \citet{MandelAgol2002apjLightcurves} model for the transit
light curve.  The transit model fit the planet-to-star radius ratio
($R\sb{p}/R\sb{s}$), transit midpoint time, cosine of inclination, and
semi-major axis-to-stellar radius ratio ($a/R\sb{s}$).  We adopted the same
limb-darkening parameters as \citetalias{FraineEtal2014natHATP11bH2O}.
We set the BLISS map grid size to the RMS of the frame-to-frame
pointing jitter; changing the grid size did not impact the transit
depth.

We determined the best-fitting aperture by minimizing the standard
deviation of the normalized residuals (SDNR) between the data and the
best-fitting model.  Both 3.6 {\micron} datasets showed a clear
SDNR minimum at an aperture radius of 3.0 pixels.  Similarly, both 4.5
{\micron} datasets showed a clear SDNR minimum at 2.5 pixels.  In
all cases, the transit depth remained consistent (within $1 \sigma$)
across the apertures.


We determined the best-fitting ramp model by minimizing the Bayesian
Information Criterion (BIC).  The value of BIC between
two competing models (${\cal M}\sb{1}$ and ${\cal M}\sb{2}$) indicates the fractional probability,
$p({\cal M}\sb{2}|D)$, of being the correct model
\citep[see][]{CubillosEtal2014apjTrES1}.  Tables \ref{table:ha011bp11} --
\ref{table:ha011bp22} show the best-fitting ramps for each dataset.

\begin{table}[ht]
\centering
\caption[3.6 {\microns} Visit 1 - Ramp Model Fits]
        {3.6 {\microns} Visit 1 - Ramp Model Fits}
\label{table:ha011bp11}
\strut\hfill\begin{tabular}{cccc}
\hline
\hline
Ramp     & $R\sb{p}/R\sb{s}$\sp{a} & $\Delta$BIC & $p({\cal M}\sb{2}|D)$ \\
\hline
exponential  & 0.05835(19)  &    0.0 &  $\cdots$ \\
linear       & 0.05767(24)  &    2.3 &  0.24     \\
quadratic    & 0.05839(35)  &    2.9 &  0.19     \\
logarithmic  & 0.05766(22)  &   13.4 & $1.2\tttt{-3}$   \\
\hline

\multicolumn{4}{l}{\footnotesize {\bf Note.} \sp{a} For this and the following tables, the values quoted in} \\
\multicolumn{4}{l}{\footnotesize \hspace{0.1cm} parentheses indicate the 1$\sigma$ uncertainty corresponding to the }  \\
\multicolumn{4}{l}{\footnotesize \hspace{0.1cm} least significant digits.}
\end{tabular}\hfill\strut
\end{table}

\begin{table}[ht]
\centering
\caption[3.6 {\microns} Visit 2 - Ramp Model Fits]
        {3.6 {\microns} Visit 2 - Ramp Model Fits}
\label{table:ha011bp12}
\strut\hfill\begin{tabular}{cccc}
\hline
\hline
Ramp     & $R\sb{p}/R\sb{s}$ & $\Delta$BIC & $p({\cal M}\sb{2}|D)$\\
\hline
exponential          & 0.05691(28)  &   0.0 & $\cdots$        \\
exponential + linear & 0.05687(26)  &  10.9 & $4.3\tttt{-3}$  \\
logarithmic          & 0.05713(32)  &  20.2 & $4.1\tttt{-5}$  \\
quadratic            & 0.05729(30)  & 105.8 & $1.1\tttt{-23}$ \\
\hline
\end{tabular}\hfill\strut
\end{table}

\begin{table}[ht]
\centering
\caption[4.5 {\microns} visit 1 - Ramp Model Fits]
        {4.5 {\microns} visit 1 - Ramp Model Fits}
\label{table:ha011bp21}
\strut\hfill\begin{tabular}{cccc}
\hline
\hline
Ramp     & $R\sb{p}/R\sb{s}$ & $\Delta$BIC & $p({\cal M}\sb{2}|D)$\\
\hline
no ramp     & 0.05798(33) &   0.0  & $\cdots$ \\
linear      & 0.05813(31) &   3.0  & 0.18 \\         
quadratic   & 0.05807(36) &  13.9  & $9.6\tttt{-4}$ \\         
exponential & 0.05814(38) &  14.0  & $9.1\tttt{-4}$ \\  
\hline
\end{tabular}\hfill\strut
\end{table}

\begin{table}[ht]
\centering
\caption[4.5 {\microns} visit 2 - Ramp Model Fits]
        {4.5 {\microns} visit 2 - Ramp Model Fits}
\label{table:ha011bp22}
\strut\hfill\begin{tabular}{cccc}
\hline
\hline
Ramp     & $R\sb{p}/R\sb{s}$ & $\Delta$BIC & $p({\cal M}\sb{2}|D)$\\
\hline
no ramp     & 0.05814(35) &    0.0  & $\cdots$ \\
linear      & 0.05808(33) &   10.9  & $4.3\tttt{-3}$ \\         
quadratic   & 0.05812(35) &   21.9  & $1.8\tttt{-5}$ \\         
exponential & 0.05814(29) &   22.1  & $1.6\tttt{-5}$ \\         
\hline
\end{tabular}\hfill\strut
\end{table}

\begin{figure}[htb]
\centering
\includegraphics[width=0.7\linewidth]{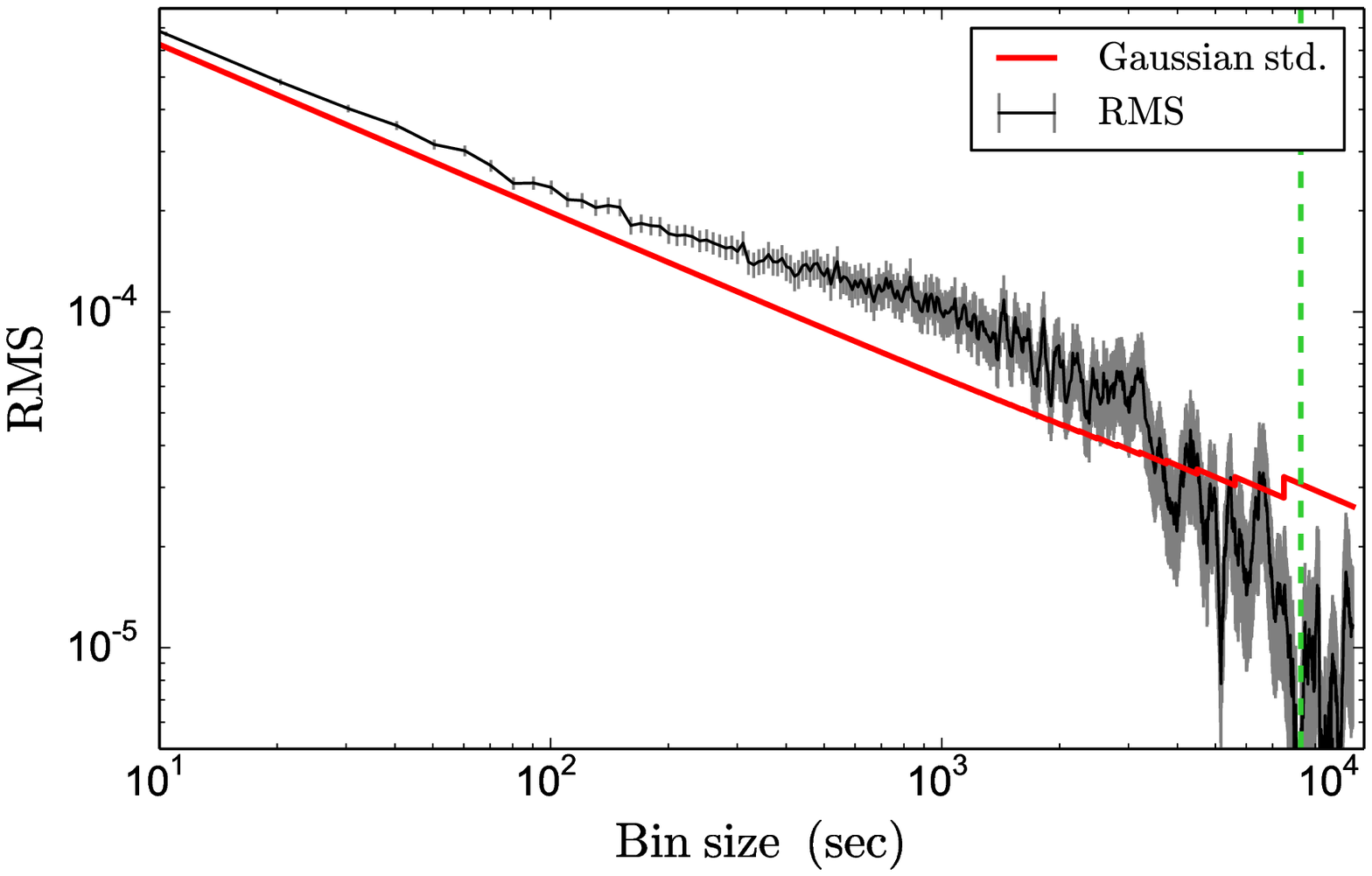}
\caption[Fit residuals' rms {\vs} bin size for second visit at 3.6 {\microns}]
 {Fit residuals' rms (black curve with 1$\sigma$ uncertainties)
  {\vs} bin size for the second visit at 3.6 {\microns}.  The red curve
  shows the expected rms for Gaussian (uncorrelated) noise.  The green
  dashed vertical line marks the transit duration time.}
\label{fig:RMS}
\end{figure}

At 3.6 {\microns}, the rising-exponential ramp outperformed the other
models in both visits.  Since the second visit at 3.6 {\microns}
showed evidence of correlated noise (Fig.\ \ref{fig:RMS}), we applied the
time-averaging correction factor, and increased the data uncertainties
by a factor of 1.5 \citep[see][]{CubillosEtal2015apjRednoise}.
At 4.5 {\microns}, the no-ramp model outperformed the other models in
both visits.  However, for the first visit, we noted a slight linear
trend in the residuals for the no-ramp model; thus, we adopted the
linear model.

\begin{figure*}[htb]
\centering
\includegraphics[width=\textwidth]{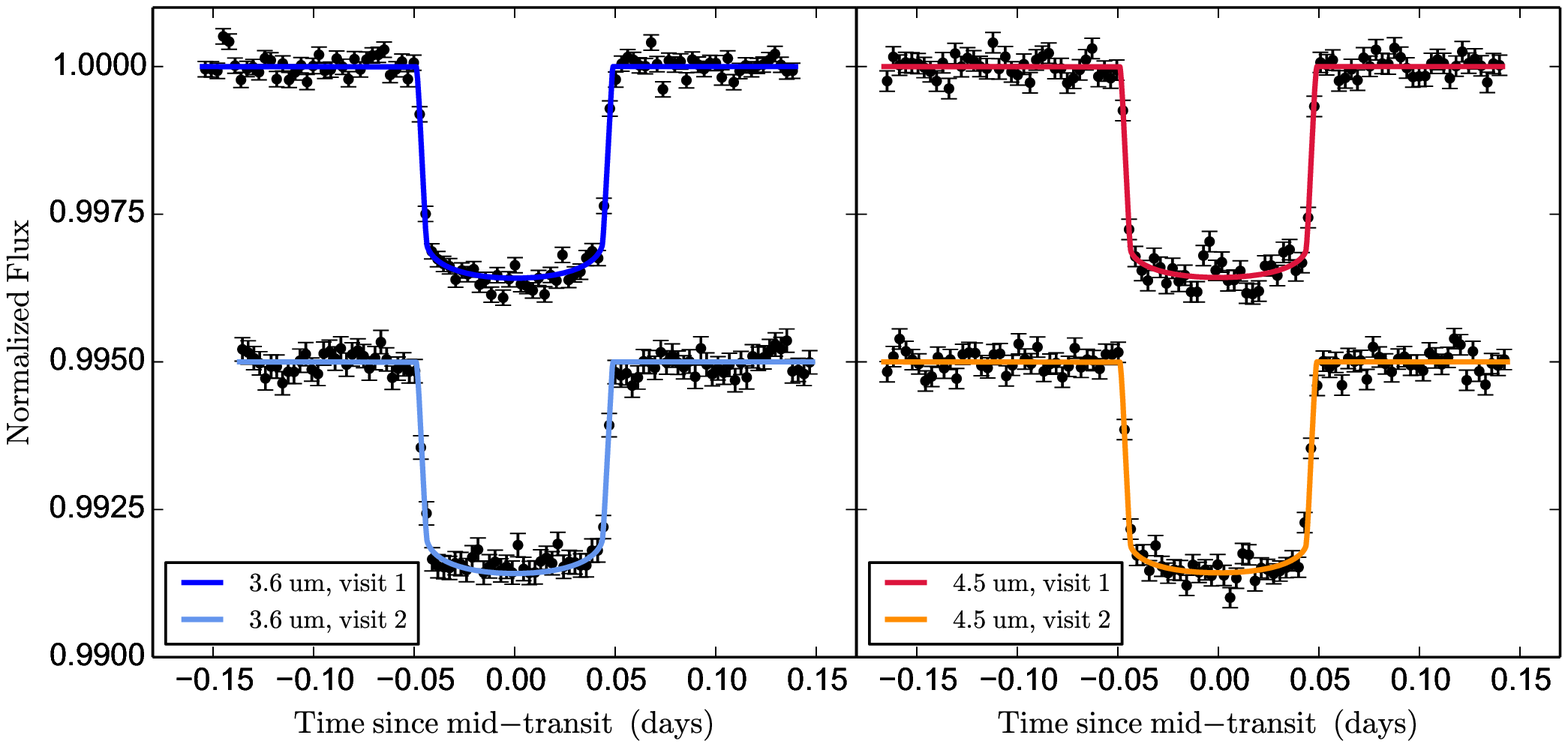}
\caption[Normalized, systematics-corrected Spitzer HAT-P-11b transit
         light curves] {Normalized, systematics-corrected Spitzer
        HAT-P-11b transit light curves (black points) with the
        best-fitting models (colored solid curves).  The error bars
        denote the 1$\sigma$ uncertainties.  For clarity, we binned the
        data points and vertically shifted the curves.}
\label{fig:c5lightcurves}
\end{figure*}

To obtain final transit depths we ran a joint fit combining all four
events.  In this fit, we shared the cosine of inclination and $a/R\sb{s}$
parameters among all events.  We also shared the $R\sb{p}/R\sb{s}$ parameter
between the events in the same wavelength band.  Figure
\ref{fig:c5lightcurves} shows the systematics-corrected light-curve
data and joint best-fitting model.  Table \ref{table:c5jointfits}
summarizes the joint-fit model setup and results.

\begin{table*}[ht]
\footnotesize
\centering
\caption[HAT-P-11b Best Joint-fit Eclipse Light-curve Parameters]
        {HAT-P-11b Best Joint-fit Eclipse Light-curve Parameters}
\label{table:c5jointfits}
\begin{tabular}{rcccc}
\hline
\hline
Waveband (\micron)                              & 3.6 (visit 1)   & 3.6 (visit 2)   & 4.5 (visit 1)   & 4.5  (visit 2)   \\ 
\hline
Mean $x$ position (pix)                         &  14.91          &  14.89          &  14.61          &  14.74           \\ 
Mean $y$ position (pix)                         &  15.13          &  15.12          &  15.06          &  15.02           \\ 
$x$-position consistency\sp{a} (pix)           & 0.004           & 0.005           & 0.012           & 0.010            \\ 
$y$-position consistency\sp{a} (pix)           & 0.011           & 0.006           & 0.005           & 0.004            \\ 
Aperture photometry radius (pixels)             & 3.0             & 3.0             & 2.5             & 2.5              \\ 
System flux \math{F\sb{s}} (\micro Jy)          & 469650(13)      & 468700(17)      & 285088.9(7.3)   & 285512.3(5.3)    \\ 
Transit midpoint (MJD\sb{UTC})\sp{b}           & 3630.7152(16)   & 3309.5283(16)   & 3630.7152(16)   & 3309.5283(16)    \\ %
Transit midpoint (MJD\sb{TDB})\sp{b}           & 3630.7159(16)   & 3309.5290(16)   & 3630.7159(16)   & 3309.5290(16)    \\ %
Transit duration (\math{t\sb{\rm 4-1}}, hrs)    & 2.351(5)        & 2.351(5)        & 2.351(5)        & 2.351(5)         \\ 
$R\sb{p}/R\sb{\star}$                           & 0.05791(22)     & 0.05791(22)     & 0.05816(25)     & 0.05816(25)      \\ 
$\cos(i)$ (deg)                                 & 89.52(12)       & 89.52(12)       & 89.52(12)       & 89.52(12)        \\ 
$a/R\sb{\star}$                                 & 16.664(89)      & 16.664(89)      & 16.664(89)      & 16.664(89)       \\ 
Limb darkening coefficient, $c1$                &  0.5750         &  0.5750         &  0.6094         &  0.6094          \\ 
Limb darkening coefficient, $c2$                & $-0.3825$       & $-0.3825$       & $-0.7325$       & $-0.7325$        \\ 
Limb darkening coefficient, $c3$                &  0.3112         &  0.3112         &  0.7237         &  0.7237          \\ 
Limb darkening coefficient, $c4$                & $-0.1099$       & $-0.1099$       & $-0.2666$       & $-0.2666$        \\ 
Ramp equation ($R(t)$)                          & Rising exp.     & Rising exp.     & Linramp         & None             \\ 
Ramp, linear term ($r\sb{1}$)                   & $\cdots$        & $\cdots$        & 0.00077(27)     & $\cdots$         \\ 
Ramp, exponential term ($t\sb{0}$)              & $-13.7(3.1)$    & 29.8(2.6)       & $\cdots$        & $\cdots$         \\ 
Ramp, exponential term ($r\sb{1}$)              & $-18.3(2.4)$    & 11.5(1.6)       & $\cdots$        & $\cdots$         \\ 
Number of free parameters\sp{c}                & 7               & 7               & 6               & 5                \\ 
Total number of frames                          & 62592           & 62592           & 62592           & 62592            \\ 
Frames used\sp{d}                              & 57999           & 56150           & 60962           & 62229            \\ 
Rejected frames (\%)                            & 7.34            & 10.29           & 2.60            & 0.58             \\ 
BIC                                             & 206372.4        & 206372.4        & 206372.4        & 206372.4         \\ 
SDNR                                            & 0.0031105       & 0.0031215       & 0.0042603       & 0.0042751        \\ 
Uncertainty scaling factor                      & 0.980           & 1.477           & 1.083           & 1.086            \\ 
Photon-limited S/N (\%)                         & 85.31           & 56.99           & 85.95           & 85.61            \\ 
\hline
\multicolumn{5}{l}{{\bf Notes.}} \\
\multicolumn{5}{l}{\sp{a} rms frame-to-frame position difference.} \\
\multicolumn{5}{l}{\sp{b} MJD = BJD $-$ 2,450,000.} \\
\multicolumn{5}{l}{\sp{c} In the individual fits.} \\
\multicolumn{5}{l}{\sp{d} We exclude frames during instrument/telescope
  settling, for insufficient points at a given BLISS bin, and} \\
\multicolumn{5}{l}{\hspace{0.1cm} for bad pixels in the photometry aperture.}
\end{tabular}
\end{table*}

The Spitzer transit depth of POET and
\citetalias{FraineEtal2014natHATP11bH2O} are consistent to each other
within $1 \sigma$.  Furthermore, the depth uncertainties also agreed,
suggesting that both reduction pipelines are statistically robust.
This is relevant, considering that there have been disagreements when
different groups analyze the same exoplanet light curve \citep{HansenEtal2014mnrasSpitzerFeatureless}.

\subsection{Atmospheric Retrieval}
\label{sec:retrieval}

We modeled the HAT-P-11b infrared spectra from 1.0 to 5.5 {\microns}.
The atmospheric model included HITEMP opacities for {\water}, CO, and
{\carbdiox} \citep{RothmanEtal2010jqsrtHITEMP}, and HITRAN opacities
for {\methane}, NH$\sb{3}$, C$\sb{2}$H$\sb{2}$, and HCN
\citep{Rothman2013JqsrtHITRAN}.  The model also included CIA opacities
for {\molhyd}-{\molhyd} and {\molhyd}-He
\citep{RichardEtal2012jqsrtCIA}.

\begin{figure}[ht]
\centering
\includegraphics[width=0.8 \linewidth]{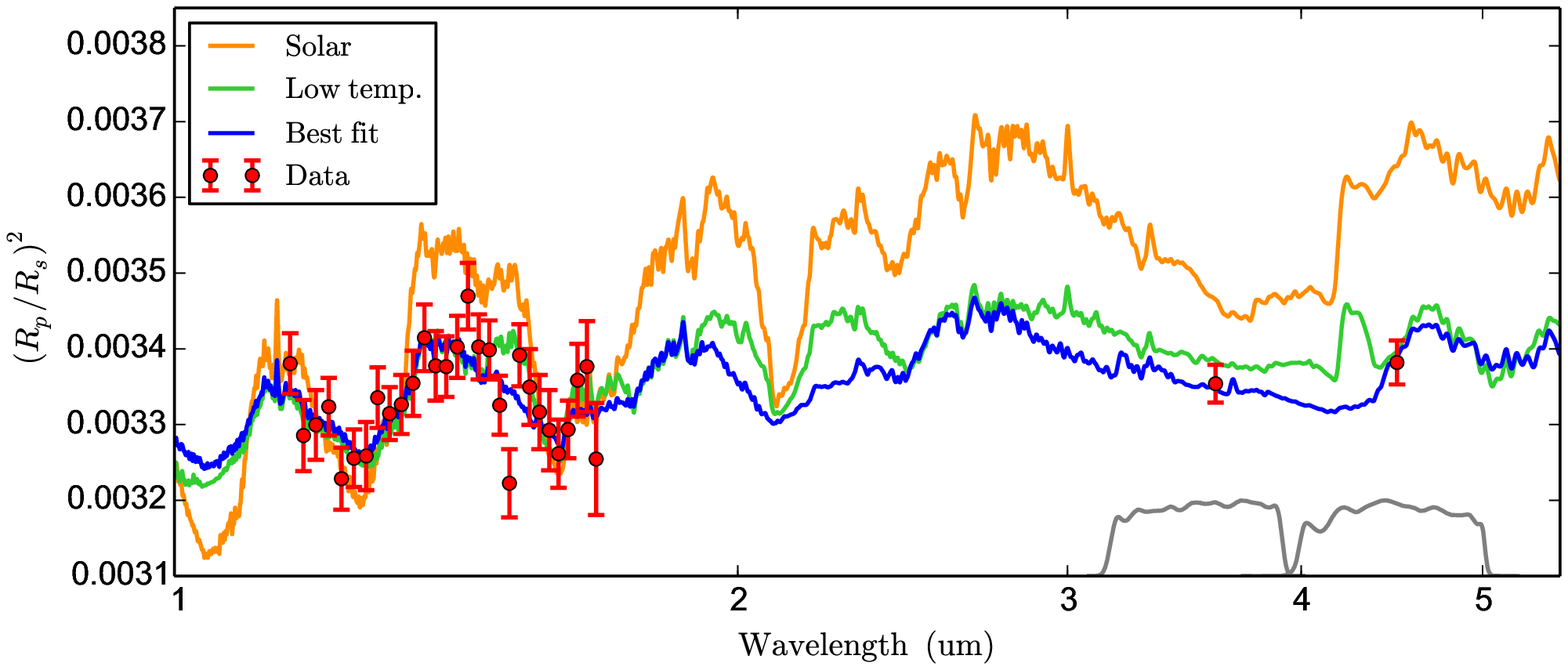}
\includegraphics[width=0.8\linewidth]{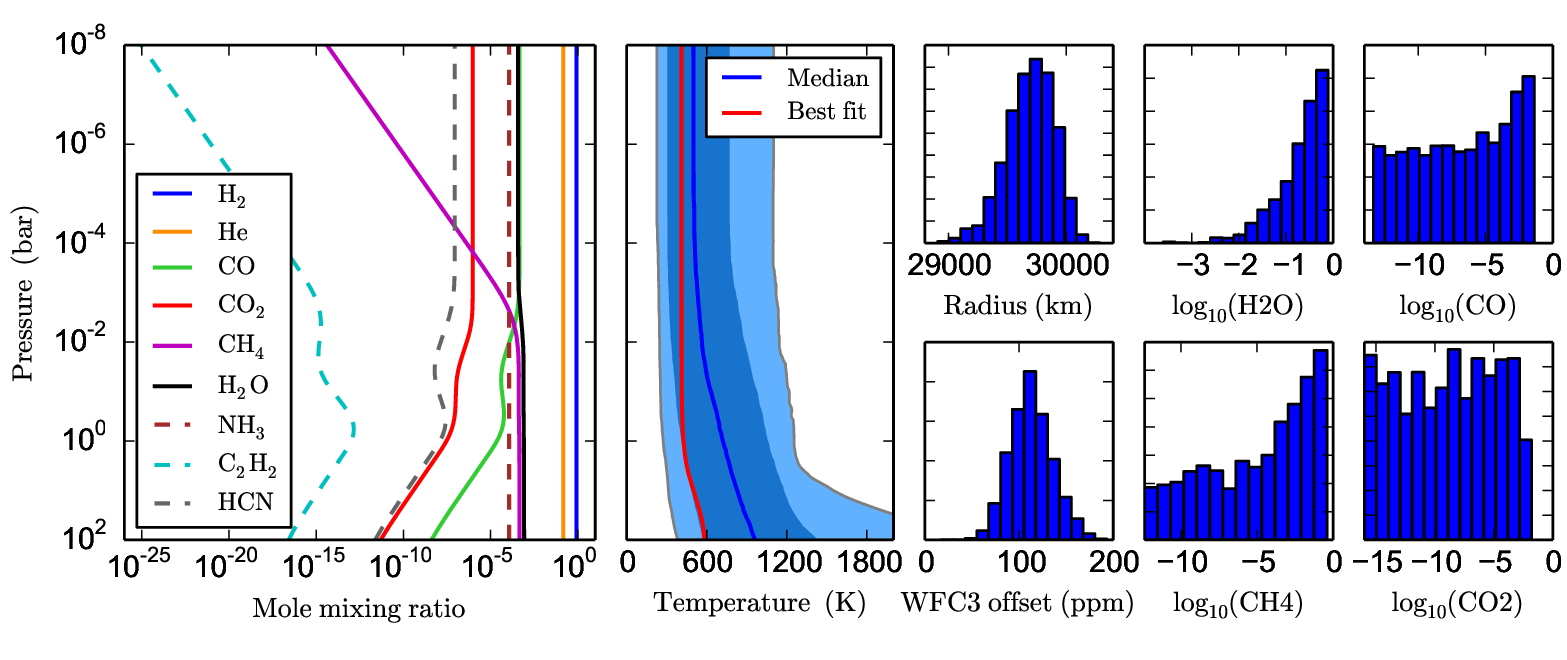}
\caption[HAT-P-11b retrieval transmission spectra and atmosphere]
{\small {\bf Top:} HAT-P-11b transmission spectra.  The red points
  with error bars denote the HST WFC3 ($1-2$ {\microns}) and
  Spitzer IRAC (3.6 and 4.5 {\microns}) data with $1\sigma$
  uncertainties.  The WFC3 points were adjusted downwards according to
  the retrieved best-fitting offset (121 p.p.m.).  The blue curve
  shows the BART-retrieved best-fitting spectrum.  The orange and
  green curves show sample models with solar abundances at equilibrium
  ($\sim 900$ K) and low ($\sim 400$ K) temperatures, respectively.
  The gray curves at the bottom show the Spitzer transmission filters.
  {\bf Bottom:} The left panel shows the initial
  thermochemical-equilibrium abundances of the atmospheric model. The
  central panel shows the posterior distribution of the temperature
  profile.  The dark and light regions denote the 68\% and 95\%
  credible regions (1 and 2$\sigma$).  The right panels show the
  posterior histograms for the 0.1 bar radius, the WFC3
  offset, and the 0.1 bar abundances of {\water}, {\methane}, CO, and
  {\carbdiox}.}
\label{fig:HATbart}
\end{figure}

We modeled the atmosphere with a set of 150 layers, equi-spaced in
log-pressure, ranging from $\ttt{-2}$ to $\ttt{-8}$ bar.  We produced an
initial temperature profile ranging from 785 K at the top of the
atmosphere to 1030 K at the bottom of the atmosphere.  TEA calculated
the initial thermochemical-equilibrium abundances (Fig.\
\ref{fig:HATbart}, bottom left) for this atmospheric model, assuming
solar elemental abundances \citep{AsplundEtal2009araSolarComposition}.
{\molhyd} and He represent the bulk of the atmospheric composition,
with abundances of $\sim85$\% and $\sim15$\%, respectively.

For the calculation of the line-transition opacities, we set a cutoff
at $\ttt{-6}$ times the strongest lines.  This threshold preserves over
$\sim 99$\% of the total opacity per channel, while significantly speeding up
the line-by-line calculations. 

The retrieval model included a total of nine free parameters.  Three
parameters, $\log\kappa$, $\log\gamma\sb{1}$, and $\beta$ of the
three-stream Eddington approximation model, fit the temperature
profile (while fixing $\alpha$ = 0.0).  Four additional
parameters scaled the log--abundance profiles of {\water}, {\methane},
CO, and {\carbdiox} (while keeping the abundances of NH$\sb{3}$,
C$\sb{2}$H$\sb{2}$, and HCN fixed at their initial values).  One
parameter fit the planet radius at 0.1 bar.  Finally, following
\citetalias{FraineEtal2014natHATP11bH2O}, a free parameter fit the
offset of the WFC3 transit-depth data relative to the Spitzer data.

Figure \ref{fig:HATbart} shows the atmospheric retrieval results.  Our
best-fitting model provided a good description of the dataset.  BART
estimated a WFC3 offset of $120 \pm 20$ p.p.m (shifted downwards)
between the WFC3 and Spitzer data.  This agrees at the 1-2$\sigma$
level with the best-fitting offset of
\citetalias{FraineEtal2014natHATP11bH2O}.  BART found a best-fitting
0.1 bar planetary radius of $29,750 \pm 200$ km.  The CO and
{\carbdiox} abundances remained unconstrained (flat posterior
histograms).  This is consistent with the weak absorption pattern from
these molecules at wavelengths shorter than 4 {\microns}.  The best
option to detect CO and {\carbdiox} would be through their strong
absorption bands in the 4.5 {\micron} filter.  However, our two
Spitzer channels have nearly equal depths.  The {\methane} posterior
histogram is nearly flat, showing a slight, insignificant peak
above $\ttt{-4}$.

The constraint on the {\water} abundance is the most interesting
case. The {\water} abundance posterior suggests an enhancement above
$100$ times the solar abundance.  The best-fitting and mean abundances
are $120$ times and $230$ times the solar abundance.  Our estimated
atmospheric compositions agree well with the results from
\citetalias{FraineEtal2014natHATP11bH2O}, finding an
enhancement in heavy elements and a robust detection of {\water}.

The key to constraining the water abundance is the amplitude of the
absorption features, which is related to the atmospheric scale height:
\begin{equation}
  H = \frac{k\sb{B}T}{\mu g},
\end{equation}
where $\mu$ is the mean molecular mass.  The scale height is the
characteristic (vertical) distance over which the pressure changes by
a factor of $e$.  For a smaller scale height, the atmosphere changes
from optically thin to optically thick over a smaller altitude range.

A higher abundance of heavy elements implies a higher mean molecular
mass, and hence a smaller scale height.  Then, the amplitude of the
modulation spectrum is smaller (compare, for example, the amplitudes
of the solar and best-fitting models in Fig.\ \ref{fig:HATbart}, top
panel).  Equivalently, lower temperatures would also decrease the
scale height, as shown in Fig.\ \ref{fig:HATbart}.  The higher {\water}
abundance, thus, allows BART to match the amplitude of the spectrum
between 1 and 2 {\microns}.

\section{Discussion}
\label{sec:discusison}

\subsection{Open Source, Open Development}
\label{sec:open}

We designed BART as an open-source, open-development project.  The
project's license will let anybody access, copy, modify, and
redistribute the code.  Researchers can freely use the BART code for
publications, as long as they make public any modification to the code
along with the scientific article.  With each publication, we will
include a compendium to reproduce all of the BART analyses from the
scientific article.  We believe that reproducibility is one of the
main principles of the scientific method, thus we encourage the community to:
\begin{enumerate}
\item Promote `Reproducible Research'
  \citep{StoddenEtal2009ciseRRlegal}.  As computer programs increase
  in complexity, the ability to reproduce an experiment solely from an
  article's description demands a significant time investment, or worse,
  the article may not contain all relevant information.  The idea of
  reproducible research is to allow anyone to reproduce any published
  experiment.

\item  Make exoplanet characterization a community effort, rather
  than individual attempts.  Planetary atmospheres modeling is a
  multi-disciplinary endeavor involving: radiative processes,
  atmospheric chemistry, circulation dynamics, cloud physics,
  and more.  The modular design of BART allows members of the scientific
  community to directly contribute their expertise with ideas and
  code.

\item Promote code verification.  Although the independent
  groups properly test their own models, there is a lack of code
  verification between competing models.  By making the inputs
  and outputs public, members of the community can directly compare the
  results from different codes for the same input.  By making the
  source code public, we facilitate code debugging and verification.
\end{enumerate}

\subsection{Conclusions}
\label{sec:c5conclusions}

Here we presented the open-source, open-development BART code to
characterize exoplanet atmospheres in a statistically-robust manner.
The project's source code and documentation are available at
\href{https://github.com/exosports/BART} {https://github.com/exosports/BART}.
The BART package includes three self-sufficient submodules: (1) the
Transit radiative-transfer code
(\href{https://github.com/exosports/transit}
{https://github.com/exosports/transit}), (2) the TEA
thermochemical-equilibrium abundances code
(\href{https://github.com/dzesmin/TEA} {https://github.com/dzesmin/TEA}), and
(3) the {\mcc} Bayesian statistical module
(\href{https://github.com/pcubillos/MCcubed}
{https://github.com/pcubillos/MCcubed}).
The documentation includes a user manual describing the inputs,
outputs, and usage of the BART code. A second, more-detailed code
manual describes in depth the routines and file structures of the
code.

The Transit module solves the radiative-transfer equation to calculate
transmission or emission spectra.  It accounts for line-transition and
collision-induced absorption opacities, which can be provided as
line-by-line or cross-section files.  The Transit spectra agree well with 
 C. Morley's
model emission spectra and \citet{SharpBurrows2007apjOpacities} opacity
spectra.  The TEA module is further described in
\citet{BlecicEtal2015apsjTEA}.  The {\mcc} statistics module is a
general-purpose model-fitting package written in Python and C
\citep[][]{CubillosEtal2015apjRednoise}.  {\mcc} provides least-square
minimization, advanced MCMC sampling, and correlated-noise estimation
algorithms.

The BART project is the outcome of the collaborative effort by the UCF
exoplanet group members and is jointly presented in this work, in
\citet{HarringtonEtal2015apjBART}, and in
\citet{BlecicEtal2015apjBART}.  The following list details the tasks
where I contributed substantially.  I revised and modified the
complete Transit code from P. Rojo to improve the code performance,
improve the code design, add comments, and remove obsolete functions.
I implemented Transit's opacity-grid calculations and
interpolation, Simpson integration, opacity threshold, and Voigt-profile
precalculation.  I refactored the CIA code into a general code that
handles cross-section absorption data for single or pairs of species. I
reimplemented the {\water} line reader for the
\citeauthor{PartridgeSchwenke1997jcpH2O} database from C to Python and
added support for the HITRAN, TiO \citeauthor{Schwenke1998TiO}, and VO
\citeauthor{Plez1998aaTiOLineList} databases.  I implemented the CTIPS
submodule to calculate the partition function for the HITRAN species.  For
the BART module, I implemented the three-stream Eddington
temperature-profile model, developed the MPI communication framework,
and implemented the spectrum integration over the filter transmission.
I wrote the {\mcc} statistical module and documentation.

The diversity of physical processes involved in atmospheric modeling
calls for the development of collaborative projects.  By releasing our
code to the community, we hope to (1) allow reproducible research, (2)
provide access to the routines discussed here, (3) promote open
development and cross validation of the software tools used in the
field, and (4) promote exoplanet characterization as a
collaborative effort.

We further applied the BART atmospheric analysis to the Spitzer and
Hubble transit observations of HAT-P-11b.  Our results agree well with
those of \citet{FraineEtal2014natHATP11bH2O}.  Both analyses detected
{\water} absorption and estimated an enhancement in heavy elements of
a few hundred times the solar composition.  Other species abundances
({\methane}, CO, and {\carbdiox}) remained unconstrained.
The low signal-to-noise ratio and limited spectral coverage of present
measurements limit the scientific yield from these observations.  The
analysis of HAT-P-11b is an example of these difficulties.  The
unknown absolute calibration of the system flux introduces an
uncertainty in the transit-depth measurements between non-simultaneous
observations.  This extra degree of freedom can mask
spectral features, limiting our capacity to lay atmospheric constraints.

Fortunately, the field has motivated new efforts to calculate
laboratory opacity data at higher temperatures
\citep[e.g.,][]{TennysonYurchenko2012mnrasExoMol,
  HargreavesEtal2012apjHotMethane, HargreavesEtal2015apjHotCH4, HargreavesEtal2015HotC2H6} and build dedicated
instrumentation.  The laboratory advancements complement the
development of next-generation telescopes; for example, JWST's large
collecting area ($\sim 6.5$ m primary diameter), fine resolving power
($R=4-3000$), and broad spectral coverage ($\sim 0.6-28.0$
{\microns}) will unveil unprecedented details of these worlds.  The
tools we have presented here prepare us to better study the diversity
of exoplanet atmospheres that will be observed in the future.

\section{Acknowledgements}

We thank Caroline Morley and Jonathan Fortney for useful conversations
and for providing radiative\-/transfer spectra for comparison.  We thank
contributors to SciPy, Matplotlib, and the Python Programming
Language; the open-source development website GitHub.com; and
contributors to the free and open-source community.
PC is supported by the Fulbright Program for Foreign Students.
Support for this work was provided by NASA through the Science Mission
Directorate's Planetary Atmospheres Program, grant NNX12AI69G, and the
Astrophysics Data Analysis Program, grant NNH12ZDA001N.  Part of this
work is based on observations made with the Spitzer Space
  Telescope, which is operated by the Jet Propulsion Laboratory,
California Institute of Technology under a contract with NASA.

\bibliographystyle{apj}
\bibliography{chap5-bart}

\end{document}